\documentclass{article}
\usepackage{graphicx} 
\usepackage{jheppub}
\usepackage{macros_JHEP}
\usetikzlibrary{arrows.meta,positioning,calc}

\usepackage[table]{xcolor}
\usepackage{subcaption}

\tikzset{
    conifoldzigzag/.style={
        line width=1.5pt,
        preaction={
            draw=white,
            line width=3.2pt
        },
        postaction={decorate},
        decoration={
            markings,
            mark=at position 0.27 with {
                \arrow{Stealth[length=4.5pt,width=3.5pt]}
            },
            mark=at position 0.68 with {
                \arrow{Stealth[length=4.5pt,width=3.5pt]}
            }
        }
    },
    zigzagnode/.style={
        circle,
        draw=black,
        thick,
        minimum size=4.5pt,
        inner sep=0pt
    },
    zigzaglabel/.style={
        fill=white,
        fill opacity=0.85,
        text opacity=1,
        inner sep=1pt,
        rounded corners=1pt
    }
}

\newcommand{\DrawConifoldTiles}{
    \foreach \x in {0,...,3}{
        \foreach \y in {0,...,3}{
            \pgfmathtruncatemacro{\parity}{mod(\x+\y,2)}

            \ifnum\parity=1
                \fill[blue!15]
                (\x,\y) rectangle ({\x+1},{\y+1});
            \else
                \fill[red!15]
                (\x,\y) rectangle ({\x+1},{\y+1});
            \fi
        }
    }

    \foreach \x in {0,...,4}{
        \draw[black,thin] (\x,0)--(\x,4);
    }
    \foreach \y in {0,...,4}{
        \draw[black,thin] (0,\y)--(4,\y);
    }
}

\newcommand{\DrawConifoldNodes}{
    \foreach \x in {0,...,4}{
        \foreach \y in {0,...,4}{
            \pgfmathtruncatemacro{\parity}{mod(\x+\y,2)}

            \ifnum\parity=0
                \node[
                    zigzagnode,
                    fill=black
                ] at (\x,\y) {};
            \else
                \node[
                    zigzagnode,
                    fill=white
                ] at (\x,\y) {};
            \fi
        }
    }
}

\usetikzlibrary{decorations.pathreplacing,decorations.markings}
\tikzset{
 segment/.style={
    decorate,
    decoration={
      show path construction,
      moveto code={},
      lineto code={
        \path [#1]
        (\tikzinputsegmentfirst) -- (\tikzinputsegmentlast);
      },
      curveto code={
        \path [#1] (\tikzinputsegmentfirst)
        .. controls
        (\tikzinputsegmentsupporta) and (\tikzinputsegmentsupportb)
        ..
        (\tikzinputsegmentlast);
      },
      closepath code={
        \path [#1]
        (\tikzinputsegmentfirst) -- (\tikzinputsegmentlast);
      },
    },
  },
  mid arrow/.style={postaction={decorate,decoration={
        markings,
        mark=at position .5 with {\arrow[#1]{stealth}}
      }}},
}

\definecolor{pyramidblue}{RGB}{0,0,255}
\definecolor{pyramidred}{RGB}{255,0,0}
\newcommand{\ConifoldPyramid}[1]{%
\begingroup
\def\R{0.55cm}
\begin{tikzpicture}[x=\R,y=\R]

  \pgfmathtruncatemacro{\maxlevel}{#1}

  \foreach \ell in {\maxlevel,...,0} {

    \pgfmathtruncatemacro{\parity}{mod(\ell,2)}

    \ifnum\parity=0

      \pgfmathtruncatemacro{\m}{\ell/2}

      \foreach \i in {0,...,\m} {
        \foreach \j in {0,...,\m} {

          \pgfmathsetmacro{\xx}{2*\i-\m}
          \pgfmathsetmacro{\yy}{2*\j-\m}

          \filldraw[
            fill=pyramidblue,
            draw=black,
            line width=0.5pt
          ]
          (\xx,\yy) circle[radius=\R];
        }
      }

    \else

      \pgfmathtruncatemacro{\m}{(\ell-1)/2}
      \pgfmathtruncatemacro{\imax}{\m+1}

      \foreach \i in {0,...,\imax} {
        \foreach \j in {0,...,\m} {

          \pgfmathsetmacro{\xx}{2*\i-(\m+1)}
          \pgfmathsetmacro{\yy}{2*\j-\m}

          \filldraw[
            fill=pyramidred,
            draw=black,
            line width=0.5pt
          ]
          (\xx,\yy) circle[radius=\R];
        }
      }

    \fi
  }

\end{tikzpicture}
\endgroup
}

\newcommand{\ConifoldAtoms}[1]{%
\begingroup
\def\R{0.55cm}

\begin{tikzpicture}[x=\R,y=\R]

  \foreach \level/\xx/\yy in {#1} {

    \pgfmathtruncatemacro{\parity}{mod(\level,2)}

    \ifnum\parity=0
      \def\atomcolor{pyramidblue}
    \else
      \def\atomcolor{pyramidred}
    \fi

    \filldraw[
      fill=\atomcolor,
      draw=black,
      line width=0.5pt
    ]
    (\xx,\yy) circle[radius=\R];
  }

\end{tikzpicture}
\endgroup
}

\newcommand{\FullBox}[2]{%
  \draw[fill=black!2]
    (#1,#2) rectangle ({#1+1},{#2+1});
}

\newcommand{\HalfBox}[2]{%
  \filldraw[fill=black!2]
    (#1,#2) --
    (#1,{#2+1}) --
    ({#1+1},#2) -- cycle;
}

\usepackage{tikz-3dplot}

\newcommand{\BlueLowerTri}{%
  \raisebox{-0.12em}{%
    \tikz[x=0.65em,y=0.65em,line join=round]
      \filldraw[fill=blue!80,line width=0.35pt]
        (0,0)--(0,1)--(1,0)--cycle;
  }%
}

\newcommand{\RedUpperTri}{%
  \raisebox{-0.12em}{%
    \tikz[x=0.65em,y=0.65em,line join=round]
      \filldraw[fill=red!80,line width=0.35pt]
        (0,1)--(1,1)--(1,0)--cycle;
  }%
}

\newcommand{\ElementaryDsixQuiver}[3]{%
\begin{tikzpicture}[
    baseline=(current bounding box.center),
    x=0.72cm,
    y=0.72cm,
    d6arrow/.style={
        black,
        thick,
        postaction={decorate},
        decoration={
            markings,
            mark=at position 0.55 with
            {\arrow[scale=0.75]{latex}}
        }
    },
    d6fermi/.style={
        red,
        thick,
        postaction={decorate},
        decoration={
            markings,
            mark=at position 0.55 with
            {\arrow[scale=0.75]{latex}}
        }
    },
    d6gauge/.style={
        circle,
        draw=black,
        thick,
        minimum size=0.46cm,
        inner sep=0pt,
        text=white,
        font=\scriptsize
    },
    d6flavor/.style={
        rectangle,
        draw=black,
        thick,
        fill=black!8,
        minimum width=0.38cm,
        minimum height=0.38cm,
        inner sep=0pt
    },
    d6label/.style={
        fill=white,
        inner sep=0.8pt,
        font=\scriptsize
    }
]

\coordinate (g1) at (-0.85,0);
\coordinate (g2) at ( 0.85,0);

\ifnum#1=1\relax
    \coordinate (f)      at (-0.85,1.25);
    \coordinate (src)    at (g1);
    \coordinate (Istart) at ($(f)+(-0.10,-0.18)$);
    \coordinate (Iend)   at ($(g1)+(-0.08,0.18)$);
    \coordinate (Ilabel) at ($(f)+(-0.34,-0.58)$);
\else
    \coordinate (f)      at (0.85,1.25);
    \coordinate (src)    at (g2);
    \coordinate (Istart) at ($(f)+(0.10,-0.18)$);
    \coordinate (Iend)   at ($(g2)+(0.08,0.18)$);
    \coordinate (Ilabel) at ($(f)+(0.34,-0.58)$);
\fi

\ifnum#2=1\relax
    \coordinate (tgt) at (g1);
\else
    \coordinate (tgt) at (g2);
\fi

\ifnum#1=#2\relax
    \ifnum#1=1\relax
        \coordinate (Lstart) at ($(f)+(0.10,-0.18)$);
        \coordinate (Lend)   at ($(g1)+(0.08,0.18)$);
        \coordinate (Llabel) at ($(f)+(0.36,-0.58)$);
    \else
        \coordinate (Lstart) at ($(f)+(-0.10,-0.18)$);
        \coordinate (Lend)   at ($(g2)+(-0.08,0.18)$);
        \coordinate (Llabel) at ($(f)+(-0.36,-0.58)$);
    \fi
\else
    \ifnum#1=1\relax
        \coordinate (Lstart) at ($(f)+(0.10,-0.18)$);
        \coordinate (Lend)   at ($(g2)+(-0.08,0.18)$);
        \coordinate (Llabel) at (0.25,0.93);
    \else
        \coordinate (Lstart) at ($(f)+(-0.10,-0.18)$);
        \coordinate (Lend)   at ($(g1)+(0.08,0.18)$);
        \coordinate (Llabel) at (-0.25,0.93);
    \fi
\fi

\ifnum#1=#2\relax
    \ifnum#1=1\relax
        \draw[d6arrow]
        ($(g1)+(-0.18,0.08)$)
        .. controls
        ($(g1)+(-0.68,0.46)$)
        and
        ($(g1)+(-0.68,-0.46)$)
        ..
        ($(g1)+(-0.18,-0.08)$);
\node[d6label]
at ($(g1)+(-0.9,-0.)$)
{$#3$};
        
    \else
        \draw[d6arrow]
        ($(g2)+(0.18,0.08)$)
        .. controls
        ($(g2)+(0.68,0.46)$)
        and
        ($(g2)+(0.68,-0.46)$)
        ..
        ($(g2)+(0.18,-0.08)$);

        \node[d6label]
at ($(g2)+(1,-0.1)$)
{$#3$};
    \fi
\else
    \ifnum#1=1\relax
        \draw[d6arrow]
        (g1)--(g2);
    \else
        \draw[d6arrow]
        (g2)--(g1);
    \fi

    \node[d6label]
    at (0,-0.30)
    {$#3$};
\fi

\draw[d6arrow] (Istart)--(Iend);
\draw[d6fermi] (Lstart)--(Lend);

\node[d6label] at (Ilabel) {$\mathsf I$};
\node[d6label,text=red] at (Llabel) {$\Lambda$};

\node[d6flavor] at (f) {};
\node[d6gauge,fill=blue!100!black] at (g1) {$1$};
\node[d6gauge,fill=red!100!black] at (g2) {$2$};

\end{tikzpicture}%
}

\title{\boldmath Gauge Origami of the Conifold $\times$ $\mathbb{C}$ }

\author[a,b]{Taro Kimura,}
\author[c]{Go Noshita}
\date{}

\affiliation[a]{Université Bourgogne Europe, CNRS, IMB UMR 5584, 21000 Dijon, France}
\affiliation[b]{Institut Universitaire de France (IUF), France}
\affiliation[c]{Center for High Energy Physics, Peking University, Beijing, China}

\emailAdd{taro.kimura@ube.fr}
\emailAdd{gnoshita969hep@gmail.com}

\abstract{ 
We construct gauge-origami systems on the toric Calabi--Yau fourfold
$\mathcal{C}\times\mathbb{C}$ from its brane brick model. Brick
matchings and phase boundaries determine elementary D6- and D4-brane
framings, respectively, realizing the magnificent-four,
tetrahedron-instanton, and spiked-instanton sectors in a unified
framed quiver quantum mechanics. We define their partition functions
by Jeffrey--Kirwan residues and describe the fixed points in terms of
colored BPS crystals. General gauge-origami configurations are
obtained by combining these elementary sectors through their common
fractional D0-brane quiver. This construction provides a first step
toward gauge origami on general toric Calabi--Yau fourfolds.
}

\begin{document}
\allowdisplaybreaks

\maketitle


\section{Introduction}\label{sec:intro}
The counting of BPS bound states in D-brane systems brings together
supersymmetric gauge theory, enumerative geometry, representation
theory, and combinatorics. For D-branes probing toric
Calabi--Yau threefolds, brane tilings provide a direct correspondence
between toric geometry and the associated quiver gauge theories
\cite{Hanany:2005ve,Franco:2005rj,Franco:2005sm,
Kennaway:2007tq,Yamazaki:2008bt}. In suitable stability chambers, the
torus-fixed points of the corresponding moduli spaces admit
crystal-melting descriptions
\cite{Szendroi:2007nu,Ooguri:2009ijd,Mozgovoy2008OnTN,
Nagao:2009rq,Yamazaki:2010fz,Bao:2022oyn}. Their chamber dependence
is closely related to wall crossing in Donaldson--Thomas theory
\cite{Nagao:2009ky,Nagao:2009rq,Sulkowski:2010eg,
Yamazaki:2010fz}. Framed quivers describing D4--D2--D0 systems can
similarly be constructed from perfect matchings and zig-zag paths of
the brane tiling
\cite{Nishinaka:2011sv,Nishinaka:2011is,
Nishinaka:2013mba,Nishinaka:2013pua}.

For toric Calabi--Yau fourfolds, the corresponding quiver theories
contain both chiral and Fermi multiplets, whose interactions are
encoded by $J$- and $E$-terms. Brane brick models realize this data
periodically on a three-torus and extend the brane-tiling
correspondence to toric Calabi--Yau fourfolds
\cite{Franco:2015tna,Franco:2015tya,Franco:2016qxh} (see also \cite{Franco:2024lxs} and the references there). Their
characteristic combinatorial objects include brick matchings and
phase boundaries, which encode lattice points and edges of the toric
diagram, respectively. Four-dimensional analogues of crystal melting
and their relation to brane brick models have also begun to be
explored
\cite{Bao:2024ygr,Franco:2023tly,Bao:2025hfu,
Bao:2025dqs,Bao:2026eym,Carcamo:2026yqu}. The brane-brick data should therefore contain information not only about the unframed quiver, but also about its framed D-brane sectors. 

Gauge origami provides a natural framework for addressing this
question. It was introduced within the BPS/CFT correspondence to
describe instantons supported on intersecting complex subspaces of
$\mathbb{C}^{4}$
\cite{Nekrasov:2016qym,Nekrasov:2016ydq}. Its principal D-brane
sectors include the magnificent four
\cite{Nekrasov:2017cih,Nekrasov:2018xsb}, tetrahedron
instantons
\cite{Pomoni:2021hkn,Pomoni:2023nlf}, and spiked instantons
\cite{Nekrasov:2016qym,Nekrasov:2016gud}. These systems arise from
D-branes supported on complex subspaces of different dimensions and
interact through a common point-like brane sector.

Most explicit gauge-origami constructions have so far been developed
for $\mathbb{C}^{4}$ and its orbifolds. Orbifold and quiver versions,
including gauge theories on $\mathbb{C}^{2}/\Gamma$ and surfaces
inside orbifolds of $\mathbb{C}^{4}$, already appear in Nekrasov's
original construction
\cite{Nekrasov:2016qym,Nekrasov:2016ydq}. Tetrahedron
instantons and their orbifold generalizations have likewise been
studied from both physical and mathematical viewpoints
\cite{Pomoni:2021hkn,Pomoni:2023nlf,Fasola:2023ypx,
Szabo:2023ixw,Szabo:2024lcp}. More generally, the associated
counting problems have been investigated in Donaldson--Thomas theory
on Calabi--Yau fourfolds
\cite{Cao:2017swr,Cao:2019tvv,Bojko:2021cih,
Cao:2023gvn,Arbesfeld:2026zrf}. A direct construction of these systems from brane brick models has not been given for non-orbifold toric CY4s. 

In this paper, we take a first step toward such a construction by
studying the simplest non-orbifold example,
$\mathcal{C}\times\mathbb{C}$, where $\mathcal{C}$ is the conifold. Our main observation is that the combinatorial data of the brane
brick model determine the elementary gauge-origami framings. Brick
matchings determine elementary D6-brane framings, whereas phase
boundaries determine elementary D4-brane framings. Together with the
D8--$\overline{\D8}$ system, these constructions realize the
magnificent-four, tetrahedron-instanton, and spiked-instanton sectors
within a common framed quiver quantum mechanics.

The D8--$\overline{\D8}$ system gives rise to the magnificent-four sector, whose fixed points are described by solid pyramid partitions.
Elementary D6-brane framings associated with brick matchings produce
inequivalent tetrahedron-instanton sectors whose fixed points are
described by pyramid partitions and super plane partitions. Their
combinations define the general tetrahedron-instanton system on
$\mathcal{C}\times\mathbb{C}$.

For D4-branes, we propose a phase-boundary prescription that
constructs the framed quiver directly from the brane brick model. In
a four-supercharge subsector, the resulting theory reproduces the
framed potential of the Nishinaka--Yamaguchi--Yoshida construction
\cite{Nishinaka:2013mba} and is closely related to the
perverse-coherent-extension construction of Butson and Rap\v{c}ák
\cite{Butson:2023eid}. Generic phase boundaries instead produce
genuinely two-supercharge framing data. Their elementary fixed points
are described by super partitions and
$\mathbb{Z}_{2}$-colored Young diagrams, and their combinations define
the general spiked-instanton system on
$\mathcal{C}\times\mathbb{C}$.

All these sectors are treated uniformly through Jeffrey--Kirwan
residues of the corresponding framed quiver quantum mechanics
\cite{Hori:2014tda,Cordova:2014oxa}. The residue prescription encodes
both the stability condition and the interactions among the framing
branes, while its contributing poles admit a direct interpretation as
colored and higher-dimensional crystal configurations
\cite{Bao:2024ygr,Bao:2025hfu}; see also
\cite{Franco:2023tly} for a complementary approach to
four-dimensional crystal melting.

Our results provide an explicit brane-brick realization of the
principal gauge-origami sectors on a non-orbifold toric
Calabi--Yau fourfold. 

The remainder of this paper is organized as follows. In
section~\ref{sec:D-brane-SQM}, we review the quiver quantum mechanics
and the brane brick model of $\mathcal{C}\times\mathbb{C}$. In
section~\ref{sec:magnificent-four}, we construct the
magnificent-four system and its solid-pyramid crystal. In
section~\ref{sec:tetrahedron-conifold}, we associate D6-brane
framings with brick matchings and construct the
tetrahedron-instanton systems. In section~\ref{sec:spikedinst}, we
formulate the phase-boundary prescription for D4-brane framings and
construct the spiked-instanton systems. We conclude with a summary
and discussion in section~\ref{sec:summary}.
Appendix~\ref{app:branetiling-conifold} reviews the brane tiling of
the conifold, while
Appendix~\ref{app:JEterms-phaseboundary} collects the superpotential
data for the D4-brane framings.

\section{D-branes and quiver quantum mechanics of \texorpdfstring{$\mathcal{C}\times \mathbb{C}$}{CC}}\label{sec:D-brane-SQM}
We summarize the notations of quivers and superpotentials of $\mathcal{C}\times\mathbb{C}$, where $\mathcal{C}$ is the conifold, in section~\ref{sec:quivers}. Basic objects and properties of the associated brane brick model are reviewed in section~\ref{sec:branebrickmodel}.

\subsection{Quivers and superpotentials}\label{sec:quivers}
In this section, we review the $\mathcal{N}=2$ supersymmetric quiver quantum mechanics associated with a toric Calabi-Yau (CY) fourfold in type IIA string theory. In particular, we will focus on the toric CY$_4$ $Z=\mathcal{C}\times \mathbb{C}$, where $\mathcal{C}$ is the conifold. The toric diagrams are
\bea\label{eq:toric-diagram}
\adjustbox{valign=c}{
\tdplotsetmaincoords{60}{110}
	\begin{tikzpicture}[scale=1.5,tdplot_main_coords]
	\draw[thick,dashed,-Triangle] (0,0,0) -- node[left,pos=1] {$1$} (1.9,0,0);
	\draw[thick,dashed,-Triangle] (0,0,0) -- node[right,pos=1] {$2$} (0,1.6,0);
	\draw[thick,dashed,-Triangle] (0,0,0) -- node[above,pos=1] {$3$} (0,0,1.5);
	\node[draw=black,line width=1pt,circle,fill=black,minimum width=0.2cm,inner sep=1pt] (O) at (0,0,0) {};
   
	\node[draw=black,line width=1pt,circle,fill=black,minimum width=0.2cm,inner sep=1pt] (X1) at (1,0,0) {};
	\node[draw=black,line width=1pt,circle,fill=black,minimum width=0.2cm,inner sep=1pt] (Z) at (0,0,1) {};
	\node[draw=black,line width=1pt,circle,fill=black,minimum width=0.2cm,inner sep=1pt] (Y1) at (0,1,0) {};
	\node[draw=black,line width=1pt,circle,fill=black,minimum width=0.2cm,inner sep=1pt] (XY) at (1,1,0) {};
	\draw[line width=1pt] (O)--(X1)--(XY)--(Y1)--(O);
	\draw[line width=1pt] (Z)--(O);
	\draw[line width=1pt] (Z)--(X1);
	\draw[line width=1pt] (Z)--(Y1);
	\draw[line width=1pt] (Z)--(XY);
     \node at (0,-0.4,0){$\boldsymbol{m}_{1}$};
     \node at (1,-0.3,0){$\boldsymbol{m}_{2}$};
      \node at (1,1.3,0){$\boldsymbol{m}_{3}$};
       \node at (0,1.3,0.2){$\boldsymbol{m}_{4}$};
        \node at (0,-0.3,1.2){$\boldsymbol{m}_{0}$};
	\end{tikzpicture}}\qquad \qquad 
\adjustbox{valign=c}{\begin{tikzpicture}[scale=1]
\draw[thick,dashed,-Triangle] (0,0,0) -- node[right,pos=1] {$1$} (2,0);
\draw[thick,dashed,-Triangle] (0,0,0) -- node[above,pos=1] {$2$} (0,2);
     \node[draw=black, circle, fill=black,minimum width=0.2cm,inner sep=1pt] (O) at (0,0){};
      \node[draw=black, circle, fill=black,minimum width=0.2cm,inner sep=1pt] (A) at (1,0){};
      \node[draw=black, circle, fill=black,minimum width=0.2cm,inner sep=1pt] (C) at (0,1){};
      \node[draw=black, circle, fill=black,minimum width=0.2cm,inner sep=1pt] (E) at (1,1){};
      \draw[line width=1pt] (O)--(A)--(E)--(C)--(O);
      \node[below left] at (0,0){$\boldsymbol{n}_{1}$};
      \node[below right] at (1,0){$\boldsymbol{n}_{2}$};
      \node[above right] at (1,1){$\boldsymbol{n}_{3}$};
      \node[above left] at (0,1){$\boldsymbol{n}_{4}$};
 \end{tikzpicture}}
\eea
and for later use we denote the lattice points by
\bea\label{eq:3dlatticept}
\boldsymbol{m}_0=(0,0,1),\quad \boldsymbol{m}_1=(0,0,0),\quad \boldsymbol{m}_2=(1,0,0),\quad \boldsymbol{m}_3=(1,1,0),\quad \boldsymbol{m}_4=(0,1,0)
\eea
and
\bea\label{eq:2dlatticept}
\boldsymbol{n}_1=(0,0),\quad \boldsymbol{n}_2=(1,0),\quad \boldsymbol{n}_3=(1,1),\quad \boldsymbol{n}_4=(0,1).
\eea

Let $\mathbb{R}\times S^{1}\times Z$ be the ten-dimensional spacetime and place D0-branes\footnote{When we say D0-branes, we are also including fractional D0-branes in this paper. Roughly speaking, fractional D0-branes are D-brane states localized at the singularity that behave as point-like objects in the non-compact spacetime. They may carry D-brane charges associated with compact cycles of the resolved geometry and can therefore be viewed, in an appropriate geometric phase, as D-branes wrapping such cycles. Different types of fractional D0-branes correspond to different gauge nodes of the quiver, and an appropriate bound state of them forms a regular D0-brane.} probing $Z$: 
\bea  \label{eq:typeIIA-CY4}
\renewcommand{\arraystretch}{1.05}
\begin{tabular}{|c|c|c|c|c|c|c|c|c|c|c|}
\hline
& \multicolumn{6}{c|}{$\mathcal{C}$} & \multicolumn{2}{c|}{$\mathbb{C}$} & \multicolumn{2}{c|}{$\mathbb{R}\times S^{1}$} \\
\cline{2-11} \raisebox{-0mm}{}  & 1 & 2 & 3 & 4& 5 & 6 & 7 & 8 & 9& 0\\[-2pt]
\hline D0& $\bullet$ & $\bullet$  & $\bullet$  & $\bullet$  & $\bullet$  & $\bullet$   & $\bullet$  & $\bullet$  & $\bullet$   & $-$\\
\hline
\end{tabular}
\eea
where $-$ means the brane extends in the corresponding direction and $\bullet$ means it is point-like. We may take the T-duality along the $\mathbb{R}\times S^{1}$ and instead consider D$(-1)$-branes or D$1$-branes in type IIB string theory. In this paper, we will mainly focus on the type IIA description.\footnote{The low energy effective theory of the D$(-1)$, D0, D1 are 0d, 1d, 2d and the partition functions will be rational, trigonometric, elliptic, respectively. }

The low energy theory on the D0-branes is an $\mathcal{N}=2$ quiver quantum mechanics. When the CY$_4$ is of the form CY$_3\times \mathbb{C}$, there is enhancement of SUSY to $\mathcal{N}=4$. In other words, we may place D3-branes on $\mathbb{C}\times\mathbb{R}\times S^{1}$, which gives a 4d $\mathcal{N}=1$ quiver gauge theory. The arising theory on the D0-branes is then understood as a dimensional reduction of it.

\paragraph{$\mathcal{N}=2$ quiver}
We briefly review basic aspects of $\mathcal{N}=2$ quiver quantum mechanics first.  Given a quiver $\overbar{Q}=(\overbar{Q}_{0},\overbar{Q}_{1})$, where $\overbar{Q}_{0}$ is a set of nodes and $\overbar{Q}_{1}$ is a set of edges, the information of the gauge symmetry and matter fields are obtained as follows.
\begin{itemize}[topsep=0pt, partopsep=0pt, itemsep=0pt]
    \item For each node $a$, a group $\U(k_{a})$ is assigned.
    \item Edges connect the nodes and there are two sets of oriented edges colored in black and red. We denote the set of black edges as $\overline{Q}_{1}^{(0)}$ while set of red edges are denoted as $\overline{Q}_{1}^{(1)}$.
    \item A black oriented edge from a source node $a$ to a terminal node $b$ represents an $\mathcal{N}=2$ chiral superfield $\Phi_{a\rightarrow b}$ transforming in the bifundamental representation $(\overline{k}_{a},k_{b})$ of $\U(k_{a})\times \U(k_{b})$.
    \item A red oriented edge connecting two nodes $a,b$ represents a Fermi superfield $\Lambda_{a\rightarrow b}$ transforming in the bifundamental representation $(\overline{k}_{a},k_{b})$ of $\U(k_{a})\times \U(k_{b})$. For the Fermi fields, we always have the conjugate $\overline{\Lambda}_{a\rightarrow b}$ transforming in $(k_{a},\overline{k}_{b})$ of $\U(k_{a})\times \U(k_{b})$, where we formally interpret the conjugate as reversing the orientation. 
\end{itemize}

Since we have both fields $(\Lambda_{a\rightarrow b},\overline{\Lambda}_{a\rightarrow b})$ in the theory, usually the Fermi superfields are written as unoriented edges in the literature. In this paper, the reference orientation of a Fermi multiplet will be kept whenever gauge representations, $J,E$-terms, or equivariant weights are relevant. In the periodic-quiver and brane-brick descriptions of the next subsection, we will occasionally suppress this orientation and regard
$\Lambda_{a\to b}$ and $\bar\Lambda_{a\to b}$ as the two oriented lifts of a single underlying Fermi edge.

Given a Fermi superfield $\Lambda_{ab}$, we additionally have a pair of holomorphic functions of chiral fields called the $E_{a\rightarrow b},J_{b\rightarrow a}$-terms. In our notation, the $E$-term is the last component of $\Lambda_{a\rightarrow b}=\cdots +\bar{\theta}^{+}E_{a\rightarrow b}(\Phi)$. Moreover, the $J$-term is coupled with $\Lambda_{a\rightarrow b}$ as
\bea
L_{J\text{-term}}=\int d\theta^{+} \Lambda_{a\rightarrow b}J_{b\rightarrow a}+\text{c.c.}.
\eea
The $E,J$-terms in quiver gauge theories are generally products of chiral fields taking the form of\footnote{The products of the chiral fields are written from the right to the left.  }
\bea
E_{ab}(\Phi)\sim \Phi_{b\leftarrow i_{n}}\Phi_{i_{n}\leftarrow i_{n-1}}\cdots \Phi_{i_{1}\leftarrow i_{0}}      \Phi_{i_{0}\leftarrow a},\\
J_{ba}(\Phi)\sim \Phi_{a\leftarrow j_{m}}\Phi_{j_{m}\leftarrow j_{m-1}}\cdots \Phi_{j_{1}\leftarrow j_{0}}      \Phi_{j_{0}\leftarrow b}
\eea
where we kept track of the arrows. Note that $\Lambda_{a\rightarrow b}$ has the same quantum numbers with $E_{a\rightarrow b}$, while $\overline{\Lambda}_{a\rightarrow b}$ has the same quantum numbers with $J_{b\rightarrow a}$. For later use, we sometimes omit the arrow and simply write $E_{ab}=E_{a\rightarrow b},J_{ba}=J_{b\rightarrow a}$ and $\Lambda_{ab}=\Lambda_{a\rightarrow b}, \overline{\Lambda}_{ab}=\overline{\Lambda}_{a\rightarrow b}$ in later discussions.


To preserve $\mathcal{N}=2$ supersymmetry, the $J,E$-terms need to obey the traceless condition
\bea\label{eq:JE-traceless}
\sum_{\alpha} \Tr \,(E_{\alpha}J_{\alpha})=0.
\eea
In particular, for a toric CY$_{4}$, the $J,E$-terms additionally obey the toric condition:
\bea\label{eq:JE-toriccond}
J_{b\rightarrow a}=J_{b\rightarrow a}^{+}-J_{b\rightarrow a}^{-},\quad E_{a\rightarrow b}=E_{a\rightarrow b}^{+}-E_{a\rightarrow b}^{-}
\eea
where $J_{b\rightarrow a }^{\pm},E_{a\rightarrow b}^{\pm}$ are monomials of chiral fields.

The $D$-term conditions for the node $a$ is
\bea\label{eq:D-term}
\mu_{a}-\zeta_{a}\mathbf{1}_{k_{a}}=0,\quad \mu_{a}\coloneqq\sum_{b:b\rightarrow a}\Phi_{b\rightarrow a}\Phi_{b\rightarrow a}^{\dagger}-\sum_{b:a\rightarrow b}\Phi_{a\rightarrow b}^{\dagger}\Phi_{a\rightarrow b}
\eea
where $\zeta_{a}$ is the FI-parameter. The vacuum moduli space is then
\bea
\mathcal{M}_{\mathcal{N}=2}=\left\{\Phi\mid \text{$D$-term}+\text{$J,E$-term}\right\}/\prod_{a}U(k_{a})
\eea

In our setup $\mathcal{C}\times \mathbb{C}$, the quiver is
\bea\label{eq:conifold-2susyquiver}
\adjustbox{valign=c}{
\begin{tikzpicture}[decoration={markings,mark=at position \arrowHeadPosition with {\arrow{latex}}}]
 \tikzset{
        box/.style={draw, minimum width=0.6cm, minimum height=0.6cm, text centered,thick},
        ->-/.style={decoration={
        markings,mark=at position #1 with {\arrow[scale=1.5]{>}}},postaction={decorate},line width=0.5mm},
        -<-/.style={decoration={
        markings,
        mark=at position #1 with {\arrow[scale=1.5]{<}}},postaction={decorate},line width=0.5mm}    
    }
\begin{scope}[xshift=4cm]
    \draw[postaction={decorate}, black,thick,scale=1.3] (0.65,0) arc(0:-180:0.65 and 0.1) ;
    \draw[postaction={decorate}, black,thick,scale=1.3] (0.75,0) arc(0:-180:0.75 and 0.2) ;
    \draw[postaction={decorate}, black,thick,scale=1.3] (-0.65,0) arc(180:0:0.65 and 0.1) ;
    \draw[postaction={decorate}, black,thick,scale=1.3] (-0.75,0) arc(180:0:0.75 and 0.2) ;
    \draw[postaction={decorate}, black,thick,scale=1.3] (-0.8,0) arc(360:0:0.4 and 0.3) ;
    \draw[postaction={decorate}, black,thick,scale=1.3] (0.8,0) arc(-180:180:0.4 and 0.3) ;
    
    \draw[red,thick,postaction={decorate},scale=1.3] (0.65,0) arc(0:-180:0.65 and 0.3) ;
    \draw[red,thick,postaction={decorate},scale=1.3] (0.75,0) arc(0:-180:0.75 and 0.4) ;
    \draw[red,thick,postaction={decorate},scale=1.3] (-0.65,0) arc(180:0:0.65 and 0.3) ;
    \draw[red,thick,postaction={decorate},scale=1.3] (-0.75,0) arc(180:0:0.75 and 0.4) ;
    \draw[fill=blue!100!white,thick](-0.95,0) circle(0.3cm);
    \draw[fill=red!100!white,thick](0.9,0) circle(0.3cm);
    
    \node at (-0.2,0.8){$\mathsf{A}_{1,2},\textcolor{red}{\Lambda^{1,2}_{1\rightarrow 2}}$};
    \node at (0,-0.8){$\mathsf{B}_{1},\mathsf{B}_{2},\textcolor{red}{\Lambda^{1,2}_{2\rightarrow 1}}$};
    \node at (-2.4,0.05){$\mathsf{C}_{1}$};
    \node at (2.5,0.05){$\mathsf{C}_{2}$};
\end{scope}
\end{tikzpicture}}
\eea
where $\mathsf{A}_{1,2},\mathsf{B}_{1,2},\mathsf{C}_{1,2}$ are $\mathcal{N}=2$ chiral fields and $\Lambda^{1,2}_{1\rightarrow 2},\Lambda^{1,2}_{2\rightarrow 1}$ are $\mathcal{N}=2$ Fermi fields. The $J,E$-terms are
\bea\label{eq:conifold-JEterms}
\begin{tabular}{c|cc}
 &\text{$J$-term}&  \text{$E$-term}\\ \hline
  $\Lambda^{1}_{12}$   & $\mathsf{B}_{1}\mathsf{A}_{2}\mathsf{B}_{2}-\mathsf{B}_{2}\mathsf{A}_{2}\mathsf{B}_{1}$  &  $\mathsf{C}_{2}\mathsf{A}_{1}-\mathsf{A}_{1}\mathsf{C}_{1}$\\
  $\Lambda^{2}_{12}$   &  $\mathsf{B}_{2}\mathsf{A}_{1}\mathsf{B}_{1}-\mathsf{B}_{1}\mathsf{A}_{1}\mathsf{B}_{2}$  & $\mathsf{C}_{2}\mathsf{A}_{2}-\mathsf{A}_{2}\mathsf{C}_{1}$ \\
  $\Lambda^{1}_{21}$  &   $\mathsf{A}_{2}\mathsf{B}_{2}\mathsf{A}_{1}-\mathsf{A}_{1}\mathsf{B}_{2}\mathsf{A}_{2}$& $\mathsf{C}_{1}\mathsf{B}_{1}-\mathsf{B}_{1}\mathsf{C}_{2}$\\
  $\Lambda^{2}_{21}$  &  $\mathsf{A}_{1}\mathsf{B}_{1}\mathsf{A}_{2}-\mathsf{A}_{2}\mathsf{B}_{1}\mathsf{A}_{1}$& $\mathsf{C}_{1}\mathsf{B}_{2}-\mathsf{B}_{2}\mathsf{C}_{2}$
\end{tabular}
\eea
where we shortly wrote $\Lambda^{1,2}_{1\rightarrow 2}=\Lambda^{1,2}_{12},\Lambda^{1,2}_{2\rightarrow 1}=\Lambda^{1,2}_{21}$.

\paragraph{$\mathcal{N}=4$ quiver}
When we have supersymmetry enhancement to $\mathcal{N}=4$, the matter fields are conveniently summarized in $\mathcal{N}=4$ vector and chiral fields.  Let $Q=(Q_{0},Q_{1})$ be an $\mathcal{N}=4$ quiver, where $Q_{0}$ is a set of nodes and $Q_{1}$ is a set of edges. The matter fields are described as follows.
\begin{itemize}[topsep=0pt, partopsep=0pt, itemsep=0pt]
    \item For each node $a$, a group $\U(k_{a})$ is assigned.
    \item There is only one type of edges connecting the nodes. 
    \item An oriented edge from a source node $a$ to a terminal node $b$ represents an $\mathcal{N}=4$ chiral superfield $\Phi_{ab}=\Phi_{a\rightarrow b}$ transforming in the bifundamental representation $(\overline{k}_{a},k_{b})$ of $\U(k_{a})\times \U(k_{b})$. We will abuse the notation and denote the $\mathcal{N}=4$ chiral superfield similar to the $\mathcal{N}=2$ chiral superfield.
\end{itemize}
An additional data is the superpotential $W$, which is a holomorphic function of the $\mathcal{N}=4$ chiral superfields. We also have the $D$-terms but we omit the discussion.

The $\mathcal{N}=4$ fields are decomposed into $\mathcal{N}=2$ fields as follows:
\begin{itemize}
    \item $\mathcal{N}=4$ vector $V^{\mathcal{N}=4}$: $\mathcal{N}=2$ vector $V^{\mathcal{N}=2}$ $+$ $\mathcal{N}=2$ adjoint chiral field $\Sigma^{\mathcal{N}=2}$
    \item $\mathcal{N}=4$ chiral $\Phi^{\mathcal{N}=4}$: $\mathcal{N}=2$ chiral $\Phi^{\mathcal{N}=2}$ $+$ $\mathcal{N}=2$ Fermi field $\Lambda^{\mathcal{N}=2}$
\end{itemize}
The $J,E$-terms of the $\mathcal{N}=4$ theory are
\bea\label{eq:4susyEJterm}
E=\Sigma^{(0,2)}\cdot  \Phi^{(0,2)},\quad J=\partial W.
\eea

For the $\mathcal{C}\times \mathbb{C}$, the $\mathcal{N}=4$ quiver and the superpotential are
\bea\label{eq:conifold-4susyquiver}
\adjustbox{valign=c}{\begin{tikzpicture}[decoration={markings,mark=at position \arrowHeadPosition with {\arrow{latex}}}]
 \tikzset{
        box/.style={draw, minimum width=0.6cm, minimum height=0.6cm, text centered,thick},
        ->-/.style={decoration={
        markings,mark=at position #1 with {\arrow[scale=1.5]{>}}},postaction={decorate},line width=0.5mm},
        -<-/.style={decoration={
        markings,
        mark=at position #1 with {\arrow[scale=1.5]{<}}},postaction={decorate},line width=0.5mm}    
    }
\begin{scope}[xshift=4cm]
    \draw[postaction={decorate}, black,thick,scale=1.3] (0.65,0) arc(0:-180:0.65 and 0.25) ;
    \draw[postaction={decorate}, black,thick,scale=1.3] (0.75,0) arc(0:-180:0.75 and 0.4) ;
    \draw[postaction={decorate}, black,thick,scale=1.3] (-0.65,0) arc(180:0:0.65 and 0.25) ;
    \draw[postaction={decorate}, black,thick,scale=1.3] (-0.75,0) arc(180:0:0.75 and 0.4) ;
    \node at (0,0.8){$\mathsf{A}_{1},\mathsf{A}_{2}$};
    \node at (0,-0.8){$\mathsf{B}_{1},\mathsf{B}_{2}$};
   
    \node at (-1,-0.5) {1};
    \node at (1,-0.5) {2};
    \draw[fill=blue!100!white,thick](-0.95,0) circle(0.3cm);
    \draw[fill=red!100!white,thick](0.9,0) circle(0.3cm);
   
\end{scope}
\end{tikzpicture}}\qquad \mathsf{W}=\Tr\left(\mathsf{A}_{1}\mathsf{B}_{1}\mathsf{A}_{2}\mathsf{B}_{2}-\mathsf{A}_{1}\mathsf{B}_{2}\mathsf{A}_{2}\mathsf{B}_{1}\right)
\eea
where we identified the $\mathcal{N}=4$ chiral superfield and $\mathcal{N}=2$ chiral superfield. The decomposition into $\mathcal{N}=2$ is
\bea
\mathcal{N}=4\text{ vector}&\rightarrow \mathcal{N}=2\text{ vector} +\mathsf{C}_{1,2},\\
\mathsf{A}_{1,2} &\rightarrow (\mathsf{A}_{1,2},\Lambda^{1,2}_{1 2})\\
\mathsf{B}_{1,2} &\rightarrow (\mathsf{B}_{1,2},\Lambda^{1,2}_{2 1}).
\eea
The $J$-term for example for $\mathsf{A}_1$ is obtained from
\bea
\frac{\partial \mathsf{W}}{\partial \mathsf{A}_1}=\mathsf{B}_{1}\mathsf{A}_{2}\mathsf{B}_{2}-\mathsf{B}_{2}\mathsf{A}_{2}\mathsf{B}_{1}.
\eea

\subsection{Brane brick models, brick matchings, and phase boundaries}\label{sec:branebrickmodel}

We briefly review the periodic quiver and brane brick model associated
with the quiver in \eqref{eq:conifold-2susyquiver}. Brane brick models
were introduced in \cite{Franco:2015tna} as type IIA brane
configurations encoding the two-dimensional $\mathcal N=(0,2)$ quiver
gauge theories on D1-branes probing toric Calabi--Yau fourfolds, and
were systematically developed in \cite{Franco:2015tya}. See also \cite{Franco:2016qxh} for a complementary mirror-symmetric
description of brane brick models. We also
summarize their string-theoretic realization and the relevant
combinatorial structures, including brick matchings and phase
boundaries, following these references.

\begin{figure}[t]
  \centering

  \begin{minipage}[c]{0.48\textwidth}
    \centering
   \begin{tikzpicture}[
  x=1.4cm,
  y=1.4cm,
  vertex1/.style={
    circle,
    draw=black,
    very thick,
    fill=blue!100!white,
    minimum size=7mm,
    inner sep=0pt
  },
  vertex2/.style={
    circle,
    draw=black,
    very thick,
    fill=red!100!white,
    minimum size=7mm,
    inner sep=0pt
  },
edge/.style={
  black,
  very thick,
  preaction={
    draw=white,
    line width=2pt
  },
  postaction={
    decorate
  },
  decoration={
    markings,
    mark=at position 0.7 with {
      \arrow{Stealth[length=3mm,width=2.5mm]}
    }
  }
},
rededge/.style={
  red!85!orange,
  very thick,
  preaction={
    draw=white,
    line width=2pt
  }
}
]

\begin{scope}[scale=0.7]

\node[vertex1,label=above left:$ $]  (a1) at (0,3) {};
\node[vertex1,label=below left:$ $]  (a0) at (0,0) {};

\node[vertex2,label=above right:$ $] (b1) at (1,4) {};
\node[vertex2,label=below right:$ $] (b0) at (1,1) {};

\node[vertex2,label=above left:$ $]  (c1) at (2,3) {};
\node[vertex2,label=below right:$ $] (c0) at (2,0) {};

\node[vertex1,label=above right:$ $] (d1) at (3,4) {};
\node[vertex1,label=below right:$ $] (d0) at (3,1) {};

\node[vertex1,label=above left:$ $]  (e1) at (4,3) {};
\node[vertex1,label=below right:$ $] (e0) at (4,0) {};

\node[vertex2,label=above right:$ $] (f1) at (5,4) {};
\node[vertex2,label=below right:$ $] (f0) at (5,1) {};

\node[right] at (4.5,0.2){\large$\mathsf{B}_{2}$};
\node[above] at (2,4.2){\large$\mathsf{A}_{2}$};
\node[above] at (3.8,4.2){\large$\textcolor{red}{\Lambda^{1}_{12}}$};
\node[right ]at (5,2.25){\large$\mathsf{C}_{2}$};
\node[left ]at (0,1.5){\large$\mathsf{C}_{1}$};
\node[right] at (4.2,2.5){\large $\textcolor{red}{\Lambda^{2}_{21}}$};
\node at (1.3,1.7){\large$\mathsf{A}_{1}$};

\draw[rededge](b1)--(d0);
\draw[edge](d1)--(f0);

\draw[edge] (f1)--(e1);

\draw[edge](c1)--(d1);
\draw[edge](d1)--(b1);
\draw[edge](b1)--(a1);

\draw[edge] (f0)--(e0);
\draw[edge] (e0)--(c0);
\draw[edge](c0)--(d0);
\draw[edge](d0)--(b0);
\draw[edge](b0)--(a0);

\draw[edge](a0)--(a1);
\draw[edge](b0)--(b1);
\draw[edge](c0)--(c1);
\draw[edge](d0)--(d1);

\draw[edge](f0)--(f1);

\draw[rededge](a1)--(b0);

\draw[edge](a1)--(c0);

\draw[rededge](d1)--(f1);
\draw[rededge](d0)--(f0);
\draw[rededge](a1)--(c1);
\draw[rededge](a0)--(c0);

\draw[rededge](e1)--(f0);

\draw[rededge](c0)--(d1);

\draw[rededge](c1)--(e0);

\draw[edge](e0)--(e1);
\draw[edge] (e1)--(c1);

\node[vertex1,label=above left:$ $]   at (0,3) {};
\node[vertex1,label=below left:$ $]  at (0,0) {};

\node[vertex2,label=above right:$ $]  at (1,4) {};

\node[vertex2,label=above left:$ $]  (c1) at (2,3) {};
\node[vertex2,label=below right:$ $] (c0) at (2,0) {};

\node[vertex1,label=above right:$ $] (d1) at (3,4) {};
\node[vertex1,label=below right:$ $] (d0) at (3,1) {};

\node[vertex1,label=above left:$ $]  (e1) at (4,3) {};
\node[vertex1,label=below right:$ $] (e0) at (4,0) {};

\node[vertex2,label=above right:$ $] (f1) at (5,4) {};
\node[vertex2,label=below right:$ $] (f0) at (5,1) {};

\draw[rededge](c1)--(e0);
\node at (3.1,0.3){\large $\textcolor{red}{\Lambda^{2}_{12}}$};
\node at (3,2.4){\large $\textcolor{red}{\Lambda^{1}_{21}}$};
\end{scope}

\end{tikzpicture}
\end{minipage}
  \begin{minipage}[c]{0.48\textwidth}
    \centering
    \includegraphics[width=0.8\linewidth]{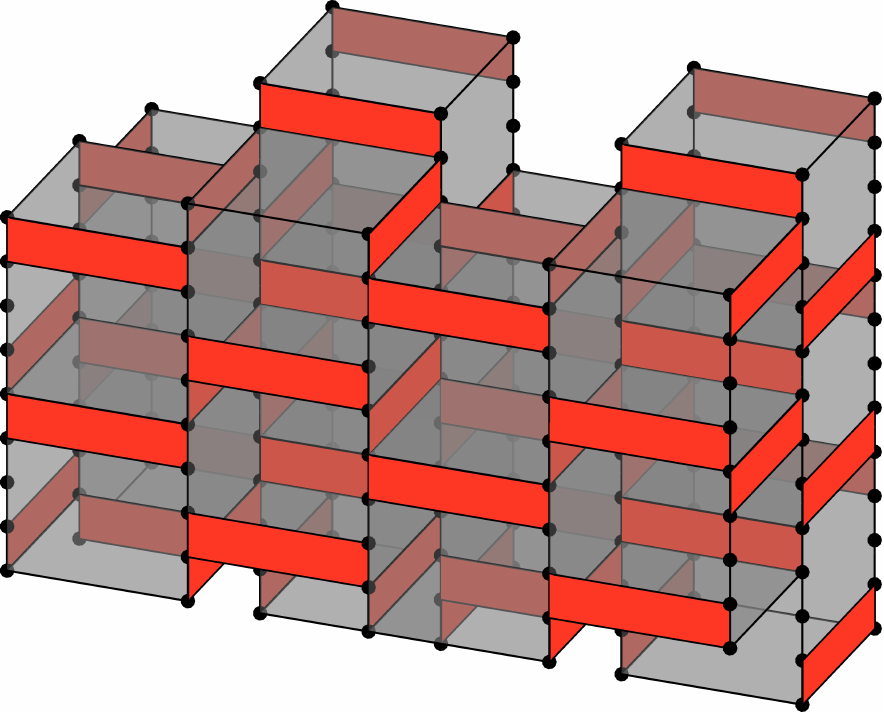}
  \end{minipage}
  \hfill
  \caption{Periodic quiver and brane brick model of $\mathcal{C}\times \mathbb{C}$}
  \label{fig:ConifoldBraneBrickQuiver}
\end{figure}
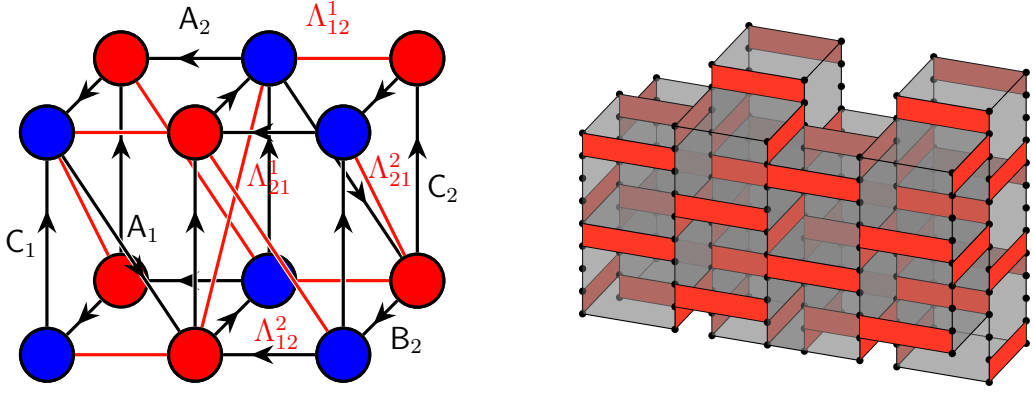

\paragraph{Periodic Quiver}
For a toric Calabi--Yau fourfold, the quiver and its $J$- and
$E$-terms are encoded in a \emph{periodic quiver} on $T^{3}$. Each
monomial terms $J_{\Lambda}^{\pm}$ or $E_{\Lambda}^{\pm}$ determines
a minimal plaquette, namely a closed path consisting of one
Fermi edge and an arbitrary number of chiral edges. For
combinatorial purposes, we package these plaquettes into the
formal quiver potential
\bea
\adjustbox{valign=c}{
\begin{tikzpicture}[
    scale=1.15,
    line cap=round,
    line join=round,
    chiral/.style={
        black,
        thick,
        postaction={
            decorate
        },
        decoration={
            markings,
            mark=at position 0.52 with {\arrow{Triangle}}
        }
    },
    fermi/.style={
        red,
        very thick
    },
    qnode/.style={
        circle,
        draw=black,
        thick,
        fill=yellow!85!orange,
        inner sep=4pt
    }
]

\coordinate (j)  at (0, 1.35);
\coordinate (i)  at (0,-1.35);

\coordinate (jt) at (-1.85, 0.75);
\coordinate (jb) at (-1.75,-1.00);

\coordinate (er1) at (1.55, 1.05);
\coordinate (er2) at (1.70, 0.05);
\coordinate (jr)  at (1.55,-1.05);

\fill[orange!35,opacity=0.78]
    (j)--(jt)--(i)--cycle;

\fill[cyan!35,opacity=0.78]
    (j)--(jb)--(i)--cycle;

\fill[cyan!35,opacity=0.78]
    (j)--(er1)--(er2)--(i)--cycle;

\fill[orange!35,opacity=0.78]
    (j)--(jr)--(i)--cycle;

\draw[chiral] (j)--(jt);
\draw[chiral] (jt)--(i);

\draw[chiral] (j)--(jb);
\draw[chiral] (jb)--(i);

\draw[chiral] (er1)--(j);
\draw[chiral] (er2)--(er1);
\draw[chiral] (i)--(er2);

\draw[chiral] (i)--(jr);
\draw[chiral] (jr)--(j);

\draw[fermi] (j)--(i);

\node[qnode] at (j)  {};
\node[qnode] at (i)  {};
\node[qnode] at (jt) {};
\node[qnode] at (jb) {};
\node[qnode] at (er1) {};
\node[qnode] at (er2) {};
\node[qnode] at (jr) {};

\node[above left=1pt] at (j) {$b$};
\node[below left=1pt] at (i) {$a$};

\node[orange!80!black] at (-1.1, 0.55)
    {$J_{b\rightarrow a}^{+}$};

\node[cyan!60!black] at (-1.00,-0.65)
    {$E_{a\rightarrow b}^{-}$};

\node[cyan!60!black] at (1.05,0.62)
    {$E_{a\rightarrow b}^{+}$};

\node[orange!80!black] at (0.6,-0.3)
    {$J_{b\rightarrow a}^{-}$};

\node[fill=white,
    fill opacity=1,
    text opacity=1,
    rounded corners=1pt,
    inner sep=1.5pt.
    right,
    text=red,
    inner sep=1.5pt
] at (0,0) {$\Lambda_{ab}$};

\end{tikzpicture}}\qquad \mathcal{W}= \Tr\,(\Lambda_{a \rightarrow b} J_{b\rightarrow a} )+\Tr\,(\overline{\Lambda}_{a\rightarrow b} E_{a\rightarrow b})
\eea
In passing from the quiver representation to the periodic quiver, it is useful to distinguish a Fermi multiplet from its graphical realization. Although every Fermi multiplet has been assigned a reference orientation in the previous subsection, its underlying edge in the periodic quiver will be drawn as a red unoriented edge. Thus,
each Fermi multiplet determines four plaquettes
$(\Lambda, J_{\Lambda}^{\pm})$ and
$(\overline{\Lambda},E_{\Lambda}^{\pm})$, sharing the same
underlying Fermi edge. Since $E_{\Lambda}$ transforms
in the same representation as $\Lambda$, while $J_{\Lambda}$
transforms in the conjugate representation, all four plaquettes
are gauge-invariant closed paths. The periodic quiver of the $\mathcal{C}\times \mathbb{C}$ is illustrated in Figure~\ref{fig:ConifoldBraneBrickQuiver}.

\begin{figure}
\centering
\adjustbox{valign=c}{\includegraphics[width=3cm]{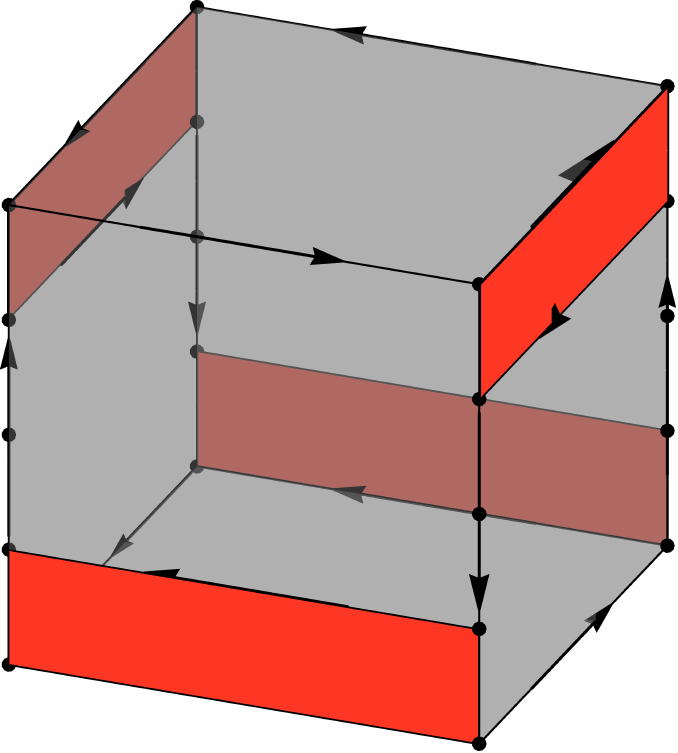}}\quad \qquad \qquad   \adjustbox{valign=c}{\begin{tikzpicture}[decoration={markings,mark=at position \arrowHeadPosition with {\arrow{latex}}}]
 \tikzset{
        box/.style={draw, minimum width=0.6cm, minimum height=0.6cm, text centered,thick},
        ->-/.style={decoration={
        markings,mark=at position #1 with {\arrow[scale=1.5]{>}}},postaction={decorate},line width=0.5mm},
        -<-/.style={decoration={
        markings,
        mark=at position #1 with {\arrow[scale=1.5]{<}}},postaction={decorate},line width=0.5mm}    
    }
\begin{scope}[xshift=4cm,scale=1.3]
   
   \draw[postaction={decorate},black, thick] (0,0)--(0,1);
   \node[above] at (0,1){$\mathsf{C}_{1}$};
   \draw[postaction={decorate},black, thick] (0,-1)--(0,0);
   \draw[postaction={decorate},black, thick] (0,0)--(1,-0.2);
    \node[right] at (1,-0.2){$\mathsf{A}_{1}$};
   \draw[postaction={decorate},black, thick] (0,0)--(-1,+0.2);
    \node[left] at (-1,+0.2){$\mathsf{A}_{2}$};
   \draw[postaction={decorate},black, thick] (0.7,0.7)--(0,0);
    \node[above right] at (0.7,0.7){$\mathsf{B}_{2}$};
     \draw[postaction={decorate},black, thick] (-0.7,-0.7)--(0,0);
        \node[below left] at (-0.7,-0.7){$\mathsf{B}_{1}$};
       \draw[red, thick] (0,0)--(1,0.4);
          \node[right,red] at (1,0.4){$\Lambda^{1}_{12}$};
       \draw[red, thick] (0,0)--(-1,0.7);
        \node[above left,red] at (-1,0.7){$\Lambda^{2}_{1 2}$};
       
       \draw[red, thick] (-0.7,-1.2)--(0,0);
       \node[below ,red] at (-0.7,-1.2){$\Lambda^{1}_{21}$};
        \draw[red, thick] (0.7,-0.6)--(0,0);
        \node[below right,red] at (0.7,-0.6){$\Lambda^{2}_{21}$};
    \draw[fill=blue!100!white,thick](0,0) circle(0.2cm);
   
\end{scope}
\end{tikzpicture}} \\
\adjustbox{valign=c}{\includegraphics[width=3cm]{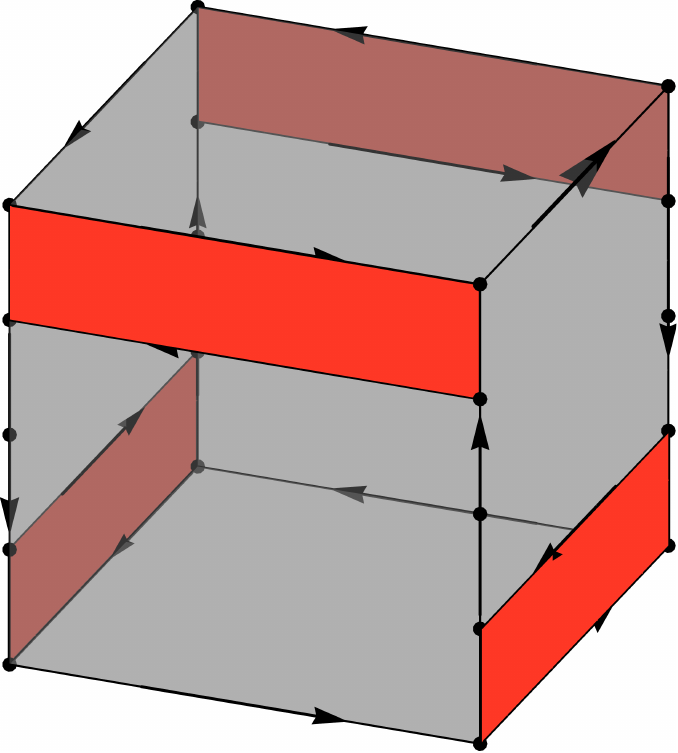}}\qquad \qquad   \adjustbox{valign=c}{\begin{tikzpicture}[decoration={markings,mark=at position \arrowHeadPosition with {\arrow{latex}}}]
 \tikzset{
        box/.style={draw, minimum width=0.6cm, minimum height=0.6cm, text centered,thick},
        ->-/.style={decoration={
        markings,mark=at position #1 with {\arrow[scale=1.5]{>}}},postaction={decorate},line width=0.5mm},
        -<-/.style={decoration={
        markings,
        mark=at position #1 with {\arrow[scale=1.5]{<}}},postaction={decorate},line width=0.5mm}    
    }
\begin{scope}[xshift=4cm,scale=1.3]
   
   \draw[postaction={decorate},black, thick] (0,0)--(0,1);
   \node[above] at (0,1){$\mathsf{C}_{2}$};
   \draw[postaction={decorate},black, thick] (0,-1)--(0,0);

   \draw[postaction={decorate},black, thick] (1,-0.2)--(0,0);
    \node[right] at (1,-0.2){$\mathsf{A}_{2}$};
   \draw[postaction={decorate},black, thick] (-1,+0.2)--(0,0);
    \node[left] at (-1,+0.2){$\mathsf{A}_{1}$};
    
   \draw[postaction={decorate},black, thick] (0,0)--(0.7,0.7);
    \node[above right] at (0.7,0.7){$\mathsf{B}_{1}$};
     \draw[postaction={decorate},black, thick] (0,0)--(-0.7,-0.7);
        \node[below left] at (-0.7,-0.7){$\mathsf{B}_{2}$};

       \draw[red, thick](1,-1.1)--(0,0);
          \node[right,red] at (1,-1.1){$\Lambda^{2}_{12}$};
          
       \draw[red, thick] (-1,-0.6)--(0,0);
        \node[left,red] at (-1,-0.6){$\Lambda^{1}_{12}$};
       
       \draw[red, thick] (0,0)--(-0.7,0.7);
       \node[left,red] at (-0.7,0.7){$\Lambda^{2}_{21}$};
        \draw[red, thick] (0,0)--(0.7,1.5);
        \node[below right,red] at (0.7,1.5){$\Lambda^{1}_{21}$};
    \draw[fill=red!100!white,thick](0,0) circle(0.2cm);
   
\end{scope}
\end{tikzpicture}}
\caption{Two fundamental bricks and the corresponding fields of the gauge theory associated with $\mathcal{C}\times \mathbb{C}$. The gray face and red face correspond to the $\mathcal{N}=2$ chiral and Fermi fields, respectively. The edges of the brick correspond to the monomial terms of the $J,E$-terms.}
\label{fig:fundamentalbrick}
\end{figure}
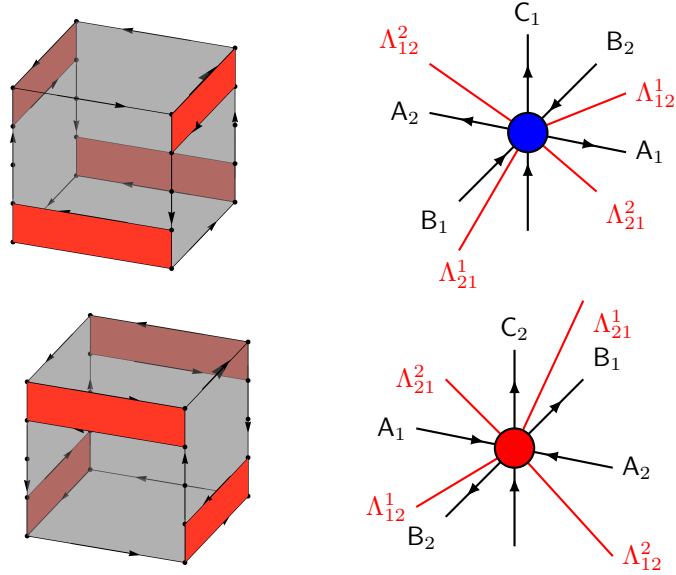

\paragraph{Brane Brick Model}
The \emph{brane brick model} is the cellular decomposition of $T^3$ dual to the periodic quiver. The correspondence between the quiver gauge theory, periodic quiver, and the brane brick model is as follows.
\bea
\renewcommand{\arraystretch}{1.45}
\begin{tabular}{
    >{\centering\arraybackslash}m{0.30\textwidth}|
    >{\centering\arraybackslash}m{0.25\textwidth}|
    >{\centering\arraybackslash}m{0.25\textwidth}
}
\hline
Gauge theory
&
Periodic quiver
&
Brane brick model
\\
\hline\hline
Gauge factor $U(k_a)$
&
Node $a$
&
Brick
\\
\hline
Chiral $\Phi_{a\to b}$
&
\parbox[c][1.05cm][c]{3.2cm}{
    \centering
    Black oriented edge\\
    $a\to b$
}
&
Oriented gray face
\\
\hline
Fermi $(\Lambda_{a\rightarrow b},\overline{\Lambda}_{a\rightarrow b})$
&
\parbox[c][1.05cm][c]{3.2cm}{
    \centering
    Red unoriented edge\\
    between $a$ and $b$
}
&
Red unoriented face
\\
\hline
Monomial $J^{\pm},E^{\pm}$
&
Minimal plaquette
&
Brick edge
\\
\hline
\end{tabular}
\label{eq:brick-quiver-dictionary}
\eea
Under the duality between the periodic quiver and the brane brick model, gauge nodes become bricks, chiral arrows become oriented gray faces, and the underlying edge of a Fermi multiplet becomes an unoriented red face separating two bricks. The boundary edges of this Fermi face encode the four $J$- and $E$-plaquettes above. Accordingly, the brane brick model treats a Fermi face as geometrically unoriented, while its oriented lifts are restored whenever the individual $J$- and $E$-terms must be distinguished.

The brane brick model for the $\mathcal{C}\times\mathbb{C}$ is illustrated in Figure~\ref{fig:ConifoldBraneBrickQuiver}, where there are two fundamental bricks corresponding to the two gauge nodes as in Figure~\ref{fig:fundamentalbrick}. 

Generally, for the toric CY$_{4}$ taking the form of $Z=X\times \mathbb{C}$, the brane brick model can be obtained by a \textit{lifting algorithm} of the brane tiling associated with the toric CY$_{3}$ $X$ \cite{Franco:2015tya}. This is a similar procedure of the decomposition of the $\mathcal{N}=4$ quivers into $\mathcal{N}=2$ quivers. We summarize the notations for the brane tiling of the conifold $\mathcal{C}$ in Appendix~\ref{app:branetiling-conifold}.

\paragraph{String-theoretic realization.}
The brane brick model admits a direct string-theoretic
realization. We choose logarithmic coordinates
\bea
z_1
&=
e^{x^2+i x^1},
\qquad
z_2
=
e^{x^4+i x^3},
\qquad
z_3
=
e^{x^6+i x^5},
\eea
so that $x^1,x^3,x^5$ parametrize the angular directions of
the toric $T^3$ fiber. Performing T-duality along these three
circles, which we denote by $T_{135}$, maps the Type IIA
configuration \eqref{eq:typeIIA-CY4} to the Type IIB brane
system
\bea
\label{eq:branebrick-web}
\renewcommand{\arraystretch}{1.05}
\begin{tabular}{|c|c|c|c|c|c|c|c|c|c|c|}
\hline
&1&2&3&4&5&6&7&8&9&0
\\[-2pt]
\hline
D3
&$-$&$\bullet$&$-$&$\bullet$&$-$&$\bullet$
&$\bullet$&$\bullet$&$\bullet$&$-$
\\
\hline
NS5
&\multicolumn{6}{c|}{$\Sigma$}
&$\bullet$&$\bullet$&$-$&$-$
\\
\hline
\end{tabular}
\eea
where we use the same labels for the T-dual coordinates.
The D0-branes become D3-branes wrapping the dual $T^3$, while
the degeneration data of the toric fibration are encoded by an
NS5-brane wrapping the holomorphic surface
\bea
\Sigma
&=
\left\{
P(z_1,z_2,z_3)=0
\right\}
\subset
(\mathbb C^\times)^3,
\quad
P(z_1,z_2,z_3)
=
\sum_{\boldsymbol v\in\Delta\cap\mathbb Z^3}
c_{\boldsymbol v}
z_1^{v_1}z_2^{v_2}z_3^{v_3},
\eea
where $\Delta$ is the toric diagram of the Calabi--Yau
fourfold.\note{The conventional brane brick construction starts
from D1-branes and produces D4-branes after the same triple
T-duality. Our D3--NS5 system is its dimensional
reduction along the common spatial direction.}

The brane brick model is obtained by taking the tropical
skeleton of the argument projection of $\Sigma$ onto $T^3$.
The resulting two-dimensional sheets divide $T^3$ into the
three-dimensional regions called bricks. The D3-branes inside the bricks are the gauge nodes of the associated quiver and the open strings connecting between different bricks give the chiral and Fermi fields of the theory.

\paragraph{Brick matchings}For later use, let us introduce a combinatorial object called the \emph{brick matching}, which is a generalization of the perfect matching of the brane tiling. As mentioned above, for each Fermi field $(\Lambda,\overline{\Lambda})$, we have four plaquettes. The brick matchings are defined as a special collection of chiral, Fermi, and conjugate Fermi fields covering every plaquette exactly once obeying the following conditions.
\begin{itemize}
    \item The chiral fields in the brick matching cover either the plaquettes $(\Lambda \cdot J_{\Lambda}^{+},\Lambda \cdot J^{-}_{\Lambda})$ or $(\overline{\Lambda}\cdot E_{\Lambda}^{+},\overline{\Lambda}\cdot E_{\Lambda}^{-})$ exactly once.
    \item If the chiral fields in the brick matching cover the plaquettes associated with the $J_{\Lambda}$-terms, then $\overline{\Lambda}$ is included in the brick matching.
    \item If the chiral fields in the brick matching cover the plaquettes associated with the $E_{\Lambda}$-terms, then $\Lambda$ is included in the brick matching.
\end{itemize}

The fields in the brick matching $\mathfrak{m}_{i}$ can be summarized in a brick matching matrix $P$, whose entries take the form as
\bea
P_{\alpha,i}=\begin{dcases}
    1 \qquad \Phi_\alpha,\Lambda_\alpha\in\mathfrak{m}_{i}\\
    0 \qquad \Phi_\alpha,\Lambda_\alpha\notin\mathfrak{m}_{i}
\end{dcases}
\eea
where $\alpha$ is the label of the chiral and Fermi fields and $i$ is the label of the brick matching.

For the $\mathcal{C}\times \mathbb{C}$ case, we have five brick matchings in total:
\bea
\mathfrak{m}_0&=\{\textcolor{purple}{\mathsf{C}_{1}},\textcolor{purple}{\mathsf{C}_{2}},\textcolor{darkgreen}{\Lambda^{1}_{1 2}},\textcolor{darkgreen}{\Lambda^{2}_{1 2}},\textcolor{darkgreen}{\Lambda^{1}_{2 1}},\textcolor{darkgreen}{\Lambda^{2}_{2 1}}\},\\
\mathfrak{m}_{1}&=\{\textcolor{purple}{\mathsf{B}_{1}},\textcolor{darkgreen}{\bar{\Lambda}^{1}_{1 2}},\textcolor{darkgreen}{\bar{\Lambda}^{2}_{1 2}},\textcolor{darkgreen}{\Lambda^{1}_{2 1}},\textcolor{darkgreen}
{\bar{\Lambda}^{2}_{2 1}}\},\\
\mathfrak{m}_{2}&=\{\textcolor{purple}{\mathsf{A}_{2}},\textcolor{darkgreen}{\bar{\Lambda}^{1}_{1 2}},\textcolor{darkgreen}{\Lambda^{2}_{1 2}},\textcolor{darkgreen}{\bar{\Lambda}^{1}_{2 1}},\textcolor{darkgreen}
{\bar{\Lambda}^{2}_{2 1}}\},\\
\mathfrak{m}_{3}&=\{\textcolor{purple}{\mathsf{B}_{2}},\textcolor{darkgreen}{\bar{\Lambda}^{1}_{1 2}},\textcolor{darkgreen}{\bar{\Lambda}^{2}_{1 2}},\textcolor{darkgreen}{\bar{\Lambda}^{1}_{2 1}},\textcolor{darkgreen}
{\Lambda^{2}_{2 1}}\},\\
\mathfrak{m}_{4}&=\{\textcolor{purple}{\mathsf{A}_{1}},\textcolor{darkgreen}{\Lambda^{1}_{1 2}},\textcolor{darkgreen}{\bar{\Lambda}^{2}_{1 2}},\textcolor{darkgreen}{\bar{\Lambda}^{1}_{2 1}},\textcolor{darkgreen}
{\bar{\Lambda}^{2}_{2 1}}\}
\eea
giving the brick matching matrix
\bea\label{eq:conifold_brickmatchingmatrix}
P_{\Lambda\bar{\Lambda}}
=
\left(
\begin{array}{c|ccccc}
     & \mathfrak{m}_1 & \mathfrak{m}_2 & \mathfrak{m}_3 & \mathfrak{m}_4 & \mathfrak{m}_0 \\ \hline
    \mathsf{A}_1 & 0&0&0&1&0\\
    \mathsf{A}_2 & 0&1&0&0&0\\
    \mathsf{B}_1 & 1&0&0&0&0\\
    \mathsf{B}_2 & 0&0&1&0&0\\
    \mathsf{C}_{1} & 0&0&0&0&1\\
    \mathsf{C}_{2} & 0&0&0&0&1\\ \hline
    \Lambda^{1}_{1 2} & 0&0&0&1&1\\
    \bar{\Lambda}^{1}_{1 2} & 1&1&1&0&0\\
    \Lambda^{2}_{1 2} & 0&1&0&0&1\\
    \bar{\Lambda}^{2}_{1 2} & 1&0&1&1&0\\
    \Lambda^{1}_{2 1} & 1&0&0&0&1\\
    \bar{\Lambda}^{1}_{2 1} & 0&1&1&1&0\\
    \Lambda^{2}_{2 1} & 0&0&1&0&1\\
    \bar{\Lambda}^{2}_{2 1} & 1&1&0&1&0
\end{array}
\right).
\eea
For example, the brick matching $\mathfrak{m}_{2}$ comes from 
\bea\label{eq:brickmatching-m2}
\begin{tabular}{cc}
 \text{$J$-term}&  \text{$E$-term}\\ 
  ${\Lambda_{1 2}^{1} }(\mathsf{B}_{1}\textcolor{purple}{\mathsf{A}_{2}}\mathsf{B}_{2}-\mathsf{B}_{2}\textcolor{purple}{\mathsf{A}_{2}}\mathsf{B}_{1})$  &  $\textcolor{darkgreen}{\bar{\Lambda}^{1}_{1 2}}(\mathsf{C}_{2}{\mathsf{A}_{1}}-{\mathsf{A}_{1}}\mathsf{C}_{1})$\\
  $\textcolor{darkgreen}{\Lambda^{2}_{1 2}}(\mathsf{B}_{2}{\mathsf{A}_{1}}\mathsf{B}_{1}-\mathsf{B}_{1}{\mathsf{A}_{1}}\mathsf{B}_{2})$  & ${\bar{\Lambda}^{2}_{1 2}}(\mathsf{C}_{2}\textcolor{purple}{\mathsf{A}_{2}}-\textcolor{purple}{\mathsf{A}_{2}}\mathsf{C}_{1})$ \\
  $\Lambda^{1}_{2 1}(\textcolor{purple}{\mathsf{A}_{2}}\mathsf{B}_{2}{\mathsf{A}_{1}}-{\mathsf{A}_{1}}\mathsf{B}_{2}\textcolor{purple}{\mathsf{A}_{2}})$& $\textcolor{darkgreen}{\bar{\Lambda}^{1}_{2 1}}(\mathsf{C}_{1}\mathsf{B}_{1}-\mathsf{B}_{1}\mathsf{C}_{2})$\\
  $\Lambda^{2}_{2 1}({\mathsf{A}_{1}}\mathsf{B}_{1}\textcolor{purple}{\mathsf{A}_{2}}-\textcolor{purple}{\mathsf{A}_{2}}\mathsf{B}_{1}{\mathsf{A}_{1}})$& $\textcolor{darkgreen}{\bar{{\Lambda}}^{2}_{2 1}}(\mathsf{C}_{1}\mathsf{B}_{2}-\mathsf{B}_{2}\mathsf{C}_{2})$
\end{tabular}
\eea
while the brick matching $\mathfrak{m}_{0}$ comes from
\bea\label{eq:brickmatching-m0}
\begin{tabular}{cc}
 \text{$J$-term}&  \text{$E$-term}\\ 
  $\textcolor{darkgreen}{\Lambda^{1}_{1 2}} (\mathsf{B}_{1}\mathsf{A}_{2}\mathsf{B}_{2}-\mathsf{B}_{2}\mathsf{A}_{2}\mathsf{B}_{1})$  &  $\bar{\Lambda}^{1}_{1 2}(\textcolor{purple}{\mathsf{C}_{2}}\mathsf{A}_{1}-\mathsf{A}_{1}\textcolor{purple}{\mathsf{C}_{1}})$\\
  $\textcolor{darkgreen}{\Lambda^{2}_{1 2}}(\mathsf{B}_{2}\mathsf{A}_{1}\mathsf{B}_{1}-\mathsf{B}_{1}\mathsf{A}_{1}\mathsf{B}_{2})$  & $\bar{\Lambda}^{2}_{1 2}(\textcolor{purple}{\mathsf{C}_{2}}\mathsf{A}_{2}-\mathsf{A}_{2}\textcolor{purple}{\mathsf{C}_{1}})$ \\
  $\textcolor{darkgreen}{\Lambda^{1}_{2 1}}(\mathsf{A}_{2}\mathsf{B}_{2}\mathsf{A}_{1}-\mathsf{A}_{1}\mathsf{B}_{2}\mathsf{A}_{2})$& $\bar{\Lambda}^{1}_{2 1}(\textcolor{purple}{\mathsf{C}_{1}}\mathsf{B}_{1}-\mathsf{B}_{1}\textcolor{purple}{\mathsf{C}_{2}})$\\
  $\textcolor{darkgreen}{\Lambda^{2}_{2 1}}(\mathsf{A}_{1}\mathsf{B}_{1}\mathsf{A}_{2}-\mathsf{A}_{2}\mathsf{B}_{1}\mathsf{A}_{1})$& $\bar{{\Lambda}}^{2}_{2 1}(\textcolor{purple}{\mathsf{C}_{1}}\mathsf{B}_{2}-\mathsf{B}_{2}\textcolor{purple}{\mathsf{C}_{2}})$
\end{tabular}
\eea

An interesting property of the brick matching is that they correspond to the lattice points of the toric diagram \eqref{eq:toric-diagram}. Generally, the brick matching associated with a lattice point is not unique, but they are unique for extremal lattice points. In our notation, the five brick matchings $\mathfrak{m}_{i}\,(i=0,1,2,3,4)$ correspond to the five lattice points in \eqref{eq:3dlatticept}. Since the lattice points of the toric diagram also correspond to the toric divisor, we have the correspondence
\bea
\boldsymbol{m}_{i}\quad\longleftrightarrow \quad \mathfrak{m}_{i}\quad \longleftrightarrow \quad  D_{\mathfrak{m}_{i}}.
\eea
For later use, we denote the set of extremal toric divisors (lattice points) and their corresponding brick matchings as
\bea\label{eq:fiveset}
\five=\{0,1,2,3,4\}.
\eea

\paragraph{Phase boundaries}
While a brick matching encodes an individual lattice point of
the toric diagram, the relative data between two brick
matchings associated with extremal lattice points of the toric diagram are encoded by a \emph{phase boundary}. A phase
boundary is a closed oriented surface in the brane brick model
and is the three-dimensional analogue of a zig-zag path in a
brane tiling.

Let $\mathfrak m_i$ and $\mathfrak m_j$ be two brick
matchings associated with two extremal lattice points of the toric diagram. Their formal difference determines the phase
boundary
\bea
\eta_{ij}
&=
\mathfrak m_i-\mathfrak m_j,
\qquad
\eta_{ji}=-\eta_{ij}.
\eea
The fields contributing to $\eta_{ij}$ and their orientations
are conveniently recorded in the phase-boundary matrix $H$.
For a chiral field $\Phi_\alpha$ and a Fermi field
$\Lambda_\alpha$, its entries are defined by
\bea
H_{\Phi_\alpha,\eta_{ij}}&=P_{\Phi_\alpha,i}-P_{\Phi_\alpha,j}\\
H_{\Lambda_\alpha,\eta_{ij}}&=\frac{1}{2}\left[(P_{\Lambda_\alpha,i}-P_{\bar{\Lambda}_\alpha,i})-(P_{\Lambda_\alpha,j}-P_{\bar{\Lambda}_\alpha,j})\right].
\eea
A vanishing entry means that the face associated with the
corresponding field does not belong to the phase boundary,
while a nonzero entry specifies its orientation. The
brick-matching condition guarantees that these oriented faces
glue together into a closed surface in $T^3$.

In the tropical description, the component of the NS5-brane skeleton
associated with a toric edge
$[\boldsymbol m_i,\boldsymbol m_j]$ is identified with the phase
boundary $\eta_{ij}$. Its homology class is determined by
$\boldsymbol m_i-\boldsymbol m_j$, consistently with the combinatorial
relation between the corresponding brick matchings.

\begin{figure}[t]
\centering
\begin{minipage}[c]{0.48\textwidth}
    \centering
    \includegraphics[width=0.8\linewidth]{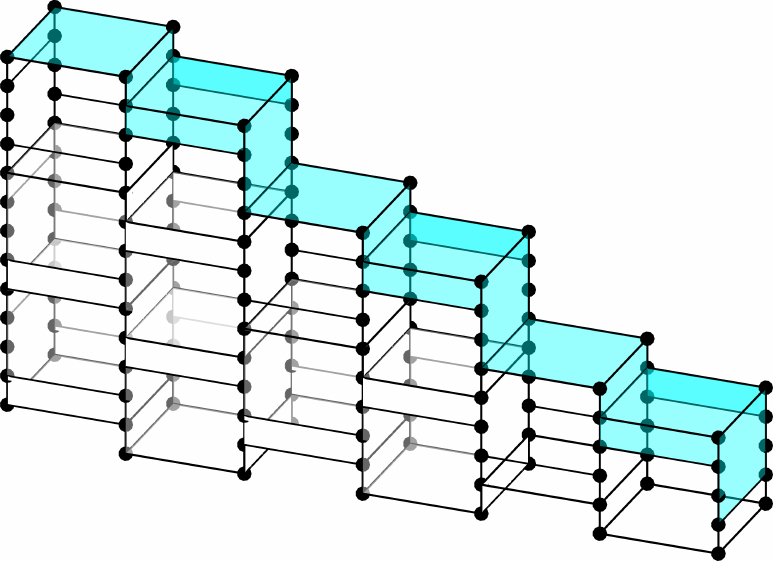}
  \end{minipage}
\begin{minipage}[c]{0.48\textwidth}
\includegraphics[width=0.8\linewidth]{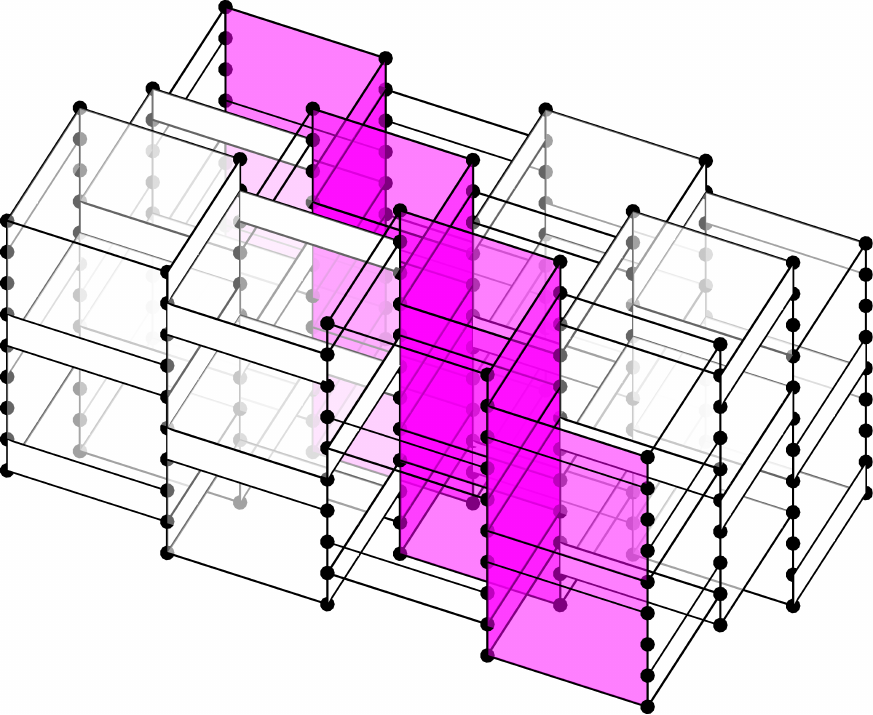} 
\end{minipage}
\caption{Phase boundaries of $\textcolor{cyan}{\eta_{02}}$ and $\textcolor{magenta}{\eta_{34}}$ in the brane brick model.}
\label{fig:phaseboundary}
\end{figure}

For $\mathcal C\times\mathbb C$, the signed field content of
the eight elementary phase boundaries is summarized by
\bea
\adjustbox{valign=c}{
\tdplotsetmaincoords{60}{110}
	\begin{tikzpicture}[scale=1.5,tdplot_main_coords]
	\draw[thick,dashed,-Triangle] (0,0,0) -- node[left,pos=1] {$ $} (1.9,0,0);
	\draw[thick,dashed,-Triangle] (0,0,0) -- node[right,pos=1] {$ $} (0,1.6,0);
	\draw[thick,dashed,-Triangle] (0,0,0) -- node[above,pos=1] {$ $} (0,0,1.5);
	\node[draw=black,line width=1pt,circle,fill=black,minimum width=0.2cm,inner sep=1pt] (O) at (0,0,0) {};
   
	\node[draw=black,line width=1pt,circle,fill=black,minimum width=0.2cm,inner sep=1pt] (X1) at (1,0,0) {};
	\node[draw=black,line width=1pt,circle,fill=black,minimum width=0.2cm,inner sep=1pt] (Z) at (0,0,1) {};
	\node[draw=black,line width=1pt,circle,fill=black,minimum width=0.2cm,inner sep=1pt] (Y1) at (0,1,0) {};
	\node[draw=black,line width=1pt,circle,fill=black,minimum width=0.2cm,inner sep=1pt] (XY) at (1,1,0) {};
	\draw[line width=1pt] (O)--(X1)--(XY)--(Y1)--(O);
	\draw[line width=1pt] (Z)--(O);
	\draw[line width=1pt] (Z)--(X1);
	\draw[line width=1pt] (Z)--(Y1);
	\draw[line width=1pt] (Z)--(XY);

    \draw[line width=1pt, cyan] (Z)--(X1);
    \node at (1,-0.3,0.5){\textcolor{cyan}{$\eta_{02}$}};
    \draw[line width=1pt, magenta] (XY)--(Y1);
    \node at (0.5,1.3,0){\textcolor{magenta}{$\eta_{34}$}};
	\end{tikzpicture}}\qquad H=
\left(
\begin{array}{c|cc>{\columncolor{magenta!30}}cc|c>{\columncolor{cyan!30}}ccc}
     & \eta_{12} & \eta_{23} & \eta_{34} & \eta_{41}
     & \eta_{01} & \eta_{02} & \eta_{03} & \eta_{04} \\ \hline
   \mathsf{A}_1
     & 0&0&-1&1 & 0&0&0&-1 \\
   \mathsf{A}_2
     & -1&1&0&0 & 0&-1&0&0 \\
   \mathsf{B}_1
     & 1&0&0&-1 & -1&0&0&0 \\
   \mathsf{B}_2
     & 0&-1&1&0 & 0&0&-1&0 \\
   \mathsf{C}_1
     & 0&0&0&0 & 1&1&1&1 \\
   \mathsf{C}_2
     & 0&0&0&0 & 1&1&1&1 \\ \hline
   \Lambda^{1}_{12}
     & 0&0&-1&1 & 1&1&1&0 \\
   \Lambda^{2}_{12}
     & -1&1&0&0 & 1&0&1&1 \\
   \Lambda^{1}_{21}
     & 1&0&0&-1 & 0&1&1&1 \\
   \Lambda^{2}_{21}
     & 0&-1&1&0 & 1&1&0&1
\end{array}
\right)
\eea
The phase boundaries $\eta_{02}$ and $\eta_{34}$ are
illustrated in Figure~\ref{fig:phaseboundary}. Since extremal
brick matchings correspond to toric lattice points, the
homology class of a phase boundary is determined by the
difference of the corresponding lattice points:
\bea
\,[\eta_{ij}]
&=
\boldsymbol m_i-\boldsymbol m_j
\in
H_2(T^3,\mathbb Z)
\cong
\mathbb Z^3.
\label{eq:phase-boundary-homology}
\eea
In particular, an elementary phase boundary is associated with
an edge joining $\boldsymbol m_i$ and $\boldsymbol m_j$ in the
toric diagram.

For later use, we introduce the index sets
\bea\label{eq:eight-four-set}
\eight
&=
\{01,02,03,04,12,23,34,41\},
\qquad
\eight
=
\four\sqcup\boldsymbol \four^\vee,
\\
\four
&=
\{01,02,03,04\},
\qquad
\four^\vee
=
\{12,23,34,41\},
\eea
and denote the phase boundary labeled by
$A\in\boldsymbol \eight$ by $\eta_A$. The toric edge joining $\boldsymbol m_i$ and $\boldsymbol m_j$ also determines the intersection of the corresponding toric divisors,
\bea
S_{\eta_{ij}}
&:=
D_{\mathfrak m_i}
\cap
D_{\mathfrak m_j}.
\eea
We therefore obtain the correspondence
\bea
\text{toric edge }
\,[\boldsymbol m_i,\boldsymbol m_j]
\quad
&\longleftrightarrow
\quad
\eta_{ij}
\quad
\longleftrightarrow
\quad
S_{\eta_{ij}}.
\eea
Here $\eta_{ij}$ is a real two-dimensional surface in the
brane brick model on $T^3$, whereas
$S_{\eta_{ij}}$ is the associated complex two-dimensional
toric subvariety of $\mathcal C\times\mathbb C$.

\section{Magnificent four of \texorpdfstring{$\mathcal{C}\times\mathbb{C}$}{C x C}}\label{sec:magnificent-four}

Although the brane brick model determines the unframed quiver describing
the D0--D0 sector, it does not by itself specify the additional framing
data produced by non-compact flavor branes wrapping cycles of the
Calabi--Yau fourfold. Such data require a choice of flavor branes and of
their positions in the T-dual brane configuration.

The D8--$\overline{\mathrm{D8}}$--D0 system on $\mathbb{C}^{4}$ was
introduced as the magnificent four in
\cite{Nekrasov:2017cih}, and its higher-rank generalizations were studied
in \cite{Nekrasov:2018xsb}. Related mathematical formulations in terms
of zero-dimensional Donaldson--Thomas theory on Calabi--Yau fourfolds,
including the equivariant and $K$-theoretic theories on $\mathbb{C}^{4}$,
have been developed in
\cite{Cao:2017swr,Cao:2019tvv,Bojko:2021cih}.
Mathematical aspects of Donaldson--Thomas counting and wall crossing for
the $\mathcal{C}\times\mathbb{C}$ have also been investigated in
\cite{Cao:2020huo,Cao:2020otr}.

In this section, we describe the corresponding framed system for $\mathcal{C}\times\mathbb{C}$ and study its partition
function. We refer to this system as the magnificent four of $\mathcal{C}\times\mathbb{C}$. We include this case mainly for completeness and to provide a reference point for the more general
flavor-brane configurations considered in the subsequent sections.

The BPS partition function is computed as the Witten index of the corresponding supersymmetric quiver quantum mechanics. Supersymmetric localization expresses this index as a Jeffrey--Kirwan residue \cite{Hori:2014tda,Cordova:2014oxa}. The contributing poles can be organized combinatorially in terms of four-dimensional molten crystals, as explained for D-brane systems on toric Calabi--Yau fourfolds in \cite{Bao:2024ygr}; see also \cite{Franco:2023tly} for a related crystal construction based on brane brick models. We will use the Jeffrey--Kirwan residue prescription to compute the partition function and the associated crystal description to characterize its contributing fixed points.

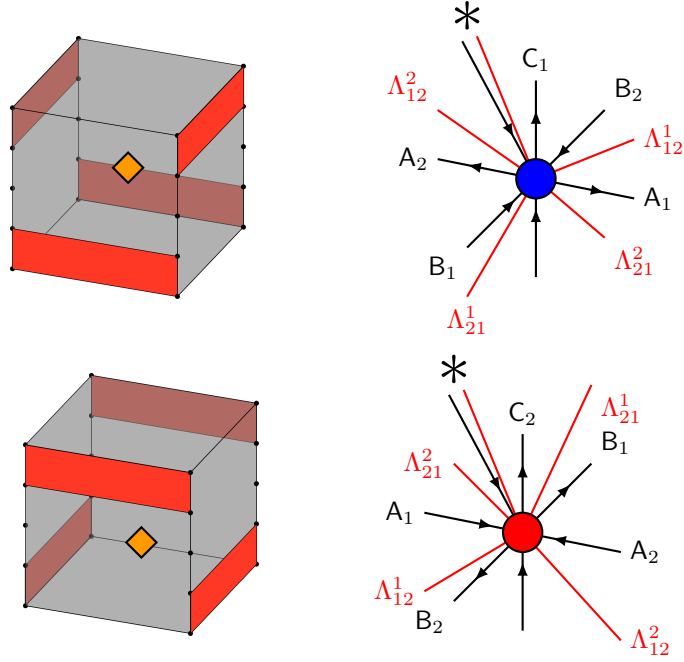
\begin{figure}
\centering
\adjustbox{valign=c}{\begin{tikzpicture}
\node[inner sep=0] (brick) at (0,0)
{\includegraphics[width=0.2\textwidth]{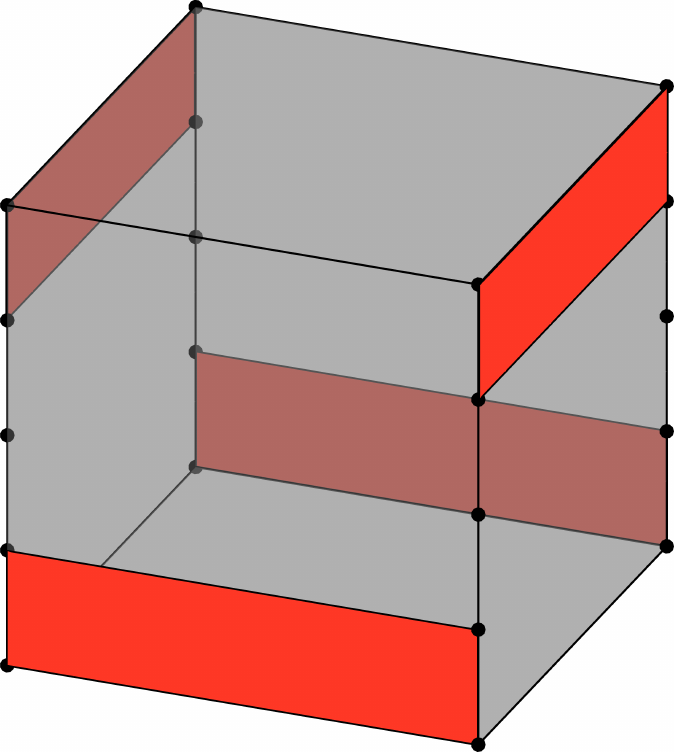}};
\filldraw[
fill=orange!80!yellow,
draw=black,
thick]
(0,0+0.2)
--(0+0.2,0)
--(0,0-0.2)
--(0-0.2,0)
--cycle;



\end{tikzpicture}}\quad \qquad \qquad   \adjustbox{valign=c}{\begin{tikzpicture}[decoration={markings,mark=at position \arrowHeadPosition with {\arrow{latex}}}]
 \tikzset{
        box/.style={draw, minimum width=0.6cm, minimum height=0.6cm, text centered,thick},
        ->-/.style={decoration={
        markings,mark=at position #1 with {\arrow[scale=1.5]{>}}},postaction={decorate},line width=0.5mm},
        -<-/.style={decoration={
        markings,
        mark=at position #1 with {\arrow[scale=1.5]{<}}},postaction={decorate},line width=0.5mm}    
    }
\begin{scope}[xshift=4cm,scale=1.3]
   
   \draw[postaction={decorate},black, thick] (0,0)--(0,1);
   \node[above] at (0,1){$\mathsf{C}_{1}$};
   \draw[postaction={decorate},black, thick] (0,-1)--(0,0);
   \draw[postaction={decorate},black, thick] (0,0)--(1,-0.2);
    \node[right] at (1,-0.2){$\mathsf{A}_{1}$};
   \draw[postaction={decorate},black, thick] (0,0)--(-1,+0.2);
    \node[left] at (-1,+0.2){$\mathsf{A}_{2}$};
   \draw[postaction={decorate},black, thick] (0.7,0.7)--(0,0);
    \node[above right] at (0.7,0.7){$\mathsf{B}_{2}$};
     \draw[postaction={decorate},black, thick] (-0.7,-0.7)--(0,0);
        \node[below left] at (-0.7,-0.7){$\mathsf{B}_{1}$};
       \draw[red, thick] (0,0)--(1,0.4);
          \node[right,red] at (1,0.4){$\Lambda^{1}_{12}$};
       \draw[red, thick] (0,0)--(-1,0.7);
        \node[above left,red] at (-1,0.7){$\Lambda^{2}_{1 2}$};
       
       \draw[red, thick] (-0.7,-1.2)--(0,0);
       \node[below ,red] at (-0.7,-1.2){$\Lambda^{1}_{21}$};
        \draw[red, thick] (0.7,-0.6)--(0,0);
        \node[below right,red] at (0.7,-0.6){$\Lambda^{2}_{21}$};
    
    \node [above] at (-0.7,1.4){\Huge $\ast$};
    \draw[postaction={decorate}, thick] (-0.75,1.4)--(0,0);
    \draw[red, thick] (-0.6,1.45)--(0,0);

   \draw[fill=blue!100!white,thick](0,0) circle(0.2cm);
   
\end{scope}
\end{tikzpicture}} \\
\adjustbox{valign=c}{\begin{tikzpicture}
\node[inner sep=0] (brick) at (0,0)
{\includegraphics[width=0.2\textwidth]{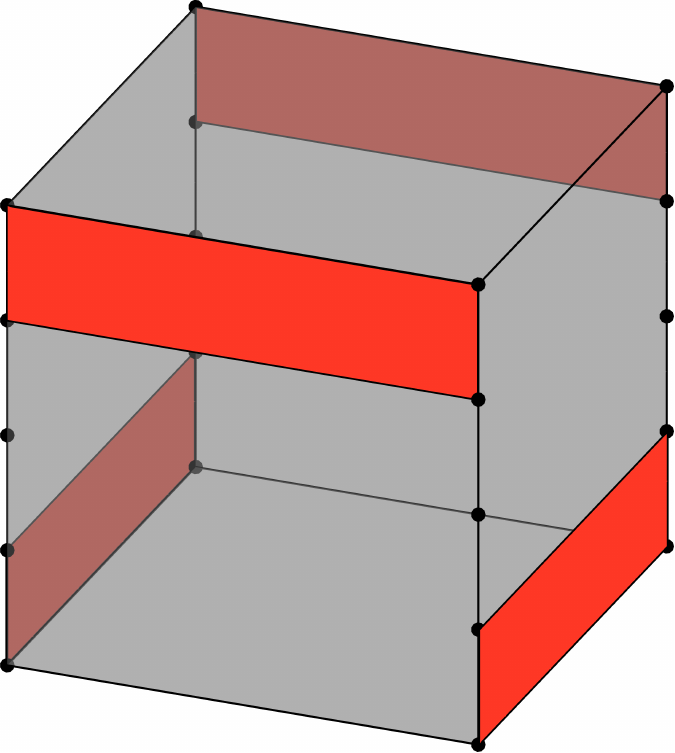}};
\filldraw[
fill=orange!80!yellow,
draw=black,
thick,yshift=-0.5cm]
(0,0+0.2)
--(0+0.2,0)
--(0,0-0.2)
--(0-0.2,0)
--cycle;

\end{tikzpicture}}\qquad \qquad   \adjustbox{valign=c}{\begin{tikzpicture}[decoration={markings,mark=at position \arrowHeadPosition with {\arrow{latex}}}]
 \tikzset{
        box/.style={draw, minimum width=0.6cm, minimum height=0.6cm, text centered,thick},
        ->-/.style={decoration={
        markings,mark=at position #1 with {\arrow[scale=1.5]{>}}},postaction={decorate},line width=0.5mm},
        -<-/.style={decoration={
        markings,
        mark=at position #1 with {\arrow[scale=1.5]{<}}},postaction={decorate},line width=0.5mm}    
    }
\begin{scope}[xshift=4cm,scale=1.3]
   
   \draw[postaction={decorate},black, thick] (0,0)--(0,1);
   \node[above] at (0,1){$\mathsf{C}_{2}$};
   \draw[postaction={decorate},black, thick] (0,-1)--(0,0);

   \draw[postaction={decorate},black, thick] (1,-0.2)--(0,0);
    \node[right] at (1,-0.2){$\mathsf{A}_{2}$};
   \draw[postaction={decorate},black, thick] (-1,+0.2)--(0,0);
    \node[left] at (-1,+0.2){$\mathsf{A}_{1}$};
    
   \draw[postaction={decorate},black, thick] (0,0)--(0.7,0.7);
    \node[above right] at (0.7,0.7){$\mathsf{B}_{1}$};
     \draw[postaction={decorate},black, thick] (0,0)--(-0.7,-0.7);
        \node[below left] at (-0.7,-0.7){$\mathsf{B}_{2}$};

       \draw[red, thick](1,-1.1)--(0,0);
          \node[right,red] at (1,-1.1){$\Lambda^{2}_{12}$};
          
       \draw[red, thick] (-1,-0.6)--(0,0);
        \node[left,red] at (-1,-0.6){$\Lambda^{1}_{12}$};
       
       \draw[red, thick] (0,0)--(-0.7,0.7);
       \node[left,red] at (-0.7,0.7){$\Lambda^{2}_{21}$};
        \draw[red, thick] (0,0)--(0.7,1.5);
        \node[below right,red] at (0.7,1.5){$\Lambda^{1}_{21}$};
            \node [above] at (-0.7,1.4){\Huge $\ast$};
    \draw[postaction={decorate}, thick] (-0.75,1.4)--(0,0);
    \draw[red, thick] (-0.6,1.45)--(0,0);
    \draw[fill=red!100!white,thick](0,0) circle(0.2cm);
   
\end{scope}
\end{tikzpicture}}
\caption{ The diamonds indicate possible locations of T-dualized D8 or $\overline{\D8}$-branes in the bricks. The framed node in the periodic quiver is denoted as a marked point $\ast$ and their associated chiral and Fermi
framing multiplets are displayed in the framed quiver as the right figure.}
\label{fig:D8framingbrick}
\end{figure}

\paragraph{D8 flavor branes and framing nodes}In the magnificent four system on $\mathbb{C}^4$, the D0--D8 open strings provide a fundamental $\mathcal{N}=2$ chiral multiplet, whereas the D0--$\overline{\mathrm{D8}}$ open strings provide a fundamental $\mathcal{N}=2$ Fermi multiplet. We adopt the same assignment as the definition of the framed quiver on $\mathcal{C}\times\mathbb{C}$. A microscopic derivation of this open-string spectrum in the singular background will not be needed for the partition-function analysis below.

Let us therefore introduce D8- and $\overline{\mathrm{D8}}$-branes wrapping the entire CY$_4$, with the same worldvolume directions. The brane configuration for a representative D8-brane is
\bea  \label{eq:conifold-D8}
\renewcommand{\arraystretch}{1.05}
\begin{tabular}{|c|c|c|c|c|c|c|c|c|c|c|}
\hline
& \multicolumn{6}{c|}{$\mathcal{C}$} & \multicolumn{2}{c|}{$\mathbb{C}$} & \multicolumn{2}{c|}{$\mathbb{R}\times S^{1}$} \\
\cline{2-11} \raisebox{-0mm}{}  & 1 & 2 & 3 & 4& 5 & 6 & 7 & 8 & 9& 0\\[-2pt]\hline
 D0& $\bullet$ & $\bullet$  & $\bullet$  & $\bullet$  & $\bullet$  & $\bullet$   & $\bullet$  & $\bullet$  & $\bullet$   & $-$\\
\hline D8& $-$ & $-$  & $-$  & $-$  & $-$  & $-$   & $-$  & $-$  & $\bullet$   & $-$\\
\hline
\end{tabular}
\eea
To identify the corresponding framing node in the brane brick model, we perform the three T-dualities along the directions $x^1$, $x^3$, and $x^5$, as in section~\ref{sec:branebrickmodel}. A D8-brane becomes a D5-brane, while an $\overline{\mathrm{D8}}$-brane becomes an $\overline{\mathrm{D5}}$-brane. The resulting configuration for the D8-brane is
\bea
\label{eq:branebrick-D8}
\renewcommand{\arraystretch}{1.05}
\begin{tabular}{|c|c|c|c|c|c|c|c|c|c|c|}
\hline
&1&2&3&4&5&6&7&8&9&0
\\[-2pt]
\hline
D3
&$-$&$\bullet$&$-$&$\bullet$&$-$&$\bullet$
&$\bullet$&$\bullet$&$\bullet$&$-$
\\
\hline
NS5
&\multicolumn{6}{c|}{$\Sigma$}
&$\bullet$&$\bullet$&$-$&$-$
\\
\hline
D5
&$\bullet$&$-$&$\bullet$&$-$&$\bullet$&$-$
&$-$&$-$&$\bullet$&$-$
\\
\hline
\end{tabular}
\eea
The D5-brane is localized at a point of the dual three-torus $T^3$, whereas the D3-branes wrap $T^3$ and are divided into gauge sectors by the NS5-brane. We place the D5-brane at a generic point away from the NS5-brane locus. Such a point belongs to the interior of a unique brick, and the D5-brane therefore couples to the D3-brane gauge sector associated with that brick. Since the D5-brane extends along non-compact directions transverse to $T^3$, its worldvolume symmetry is non-dynamical from the viewpoint of the D3-brane quantum mechanics and becomes a flavor symmetry. The corresponding D3--D5 open strings provide the framing multiplets.

Consequently, a single D8- or $\overline{\mathrm{D8}}$-brane has $|Q_0|$ possible generic locations, one for each brick of the brane brick model. Choosing its location determines the gauge node to which the associated framing node is attached. Moving the flavor brane across an NS5-brane phase boundary changes this choice and hence changes the framed quiver.

For $\mathcal{C}\times\mathbb{C}$, the brane brick model contains two bricks associated with the two gauge nodes of the quiver. For each brick $a=1,2$, let $N_a$ and $M_a$ denote the numbers of D8- and $\overline{\mathrm{D8}}$-branes placed in its interior, respectively. See Figure~\ref{fig:D8framingbrick} for example. They give rise to the flavor symmetry
\bea
\prod_{a=1}^{2}U(N_a)\times U(M_a).
\eea
We use $N_a|M_a$ as a compact notation for the corresponding pair of D8- and $\overline{\mathrm{D8}}$-flavor nodes. The D8-branes provide chiral framing multiplets $\mathsf{I}_a$, while the $\overline{\mathrm{D8}}$-branes provide Fermi framing multiplets $\Lambda_a$, both connected to the gauge node associated with the $a$-th brick. The most general framing obtained by distributing D8- and $\overline{\mathrm{D8}}$-branes among the two bricks is represented by the following quiver:
\bea
\adjustbox{valign=c}{
\begin{tikzpicture}[decoration={markings,mark=at position \arrowHeadPosition with {\arrow{latex}}}]
 \tikzset{
        box/.style={draw, minimum width=0.6cm, minimum height=0.6cm, text centered,thick},
        ->-/.style={decoration={
        markings,mark=at position #1 with {\arrow[scale=1.5]{>}}},postaction={decorate},line width=0.5mm},
        -<-/.style={decoration={
        markings,
        mark=at position #1 with {\arrow[scale=1.5]{<}}},postaction={decorate},line width=0.5mm}    
    }
\begin{scope}[xshift=4cm]
    \draw[postaction={decorate}, black,thick,scale=1.3] (0.65,0) arc(0:-180:0.65 and 0.1) ;
    \draw[postaction={decorate}, black,thick,scale=1.3] (0.75,0) arc(0:-180:0.75 and 0.2) ;
    \draw[postaction={decorate}, black,thick,scale=1.3] (-0.65,0) arc(180:0:0.65 and 0.1) ;
    \draw[postaction={decorate}, black,thick,scale=1.3] (-0.75,0) arc(180:0:0.75 and 0.2) ;
    \draw[postaction={decorate}, black,thick,scale=1.3] (-0.8,0) arc(360:0:0.4 and 0.3) ;
    \draw[postaction={decorate}, black,thick,scale=1.3] (0.8,0) arc(-180:180:0.4 and 0.3) ;
    
    \draw[red,thick,postaction={decorate},scale=1.3] (0.65,0) arc(0:-180:0.65 and 0.3) ;
    \draw[red,thick,postaction={decorate},scale=1.3] (0.75,0) arc(0:-180:0.75 and 0.4) ;
    \draw[red,thick,postaction={decorate},scale=1.3] (-0.65,0) arc(180:0:0.65 and 0.3) ;
    \draw[red,thick,postaction={decorate},scale=1.3] (-0.75,0) arc(180:0:0.75 and 0.4) ;
    
    \node[box,fill=black!10!white] at (-0.95,2.0){$ N_{1}|M_{1} $};
    \node[box,fill=black!10!white] at (0.95,2.0){$ N_{2}|M_{2} $};
    \draw[fill=blue!100!white,thick](-0.95,0) circle(0.3cm);
    \draw[fill=red!100!white,thick](0.9,0) circle(0.3cm);
   \draw[postaction={decorate},black,thick, scale=1.3] (-0.8,1.3)--(-0.8,0.2);
    \draw[postaction={decorate},red,thick, scale=1.3] (-0.7,1.3)--(-0.7,0.24);
     \draw[postaction={decorate},red,thick, scale=1.3] (0.8,1.3)--(0.8,0.2);
    \draw[postaction={decorate},black,thick, scale=1.3] (0.7,1.3)--(0.7,0.24);
\end{scope}
\end{tikzpicture}}
\eea

No additional $J$- or $E$-term is introduced for the framing Fermi multiplets. Indeed, the framed quiver contains no chiral path with the flavor and gauge quantum numbers required to form a gauge-invariant closed path in the periodic quiver involving a framing Fermi multiplet. Equivalently, no chiral monomial with the appropriate representation can be assigned to either its $J$- or $E$-term. We therefore set
\bea
J_{\Lambda_a}=0,
\qquad
E_{\Lambda_a}=0.
\eea
Here we treat the D8- and $\overline{\mathrm{D8}}$-flavor symmetries as independent and do not introduce additional D8--$\overline{\mathrm{D8}}$ open-string fields, such as tachyon or mass parameters.

\begin{remark}
\label{rem:framing-toric-condition}
The toric condition \eqref{eq:JE-toriccond} reviewed in
section~\ref{sec:branebrickmodel} applies only to the unframed
D0--D0 quiver. The flavor extension introduced by non-compact
D-branes need not satisfy the toric condition. In particular, a
framing Fermi multiplet is not required to be associated with the four
minimal plaquettes of a toric Fermi multiplet: it may have only one
non-vanishing $J$- or $E$-term, or it may have
\bea
J_{\Lambda}
=
E_{\Lambda}
=
0.
\eea
This does not violate $\mathcal{N}=2$ supersymmetry, provided that the
full framed quiver satisfies \eqref{eq:JE-traceless}. Throughout this paper, brick matchings and the brick-matching matrix
$P$ always refer to the original unframed toric quiver. The framing
fields are not included in $P$.
\end{remark}

\paragraph{Witten index}
Let us briefly review the flavoured Witten index and the Jeffrey--Kirwan (JK) residue prescription. For details, see the original papers \cite{Benini:2013xpa,Benini:2013nda, Hori:2014tda,Cordova:2014oxa,Ohta:2014ria}. Given the 1d $\mathcal{N}=2$ supersymmetric quantum mechanics, the Witten index is defined as
\bea
\mathcal{Z}(y)=\Tr\left[(-1)^{F}e^{-\beta \{Q,Q^{\dagger}\}}\prod_{a}y_{a}^{T_{a}}\right]
\eea
where $F$ is the fermion number, $\{y_{a}=e^{u_{a}}\}$ are the fugacities of the flavor symmetry, $\{T_{a}\}$ are the Cartan generators of the flavor symmetry group, and $\beta$ is the circumference of the ${S}^{1}$. Thanks to supersymmetric localization, the Witten index reduces to a finite dimensional contour integral. For a one-dimensional system described by a vector multiplet $V$ with gauge group $G$, chiral multiplets $\Phi_{i}$ transforming in a representation $V_{\text{chiral}}$ of $G\times G_{f}$, and Fermi multiplets $\Lambda_{\alpha}$ transforming in the representation $V_{\text{Fermi}}$ of $G\times G_{f}$, the contour integral schematically takes the form as
\bea
\mathcal{Z}(y)=\frac{1}{|W_{G}|}\oint _{\text{JK}}Z_{\text{V}}\prod_{i}Z_{\Phi_{i}}\prod_{\alpha}Z_{\Lambda_{\alpha}}
\eea
where
\bea\label{eq:2susylocalization}
Z_{\text{V}}&=\prod_{I}\frac{d\phi_{I}}{2\pi i}\prod_{\alpha\in G}\sh\left(\beta \alpha\cdot \phi\right),\quad Z_{\Phi_{i}}=\prod_{\rho\in V_{\text{chiral}}}\frac{1}{\sh(\beta\rho\cdot \mu)},\\
Z_{\Lambda_{\alpha}}&=\prod_{\rho\in V_{\text{Fermi}}}\sh(\beta \rho\cdot \mu),\quad \sh(x)=2\sinh (\frac{x}{2})
\eea
and $|W_{G}|$ is the order of the Weyl group of $G$. The variable $\phi$ takes a value in the Cartan subalgebra of $G$ and $\mu$ contains $\phi$ and $u_{a}$. The products over $G$ and $V$ mean those over the associated roots and the weights. It is known that the correct way to evaluate the contour integral formula is the JK-residue formalism which we will explain later. Note that the rational and elliptic version is obtained by replacement
\bea\label{eq:rat-trig-ell}
\sh(x)\rightarrow\begin{dcases}
    x,\quad & \text{rational}\\
    \frac{i\theta_{1}(\tau,x)}{\eta(\tau)},\quad & \text{elliptic}
\end{dcases}
\eea
where $\eta(\tau)$ is the Dedekind eta function and $\theta_{1}(\tau,x)$ is the Jacobi theta function.

In particular, we are interested in the Witten index of 1d $\mathcal{N}=2$ quiver quantum mechanics with unitary groups.  We denote the flavor charges of the chiral, Fermi superfields as $p(\Phi)=e^{\epsilon(\Phi)}$, $p(\Lambda)=e^{\epsilon(\Lambda)}$. We further have flavor nodes and we denote the fugacities of them collectively as $\{v_{i}=e^{\mathfrak{a}_{i}}\}$. On the other hand, the exponent of the root of the gauge group will be denoted as $x=e^{\phi}$ and at the end the integral over it will be taken. Under this notations, the basic factors appearing in the integrand are determined as follows.
\begin{itemize}[topsep=0pt, partopsep=0pt, itemsep=0pt]
    \item For each gauge group (circle node) of the theory, we have 
    \bea
\adjustbox{valign=c}{\begin{tikzpicture}[decoration={markings,mark=at position \arrowHeadPosition with {\arrow{latex}}}]
 \tikzset{
        box/.style={draw, minimum width=0.6cm, minimum height=0.6cm, text centered,thick},
        ->-/.style={decoration={
        markings,mark=at position #1 with {\arrow[scale=1.5]{>}}},postaction={decorate},line width=0.5mm},
        -<-/.style={decoration={
        markings,
        mark=at position #1 with {\arrow[scale=1.5]{<}}},postaction={decorate},line width=0.5mm}    
    }
\begin{scope}{xshift=0cm}
    \draw[fill=black!10!white,thick](0,0) circle(0.4cm);
    \node at (0,0){$k_{a}$};
\end{scope}
\end{tikzpicture}}\quad {\rightsquigarrow} \quad \prod_{I=1}^{k_{a}}\frac{d\phi_{I}^{(a)}}{2\pi i } \prod_{I\neq J}\sh(\phi_{I}^{(a)}-\phi_{J}^{(a)})
    \eea
where we simply set $\beta=1$.
\item The chiral superfield corresponding to an arrow from $\U(k_{a})$ to $\U(k_{b})$ gives the contribution
    \bea
   \adjustbox{valign=c}{ \begin{tikzpicture}[decoration={markings,mark=at position \arrowHeadPosition with {\arrow{latex}}}]
 \tikzset{
        box/.style={draw, minimum width=0.6cm, minimum height=0.6cm, text centered,thick},
        ->-/.style={decoration={
        markings,mark=at position #1 with {\arrow[scale=1.5]{>}}},postaction={decorate},line width=0.5mm},
        -<-/.style={decoration={
        markings,
        mark=at position #1 with {\arrow[scale=1.5]{<}}},postaction={decorate},line width=0.5mm}    
    }

\begin{scope}{}
\draw[fill=black!10!white,thick](4.7,-2) circle(0.4cm);
\node at (4.7,-2){$k_{b}$};
\draw[fill=black!10!white,thick](2.1,-2) circle(0.4cm);
\node at (2.1,-2){$k_{a}$};
\draw[postaction={decorate}, thick](2.5,-2)--(4.3,-2);
\node[above] at (3.4,-2){$p(\Phi_{a\rightarrow b})$};
\end{scope}
\end{tikzpicture}}\quad \rightsquigarrow\quad \prod_{I=1}^{k_{a}}\prod_{J=1}^{k_{b}}\frac{1}{\sh(\phi_{J}^{(b)}-\phi_{I}^{(a)}-\epsilon(\Phi_{a\rightarrow b}))}.
    \eea
    \item The Fermi superfield corresponding to an arrow from $\U(k_{a})$ to $\U(k_{b})$ gives the contribution 
    \bea
   \adjustbox{valign=c}{ \begin{tikzpicture}[decoration={markings,mark=at position \arrowHeadPosition with {\arrow{latex}}}]
 \tikzset{
        box/.style={draw, minimum width=0.6cm, minimum height=0.6cm, text centered,thick},
        ->-/.style={decoration={
        markings,mark=at position #1 with {\arrow[scale=1.5]{>}}},postaction={decorate},line width=0.5mm},
        -<-/.style={decoration={
        markings,
        mark=at position #1 with {\arrow[scale=1.5]{<}}},postaction={decorate},line width=0.5mm}    
    }

\begin{scope}{}
\draw[fill=black!10!white,thick](4.7,-2) circle(0.4cm);
\node at (4.7,-2){$k_{b}$};
\draw[fill=black!10!white,thick](2.1,-2) circle(0.4cm);
\node at (2.1,-2){$k_{a}$};
\draw[red, postaction={decorate}, thick](2.5,-2)--(4.3,-2);
\node[above] at (3.4,-2){\textcolor{red}{$p(\Lambda_{a\rightarrow b})$}};
\end{scope}
\end{tikzpicture}}\quad \rightsquigarrow\quad \prod_{I=1}^{k_{a}}\prod_{J=1}^{k_{b}}\sh(\phi_{J}^{(b)}-\phi_{I}^{(a)}-\epsilon(\Lambda_{a\rightarrow b})).
    \eea
    If one uses the conjugate Fermi superfield, the contribution to the Witten index is
    \bea
   \adjustbox{valign=c}{ \begin{tikzpicture}[decoration={markings,mark=at position \arrowHeadPosition with {\arrow{latex}}}]
 \tikzset{
        box/.style={draw, minimum width=0.6cm, minimum height=0.6cm, text centered,thick},
        ->-/.style={decoration={
        markings,mark=at position #1 with {\arrow[scale=1.5]{>}}},postaction={decorate},line width=0.5mm},
        -<-/.style={decoration={
        markings,
        mark=at position #1 with {\arrow[scale=1.5]{<}}},postaction={decorate},line width=0.5mm}    
    }

\begin{scope}{}
\draw[fill=black!10!white,thick](4.7,-2) circle(0.4cm);
\node at (4.7,-2){$k_{b}$};
\draw[fill=black!10!white,thick](2.1,-2) circle(0.4cm);
\node at (2.1,-2){$k_{a}$};
\draw[red, postaction={decorate}, thick](4.3,-2)--(2.5,-2);
\node[above] at (3.4,-2){\textcolor{red}{$p(\overline{\Lambda}_{a\rightarrow b})$}};
\end{scope}
\end{tikzpicture}}\quad \rightsquigarrow\quad 
\prod_{I=1}^{k_{a}}\prod_{J=1}^{k_{b}}\sh(\phi_{I}^{(a)}-\phi_{J}^{(b)}-\epsilon(\bar{\Lambda}_{a\rightarrow b})).
    \eea
  Since we have the relation
  \bea
    \eps(\bar{\Lambda}_{a\rightarrow b})=-\eps(\Lambda_{a\rightarrow b})
  \eea
  note that the integrand differs by a sign factor $(-1)^{k_{a}k_{b}}$.

\end{itemize}

A useful remark concerns the sign associated with the choice of orientation for the Fermi multiplets. For a fixed dimension vector of the gauge group, changing the orientation convention multiplies the Witten index by an overall sign. This sign may be absorbed into the choice of normalization, or equivalently the orientation convention, for that particular sector and does not affect the pole structure or the relative weights of its fixed points.

In BPS counting, however, we consider the generating function over all gauge-group dimension vectors. Replacing a bifundamental Fermi multiplet between gauge nodes $a$ and $b$ by its conjugate can produce a dimension-dependent sign of the form $(-1)^{k_{a}k_{b}}$. Such a quadratic sign cannot, in general, be absorbed into a redefinition of the topological fugacities. A globally consistent choice between each Fermi multiplet and its conjugate is therefore required in defining the BPS partition function.

The situation is different for a framing Fermi multiplet connecting a gauge node $a$ of rank $k_a$ to a flavor node of fixed rank $N_a$. Replacing the framing Fermi multiplet by its conjugate changes its one-loop contribution only by $(-1)^{k_{a}N_{a}}$. Since the flavor rank $N_a$ is fixed, this sign is linear in the gauge-group dimension vector and can be absorbed into a redefinition of the corresponding topological fugacity. Therefore, the orientation ambiguity of the framing Fermi multiplets does not affect the relative weights of the fixed points or the resulting BPS partition function, up to a conventional redefinition of the topological parameters. 

It was argued in \cite{Bao:2024ygr} that, for a toric CY$_4$ quiver admitting a brane brick model, such a choice of Fermi fields for gauge nodes can be fixed by selecting a column of the brick-matching matrix. The selected brick matching simultaneously determines, for every unoriented Fermi edge, whether $\Lambda$ or $\bar\Lambda$ should be used, or equivalently whether the corresponding $J$- or $E$-plaquettes are selected. The resulting choices provide consistent orientation data for the DT$_4$ moduli problem. Moreover, the resulting BPS index was argued to be independent of the brick-matching column used in this prescription.

Throughout this paper, we select the brick matching $\mathfrak m_0$. Since the $\mathfrak m_0$ column selects the unconjugated field $\Lambda$ for every Fermi edge, we consistently use the Fermi multiplets $\Lambda$, rather than their conjugates $\bar\Lambda$ for the gauge--gauge Fermi multiplets, in defining the Witten index. On the other hand, for the Fermi multiplets of the flavor nodes, we choose the orientations for convenience.

\paragraph{Flavor charges and contour integrand}
Let us apply the general formalism above to the magnificent four system of $\mathcal{C}\times \mathbb{C}$. Assume the flavor charges are
\bea
\mathsf{A}_{1}\rightarrow p_{1}\mathsf{A}_{1},\quad \mathsf{A}_{2}\rightarrow p_{2}\mathsf{A}_{2},\quad \mathsf{B}_{1}\rightarrow p_{3}\mathsf{B}_{1},\quad \mathsf{B}_{2}\rightarrow p_{4}\mathsf{B}_{2},\quad \mathsf{C}_{1}\rightarrow p_{5}\mathsf{C}_{1},\quad \mathsf{C}_{2}\rightarrow p_{6}\mathsf{C}_{2}
\eea
where $p_{a}=e^{\eps_{a}}$. The $J,E$-terms \eqref{eq:conifold-JEterms} impose the conditions
\bea\label{eq:CY4condition}
p_{1}p_{2}p_{3}p_{4}p_{5}=1,\quad p_{6}=p_{5},
\eea
which also means $\sum_{a=1}^{5}\eps_{a}=0$. The first condition is usually called the CY condition. The $J,E$-terms then transform as
\bea
\begin{tabular}{c|cc}
 &\text{$J$-term}&  \text{$E$-term}\\ \hline
  $\Lambda^{1}_{1\rightarrow 2}$   & $p_{2}p_{3}p_{4}$  &  $p_{1}p_{5}$\\
  $\Lambda^{2}_{1\rightarrow 2}$   &  $p_{1}p_{3}p_{4}$  & $p_{2}p_{5}$ \\
  $\Lambda^{1}_{2\rightarrow 1}$  &   $p_{1}p_{2}p_{4}$& $p_{3}p_{5}$\\
  $\Lambda^{2}_{2\rightarrow 1}$  &  $p_{1}p_{2}p_{3}$& $p_{5}p_{4}$
\end{tabular}
\eea
The fractional D0-brane contributions to the Witten index of the SQM of the fields are then
\bea
\text{vector}:&\quad    \prod_{I\neq J}^{k_{1}}\sh(\phi_{I}^{(1)}-\phi_{J}^{(1)})\prod_{I\neq J}^{k_{2}}\sh(\phi_{I}^{(2)}-\phi_{J}^{(2)})\\
\text{chiral}:&\quad  \prod_{I,J=1}^{k_{1}}\frac{1}{\sh(\phi_{I}^{(1)}-\phi_{J}^{(1)}-\eps_5)}\prod_{I,J=1}^{k_{2}}\frac{1}{\sh(\phi_{I}^{(2)}-\phi_{J}^{(2)}-\eps_5)}\\
&\quad \prod_{I=1}^{k_{1}}\prod_{J=1}^{k_{2}}\frac{1}{\sh(\phi_{J}^{(2)}-\phi_{I}^{(1)}-\eps_{1,2})\sh(\phi_{I}^{(1)}-\phi_{J}^{(2)}-\eps_{3,4})}\\
\text{Fermi}:&\quad \prod_{I=1}^{k_{1}}\prod_{J=1}^{k_{2}} \sh(\phi_{J}^{(2)}-\phi_{I}^{(1)}-\eps_5-\eps_{1,2})\sh(\phi_{I}^{(1)}-\phi_{J}^{(2)}-\eps_5-\eps_{3,4}),
\eea
where we chose the $E$-terms for the Fermi superfield contributions. For later use, we shortly write the fractional D0-brane contributions as
\bea
\mathcal{Z}^{\D0\tbar\D0}(\phi)&=\prod_{I,J=1}^{k_{1}}\frac{\sh(\phi_{I}^{(1)}-\phi_{J}^{(1)})'}{\sh(\phi_{I}^{(1)}-\phi_{J}^{(1)}-\eps_5)}\prod_{I,J=1}^{k_{2}}\frac{\sh(\phi_{I}^{(2)}-\phi_{J}^{(2)})'}{\sh(\phi_{I}^{(2)}-\phi_{J}^{(2)}-\eps_5)}\\
&\times \prod_{I=1}^{k_{1}}\prod_{J=1}^{k_{2}}\frac{\sh(\phi_{J}^{(2)}-\phi_{I}^{(1)}-\eps_5-\eps_{1,2})\sh(\phi_{I}^{(1)}-\phi_{J}^{(2)}-\eps_5-\eps_{3,4})}{\sh(\phi_{J}^{(2)}-\phi_{I}^{(1)}-\eps_{1,2})\sh(\phi_{I}^{(1)}-\phi_{J}^{(2)}-\eps_{3,4})}
\eea
where $\prod_{I,J}\sh(\phi_I-\phi_J)'$ means we omit the diagonal part.

The additional $\mathcal{N}=2$ chiral fields or Fermi fields arising from the D8, $\overline{\D8}$-branes do not affect the $J,E$-terms and thus we can simply set the flavor charges to be neutral. The contributions to the Witten index are
\bea
\mathcal{Z}^{\D8\tbar\D0}_{a}(\fra^{(a)},\phi)=\prod_{I=1}^{k_{a}}\frac{1}{\sh(\phi_{I}^{(a)}-\fra^{(a)})},\quad \mathcal{Z}^{\overline{\D8}\tbar\D0}_{a}(\frb^{(a)},\phi)=\prod_{I=1}^{k_{a}}\sh(\phi_{I}^{(a)}-\frb^{(a)})
\eea
where $a=1,2$. The choice of $a$ here represents the choice to which the framing node is attached.

\begin{definition}
   The magnificent four partition function of $(N_{1},N_{2})$ $\D8$-branes and $(M_{1},M_{2})$ $\overline{\D8}$-branes is defined as
    \bea
    \mathcal{Z}(\mathfrak{q}_{1},\mathfrak{q}_{2};p_{1,2,3,4,5})=\sum_{k_{1}=0}^{\infty}\sum_{k_{2}=0}^{\infty}\mathfrak{q}_{1}^{k_{1}}\mathfrak{q}_{2}^{k_{2}}\mathcal{Z}_{k_{1},k_{2}}
    \eea
    where
    \bea
    \mathcal{Z}_{k_{1},k_{2}}&=\frac{1}{k_{1}!k_{2}!}\oint_{\text{JK}} \prod_{a=1}^{2}\prod_{I=1}^{k_{a}}\frac{d \phi_I^{(a)}}{2\pi i}\prod_{a=1}^{2} \prod_{\alpha=1}^{N_{a}}\mathcal{Z}_{a}^{\D8\tbar\D0}(\fra^{(a)}_{\alpha},\phi)\prod_{\beta=1}^{M_{a}}\mathcal{Z}_{a}^{\overline{\D8}\tbar\D0}(\frb^{(a)}_{\beta},\phi)\mathcal{Z}^{\D0\tbar\D0}(\phi).
    \eea
\end{definition}

Actually, the integrand above additionally has a redundant symmetry
\bea
&\eps_{I}\rightarrow \eps_{I}+\delta_{a}\text{sign}_{a}(I),\quad \text{sign}_{a}(I)=\begin{dcases}
    +1,\quad s(I)=a,\\
    -1,\quad t(I)=a,\\
    0,\quad \text{otherwise}
\end{dcases},\\
&\phi_{I}^{(a)}\rightarrow \phi_{I}^{(a)}-\delta_{a},
\eea
where $s(I)$ ($t(I)$) is the source (target) node of the arrow $I$, respectively. In particular,
\bea
&\eps_{1,2}\rightarrow \eps_{1,2}+\delta_{1}-\delta_{2},\quad \eps_{3,4}\rightarrow \eps_{3,4}+\delta_{2}-\delta_{1},\quad \eps_5 \rightarrow \eps_5,\\
&\phi_{I}^{(1)} \rightarrow \phi_{I}^{(1)}-\delta_{1},\quad \phi_{I}^{(2)} \rightarrow \phi_{I}^{(2)}-\delta_{2}
\eea
Due to this symmetry, we can further impose one restriction to the parameters $\{p_{1,2,3,4,5}\}$ and obtain only three independent parameters. This is true for general toric CY$_{4}$ and related with the $U(1)^{3}$ isometries of the geometry. In this paper, to keep the notation in a symmetric way, we do not impose any further condition and simply use the five parameters obeying the condition \eqref{eq:CY4condition}.

\paragraph{JK-residue}
To evaluate the integral, we need to use the JK-residue prescription. Let the contour integral be
\bea
\mathcal{Z}=\oint_{\JK}\prod_{i=1}^{k}\frac{d\phi_i}{2\pi i }\mathcal{Z}(\phi)
\eea
where $\phi_i$ takes values in the Cartan subalgebra $\mathfrak{h}$ of the gauge group. The poles come from the zeros of them in the denominator:
\bea
Q_{i}(\phi)+f_{i}(\fra,\eps_{a})=0,\quad i=1,\ldots n
\eea
where $f_{i}$ is some function depending on the parameters of the theory. Each pole corresponds to a hyperplane $H_i$ in $\mathfrak{h}$ and the charge vector $Q_{i}\in\mathfrak{h}^{\ast}$ is associated with it. The integrand is singular at $\mathcal{M}_{\text{sing}}=\cup_i H_{i}$ and we denote $\mathcal{M}^{\ast}_{\text{sing}}$ to be the set of isolated points where $n\geq k$ linearly independent singular hyperplanes meet. 

To perform the JK-residue formalism, we need an additional ingredient\footnote{We hope the readers do not get confused with the phase boundary of the brane brick models.} $\eta\in\mathfrak{h}^{\ast}$, which is called the reference vector. The integral is then evaluated as
\bea
\oint _{\text{JK}}\prod_{i=1}^{k}\frac{d\phi_i}{2\pi i}\mathcal{Z}(\phi)=\sum_{\phi_{\ast}\in \mathcal{M}_{\text{sing}}^{\ast}}\underset{\phi=\phi_{\ast}}{\text{JK-Res}}(Q(\phi_{\ast}),\eta)\mathcal{Z}(\phi).
\eea
If $\phi_{\ast}$ is a non-degenerate pole associated with the charge vectors $Q_{1},\ldots,Q_{k}$, the JK-residue is
\bea
\underset{\phi=\phi_{\ast}}{\text{JK-}\text{Res}}(Q(\phi_{\ast}),\eta)\mathcal{Z}(\phi)=\delta(Q,\eta)\frac{1}{|\det Q|}\underset{\delta_{k}=0}{\Res}\cdots \underset{\delta_{1}=0}{\Res}\left.\mathcal{Z}(\phi)\right|_{Q_{i}(\phi)+f_{i}(\fra,\eps_{a})=\delta_{i}}
\eea
where 
\bea
\delta(Q,\eta)=\begin{dcases}
    1,\,\,\eta\in \text{Cone}(Q_{1},\ldots,Q_{k})\\
    0,\,\, \text{otherwise}
\end{dcases},\quad 
\text{Cone}(Q_{1},\ldots,Q_{k})=\left\{\sum_{i=1}^{k}\lambda_{i}Q_{i}\mid \lambda_{i}>0\right\}.
\eea 
When the pole is degenerate, we need more detailed analysis but we omit the discussion (see \cite{Benini:2013xpa,Benini:2013nda,Hori:2014tda}). The JK-residue generally depends on the choice of the vector $\eta$ and it is related to wall crossing. In this paper, we will always fix it to be
\bea
\eta=(1,\ldots,1),
\eea
namely the cyclic chamber. Discussion of wall crossing will be left for future work.

For the cyclic chamber, the JK condition admits a simple tree interpretation. Denoting the standard basis of $\mathfrak{h}^{\ast}$ by $e_I=(0,0,\ldots,1,\ldots,0,0)\in\mathbb{R}^{k}$, an admissible set of charge vectors $Q$ satisfying $\eta\in\operatorname{Cone}(Q) $ forms an oriented rooted tree. The root is associated with a charge vector $e_I$. Starting from the root, if $e_J$ has already appeared, the charge vector $e_K-e_J$ is allowed and corresponds to an oriented edge $J\rightarrow K$, while the opposite orientation $e_J-e_K$ is not. Repeating this procedure recursively generates the allowed poles.

\paragraph{Pyramid partitions and solid pyramid partitions}
\begin{figure}[t]
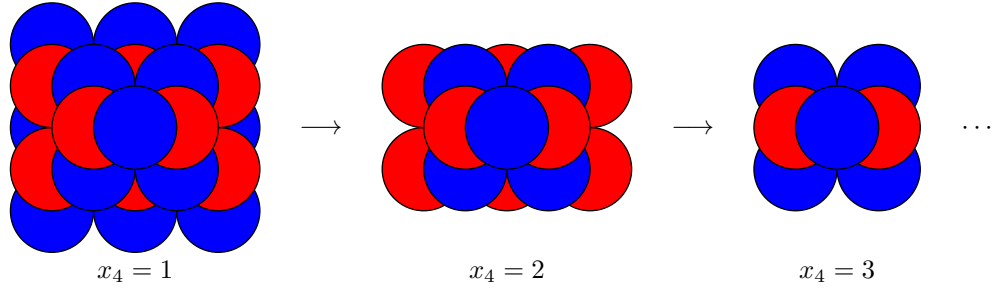

  \centering
  \begin{tabular}{c@{\hspace{0.4cm}}c@{\hspace{0.4cm}}c@{\hspace{0.4cm}}c@{\hspace{0.4cm}}cc}
    \adjustbox{valign=c}{\ConifoldPyramid{4}}
    &
    $\longrightarrow$
    &
    \adjustbox{valign=c}{\ConifoldPyramid{3}}
    &
    $\longrightarrow$
    &
    \adjustbox{valign=c}{\ConifoldPyramid{2}} & $\cdots $
    \\[2mm]
    $x_4=1$
    &&
    $x_4=2$
    &&
    $x_{4}=3$
    & 
  \end{tabular}
  \caption{A four-dimensional BPS crystal of
  $\mathcal{C}\times\mathbb{C}$ represented by its
  three-dimensional slices. 
  \label{fig:4dBPScrystal}
  }
\end{figure}
Application of the JK-residue prescription to toric CY$_{4}$ quivers were done in \cite{Bao:2024ygr} (see also \cite{Franco:2023tly}), generalizing the discussion of \cite{Nekrasov:2018xsb}. It was shown there that the genuine poles in the cyclic chamber giving non-zero JK-residues are classified by four-dimensional statistical models called 4d BPS crystals, which are natural generalizations of the well-known 3d BPS crystals \cite{Ooguri:2009ijd}. For example, for the $\mathbb{C}^{4}$ case, the 4d BPS crystal is the solid partition.

For the $\mathcal{C}\times \mathbb{C}$ case, the 4d BPS crystal is a 4d crystal whose horizontal slice is an ordinary pyramid partition. We first describe the three-dimensional pyramid crystal appearing in the most left figure in Figure~\ref{fig:4dBPScrystal}. The crystal is generated recursively from a single blue atom placed at the origin. From every blue atom, two red atoms can be generated in the left and right directions. From every red atom, two blue atoms can be generated in the front and back directions. Each elementary step lowers the position of the new atom by one level. Schematically, the local generation rule is 
\bea
\text{blue}
\quad
\xrightarrow{\ \mathrm{left},\mathrm{right}\ }
\quad
\text{red},
\qquad
\text{red}
\quad
\xrightarrow{\ \mathrm{forward},\mathrm{backward}\ }
\quad
\text{blue}.
\eea
Repeating these steps from the origin produces the infinite two-colored pyramid crystal. When different sequences of elementary steps reach the same position, they are regarded as producing the same atom.

A \textit{pyramid partition} is a finite configuration of \textit{atoms} satisfying the following predecessor condition. If an atom belongs to the configuration, then every atom encountered earlier along any directed path from the origin to that atom must also belong to the configuration. Equivalently, starting from the atom at the origin, a new atom can be added only after all its predecessor atoms have already been added. This is the growth form of the melting rule.

More formally, for two atoms $x$ and $y$, let
\bea
x\preceq y
\eea
mean that $y$ can be reached from $x$ by a sequence of the elementary steps above. Then a pyramid partition is a finite subset $\rho$ of the infinite pyramid crystal such that
\bea
y\in\rho,
\qquad
x\preceq y
\qquad
\Longrightarrow
\qquad
x\in\rho.
\eea
We denote the set of pyramid partitions by $\mathsf{Pyr}$.

A \textit{solid pyramid partition} $\boldsymbol{\rho}$ is a function
\bea
h_{\boldsymbol{\rho}}:
\mathsf{Pyr}
\longrightarrow
\mathbb Z_{\geq0}
\eea
with finite support, such that
\bea
x\preceq y
\qquad
\Longrightarrow
\qquad
h_{\boldsymbol{\pi}}(x)
\geq
h_{\boldsymbol{\pi}}(y).
\eea
Thus, the non-negative integers assigned to the atoms weakly decrease along every directed path emanating from the blue atom at the origin. The integer
$h_{\boldsymbol{\pi}}(x)$ counts the number of atoms stacked over the position $x$ along the fourth direction. We denote the set of solid pyramid partitions by
$\mathsf{SolPyr}$.

The solid pyramid partition can be visualized as a finite nested sequence of ordinary pyramid partitions (see Figure~\ref{fig:4dBPScrystal}):
\bea
\boldsymbol{\rho}=\{\rho^{(1)},\rho^{(2)},\ldots\} ,\quad \rho^{(1)}
\supseteq
\rho^{(2)}
\supseteq
\rho^{(3)}
\supseteq
\cdots .
\eea

The number of blue (red) atoms of a pyramid partition is denoted $|\rho|_{1}$ ($|\rho|_{2}$, respectively). Analogously, the number of blue (red) atoms of a solid pyramid partition is denoted $|\boldsymbol{\rho}|_{1}$ ($|\boldsymbol{\rho}|_{2}$, respectively).

\paragraph{Equivariant weights and JK-poles}
We next assign an equivariant weight to each atom of the solid pyramid crystal. The blue atom at the origin is assigned the framing weight $\mathfrak{a}$. Starting from a blue atom, the two elementary moves producing red atoms are assigned the weights $\epsilon _1$ and $\epsilon _2$, while the two moves from a red atom to blue atoms are assigned the weights $\epsilon _3$ and $\epsilon _4$. Finally, the elementary move to the next slice in the fourth direction is assigned the weight $\epsilon _5$. For an atom of a solid pyramid partition $x\in\boldsymbol{\rho}$ at layer $k$, the equivariant parameter is associated as
\bea\label{eq:equivweight-def}
\text{wt}(x)=\sum_{e\in\gamma_{x}}\text{wt}(e)+(k-1)\eps_{5},\qquad \qquad \adjustbox{valign=c}{
\begin{tikzpicture}[scale=0.5]

\filldraw [fill= blue!80!white, draw=black, thick] (1,1) circle (1cm);

\filldraw [fill= blue!80!white, draw=black, thick] (1,-1) circle (1cm);
\filldraw [fill= blue!80!white, draw=black, thick] (-1,1) circle (1cm);

\filldraw [fill= blue!80!white, draw=black, thick] (-1,-1) circle (1cm);

\filldraw [fill= red!80!white, draw=black, thick] (-1,0) circle (1cm);
\filldraw [fill= red!80!white, draw=black, thick] (1,0) circle (1cm);
\filldraw [fill= blue!80!white, draw=black, thick] (0,0) circle (1cm);
\draw[-Triangle,thick] (0,0)--(1.7,0);
\draw[-Triangle,thick](1.3,0.2)--(1.3,1.5);
\draw[-Triangle,thick](1.3,-0.2)--(1.3,-1.5);

\draw[-Triangle,thick] (0,0)--(-1.7,0);
\draw[-Triangle,thick](-1.3,0.2)--(-1.3,1.5);
\draw[-Triangle,thick](-1.3,-0.2)--(-1.3,-1.5);
\node[right] at (1.8,0){$p_{1}$};
\node[right] at (1.8,0.9){$p_{3}$};
\node[right] at (1.8,-0.9){$p_{4}$};
\node[left] at (-1.8,0){$p_{2}$};
\node[left] at (-1.8,0.9){$p_{3}$};
\node[left] at (-1.8,-0.9){$p_{4}$};
\end{tikzpicture}}
\eea
where $\gamma_{x}$ is any path from the origin blue atom of layer $k$.

\begin{theorem}
    For $(N_{1},N_{2},M_{1},M_{2})=(1,0,M_{1},M_{2})$, under the assumption that the equivariant parameters of the D8 and $\overline{\D8}$-branes are generic, the non-vanishing JK poles of the Witten index are in one-to-one correspondence with the solid pyramid partitions. Let $\boldsymbol\rho$ be a solid pyramid partition, then up to permutations, the corresponding pole configuration is
    \bea
    \{\phi_{I}^{(1)}\}_{I=1}^{k_{1}}&=\{\text{wt}(x)\mid x\in\boldsymbol{\rho},\quad \text{$x$ is a blue atom}\}\\
    \{\phi_{I}^{(2)}\}_{I=1}^{k_{2}}&=\{\text{wt}(x)\mid x\in\boldsymbol{\rho},\quad \text{$x$ is a red atom}\}
    \eea
    where
    \bea
    k_{1}=|\boldsymbol{\rho}|_{1},\quad k_{2}=|\boldsymbol{\rho}|_{2}.
    \eea
\end{theorem}

To make the discussion concrete, let us focus on the case $(N_{1},N_{2},M_{1},M_{2})=(1,0,0,0)$ whose integrand is
\bea
\mathcal{Z}_{1}^{\D8\tbar\D0}(\fra,\phi_I)\mathcal{Z}^{\D0\tbar\D0}(\phi).
\eea
The poles picked by the JK-residue prescription are
\bea\label{eq:D8-JK-poles}
\phi_{I}^{(1)}-\fra=0,\quad \phi_{I}^{(1)}-\phi_{J}^{(1)}-\eps_{5}=0,\quad \phi_{I}^{(2)}-\phi_{J}^{(2)}-\eps_{5}=0,\\
\phi_{J}^{(2)}-\phi_{I}^{(1)}-\eps_{1,2}=0,\quad \phi_{I}^{(1)}-\phi_{J}^{(2)}-\eps_{3,4}=0.
\eea

We will call the total number of the dimension vectors $(k_{1},k_{2})$ the level $k=k_{1}+k_{2}$. For level one, the only nontrivial case is when $(k_{1},k_{2})=(1,0)$ whose pole is
\bea
\phi_{1}^{(1)}-\fra=0.
\eea
This configuration is the blue atom at the origin:
\bea
 \adjustbox{valign=c}{
\begin{tikzpicture}[scale=0.5]
\filldraw [fill= blue!80!white, draw=black, thick] (0,0) circle (1cm);
\end{tikzpicture}}
\eea

For level two, we have following three configurations:
\bea
\phi_{1}^{(1)}-\fra=0,\quad \phi^{(1)}_{2}-\phi^{(1)}_{1}-\eps_{5}=0,&\qquad \qquad  \adjustbox{valign=c}{
\begin{tikzpicture}[scale=0.5]

\filldraw [fill= blue!80!white, draw=black, thick] (0,0) circle (1cm);
\filldraw [fill= blue!80!white, draw=black, thick] (2.5,0) circle (1cm);
\draw[-Triangle,thick] (0,-1.3)--(2.5,-1.3);
\node[below] at (1.25,-1.3){$p_{5}$}; 
\end{tikzpicture}},\\
\phi_{1}^{(1)}-\fra=0,\quad \phi^{(2)}_{1}-\phi^{(1)}_{1}-\eps_{1}=0,&\qquad \qquad  \adjustbox{valign=c}{
\begin{tikzpicture}[scale=0.5]

\filldraw [fill= red!80!white, draw=black, thick] (1,0) circle (1cm);
\filldraw [fill= blue!80!white, draw=black, thick] (0,0) circle (1cm);
\draw[-Triangle,thick, black] (0.1,0)--(1.7,0);
\node[right,black] at (2,-0.1){$p_{1}$};
\end{tikzpicture}}\\
\phi_{1}^{(1)}-\fra=0,\quad \phi^{(2)}_{1}-\phi^{(1)}_{1}-\eps_{2}=0,&\qquad \qquad \adjustbox{valign=c}{
\begin{tikzpicture}[scale=0.5]
\filldraw [fill= red!80!white, draw=black, thick] (-1,0) circle (1cm);
\filldraw [fill= blue!80!white, draw=black, thick] (0,0) circle (1cm);
\draw[-Triangle,thick, black] (0.1,0)--(-1.7,0);
\node[left,black] at (-2,-0.1){$p_{2}$};
\end{tikzpicture}}
\eea
where for the first configuration, the second atom is stacked in the second layer $x_{4}=2$.

For level three, vanishing terms appear from the numerators of the integrand. Naively, JK-residue picks the configuration
\bea
\phi_{1}^{(1)}-\fra=0,\quad \phi_{2}^{(1)}-\phi^{(1)}_{1}-\eps_{5}=0,\quad \phi^{(2)}_{1}-\phi^{(1)}_{2}-\eps_{1,2}=0.
\eea
but due to the numerator $\sh(\phi_{1}^{(2)}-\phi_{1}^{(1)}-\eps_{1,2}-\eps_{5})$, such contributions vanish. Generally, this means that so that one can place atoms at higher layers at the fourth direction, we need atoms at lower layers. Similarly, the configuration with no blue atom in the second layer
\bea
\,&\adjustbox{valign=c}{
\begin{tikzpicture}[scale=0.5]

\filldraw [fill= red!80!white, draw=black, thick] (1,0) circle (1cm);
\filldraw [fill= blue!80!white, draw=black, thick] (0,0) circle (1cm);

\filldraw [fill= red!80!white, draw=black] (7,0) circle (1cm);
\filldraw [fill= blue!20!white, draw=black, dashed ] (6,0) circle (1cm);

\end{tikzpicture}}\\
\phi_{1}^{(1)}-\fra=0,&\quad \phi_{1}^{(2)}-\phi^{(1)}_{1}-\eps_{1}=0,\quad \phi^{(2)}_{2}-\phi^{(2)}_{1}-\eps_{5}=0,
\eea
 also vanishes due to the numerator $\sh(\phi_{2}^{(2)}-\phi_{1}^{(1)}-\eps_{1}-\eps_{5})$. In total, the nontrivial configurations are
\bea
 \adjustbox{valign=c}{
\begin{tikzpicture}[scale=0.5]
\filldraw [fill= blue!80!white, draw=black, thick] (0,0) circle (1cm);
\node at (0,0){\Large$3$};
\end{tikzpicture}},\qquad  \adjustbox{valign=c}{
\begin{tikzpicture}[scale=0.5]
\filldraw [fill= red!80!white, draw=black, thick] (1,0) circle (1cm);
\filldraw [fill= blue!80!white, draw=black, thick] (0,0) circle (1cm);
\node at (0,0){\Large$2$};

\end{tikzpicture}},\qquad \adjustbox{valign=c}{
\begin{tikzpicture}[scale=0.5]
\filldraw [fill= red!80!white, draw=black, thick] (-1,0) circle (1cm);
\filldraw [fill= blue!80!white, draw=black, thick] (0,0) circle (1cm);
\node at (0,0){\Large$2$};

\end{tikzpicture}},\qquad \adjustbox{valign=c}{
\begin{tikzpicture}[scale=0.5]

\filldraw [fill= blue!80!white, draw=black, thick] (1,1) circle (1cm);
\filldraw [fill= red!80!white, draw=black, thick] (1,0) circle (1cm);
\filldraw [fill= blue!80!white, draw=black, thick] (0,0) circle (1cm);
\draw[-Triangle,thick] (0,0)--(1.7,0);
\draw[-Triangle,thick](1.3,0.2)--(1.3,1.5);
\node[below] at (1.3,0){$p_{1}$};
\node[right] at (1.8,0.9){$p_{3}$};
\end{tikzpicture}},\\
\adjustbox{valign=c}{
\begin{tikzpicture}[scale=0.5]

\filldraw [fill= blue!80!white, draw=black, thick] (1,-1) circle (1cm);
\filldraw [fill= red!80!white, draw=black, thick] (1,0) circle (1cm);
\filldraw [fill= blue!80!white, draw=black, thick] (0,0) circle (1cm);
\draw[-Triangle,thick] (0,0)--(1.7,0);
\draw[-Triangle,thick](1.3,-0.2)--(1.3,-1.5);
\node[above] at (1.3,0){$p_{1}$};
\node[right] at (1.8,-0.9){$p_{4}$};
\end{tikzpicture}},\qquad \adjustbox{valign=c}{
\begin{tikzpicture}[scale=0.5]

\filldraw [fill= blue!80!white, draw=black, thick] (-1,1) circle (1cm);
\filldraw [fill= red!80!white, draw=black, thick] (-1,0) circle (1cm);
\filldraw [fill= blue!80!white, draw=black, thick] (0,0) circle (1cm);
\draw[-Triangle,thick] (0,0)--(-1.7,0);
\draw[-Triangle,thick](-1.3,0.2)--(-1.3,1.5);
\node[below] at (-1.3,0){$p_{2}$};
\node[left] at (-1.8,0.9){$p_{3}$};
\end{tikzpicture}},\qquad  \adjustbox{valign=c}{
\begin{tikzpicture}[scale=0.5]

\filldraw [fill= blue!80!white, draw=black, thick] (-1,-1) circle (1cm);
\filldraw [fill= red!80!white, draw=black, thick] (-1,0) circle (1cm);
\filldraw [fill= blue!80!white, draw=black, thick] (0,0) circle (1cm);
\draw[-Triangle,thick] (0,0)--(-1.7,0);
\draw[-Triangle,thick](-1.3,-0.2)--(-1.3,-1.5);
\node[above] at (-1.3,0){$p_{2}$};
\node[left] at (-1.8,-0.9){$p_{4}$};
\end{tikzpicture}},\qquad  \adjustbox{valign=c}{
\begin{tikzpicture}[scale=0.5]

\filldraw [fill= red!80!white, draw=black, thick] (1,0) circle (1cm);
\filldraw [fill= red!80!white, draw=black, thick] (-1,0) circle (1cm);
\filldraw [fill= blue!80!white, draw=black, thick] (0,0) circle (1cm);
\end{tikzpicture}}
\eea
where the integers are the numbers of atoms stacked towards the fourth direction.

Let us list some examples for the partition functions we will use in later sections up to level two are as follows. For $(N_{1},N_{2},M_{1},M_{2})=(1,0,0,0)$, we have
\bea\label{eq:D8-noantiD8}
\mathcal{Z}(\fq_1,\fq_2;p_{1,2,3,4,5})&=1
+\frac{\sqrt{p_5}}
{1-p_{5}}\,
\mathfrak{q}_1
\nonumber\\
&+
\frac{
p_1 (p_2p_3-1)(p_2p_4-1)(p_1^2p_3p_4-1)
}{
(p_1-p_2)(p_1p_3-1)
(p_1p_4-1)(p_1p_2p_3p_4-1)
}\,
\mathfrak{q}_1\mathfrak{q}_2
\nonumber\\
&-
\frac{
p_2 (p_1p_3-1)(p_1p_4-1)(p_2^2p_3p_4-1)
}{
(p_1-p_2)(p_2p_3-1)(p_2p_4-1)(p_1p_2p_3p_4-1)
}\,
\mathfrak{q}_1\mathfrak{q}_2
\nonumber\\
&+
\frac{
p_{5}^{3/2}
}{
(1-p_{5})(1-p_{5}^{2})
}\,
\mathfrak{q}_1^2 +\cdots.
\eea
For $(N_{1},N_{2},M_{1},M_{2})=(1,0,1,0)$, we have
\bea\label{eq:D81D81}
\mathcal{Z}(\fq_1,\fq_2;p_{1,2,3,4,5})={}&1
+\frac{(1-\mu)\sqrt{p_5}}
{\sqrt{\mu}\,(1-p_5)}
\,\mathfrak{q}_1
\nonumber\\
&+
\frac{(1-\mu)p_1(p_2p_3-1)(p_2p_4-1)(p_1^2p_3p_4-1)}
{\sqrt{\mu}\,(p_1-p_2)(p_1p_3-1)(p_1p_4-1)(p_1p_2p_3p_4-1)}
\,\mathfrak{q}_1\mathfrak{q}_2
\nonumber\\
&+
\frac{(1-\mu)p_2(p_1p_3-1)(p_1p_4-1)(p_2^2p_3p_4-1)}
{\sqrt{\mu}\,(p_2-p_1)(p_2p_3-1)(p_2p_4-1)(p_1p_2p_3p_4-1)}
\,\mathfrak{q}_1\mathfrak{q}_2
\nonumber\\
&-
\frac{(1-\mu)p_5
\left(\mu -p_{5}\right)}
{\mu\,(1-p_{5})(1-p_{5}^{2})}
\,\mathfrak{q}_1^2 +\cdots 
\eea
where the flavor fugacity of the $\overline{\D8}_{1}$-brane is set $\frb=\fra+\mathfrak{m}$ and $\mu=e^{\mathfrak{m}}$. Similarly, for $(N_{1},N_{2},M_{1},M_{2})=(1,0,0,1)$, we have
\bea\label{eq:D81D82}
\mathcal{Z}(\fq_1,\fq_2;p_{1,2,3,4,5})={}&1
+\frac{\sqrt{p_5}}
{1-p_{5}}\,
\mathfrak{q}_1
\nonumber\\
&+
\frac{
\sqrt{p_1}\,(p_2p_3-1)(p_2p_4-1)(p_1^2p_3p_4-1)(p_1-\mu)
}{
\sqrt{\mu}\,(p_1-p_2)(p_1p_3-1)(p_1p_4-1)(p_1p_2p_3p_4-1)
}\,
\mathfrak{q}_2\mathfrak{q}_1
\nonumber\\
&+
\frac{
\sqrt{p_2}\,(p_1p_3-1)(p_1p_4-1)(p_2^2p_3p_4-1)(p_2-\mu)
}{
\sqrt{\mu}\,(p_2-p_1)(p_2p_3-1)(p_2p_4-1)(p_1p_2p_3p_4-1)
}\,
\mathfrak{q}_2\mathfrak{q}_1
\nonumber\\
&+
\frac{
p_{5}^{3/2}
}{
(1-p_{5})(1-p_{5}^{2})
}\,
\mathfrak{q}_1^2 +\cdots 
\eea
where the flavor fugacity of the $\overline{\D8}_{2}$-brane is set $\frb=\fra+\mathfrak{m}$ and $\mu=e^{\mathfrak{m}}$.

Although we restricted ourselves to the single-framing case $(N_{1},N_{2})=(1,0)$, the higher-rank generalization is straightforward. For generic flavor parameters $\mathfrak{a}_{\alpha}^{(c)}$, the non-vanishing JK poles are classified by collections
\bea
\left\{
\boldsymbol{\rho}^{(c)}_{\alpha}
\mathrel{\big|}
c=1,2,\
\alpha=1,\ldots,N_c
\right\},
\eea
where $\boldsymbol{\rho}^{(1)}_{\alpha}$ is rooted at a blue atom of weight $\mathfrak{a}_{\alpha}^{(1)}$, while $\boldsymbol{\rho}^{(2)}_{\alpha}$ is rooted at a red atom of weight $\mathfrak{a}_{\alpha}^{(2)}$. For the crystal whose origin is a red atom, the solid pyramid partition is obtained by simultaneously exchanging the blue and red atoms and permuting the equivariant parameters as
\bea
(\epsilon _1,\epsilon _2,\epsilon _3,\epsilon _4,\epsilon _5)
\longmapsto
(\epsilon _3,\epsilon _4,\epsilon _1,\epsilon _2,\epsilon _5).
\eea

\section{Tetrahedron instantons of \texorpdfstring{$\mathcal{C}\times \mathbb{C}$}{CC}}\label{sec:tetrahedron-conifold}
Tetrahedron instantons on $\mathbb{C}^{4}$ were introduced in \cite{Pomoni:2021hkn}, and their algebro-geometric formulation in Donaldson--Thomas theory was developed in \cite{Fasola:2023ypx}. Related extensions to Abelian orbifolds have also been considered in \cite{Cao:2023gvn,Szabo:2024lcp}. However, a systematic construction of tetrahedron instantons for general toric Calabi--Yau fourfolds beyond these quotient geometries is not yet available.

As a first step toward such a generalization, in this section we extend the tetrahedron-instanton construction to $\mathcal{C}\times\mathbb{C}$. We identify the relevant D6-brane framings from the toric data, construct the corresponding framed quivers, and compute their BPS partition functions.

We first associate a framed
D6-brane system with each corner toric divisor by choosing the
corresponding brick matching in section~\ref{sec:D6-brickmatchings}. We then study a single D6-brane
on the two inequivalent types of divisors in section~\ref{sec:D6-brick-m0} and \ref{sec:D6-brick-m2}. Their fixed points
are described by pyramid partitions and super plane partitions,
respectively. Finally, we combine these elementary D6-brane
sectors to define the general tetrahedron instanton partition
function in section~\ref{sec:tetrahedron-instanton}.

\subsection{D6-brane configurations from brick matchings}\label{sec:D6-brickmatchings}
We now turn to systems obtained by introducing D6-branes wrapping non-compact toric divisors of $\mathcal{C}\times\mathbb{C}$. In this section, we focus on setups involving a single D6-brane. Generalizations to the higher-rank case are discussed in section~\ref{sec:tetrahedron-instanton}.

Let $\mathfrak m$ be a brick matching associated with an extremal point of the toric diagram, and denote the corresponding toric divisor by $D_{\mathfrak m}$. We first place a single D6-brane on $D_{\mathfrak m}$:
\bea  \label{eq:conifold-D6-divisor}
\renewcommand{\arraystretch}{1.05}
\begin{tabular}{|c|c|c|c|c|c|c|c|c|c|c|}
\hline
& \multicolumn{6}{c|}{$\mathcal{C}$} & \multicolumn{2}{c|}{$\mathbb{C}$} & \multicolumn{2}{c|}{$\mathbb{R}\times S^{1}$} \\
\cline{2-11} \raisebox{-0mm}{}  & 1 & 2 & 3 & 4& 5 & 6 & 7 & 8 & 9& 0\\[-2pt]\hline
 D0& $\bullet$ & $\bullet$  & $\bullet$  & $\bullet$  & $\bullet$  & $\bullet$   & $\bullet$  & $\bullet$  & $\bullet$   & $-$\\
\hline D6& \multicolumn{8}{c|}{$D_{\mathfrak{m}}$}& $\bullet$   & $-$\\
\hline
\end{tabular}
\eea
Since the divisor $D_{\mathfrak m}$ is non-compact, the D6-brane behaves as a flavor brane from the viewpoint of the fractional D0-branes and introduces a framing node into the quiver quantum mechanics.

As a local model, recall the D0--D6 system on $\mathbb{C}^{3}$. In the language of one-dimensional $\mathcal{N}=2$ quiver quantum mechanics, the D0--D6 open strings give rise to a framing chiral multiplet $\mathsf{I}$ and a framing Fermi multiplet $\Lambda$. Moreover, the $E$-term of $\Lambda$ takes the form as $E_{\Lambda}=\Phi\cdot \mathsf{I}$, where $\Phi$ is some chiral field. Since the D0-brane theory on $\mathbb{C}^{3}$ has a single gauge node, both framing multiplets are associated with the same node.

For a general toric quiver, however, the chiral and Fermi multiplets may be associated with different gauge nodes. A natural generalization to a toric quiver with several gauge nodes is
\bea
\adjustbox{valign=c}{
\begin{tikzpicture}[
    scale=1.15,
    line cap=round,
    line join=round,
    chiral/.style={
        black,
        thick,
        postaction={
            decorate
        },
        decoration={
            markings,
            mark=at position 0.52 with {\arrow{Triangle}}
        }
    },
    fermi/.style={
        red,
         thick,
        postaction={
            decorate
        },
        decoration={
            markings,
            mark=at position 0.52 with {\arrow{Triangle}}
        }
    },
    qnode/.style={
        circle,
        draw=black,
        thick,
        fill=yellow!85!orange,
        inner sep=5pt
    }, box/.style={draw, minimum width=0.6cm, minimum height=0.6cm, text centered,thick}
]

\draw[chiral] (0,2)--(0,0); 
\draw[fermi] (0,2)--(2,0.5);
\draw[chiral] (0,2)--(0,0); 
\draw[thick,-Triangle] (0,0)--(0.7,-0.1); 
\draw[chiral] (1.3,0.2)--(2,0.5); 
\node[left] at (-0.05,1){$\mathsf{I}$};
\node[right] at (1,1.75){$\Lambda$};

\node[rotate=25] at (1,0.1){$\cdots $};
 
\node[qnode] at (0,0){$a$ };

\node[qnode] at (2,0.5){$b$ };
\node[box,fill=black!10!white] at (0,2){ };
\end{tikzpicture}}\qquad E_{\Lambda}=\Phi_{\gamma}\cdot \,\mathsf{I},\quad \Phi_{\gamma}=\Phi_{b\leftarrow i_{n}}\cdots \Phi_{i_{1}\leftarrow a}
\eea
where $\gamma$ is an oriented path of chiral edges from the gauge node $a$ to the gauge node $b$. In other words, the framing chiral multiplet $\mathsf{I}$ is attached to the initial node \(a\), while the framing Fermi multiplet \(\Lambda\) is
attached to the terminal node \(b\). In general, the path \(\gamma\)
may consist of several chiral arrows, and hence the two framing multiplets need not be attached to adjacent gauge nodes.


The diagram above describes a general gauge-invariant framing. Motivated by the local D0--D6 system on $\mathbb{C}^{3}$, we identify the elementary configuration of a single D6-brane with the minimal case in which \(\gamma\) consists of a single chiral arrow. We then propose the following prescription for a D6-brane wrapping a toric divisor associated with a corner of the toric diagram.
\begin{proposal}[D6-brane framing prescription]\label{prop:D6-framing}
Let $\mathfrak m_i$ be a corner brick matching and let
$D_{\mathfrak m_i}$ be the toric divisor associated with the
corresponding corner of the toric diagram. Define
\bea
\mathcal{A}_{\mathfrak m_i}
=
\left\{
\Phi_{a\rightarrow b}\in\overline{Q}_{1}^{(0)}
\mathrel{\big|}
P_{\Phi,\mathfrak m_i}=1
\right\},
\eea
where $P$ is the brick-matching matrix.

We propose that the admissible elementary framings of a single
D6-brane wrapping $D_{\mathfrak m_i}$ are in one-to-one
correspondence with the elements of
$\mathcal{A}_{\mathfrak m_i}$. For each
$\Phi_{a\rightarrow b}\in\mathcal{A}_{\mathfrak m_i}$,
we introduce
\bea
\mathsf{I}_{\mathfrak m_i}^{\Phi}
:
\ast\longrightarrow a,
\qquad
\Lambda_{\mathfrak m_i}^{\Phi}
:
\ast\textcolor{red}{\longrightarrow} b,
\eea
with the $E$-term
\bea
E_{\Lambda_{\mathfrak m_i}^{\Phi}}
=
\Phi_{a\rightarrow b}\,
\mathsf{I}_{\mathfrak m_i}^{\Phi}.
\eea
Thus, the brick matching $\mathfrak m_i$ determines the wrapped
divisor $D_{\mathfrak m_i}$, whereas the choice of
$\Phi_{a\rightarrow b}\in\mathcal{A}_{\mathfrak m_i}$
determines the color of the elementary framing. In particular, if
$\lvert\mathcal{A}_{\mathfrak m_i}\rvert>1$, a D6-brane wrapping the
same toric divisor admits more than one elementary framing choice.
\end{proposal}

This proposal is the minimal generalization of the local D0--D6
framing on \(\mathbb{C}^{3}\). Indeed, when \(a=b\),
\(\Phi_{a\rightarrow a}\) is an adjoint chiral field and the construction
reduces to the familiar one-node framing. For a general toric quiver,
the source and target of \(\Phi_{a\rightarrow b}\) determine the two gauge
nodes supporting \(\mathsf{I}\) and \(\Lambda\), respectively. Although
gauge invariance also allows the replacement of
\(X_{a\rightarrow b}\) by a composite chiral path, we do not identify
such longer-path framings with elementary configurations of a single
D6-brane. A microscopic derivation of the proposed correspondence from
the D0--D6 open-string spectrum will be left for future work.

As emphasized in Rem.~\ref{rem:framing-toric-condition}, the
framing multiplets introduced above are not required to satisfy the
toric condition \eqref{eq:JE-toriccond}. Note that since the $J$-term is $0$, the traceless condition \eqref{eq:JE-traceless} is automatically satisfied.  We also note that the matrix element
$P_{\Phi,\mathfrak m_i}$ appearing in Proposal~\ref{prop:D6-framing} refers to the
brick-matching matrix of the unframed D0--D0 quiver, not the framed quiver.

In particular, for the $\mathcal{C}\times \mathbb{C}$ case, we have five brick matchings giving
\bea
\mathcal{A}_{\mathfrak{m}_{0}}=\{\mathsf{C}_{1},\mathsf{C}_{2}\},\quad \mathcal{A}_{\mathfrak{m}_{1}}=\{\mathsf{B}_{1}\},\quad \mathcal{A}_{\mathfrak{m}_{2}}=\{\mathsf{A}_{2}\},\quad \mathcal{A}_{\mathfrak{m}_{3}}=\{\mathsf{B}_{2}\},\quad \mathcal{A}_{\mathfrak{m}_{4}}=\{\mathsf{A}_{1}\},
\eea
where for only the brick matching $\mathcal{A}_{\mathfrak{m}_{0}}$, we have two chiral fields. For simplicity, we denote the sets $\mathcal{A}_{\mathfrak{m}_{i}}$ by the starting nodes of the chiral fields, because for the $\mathcal{C}\times \mathbb{C}$ case, they uniquely determine the fields that we use:
\bea
 \mathcal{A}_{\mathfrak{m}_{0}}=\{1,2\},\quad \mathcal{A}_{\mathfrak{m}_{1}}=\{2\},\quad \mathcal{A}_{\mathfrak{m}_{2}}=\{1\},\quad \mathcal{A}_{\mathfrak{m}_{3}}=\{2\},\quad \mathcal{A}_{\mathfrak{m}_{4}}=\{1\}.
\eea
We also shortly denote $\mathcal{A}_{\mathfrak{m}_{i}}=\mathcal{A}_{i}$ for $i\in\five$.

The elementary single-D6-brane framings are summarized as follows.
The table is written in terms of $\mathcal N=2$ multiplets, including
the two $\mathcal N=4$ configurations associated with
$\mathfrak m_0$:
\bea
\renewcommand{\arraystretch}{1.25}
\begin{tabular}{c|c|c|c|c}
\hline
Brick matching
&
Toric point
&
Elementary framing
&
$E$-term
&
$\mathcal N$
\\
\hline\hline
$\mathfrak m_0$
&
$(0,0,1)$
&
\ElementaryDsixQuiver{1}{1}{\mathsf C_1}
&
$E_{\Lambda}
 =
 \mathsf C_1\mathsf I$
&
$4$
\\

$\mathfrak m_0$
&
$(0,0,1)$
&
\ElementaryDsixQuiver{2}{2}{\mathsf C_2}
&
$E_{\Lambda}
 =
 \mathsf C_2\mathsf I$
&
$4$
\\
\hline

$\mathfrak m_1$
&
$(0,0,0)$
&
\ElementaryDsixQuiver{2}{1}{\mathsf B_1}
&
$E_{\Lambda}
 =
 \mathsf B_1\mathsf I$
&
$2$
\\

$\mathfrak m_2$
&
$(1,0,0)$
&
\ElementaryDsixQuiver{1}{2}{\mathsf A_2}
&
$E_{\Lambda}
 =
 \mathsf A_2\mathsf I$
&
$2$
\\

$\mathfrak m_3$
&
$(1,1,0)$
&
\ElementaryDsixQuiver{2}{1}{\mathsf B_2}
&
$E_{\Lambda}
 =
 \mathsf B_2\mathsf I$
&
$2$
\\

$\mathfrak m_4$
&
$(0,1,0)$
&
\ElementaryDsixQuiver{1}{2}{\mathsf A_1}
&
$E_{\Lambda}
 =
 \mathsf A_1\mathsf I$
&
$2$
\\

\hline
\end{tabular}
\label{eq:elementary-D6-framings}
\eea
Note that for the $\mathcal N=4$ divisor $D_{\mathfrak m_0}$, no independent
interaction needs to be added in the $\mathcal N=4$ notation. After
decomposing the framing multiplet into $\mathcal N=2$ multiplets, the
same interaction is encoded by the framing Fermi multiplet through
\bea
E_{\Lambda_{\mathfrak m_0}^{(a)}}
=
\mathsf C_a
\mathsf I_{\mathfrak m_0}^{(a)},
\qquad
a=1,2.
\eea

\paragraph{Brane brick model and tachyon condensation}
We next provide a heuristic brane-brick interpretation of the framing
prescription proposed above. After performing the three T-dualities on \eqref{eq:conifold-D6-divisor} as used in \eqref{eq:branebrick-web}, the T-dual image of a D6-brane wrapping $D_{\mathfrak{m}}$ is expected to be a point-like object in the brane brick model. For a choice $\Phi_{a\rightarrow b}\in\mathcal A_{\mathfrak m}$, we propose that this object is localized on the chiral face separating the bricks $a$ and $b$ associated with $\Phi_{a\rightarrow b}$.

This proposal admits a natural interpretation in terms of tachyon
condensation. Consider a D8-brane and an
$\overline{\mathrm{D8}}$-brane whose T-dual images are placed inside
two adjacent bricks $a$ and $b$, respectively. Before tachyon
condensation, they give rise to a framing chiral multiplet
$\mathsf I:\ast\rightarrow a$ and a framing Fermi multiplet
$\Lambda:\ast\rightarrow b$. We then choose a tachyon profile whose
zero locus is the toric divisor $D_{\mathfrak m}$, represented in
the brane-brick model by their common chiral face
$\Phi_{a\rightarrow b}$. The D8--$\overline{\mathrm{D8}}$ pair
condenses into a D6-brane wrapping $D_{\mathfrak m_i}$, whose T-dual
image is localized on this face:
\bea\label{eq:tachyon-condense}
\begin{tikzpicture}[
    scale=0.95,
    brickedge/.style={
        draw=black,
        thick
    },
    sharedface/.style={
        fill=gray!25,
        draw=black,
        thick
    },
    dbrane/.style={
        diamond,
        draw=orange!60!black,
        fill=orange!75,
        thick,
        minimum size=0.42cm,
        inner sep=0pt
    },
    d6brane/.style={
    diamond,
    draw=violet!70!black,
    fill=violet!65,
    thick,
    minimum size=0.50cm,
    inner sep=0pt
}
]

\def\DrawTwoBricks{%
    \filldraw[sharedface]
    (1.20,0)
    --
    (1.60,0.30)
    --
    (1.60,1.50)
    --
    (1.20,1.20)
    --
    cycle;

    \draw[brickedge]
    (0,0)--(1.20,0)--(1.20,1.20)--(0,1.20)--cycle;

    \draw[brickedge]
    (0.40,0.30)--(1.60,0.30)
    --(1.60,1.50)--(0.40,1.50)--cycle;

    \draw[brickedge] (0,0)--(0.40,0.30);
    \draw[brickedge] (0,1.20)--(0.40,1.50);

    \draw[brickedge]
    (1.20,0)--(2.40,0)
    --(2.40,1.20)--(1.20,1.20);

    \draw[brickedge]
    (1.60,0.30)--(2.80,0.30)
    --(2.80,1.50)--(1.60,1.50);

    \draw[brickedge] (2.40,0)--(2.80,0.30);
    \draw[brickedge] (2.40,1.20)--(2.80,1.50);
}

\begin{scope}
    \DrawTwoBricks

    \node[dbrane] (D8) at (0.72,0.73) {};
    \node[dbrane] (antiD8) at (2.08,0.73) {};

    \node[font=\scriptsize,below=2pt]
    at (D8.south)
    {$\mathrm{D8}$};

    \node[font=\scriptsize,below=2pt]
    at (antiD8.south)
    {$\overline{\mathrm{D8}}$};

    \node[font=\scriptsize,above]
    at (1.40,1.54)
    {$\Phi_{a\rightarrow b}$};
\end{scope}

\draw[
    -{Triangle[length=3mm,width=2.2mm]},
    very thick
]
(3.15,0.75)--(4.55,0.75);

\node[
    font=\scriptsize,
    align=center
]
at (3.85,1.12)
{tachyon\\ condensation};

\begin{scope}[xshift=4.90cm]
    \DrawTwoBricks

    \node[
        d6brane,
        minimum size=0.50cm
    ]
    (D6) at (1.40,0.75) {};

    \node[font=\scriptsize,right=3pt]
    at (D6.east)
    {$\mathrm{D6}$};

    \node[font=\scriptsize,above]
    at (1.40,1.54)
    {$\Phi_{a\rightarrow b}$};
\end{scope}
\end{tikzpicture}
\eea
where the branes are labeled by their
original type IIA charges, while the diamonds represent their T-dual
images in the brane-brick model.

The three fields associated with this configuration form the
gauge-invariant closed path
\bea
\overline{\Lambda}\,
\Phi_{a\rightarrow b}\,
\mathsf I,
\eea
which gives the $E$-term
\bea
E_{\Lambda}
=
\Phi_{a\rightarrow b}\mathsf I.
\eea
This reproduces precisely the elementary framing introduced in the
proposal above. If the brick matching $\mathfrak m$ contains more
than one admissible chiral field, each choice specifies a different
chiral face and hence a different possible position of the D6-brane
in the brane-brick model. This geometrically accounts for the
different framing colors associated with the same wrapped divisor,
as occurs for $\mathfrak m_0$.

The discussion above should be regarded as a physical motivation for
our proposal rather than a microscopic derivation. A detailed
derivation from the D-brane open-string spectrum and the corresponding
tachyon profile will be left for future work.

\subsection{The \texorpdfstring{$\mathcal N=4$}{N=4} D6--D2--D0 system: the brick matching \texorpdfstring{$\mathfrak m_0$}{m0}}\label{sec:D6-brick-m0}
In this section, we will focus on the D6--D2--D0 system associated with the toric divisor $D_{\mathfrak{m}_{0}}$. For this case, we have two choices as explained in \eqref{eq:elementary-D6-framings}. The flavor charges of the chiral and Fermi fields are determined as
\bea
\mathsf{I}_{\mathfrak{m}_{0}}^{(a)}\rightarrow \mathsf{I}_{\mathfrak{m}_{0}}^{(a)},\quad \Lambda_{\mathfrak{m}_{0}}^{(a)}\rightarrow p_{5}\Lambda_{\mathfrak{m}_{0}}^{(a)}
\eea
for $a\in\mathcal{A}_{0}$. The framing node contribution to the Witten index is
\bea
\mathcal{Z}_{\mathfrak{m}_{0},a}^{\D6\tbar\D0}( \fra,\phi)=\prod_{I =1}^{k_{a}}\frac{\sh(\phi_{I}^{(a)}-\fra-\eps_5)}{\sh(\phi_{I}^{(a)}-\fra)},\qquad a\in\mathcal{A}_{0}.
\eea
Note that this system preserves $\mathcal{N}=4$ supersymmetry and the factor above is interpreted as the $\mathcal{N}=4$ chiral field.

Since for the case $a=2$, one can obtain a similar result after permuting the colors of nodes and the equivariant parameters, we focus only on the case $a=1$.

\begin{definition}\label{def:D6-index-m0}
    The BPS partition function of the system with a D6-brane wrapping the toric divisor $\mathfrak{m}_{0}$ of $\mathcal{C}\times\mathbb{C}$ is defined as
    \bea
    \mathcal{Z}^{\D6\tbar\D0}(\mathfrak{m}_{0};\mathfrak{q}_{1},\mathfrak{q}_{2};p_{1,2,3,4,5})=\sum_{k_{1}=0}^{\infty}\sum_{k_{2}=0}^{\infty}\mathfrak{q}_{2}^{k_{2}}\mathfrak{q}_{1}^{k_{1}}\mathcal{Z}^{\D6\tbar\D0}_{k_{1},k_{2}}(\mathfrak{m}_{0})
    \eea
    where
    \bea
    \mathcal{Z}^{\D6\tbar\D0}_{k_{1},k_{2}}(\mathfrak{m}_{0})&=\frac{1}{k_{2}!k_{1}!}\oint_{\text{JK}}\prod_{a=1}^{2}\prod_{I=1}^{k_{a}}\frac{d\phi_{I}^{(a)}}{2\pi i } \mathcal{Z}_{\mathfrak{m}_{0},1}^{\D6\tbar\D0}(\fra,\phi)\mathcal{Z}^{\D0\tbar\D0}(\phi).
    \eea

\end{definition}

One observation is that the D6-brane contribution can be obtained by the $\D8\tbar\overline{\D8}\tbar\D0$ system of rank $(N_{1},N_{2},M_{1},M_{2})=(1,0,1,0)$ after tuning the distance $\frb-\fra=\eps_{5}$ as
\bea
\mathcal{Z}_{\mathfrak{m}_{0},1}^{\D6\tbar\D0}(\fra,\phi)=\mathcal{Z}_{1}^{\D8\tbar\D0}(\fra,\phi)\mathcal{Z}_{1}^{\overline{\D8}\tbar\D0}(\fra+\eps_{5},\phi),
\eea
which is what we have mentioned in \eqref{eq:tachyon-condense}.

Since the numerator is not generic, the genuine poles picked up from the JK-residue prescription at the cyclic chamber $\eta=(1,\ldots,1)$ are not the same as the one from the previous section, but rather a restricted version. For example, at level two, the configuration $(\phi_{1}^{(1)},\phi_{2}^{(1)})=(\fra,\fra+\eps_5)$ is canceled by the numerator $\sh(\phi_{2}^{(1)}-\fra-\eps_{5})$. Generally, the growth of the crystal in the fourth direction with coordinates $p_{5}$ is terminated, and the poles are classified by the pyramid partitions introduced in section~\ref{sec:magnificent-four}.

\begin{theorem}
    The non-vanishing JK poles of the BPS partition function in Def.~\ref{def:D6-index-m0} are classified by the 3d pyramid partitions whose origin atom is blue and whose atoms are generated by the equivariant parameters $p_{1,2,3,4}$. Let $\rho$ be a pyramid partition, then up to permutations, the corresponding pole configuration is
    \bea
    \{\phi_{I}^{(1)}\}_{I=1}^{k_{1}}&=\{\text{wt}(x)\mid x\in{\rho},\quad \text{$x$ is a blue atom}\}\\
    \{\phi_{I}^{(2)}\}_{I=1}^{k_{2}}&=\{\text{wt}(x)\mid x\in{\rho},\quad \text{$x$ is a red atom}\}
    \eea
    where
    \bea
    k_{1}=|{\rho}|_{1},\quad k_{2}=|{\rho}|_{2}.
    \eea
    $|\rho|_{1,2}$ is the number of atoms with color $1,2$ and the weight function $\text{wt}(x)$ is obtained from \eqref{eq:equivweight-def} by setting the layer to be one.
    
\end{theorem} 
Up to level two, the partition function takes the form of
\bea
\mathcal{Z}^{\D6\tbar\D0}(\mathfrak{m}_{0};\mathfrak{q}_{1},\mathfrak{q}_{2};p_{1,2,3,4,5})&=1+\fq_{1}-\left(\frac{\sqrt{p_1} \left(p_2 p_3-1\right) \left(p_2 p_4-1\right) \left(p_1^2 p_3
   p_4-1\right)}{\sqrt{p_2} \left(p_2-p_1\right) \sqrt{p_3} \left(p_1 p_3-1\right)
   \sqrt{p_4} \left(p_1 p_4-1\right)}\right.\\
   &\left.+\frac{\sqrt{p_2} \left(p_1 p_3-1\right) \left(p_1
   p_4-1\right) \left(p_2^2 p_3 p_4-1\right)}{\sqrt{p_1} \left(p_1-p_2\right) \sqrt{p_3}
   \left(p_2 p_3-1\right) \sqrt{p_4} \left(p_2 p_4-1\right)}\right)\fq_{1}\fq_{2}+\cdots 
\eea
where one can see that the term with $\fq_{1}^{2}$ in \eqref{eq:D81D82} disappears due to the condition $\mu=p_{5}$. This partition function has a unrefined limit $p_{5}\rightarrow 1$ trivializing all the factors:
\bea
\mathcal{Z}^{\D6\tbar\D0}(\mathfrak{m}_{0};\mathfrak{q}_{1},\mathfrak{q}_{2};p_{1,2,3,4,5})\xrightarrow{p_{5}\rightarrow 1} 1+\fq_1+2\fq_{1}\fq_{2}+5\fq_1^{2}\fq_2+\cdots
\eea
and one obtains the generating function of the pyramid partitions. 

\subsection{The \texorpdfstring{$\mathcal N=2$}{N=2} D6--D2--D0 system: the brick matching \texorpdfstring{$\mathfrak m_2$}{m2}}\label{sec:D6-brick-m2}
Let us next consider the D6--D2--D0 system associated with the toric divisors $D_{\mathfrak{m}_{1,2,3,4}}$. The flavor charges of the chiral and Fermi fields are determined as
\bea
\mathsf{I}_{\mathfrak{m}_{1}}\rightarrow \mathsf{I}_{\mathfrak{m}_1},&\quad \Lambda_{\mathfrak{m}_{1}}\rightarrow p_{3}\Lambda_{\mathfrak{m}_{1}},\\
\mathsf{I}_{\mathfrak{m}_{2}}\rightarrow \mathsf{I}_{\mathfrak{m}_2},&\quad \Lambda_{\mathfrak{m}_{2}}\rightarrow p_{2}\Lambda_{\mathfrak{m}_{2}},\\
\mathsf{I}_{\mathfrak{m}_{3}}\rightarrow \mathsf{I}_{\mathfrak{m}_3},&\quad \Lambda_{\mathfrak{m}_{3}}\rightarrow p_{4}\Lambda_{\mathfrak{m}_{3}},\\
\mathsf{I}_{\mathfrak{m}_{4}}\rightarrow \mathsf{I}_{\mathfrak{m}_4},&\quad \Lambda_{\mathfrak{m}_{4}}\rightarrow p_{1}\Lambda_{\mathfrak{m}_{4}}.
\eea
The framing node contributions to the Witten index are
\bea
\mathcal{Z}_{\mathfrak{m}_{1}}^{\D6\tbar\D0}( \fra,\phi)=\frac{\prod_{J=1}^{k_{1}}\sh(\phi_{J}^{(1)}-\fra-\eps_{3})}{\prod_{I=1}^{k_{2}}\sh(\phi_{I}^{(2)}-\fra)},\\
\mathcal{Z}_{\mathfrak{m}_{2}}^{\D6\tbar\D0}( \fra,\phi)=\frac{\prod_{J=1}^{k_{2}}\sh(\phi_{J}^{(2)}-\fra-\eps_{2})}{\prod_{I=1}^{k_{1}}\sh(\phi_{I}^{(1)}-\fra)},\\
\mathcal{Z}_{\mathfrak{m}_{3}}^{\D6\tbar\D0}( \fra,\phi)=\frac{\prod_{J=1}^{k_{1}}\sh(\phi_{J}^{(1)}-\fra-\eps_{4})}{\prod_{I=1}^{k_{2}}\sh(\phi_{I}^{(2)}-\fra)},\\
\mathcal{Z}_{\mathfrak{m}_{4}}^{\D6\tbar\D0}( \fra,\phi)=\frac{\prod_{J=1}^{k_{2}}\sh(\phi_{J}^{(2)}-\fra-\eps_{1})}{\prod_{I=1}^{k_{1}}\sh(\phi_{I}^{(1)}-\fra)}.
\eea

In this section, we focus on the case $\mathfrak{m}_{2}$. Other cases can be obtained by permuting the colors and the equivariant parameters.

\begin{definition}\label{def:D6_m2}
    The BPS partition function of the system with a D6-brane wrapping the toric divisor $\mathfrak{m}_{2}$ is defined as
    \bea
    \mathcal{Z}^{\D6\tbar\D0}(\mathfrak{m}_{2};\mathfrak{q}_{1},\mathfrak{q}_{2};p_{1,2,3,4,5})=\sum_{k_{1}=0}^{\infty}\sum_{k_{2}=0}^{\infty}\mathfrak{q}_{2}^{k_{2}}\mathfrak{q}_{1}^{k_{1}}\mathcal{Z}^{\D6\tbar\D0}_{k_{1},k_{2}}(\mathfrak{m}_{2})
    \eea
    where
    \bea
    \mathcal{Z}^{\D6\tbar\D0}_{k_{1},k_{2}}(\mathfrak{m}_{2})&=\frac{1}{k_{2}!k_{1}!}\oint_{\text{JK}} \prod_{a=1}^{2}\prod_{I=1}^{k_{a}}\frac{d\phi_{I}^{(a)}}{2\pi i } \mathcal{Z}_{\mathfrak{m}_{2}}^{\D6_{1}\tbar\D0}(\fra,\phi)\mathcal{Z}^{\D0\tbar\D0}(\phi)
    \eea

\end{definition}

Similar to the previous example, the framing node contribution is obtained from the magnificent four system with $(N_{1},N_{2},M_{1},M_{2})=(1,0,0,1)$:
\bea
 \mathcal{Z}_{\mathfrak{m}_{2}}^{\D6_{1}\tbar\D0}(\fra,\phi)=\mathcal{Z}^{\D8\tbar\D0}_{1}(\fra,\phi)\mathcal{Z}^{\overline{\D8}\tbar\D0}_{2}(\frb,\phi),\quad \frb=\fra+\eps_{2}
\eea
where we tuned the parameter of the $\overline{\D8}$-brane. Again, the numerator is not generic and the genuine poles are the restricted version of \eqref{eq:D81D82}. In particular, at level two, the configuration $(\phi_{1}^{(1)},\phi_{2}^{(2)})=(\fra,\fra+\eps_2)$ gives zero residue because of the numerator $\sh(\phi^{(2)}_{1}-\phi^{(1)}_{1}-\eps_{2})$.

\paragraph{Super partitions and super plane partitions}
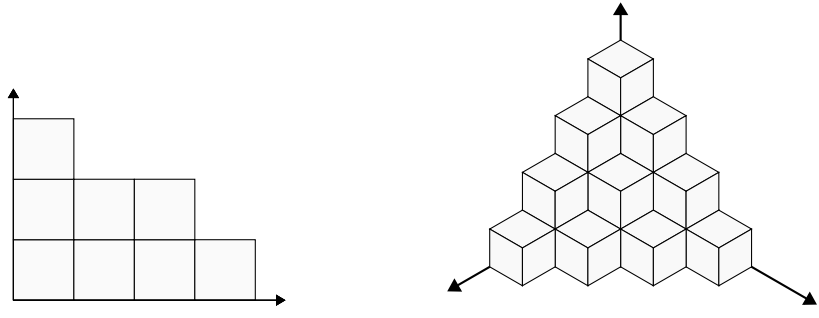
\begin{figure}
    \centering
    \begin{tikzpicture}[scale=0.8]
  \foreach \y/\length in {0/4,1/3,2/1}{
    \foreach \x in {1,...,\length}{
      \draw[fill=black!2!white]
        ({\x-1},\y) rectangle (\x,{\y+1});
    }
    \draw[-Triangle] (0,0)--(4.5,0);
     \draw[-Triangle] (0,0)--(0,3.5);
  }
\end{tikzpicture}\hspace{2cm}
       \begin{tikzpicture}[scale=0.5]
\draw[thick,-Triangle] (0,0)--(-30:6); 
    \draw[thick,-Triangle] (0,0)--(210:5.3);
    \draw[thick,-Triangle] (0,0)--(90:5);
    \planepartition{{4,3,2,1},{3,2,1},{2,1,0},{1,0,0}}
\end{tikzpicture}
    \caption{Partitions and plane partitions. The left partition is $\lambda=(3,2,2,1)$, while the right plane partition is $\pi=\{\{4,3,2,1\},\{3,2,1\},\{2,1\},\{1\}\}$.}
    \label{fig:2d3dpartition}
\end{figure}
The genuine poles actually have a nice combinatorial description so let us introduce some notations for the following discussions. Recall that a partition or a Young diagram $\lambda$ is a sequence of integers $\lambda=(\lambda_{1},\lambda_{2},\ldots ,\lambda_{\ell(\lambda)} ) $ obeying the condition
\bea\label{eq:2dpartition}
\lambda_{1}\geq \lambda_{2}\geq \lambda_{3}\geq \cdots\geq \lambda_{\ell(\lambda)}\geq 0,
\eea
where $\ell(\lambda)$ is the length of the Young diagram. It can be illustrated as a two-dimensional stack of boxes as the left of Figure~\ref{fig:2d3dpartition}. A box of the Young diagram is identified by its coordinate $(i,j)\in\mathbb{Z}^{2}_{>0}$ and obeys the condition
\bea
(i,j)\in\lambda \quad \Leftrightarrow\quad  1\leq j\leq \lambda_{i}.
\eea
The size of the Young diagram or the number of boxes in the Young diagram is defined as
\bea
|\lambda|=\sum_{i=1}\lambda_{i}.
\eea

A plane partition is a three-dimensional version of the partition and it is a two-dimensional sequence of integers $\pi=\{\pi_{i,j}\in\mathbb{Z}_{\geq 0}\}_{i,j\in\mathbb{Z}_{>0}}$ obeying the condition
\bea
\pi_{i,j}\geq \pi_{i+1,j},\quad \pi_{i,j}\geq \pi_{i,j+1}.
\eea
Similarly, we can illustrate it as a three-dimensional stack of boxes as Figure~\ref{fig:2d3dpartition} and a box is identified with its coordinate $(i,j,k)\in\mathbb{Z}^{3}_{>0}$ obeying the condition
\bea
(i,j,k)\in\pi\quad \Leftrightarrow\quad 1\leq k\leq \pi_{i,j}.
\eea
The size of the plane partition is defined as
\bea
|\pi|=\sum_{i,j}\pi_{i,j}.
\eea

A super partition is a finite sequence
\bea\label{eq:superpartition-def}
 \Lambda
 =
 (\Lambda_1,\ldots,\Lambda_{\ell(\Lambda)}),
 \qquad
 \Lambda_i\in\frac{1}{2}\mathbb Z_{>0},
\eea
satisfying
\bea
 \Lambda_1\geq\Lambda_2\geq\cdots
 \geq\Lambda_{\ell(\Lambda)}.
\eea
In addition, the subsequence consisting of the half-integer
components is strictly decreasing. We also denote the set of super partitions as $\mathsf{SP}$. We illustrate the super partition as a super Young diagram as
\bea\label{eq:superpartition}
\adjustbox{valign=c}{
\begin{tikzpicture}[
  scale=0.8,
  line width=0.45pt,
  line join=round
]

\begin{scope}

  \foreach \x in {0,1,2,3}{
    \FullBox{\x}{0}
  }

  \foreach \x in {0,1,2}{
    \FullBox{\x}{1}
  }

  \foreach \x in {0,1}{
    \FullBox{\x}{2}
  }
  \HalfBox{2}{2}

  \HalfBox{0}{3}

  \draw[-Triangle,thick] (0,0)--(5,0);
  \draw[-Triangle,thick] (0,0)--(0,5);

  \node[below] at (2,-0.5)
    {super partition};

\end{scope}

\begin{scope}[xshift=6cm]
 \draw[-Triangle,thick] (0,0)--(5,0);
    \draw[-Triangle,thick] (0,0)--(0,5);
    \draw (0,1)--(4,1);
    \foreach \x in {0,1,2,3} \filldraw[fill= blue!80!white, draw=black] (\x,0)--(\x,1)--(\x+1,0)--cycle;
    \foreach \x in {0,1,2,3} \filldraw[fill= red!80!white, draw=black] (\x,1)--(\x+1,1)--(\x+1,0)--cycle;
    \foreach \x in {1,2,3,4} \draw (\x,1)--(\x,0);
    \foreach \x in {0,1,2,3} \draw (\x,1)--(\x+1,0);

    \draw (0,2)--(3,2);
    \foreach \x in {0,1,2} \filldraw[fill= blue!80!white, draw=black] (\x,1)--(\x,2)--(\x+1,1)--cycle;
    \foreach \x in {0,1,2} \filldraw[fill= red!80!white, draw=black] (\x,2)--(\x+1,2)--(\x+1,1)--cycle;
    \foreach \x in {1,2,3} \draw (\x,2)--(\x,1);
    \foreach \x in {0,1,2} \draw (\x,2)--(\x+1,1);

    \draw (0,3)--(2,3);
    \foreach \x in {0,1,2} \filldraw[fill= blue!80!white, draw=black] (\x,2)--(\x,3)--(\x+1,2)--cycle;
    \foreach \x in {0,1} \filldraw[fill= red!80!white, draw=black] (\x,3)--(\x+1,3)--(\x+1,2)--cycle;
    \foreach \x in {1,2} \draw (\x,3)--(\x,2);
    \foreach \x in {0,1,2} \draw (\x,3)--(\x+1,2);

    \filldraw[fill= blue!80!white, draw=black] (0,4)--(1,3)--(0,3)--cycle;

  \node[below,align=center] at (2,-0.5)
    {Fully triangulated\\super partition};

\end{scope}
\end{tikzpicture}}\qquad \Lambda=\left(\frac{7}{2},3,\frac{5}{2},1\right)
\eea
where the right figure is the fully triangulated super partition. In later discussions, we will not distinguish both of them. Given the triangulation, the super partition $\Lambda$ can be also understood as a pair of partitions $\Lambda=(\lambda^{(1)},\lambda^{(2)})$ such that $\lambda^{(1)}$ is understood as the collection of the triangles $\BlueLowerTri$, while $\lambda^{(2)}$ is understood as the collection of the triangles $\RedUpperTri$. In particular, we can associate with $\Lambda$ a pair of partitions
\[
 \Lambda
 \mapsto 
 \left(\lambda^{(1)},\lambda^{(2)}\right),
 \qquad
 \lambda^{(1)}_i=\left\lceil\Lambda_i\right\rceil,
 \qquad
 \lambda^{(2)}_i=\left\lfloor\Lambda_i\right\rfloor.
\]
For the above example \eqref{eq:superpartition}, we have
\bea
\Lambda=(\lambda^{(1)},\lambda^{(2)}),\quad \lambda^{(1)}=(4,3,3,1),\quad \lambda^{(2)}=(3,3,2,1).
\eea
Conversely, for a pair of partitions $(\lambda^{(1)},\lambda^{(2)})$  satisfying
\[
 \lambda^{(1)}_i
 \geq
 \lambda^{(2)}_i
 \geq
 \lambda^{(1)}_{i+1},
 \qquad
 \lambda^{(1)}_i-\lambda^{(2)}_i\in\{0,1\}.
\]
we can associate a super partition $\Lambda$. The size of the super partition is defined as
\bea
|\Lambda|_{1}=|\lambda^{(1)}|,\quad |\Lambda|_2=|\lambda^{(2)}|,\quad |\Lambda|=2\sum_{i}\Lambda_{i}=|\lambda^{(1)}|+|\lambda^{(2)}|,
\eea
where $|\Lambda|_{1,2}$ counts the number of triangles $\BlueLowerTri,\RedUpperTri$, respectively and $|\Lambda|$ counts the total number of triangles. The coordinates of the triangles $\BlueLowerTri,\RedUpperTri$ are induced from the coordinates of the pair of partitions $(\lambda^{(1)},\lambda^{(2)})$.

A super plane partition is a pair of finitely supported arrays
\bea
 \Pi=\left(\pi^{(1)},\pi^{(2)}\right),\quad \pi^{(a)}
 =
 \left\{
 \pi^{(a)}_{i,j}\in\mathbb Z_{\geq0}
 \right\}_{i,j\geq1}
\eea
satisfying
\bea\label{eq:superplanepartition}
 \pi^{(1)}_{i,j}
 \geq
 \pi^{(2)}_{i,j}
 \geq
 \max\left\{
 \pi^{(1)}_{i+1,j},
 \pi^{(1)}_{i,j+1}
 \right\}.
\eea
From this condition, $\pi^{(a)}$ are automatically plane partitions. The set of super plane partitions is denoted as $\mathsf{SPP}$. We can illustrate the super plane partition as a three dimensional stack of triangular prisms as
\bea\label{eqfig:superplanepartition}
\adjustbox{valign=c}{\tdplotsetmaincoords{60}{125}
\begin{tikzpicture}[scale=0.7,tdplot_main_coords]
\BluePrism{0}{0}{0}
\RedPrism{0}{0}{0}
\BluePrism{0}{1}{0}
\RedPrism{0}{1}{0}
\BluePrism{0}{2}{0}
\RedPrism{0}{2}{0}
\BluePrism{0}{3}{0}

\BluePrism{1}{0}{0}
\RedPrism{1}{0}{0}
\BluePrism{1}{1}{0}
\RedPrism{1}{1}{0}
\BluePrism{1}{2}{0}

\BluePrism{2}{0}{0}
\RedPrism{2}{0}{0}

\BluePrism{3}{0}{0}

\BluePrism{0}{0}{1}
\RedPrism{0}{0}{1}
\BluePrism{0}{1}{1}
\RedPrism{0}{1}{1}
\BluePrism{1}{0}{1}
\RedPrism{1}{0}{1}
\BluePrism{2}{0}{1}

\BluePrism{0}{0}{2}
\RedPrism{0}{0}{2}
\BluePrism{0}{0}{3}
\RedPrism{0}{0}{3}

\BluePrism{0}{1}{2}
\RedPrism{0}{1}{2}

\BluePrism{0}{2}{1}

\BluePrism{1}{0}{2}
\RedPrism{1}{0}{2}

\BluePrism{0}{0}{4}

\RedPrism{0}{3}{0}

\end{tikzpicture}}
\eea
where compared to the plane partition, each unit cube is divided into two triangular prisms colored in blue and red:
\bea
\adjustbox{valign=c}{
\tdplotsetmaincoords{60}{125}
\begin{tikzpicture}[xshift=-7cm,tdplot_main_coords]
    \draw[-Triangle] (0,0,0)--(0.5,0,0);
    \draw[-Triangle] (0,0,0)--(0,0.5,0);
    \draw[-Triangle] (0,0,0)--(0,0,0.4);
    \node[below left,scale=0.8] at (0.5,0,0){$1$};
    \node[below right,scale=0.8] at (0,0.5,0){$2$};
    \node[above,scale=0.8] at (0,0,0.4){$3$};
\end{tikzpicture}
\begin{tikzpicture}[scale=1,tdplot_main_coords]
\draw [thick] (0,0,1)--(0,1,1)--(1,1,1)--(1,0,1)--cycle;
\draw[thick] (1,0,0)--(1,1,0)--(0,1,0);
\draw[thick] (1,0,1)--(1,0,0);
\draw[thick] (1,1,1)--(1,1,0);
\draw[thick] (0,1,1)--(0,1,0);
\end{tikzpicture}}\qquad  \qquad 
\adjustbox{valign=c}{
\tdplotsetmaincoords{60}{125}
\begin{tikzpicture}[xshift=-7cm,tdplot_main_coords]
    \draw[-Triangle] (0,0,0)--(0.5,0,0);
    \draw[-Triangle] (0,0,0)--(0,0.5,0);
    \draw[-Triangle] (0,0,0)--(0,0,0.4);
    \node[below left,scale=0.8] at (0.5,0,0){$1$};
    \node[below right,scale=0.8] at (0,0.5,0){$2$};
    \node[above,scale=0.8] at (0,0,0.4){$3$};
\end{tikzpicture}
\begin{tikzpicture}[scale=1,tdplot_main_coords]
\filldraw[fill=blue!70!white] (0,1,1)--(0,0,1)--(1,0,1)--cycle;
\filldraw[fill=red!70!white] (1,0,1)--(1,1,1)--(0,1,1)--cycle;
\filldraw[fill=red!70!white] (1,0,1)--(1,0,0)--(1,1,0)--(1,1,1)--cycle;
\filldraw[fill=red!70!white] (1,1,1)--(1,1,0)--(0,1,0)--(0,1,1)--cycle;
\draw[thick] (0,0,1)--(0,1,1)--(1,1,1)--(1,0,1)--cycle;
\draw[thick] (1,0,0)--(1,1,0)--(0,1,0);
\draw[thick] (1,0,1)--(1,0,0);
\draw[thick] (1,1,1)--(1,1,0);
\draw[thick] (0,1,1)--(0,1,0);
\draw[thick] (1,0,1)--(0,1,1);
\end{tikzpicture}}
\eea
Obviously, one can see that each layer is a super partition and the super plane partition is a nested sequence of super partitions. For each unit cube, we assign a lattice coordinate $(i,j,k)\in\mathbb Z_{>0}^{3}$ and both prisms inherit the
coordinate $(i,j,k)$ of the parent cube and are distinguished
only by their colors. Note that the blue/red prism with coordinate $(i,j,k)$ obeys $(i,j,k)\in\pi^{(1,2)}$. The size of the super plane partition is defined as
\bea
|\Pi|_{1}=|\pi^{(1)}|,\quad |\Pi|_2=|\pi^{(2)}|,\quad |\Pi|=|\Pi|_{1}+|\Pi|_{2}
\eea
where $|\Pi|_{1,2}$ is the number of blue and red prisms, respectively.


\begin{theorem}
The non-vanishing JK poles of the BPS partition function in Def.~\ref{def:D6_m2} are in one-to-one
correspondence with super plane partitions.  Let
\bea
 \Pi=(\pi^{(1)},\pi^{(2)})\in\mathsf{SPP}
\eea
be a super plane partition.  Then, up to permutations, the
corresponding pole configuration is
\bea
 \left\{\phi_I^{(1)}\right\}_{I=1}^{k_1}
 &=
 \left\{
 \fra
 +(\eps_1+\eps_4)(i-1)
 +(\eps_1+\eps_3)(j-1)
 +\eps_5(k-1)
 \mathrel{}\middle|\mathrel{}
 (i,j,k)\in\pi^{(1)}
 \right\},
 \\
 \left\{\phi_I^{(2)}\right\}_{I=1}^{k_2}
 &=
 \left\{
 \fra+\eps_1
 +(\eps_1+\eps_4)(i-1)
 +(\eps_1+\eps_3)(j-1)
 +\eps_5(k-1)
 \mathrel{}\middle|\mathrel{}
 (i,j,k)\in\pi^{(2)}
 \right\},
\eea
where
\bea
 k_1=|\Pi|_1=|\pi^{(1)}|,
 \qquad
 k_2=|\Pi|_2=|\pi^{(2)}|.
\eea
\end{theorem}
Starting from the blue prism in the corner cube $(1,1,1)$,
moving one unit in the $i$-, $j$-, and $k$-directions multiplies
the equivariant weight by
\[\adjustbox{valign=c}{\begin{tikzpicture}
    \draw[-Triangle] (0,0)--(0,1);
    \draw[-Triangle] (0,0)--(-1,-0.7);
    \draw[-Triangle] (0,0)--(1,-0.7);
    \node[left] at (-1,-0.7){$\eps_{1}+\eps_{4}$};
    \node[right] at (1,-0.7){$\eps_{1}+\eps_{3}$};
    \node[above] at (0,1){$\eps_{5}$};
\end{tikzpicture}}\qquad \qquad
 p_1p_4,\qquad p_1p_3,\qquad p_5,
\]
respectively. Inside each cube, the weight of the red prism is
$p_1$ times that of the blue prism. Up to level three, we have {
\bea
 \mathcal{Z}^{\D6\tbar\D0}(\mathfrak{m}_{2};\mathfrak{q}_{1},\mathfrak{q}_{2};p_{1,2,3,4,5})&=1\\
 &+\frac{\sqrt{p_{5}}}{1-p_{5}}\fq_1\qquad \adjustbox{valign=c}{\tdplotsetmaincoords{60}{125}
\begin{tikzpicture}[scale=0.5,tdplot_main_coords]
\BluePrism{0}{0}{0}
\end{tikzpicture}}\\
&+\frac{p_{5}^{3/2}}{(1-p_{5})(1-p_{5}^{2})}\fq_1^{2}\qquad \tdplotsetmaincoords{60}{125}
\adjustbox{valign=c}{\begin{tikzpicture}[scale=0.5,tdplot_main_coords]
\BluePrism{0}{0}{0}
\BluePrism{0}{0}{1}
\end{tikzpicture}}\\
&+\frac{\left(p_1^2 p_3 p_4-1\right) \left(p_1 p_3 p_5-1\right) \left(p_1 p_4 p_5-1\right)
   }{p_1 \sqrt{p_3} \left(p_1 p_3-1\right) \sqrt{p_4} \left(p_1 p_4-1\right)
   \left(p_5-1\right) \sqrt{p_5}} \fq_{1}\fq_{2}\qquad  \tdplotsetmaincoords{60}{125}
\adjustbox{valign=c}{\begin{tikzpicture}[scale=0.5,tdplot_main_coords]
\BluePrism{0}{0}{0}
\RedPrism{0}{0}{0}
\end{tikzpicture}}\\
&+\frac{p_{5}^{3}}{(1-p_{5})(1-p_{5}^{2})(1-p_{5}^{3})}\fq^{3}_{1}\qquad 
\adjustbox{valign=c}{\tdplotsetmaincoords{60}{125}\begin{tikzpicture}[scale=0.5,tdplot_main_coords]
\BluePrism{0}{0}{0}
\BluePrism{0}{0}{1}
\BluePrism{0}{0}{2}
\end{tikzpicture}}\\
&+\fq_{1}^{2}\fq_{2}\left(\frac{\left(p_1^2 p_3 p_4-p_5\right) \sqrt{p_5} \left(p_1 p_3 p_5-1\right) \left(p_1 p_4
   p_5-1\right)}{p_1 \sqrt{p_3} \sqrt{p_4} \left(p_1 p_3-p_5\right) \left(p_1
   p_4-p_5\right) \left(p_5-1\right){}^2}\qquad \right.\adjustbox{valign=c}{\tdplotsetmaincoords{60}{125}\begin{tikzpicture}[scale=0.5,tdplot_main_coords]
\BluePrism{0}{0}{0}
\BluePrism{0}{0}{1}
\RedPrism{0}{0}{0}
\end{tikzpicture}}\\
&-\frac{\left(p_1^2 p_3 p_4-1\right) \left(p_1 p_3 p_5-1\right) \left(p_3-p_4
   p_5\right)}{\sqrt{p_1} \left(p_1 p_3-1\right) \left(p_3-p_4\right) \sqrt{p_4}
   \left(p_1 p_3-p_5\right) \left(p_5-1\right)}\qquad \adjustbox{valign=c}{\tdplotsetmaincoords{60}{125}\begin{tikzpicture}[scale=0.5,tdplot_main_coords]
\BluePrism{0}{0}{0}
\RedPrism{0}{0}{0}
\BluePrism{0}{1}{0}
\end{tikzpicture}}\\
&-\left.\frac{\left(p_1^2 p_3 p_4-1\right) \left(p_3 p_5-p_4\right) \left(p_1 p_4
   p_5-1\right)}{\sqrt{p_1} \sqrt{p_3} \left(p_3-p_4\right) \left(p_1 p_4-1\right)
   \left(p_1 p_4-p_5\right) \left(p_5-1\right)}\right)\qquad \adjustbox{valign=c}{\tdplotsetmaincoords{60}{125}\begin{tikzpicture}[scale=0.5,tdplot_main_coords]
\BluePrism{0}{0}{0}
\RedPrism{0}{0}{0}
\BluePrism{1}{0}{0}

\end{tikzpicture}}\\
&+\cdots 
\eea}
It is indeed the result obtained by setting $\mu=p_{2}$ in \eqref{eq:D81D82}, where after tuning the parameter, some terms vanish.

\subsection{General tetrahedron instanton configurations}\label{sec:tetrahedron-instanton}
We now combine the elementary D6-brane systems studied in the
previous subsections and consider their higher rank generalizations
and general tetrahedron instanton configurations.

For each brick matching $\mathfrak{m}_{i}$ $(i\in\five)$ and $a\in\mathcal A_i$, we introduce a flavor space of rank $
N_{\mathfrak m_i}^{(a)}$. The total number of D6-branes wrapping the toric divisor $D_{\mathfrak m_i}$ is defined by
\bea
N_{\mathfrak m_i}
&:=
\sum_{a\in\mathcal A_i}
N_{\mathfrak m_i}^{(a)}.
\eea
In particular,
\bea
N_{\mathfrak m_0}
&=
N_{\mathfrak m_0}^{(1)}
+
N_{\mathfrak m_0}^{(2)},
\quad 
N_{\mathfrak m_1}
=
N_{\mathfrak m_1}^{(2)},
\qquad
N_{\mathfrak m_2}
=
N_{\mathfrak m_2}^{(1)},
\quad 
N_{\mathfrak m_3}
=
N_{\mathfrak m_3}^{(2)},
\qquad
N_{\mathfrak m_4}
=
N_{\mathfrak m_4}^{(1)}.
\label{eq:tetrahedron-framing-ranks}
\eea
For $i=1,\ldots,4$, we omit the gauge-node superscript whenever
there is no ambiguity.

We denote the flavor parameters of the individual D6-branes by
\bea
\mathfrak a_{\mathfrak m_i,\alpha}^{(a)},
\qquad
a\in\mathcal A_i,
\qquad
\alpha
=
1,\ldots,N_{\mathfrak m_i}^{(a)}.
\eea
Here $i$ labels the toric divisor, $a$ labels the gauge node to which
the framing is attached, and $\alpha$ labels an individual D6-brane
within the corresponding stack. We collectively denote the rank and
flavor data by
\bea
\boldsymbol{ N_{\mathfrak{m}}}
&:=
\left(
N_{\mathfrak m_i}^{(a)}
\right)_{
\substack{
i=0,\ldots,4\\
a\in\mathcal A_i
}},
\quad
\boldsymbol{\mathfrak a_{\mathfrak{m}}}
:=
\left(
\mathfrak a_{\mathfrak m_i,\alpha}^{(a)}
\right)_{
\substack{
i=0,\ldots,4,\;
a\in\mathcal A_i\\
\alpha=1,\ldots,N_{\mathfrak m_i}^{(a)}
}}.
\eea

Adding each stack of D6-branes introduce an $\mathcal{N}=2$ chiral field $\mathsf{I}^{(a)}_{\mathfrak{m}_{i}}$ and an $\mathcal{N}=2$ Fermi field $\Lambda_{\mathfrak{m}_{i}}^{(a)}$. The quiver diagram of the most generic tetrahedron instanton system is 
\bea
\begin{tikzpicture}[scale=0.8,
  >=Triangle,
  gauge/.style={
    circle,
    draw=#1!80!black,
    fill=#1!15,
    very thick,
    minimum size=13mm,
    inner sep=1pt
  },
  flavor/.style={
    rectangle,
    draw=black,
    rounded corners=2pt,
    thick,
    minimum width=20mm,
    minimum height=9mm,
    fill=black!5!white
  },
  chiral/.style={
    -{Triangle[length=2.1mm,width=1.6mm]},
    black,
    thick
  },
  fermi/.style={
    -{Triangle[length=2.1mm,width=1.6mm]},
    red!100!black,
    thick,
    preaction={draw=white,line width=3.5pt}
  },
  twotips/.style={
    decoration={
      markings,
      mark=at position .47 with {
        \arrow{Triangle[length=2.1mm,width=1.6mm]}
      },
      mark=at position .58 with {
        \arrow{Triangle[length=2.1mm,width=1.6mm]}
      }
    },
    postaction={decorate}
  },
  bulkchiral/.style={
    black,
    thick,
    twotips
  },
  bulkfermi/.style={
    red!100!black,
    thick,
    preaction={draw=white,line width=3.5pt},
    twotips
  },
  every node/.style={
    font=\small
  }
]


\node[gauge=blue] (K1) at (-2.35,0) {$k_1$};
\node[gauge=red]  (K2) at ( 2.35,0) {$k_2$};


\draw[chiral]
  (K1.145)
  .. controls +(-1.35,1.00) and +(-1.35,-1.00) ..
  node[midway,left] {$\mathsf C_1$}
  (K1.215);

\draw[chiral]
  (K2.35)
  .. controls +(1.35,1.00) and +(1.35,-1.00) ..
  node[midway,right] {$\mathsf C_2$}
  (K2.325);

%


\draw[bulkchiral]
  ([yshift=4pt]K1.east)
  --
  ([yshift=4pt]K2.west);


\draw[bulkchiral]
  ([yshift=-4pt]K2.west)
  --
  ([yshift=-4pt]K1.east);


\draw[bulkfermi]
  ([yshift=13pt]K1.east)
  --
  ([yshift=13pt]K2.west);


\draw[bulkfermi]
  ([yshift=-13pt]K2.west)
  --
  ([yshift=-13pt]K1.east);


\node[flavor] (F01) at (-2.35,3.35)
  {$N_{\mathfrak m_0}^{(1)}$};

\node[flavor] (F02) at (2.35,3.35)
  {$N_{\mathfrak m_0}^{(2)}$};


\draw[chiral]
  ([xshift=-4pt]F01.south)
  --
  node[left]
  {$\mathsf I_{\mathfrak m_0}^{(1)}$}
  ([xshift=-4pt]K1.north);

\draw[fermi]
  ([xshift=4pt]F01.south)
  --
  node[right,text=red!100!black]
  {$\Lambda_{\mathfrak m_0}^{(1)}$}
  ([xshift=4pt]K1.north);


\draw[chiral]
  ([xshift=-4pt]F02.south)
  --
  node[left]
  {$\mathsf I_{\mathfrak m_0}^{(2)}$}
  ([xshift=-4pt]K2.north);

\draw[fermi]
  ([xshift=4pt]F02.south)
  --
  node[right,text=red!100!black]
  {$\Lambda_{\mathfrak m_0}^{(2)}$}
  ([xshift=4pt]K2.north);


\node[flavor] (F2) at (-5.35,-4.75)
  {$N_{\mathfrak m_2}$};

\node[flavor] (F4) at (-2.35,-4.20)
  {$N_{\mathfrak m_4}$};

\node[flavor] (F1) at (2.35,-4.20)
  {$N_{\mathfrak m_1}$};

\node[flavor] (F3) at (5.35,-4.75)
  {$N_{\mathfrak m_3}$};


\draw[chiral]
  (F2.north east)
  to[bend left=5]
  node[
    pos=.35,
    above left,
    fill=white,
    inner sep=1pt
  ]
  {$\mathsf I_{\mathfrak m_2}$}
  (K1.south west);

\draw[chiral]
  (F4.north)
  to[bend right=5]
  node[
    pos=.48,
    left,
    fill=white,
    inner sep=1pt
  ]
  {$\mathsf I_{\mathfrak m_4}$}
  (K1.south);

\draw[chiral]
  (F1.north)
  to[bend left=5]
  node[
    pos=.48,
    right,
    fill=white,
    inner sep=1pt
  ]
  {$\mathsf I_{\mathfrak m_1}$}
  (K2.south);

\draw[chiral]
  (F3.north west)
  to[bend right=5]
  node[
    pos=.35,
    above right,
    fill=white,
    inner sep=1pt
  ]
  {$\mathsf I_{\mathfrak m_3}$}
  (K2.south east);

%

\draw[fermi]
  (F2.north)
  to[out=27,in=-145,looseness=.72]
  node[
    pos=.30,
    above,
    fill=white,
    inner sep=1pt,
    text=red!100!black
  ]
  {$\Lambda_{\mathfrak m_2}$}
  (K2.south west);

\draw[fermi]
  (F4.north east)
  to[out=52,in=-125,looseness=.70]
  node[
    pos=.30,
    below left,
    fill=white,
    inner sep=1pt,
    text=red!100!black
  ]
  {$\Lambda_{\mathfrak m_4}$}
  (K2.south west);

\draw[fermi]
  (F1.north west)
  to[out=128,in=-55,looseness=.70]
  node[
    pos=.30,
    below right,
    fill=white,
    inner sep=1pt,
    text=red!100!black
  ]
  {$\Lambda_{\mathfrak m_1}$}
  (K1.south east);

\draw[fermi]
  (F3.north)
  to[out=153,in=-35,looseness=.72]
  node[
    pos=.30,
    above,
    fill=white,
    inner sep=1pt,
    text=red!100!black
  ]
  {$\Lambda_{\mathfrak m_3}$}
  (K1.south east);

\end{tikzpicture}
\eea

We additionally need potential terms to the theory. Recall that the $E$-term does not modify the potential terms coming only from the fractional D0-branes and the traceless condition \eqref{eq:JE-traceless} is automatically satisfied. Therefore, the potential terms of the tetrahedron instanton system are just a collection of all of the $E$-terms in \eqref{eq:elementary-D6-framings}:
\bea
\begin{array}{c|cc}
 & \text{$J$-term}\quad & \quad \text{$E$-term}\\\hline
\Lambda^{(1)}_{\mathfrak{m}_0}     & 0 & \mathsf{C}_{1}\,\mathsf{I}_{\mathfrak{m}_{0}}^{(1)}\\
\Lambda^{(2)}_{\mathfrak{m}_0}     & 0 & \mathsf{C}_{2}\,\mathsf{I}_{\mathfrak{m}_{0}}^{(2)}\\
\Lambda_{\mathfrak{m}_1}     & 0 & \mathsf{B}_{1}\,\mathsf{I}_{\mathfrak{m}_{1}}\\
\Lambda_{\mathfrak{m}_2}     & 0 & \mathsf{A}_{2}\,\mathsf{I}_{\mathfrak{m}_{2}}\\
\Lambda_{\mathfrak{m}_3}     & 0 & \mathsf{B}_{2}\,\mathsf{I}_{\mathfrak{m}_{3}}\\
\Lambda_{\mathfrak{m}_4}     & 0 & \mathsf{A}_{1}\,\mathsf{I}_{\mathfrak{m}_{4}}\\
\end{array}
\eea

The flavor charges of the additional multiplets are inherited from
the corresponding elementary D6-brane systems. The total framing
contribution is therefore
\bea\label{eq:tetrahedron-framing-factor}
\mathcal{Z}_{\boldsymbol{ N_{\mathfrak{m}}}}^{\mathrm{D6-D0}}
\left(
\boldsymbol{\mathfrak a_{\mathfrak{m}}},\phi
\right)
&=
\prod_{i=0}^{4}
\prod_{a\in\mathcal A_i}
\prod_{\alpha=1}^{N_{\mathfrak m_i}^{(a)}}
\mathcal{Z}_{\mathfrak m_i,a}^{\mathrm{D6-D0}}
\left(
\mathfrak a_{\mathfrak m_i,\alpha}^{(a)},
\phi
\right),
\eea
where
$\mathcal{Z}_{\mathfrak m_i,a}^{\mathrm{D6-D0}}$
denotes the rank-one contribution derived in the previous
subsections. For $i=1,\ldots,4$, we omit the index $a$ because the
admissible framing node is unique.

\begin{definition}\label{def:tetrahedron-partition-function}
The tetrahedron instanton partition function on
$\mathcal{C}\times\mathbb{C}$ is then defined by
\bea
 \mathcal{Z}^{\mathrm{tet}}_{\boldsymbol{ N_{\mathfrak{m}}}}
 (\boldsymbol{\mathfrak{a}_{\mathfrak{m}}};\mathfrak{q}_{1},\mathfrak{q}_{2};
 p_{1,2,3,4,5})
 &=
 \sum_{k_{1}=0}^{\infty}
 \sum_{k_{2}=0}^{\infty}
 \mathfrak{q}_{1}^{k_{1}}
 \mathfrak{q}_{2}^{k_{2}}\,
 \mathcal{Z}^{\mathrm{tet}}_{k_{1},k_{2};
 \boldsymbol{N_{\mathfrak{m}}}}(\boldsymbol{\mathfrak{a}_{\mathfrak{m}
 }}),
 \\
 \mathcal{Z}^{\mathrm{tet}}_{k_{1},k_{2};
 \boldsymbol{ N_{\mathfrak{m}}}}(\boldsymbol{\mathfrak{a}_{\mathfrak{m}}})
 &=
 \frac{1}{k_{1}!k_{2}!}
 \oint_{\mathrm{JK}}
 \prod_{a=1}^{2}
 \prod_{I=1}^{k_{a}}
 \frac{d\phi_{I}^{(a)}}{2\pi i}\,
 \mathcal{Z}^{\mathrm{D6-D0}}_{\boldsymbol{N_{\mathfrak{m}}}}
 (\boldsymbol{\mathfrak{a}_{\mathfrak{m}}},\phi)\,\mathcal{Z}^{\mathrm{D0-D0}}(\phi)\,.
 \label{eq:tetrahedron-partition-function}
\eea
\end{definition}

For generic values of the flavor parameters
$\boldsymbol{\mathfrak a}$, every pole selected in the cyclic
chamber is generated from one of the framing poles
\bea
\mathfrak a_{\mathfrak m_i,\alpha}^{(a)},
\qquad
a\in\mathcal A_i,
\qquad
\alpha=1,\ldots,N_{\mathfrak m_i}^{(a)}.
\eea
Accordingly, a fixed point is labeled by a collection of crystal
configurations
\bea\label{eq:tetrahedron-fixed-points}
\boldsymbol{{C}_{\mathfrak{m}}}
=
\left(
{C}_{\mathfrak m_i,\alpha}^{(a)}
\right)_{
\substack{
i=0,\ldots,4,\;
a\in\mathcal A_i\\
\alpha=1,\ldots,N_{\mathfrak m_i}^{(a)}
}},
\qquad
\mathcal{C}_{\mathfrak m_i,\alpha}^{(a)}
\in
\operatorname{Cryst}_{\mathfrak m_i}^{(a)}.
\eea
Here
$\operatorname{Cryst}_{\mathfrak m_i}^{(a)}$
denotes the set of elementary crystal configurations generated from
a framing pole attached to the gauge node $a$.

For $\mathfrak m_0$, there are two possible colors of the initial
atom, corresponding to the two admissible framing nodes. We denote
the corresponding sets of pyramid partitions by
\bea
\operatorname{Cryst}_{\mathfrak m_0}^{(1)}
=
\mathsf{Pyr}^{(1)},
\qquad
\operatorname{Cryst}_{\mathfrak m_0}^{(2)}
=
\mathsf{Pyr}^{(2)}.
\eea
For the remaining four brick matchings, the framing node is uniquely
fixed, and the elementary configurations are super plane partitions,
up to permutations of the colors and equivariant parameters:
\bea
\operatorname{Cryst}_{\mathfrak m_i}^{(a_i)}
\simeq
\mathsf{SPP}_{\mathfrak m_i},
\qquad
(a_1,a_2,a_3,a_4)=(2,1,2,1).
\eea
Thus, for generic flavor parameters, the fixed-point data of the
general tetrahedron-instanton system consist of one elementary
crystal configuration for each individual D6-brane. No additional
melting rule is required beyond those of the elementary systems
studied in the previous subsections.

The full contribution associated with
${{C}_{\mathfrak{m}}}$ is nevertheless not, in general, a product of
independent single-D6 contributions, since all crystal configurations
couple through the common fractional D0--D0 sector. Their interaction is automatically incorporated in the JK residue of
Def.~\ref{def:tetrahedron-partition-function}.

The above description assumes generic flavor parameters. If two
framing parameters are tuned such that their equivariant orbits
overlap, degenerate poles may appear. Such non-generic configurations
require a separate JK-residue analysis and will not be considered
here.

\section{Spiked instantons of \texorpdfstring{$\mathcal{C}\times \mathbb{C}$}{CC}}\label{sec:spikedinst}
Spiked instantons on $\mathbb{C}^{4}$ were introduced in
\cite{Nekrasov:2016qym,Nekrasov:2016gud} as moduli spaces associated
with intersecting D-brane configurations. Their cyclic-orbifold
generalizations and the resulting orbifold $qq$-characters were
developed in \cite{Nekrasov:2016qym,Nekrasov:2016ydq}; see also
\cite{Jeong:2021rll} for fractional $qq$-characters obtained from
orbifold gauge origami. A recent algebro-geometric construction of the
gauge-origami moduli spaces and their virtual invariants was given in
\cite{Arbesfeld:2026zrf}.

D4--D2--D0 systems in which a D4-brane wraps a toric divisor of a
toric Calabi--Yau threefold, including the conifold, have also been
studied using brane tilings in \cite{Nishinaka:2013mba}. However, an
analogous prescription for D4-brane framings in brane brick models has
not yet been systematically developed. In this section, we take a first
step in this direction by constructing such framings for
$\mathcal{C}\times\mathbb{C}$ using phase boundaries.

We first associate a framed D4-brane
system with each toric surface by choosing the corresponding phase
boundary in section~\ref{sec:D4-phaseboundaries}. We then study a single
D4-brane associated with each of the two inequivalent types of phase
boundaries in sections~\ref{sec:D4-phase-02}
and~\ref{sec:D4-phase-34}. Their torus-fixed points are described by
super partitions and $\mathbb{Z}_{2}$-colored Young diagrams,
respectively. Finally, we combine these elementary D4-brane sectors and
define the general spiked-instanton partition function in
section~\ref{sec:spiked-instanton}.

\subsection{D4-brane configurations from phase boundaries}\label{sec:D4-phaseboundaries}
In this section, we will consider the situation when have one non-compact D4-brane wrapping a four cycle in the $\mathcal{C}\times \mathbb{C}$. Let $\eta$ be a phase boundary associated with an edge of the toric diagram connected to two extremal lattice points and denote the corresponding toric surface $S_{\eta}$. We place a single D4-brane on $S_{\eta}$:
\bea  \label{eq:conifold-D4-toricsurface}
\renewcommand{\arraystretch}{1.05}
\begin{tabular}{|c|c|c|c|c|c|c|c|c|c|c|}
\hline
& \multicolumn{6}{c|}{$\mathcal{C}$} & \multicolumn{2}{c|}{$\mathbb{C}$} & \multicolumn{2}{c|}{$\mathbb{R}\times S^{1}$} \\
\cline{2-11} \raisebox{-0mm}{}  & 1 & 2 & 3 & 4& 5 & 6 & 7 & 8 & 9& 0\\[-2pt]\hline
 D0& $\bullet$ & $\bullet$  & $\bullet$  & $\bullet$  & $\bullet$  & $\bullet$   & $\bullet$  & $\bullet$  & $\bullet$   & $-$\\
\hline D4& \multicolumn{8}{c|}{$S_{\eta}$}& $\bullet$   & $-$\\
\hline
\end{tabular}
\eea
Since the toric surface is non-compact, this D4-brane will again play the role of a flavor brane from the viewpoint of the fractional D0-branes.

As a local model, let us first recall the D0--D4 system on $\mathbb{C}^{2}\subset \mathbb{C}^{4}$. In the language of the $\mathcal{N}=2$ quiver quantum mechanics, the D4-brane adds a framing node $\ast$, one fundamental $\mathcal{N}=2$ chiral $\mathsf{I}:\ast\rightarrow  1$, one anti-fundamental $\mathcal{N}=2$ chiral $\mathsf{J}:1\rightarrow \ast$, and two $\mathcal{N}=2$ Fermi multipelts $\Lambda_{1},\Lambda_{2}$, where $1$ here denotes the unique gauge node of the $\mathbb{C}^{4}$-theory. For a general toric CY$_4$ quiver, however, the four gauge nodes involved in the framing construction need not coincide.

The general local configuration relevant for the D4-brane framing is
of the following form:
\bea
\adjustbox{valign=c}{
\begin{tikzpicture}[
    scale=1.15,
    line cap=round,
    line join=round,
    chiral/.style={
        black,
        thick,
        postaction={decorate},
        decoration={
            markings,
            mark=at position 0.52 with {\arrow{Triangle}}
        }
    },
    fermi/.style={
        red!100!black,
        thick,
        postaction={decorate},
        decoration={
            markings,
            mark=at position 0.52 with {\arrow{Triangle}}
        }
    },
    qnode/.style={
        circle,
        draw=black,
        thick,
        fill=yellow!85!orange,
        minimum size=0.72cm,
        inner sep=0pt
    },
    box/.style={
        rectangle,
        draw=black,
        thick,
        fill=black!10!white,
        minimum width=0.65cm,
        minimum height=0.65cm,
        inner sep=0pt
    }
]

\node[box] (f) at (0,3.2) {$*$};

\node[qnode] (a) at (-3.0,0) {$a$};
\node[qnode] (b) at ( 3.0,0) {$b$};

\node[qnode] (c) at (-0.8, 1.05) {$c$};
\node[qnode] (d) at ( 0.8,-1.05) {$d$};

\draw[chiral]
    ($(f.south west)+(0.08,-0.02)$)
    --
    ($(a.north)+(0,0.02)$)
    node[
        pos=0.48,
        above left=1pt,
        fill=white,
        inner sep=0.5pt
    ]
    {$\mathsf I$};

\draw[chiral]
    ($(b.north)+(0,0.02)$)
    --
    ($(f.south east)+(-0.08,-0.02)$)
    node[
        pos=0.50,
        above right=1pt,
        fill=white,
        inner sep=0.5pt
    ]
    {$\mathsf J$};

\draw[fermi]
    ($(f.south)+(-0.12,0)$)
    to[bend right=9]
    ($(c.north)+(0,0.02)$)
    node[
        pos=0.53,
        left=3pt,
        fill=white,
        inner sep=0.5pt,
        text=red!100!black
    ]
    {$ $};

\draw[chiral]
    (a)--(c)
    node[
        pos=0.46,
        above left=1pt,
        fill=white,
        inner sep=0.5pt
    ]
    {$\mathsf X$};

\draw[chiral]
    (c)--(b)
    node[
        pos=0.52,
        above right=1pt,
        fill=white,
        inner sep=0.5pt
    ]
    {$\mathsf U$};

\draw[chiral]
    (a)--(d)
    node[
        pos=0.47,
        below left=1pt,
        fill=white,
        inner sep=0.5pt
    ]
    {$\mathsf Y$};

\draw[chiral]
    (d)--(b)
    node[
        pos=0.52,
        below right=1pt,
        fill=white,
        inner sep=0.5pt
    ]
    {$\mathsf V$};

\draw[fermi]
    ($(b.west)+(-0.02,0)$)
    --
    ($(a.east)+(0.02,0)$)
    node[
        pos=0.50,
        below=3pt,
        fill=white,
        inner sep=1.5pt,
        text=red!100!black
    ]
    {$ $};

\draw[
    white,
    line width=4.5pt
]
    ($(d.north)+(0.05,0)$)
    to[bend right=15]
    ($(f.south)+(0.15,0)$);

\draw[fermi]
    ($(d.north)+(0.05,0)$)
    to[bend right=15]
    ($(f.south)+(0.15,0)$)
    node[
        pos=0.9,
        right=3pt,
        fill=white,
        inner sep=0.1pt,
        text=red!100!black
    ]
    {$ $};

\node at (0,-0.2) {\textcolor{red}{$\Lambda_{b\rightarrow a}$}};

\node at (0.9,1.7) {\textcolor{red}{$\Lambda_{2}$}};

\node at (-0.3,1.7) {\textcolor{red}{$\Lambda_{1}$}};

\node[box]  at (0,3.2) {$ $};
\node[qnode]  at ( 0.8,-1.05) {$d$};

\end{tikzpicture}
}\label{eq:D4-general-framing}
\eea
where $\mathsf{X},\mathsf{Y},\mathsf{U},\mathsf{V}$ are possibly composite paths of chiral fields and $\Lambda_{b\rightarrow a}$ is a Fermi field of the original CY$_{4}$ quiver,  while
\bea
\Lambda_{1},
\qquad
\Lambda_{2}
\eea
are the two framing Fermi multiplets introduced by the D4-brane. Since in the computations, the orientations of the framing Fermi multiplets are non-essential, we choose the orientations of the two framing
Fermi multiplets as
\bea
\Lambda_{1}:*\longrightarrow c,
\qquad
\Lambda_{2}:d\longrightarrow *.
\label{eq:D4-framing-Fermi-orientations}
\eea
for convenience. Note that their conjugate representatives have the opposite orientations.

Once an orientation of a Fermi edge is chosen, its $J$- and $E$-terms
can be read from the closed plaquettes containing that edge.  For the first Fermi multiplet $\Lambda_{1}$, the upper part of \eqref{eq:D4-general-framing} contains two chiral paths:
\bea
\mathsf{X}\mathsf{I}:*\longrightarrow c,
\qquad
\mathsf{J}\mathsf{U}:c\longrightarrow *.
\eea
They therefore determine
\bea
E_{\Lambda_{1}}
=
\mathsf{X}\mathsf{I},
\qquad
J_{\Lambda_{1}}
=
\mathsf{J}\mathsf{U}.
\label{eq:D4-framing-Lambda1}
\eea
Indeed, they are the two closed plaquettes associated with this framing Fermi edge.

Similarly, for
$\Lambda_{2}:d\rightarrow *$, the lower part of the figure contains
the oppositely oriented chiral paths
\bea
\mathsf{J}\mathsf{V}:d\longrightarrow *,
\qquad
\mathsf{Y}\mathsf{I}:*\longrightarrow d.
\eea
Consequently, we take
\bea
E_{\Lambda_{2}}
=
-\mathsf{J}\mathsf{V},
\qquad
J_{\Lambda_{2}}
=
\mathsf{Y}\mathsf{I}.
\label{eq:D4-framing-Lambda2}
\eea
Thus,
$\overline{\Lambda}_{2}\mathsf{J}\mathsf{V}$ and
$\Lambda_{2}\mathsf{Y}\mathsf{I}$ form the corresponding closed
plaquettes. The relative minus sign is fixed by the $\mathcal{N}=2$ supersymmetry
constraint \eqref{eq:JE-traceless}.

Unlike a Fermi multiplet of the unframed toric CY$_4$ quiver, a
framing Fermi multiplet need not satisfy the toric condition by
itself. In particular, $\Lambda_{1}$ and $\Lambda_{2}$ need not each
appear in four distinct minimal plaquettes. It is sufficient that
their $J$- and $E$-terms are gauge-covariant paths with the required
endpoints and that the complete framed quiver satisfies \eqref{eq:JE-traceless}.

The two chiral paths connecting $a$ and $b$ are
\bea
\mathsf{U}\mathsf{X}:a\longrightarrow b,
\qquad
\mathsf{V}\mathsf{Y}:a\longrightarrow b.
\eea
They form the two monomials in one of the $J$- or $E$-polynomials of
the bulk Fermi multiplet $\Lambda_{b\rightarrow a}$. With the
orientation shown in \eqref{eq:D4-general-framing}, we write
this polynomial as
\bea
J_{\Lambda_{b\rightarrow a}}
=
-\mathsf{U}\mathsf{X}
+\mathsf{V}\mathsf{Y}.
\label{eq:D4-bulk-Fermi-relation}
\eea
The D4-brane framing deforms the complementary polynomial by
\bea
E_{\Lambda_{b\rightarrow a}}
=
E_{\Lambda_{b\rightarrow a}}^{(0)}
+\mathsf{I}\mathsf{J},
\label{eq:D4-IJ-deformation}
\eea
where $E_{\Lambda_{b\rightarrow a}}^{(0)}$ is the $E$-term of the
unframed quiver. The complete local framing data are therefore
summarized as
\bea
\begin{array}{c|cc}
&
\text{$J$-term}
&
\text{$E$-term}
\\ \hline
\Lambda_{*\rightarrow c}
&
\mathsf{J}\mathsf{U}
&
\mathsf{X}\mathsf{I}
\\
\Lambda_{d\rightarrow *}
&
\mathsf{Y}\mathsf{I}
&
-\mathsf{J}\mathsf{V}
\\
\Lambda_{b\rightarrow a}
&
-\mathsf{U}\mathsf{X}
+\mathsf{V}\mathsf{Y}
&
E_{\Lambda_{b\rightarrow a}}^{(0)}
+\mathsf{I}\mathsf{J}
\end{array}.
\label{eq:general-D4-framing-JE}
\eea
Note that the additional fields and potential terms coming from the framing nodes obey \eqref{eq:JE-traceless} themselves.

If the polynomial
$-\mathsf{U}\mathsf{X}+\mathsf{V}\mathsf{Y}$ appears as an
$E$-term in the globally chosen Fermi orientation, the same
construction is obtained by exchanging $J$ and $E$ and reversing the
corresponding Fermi representatives.

\paragraph{Phase boundaries and D4-branes framings}
The local quiver in \eqref{eq:D4-general-framing} does not by
itself determine the gauge nodes $a,b,c,d$ or the chiral paths
$\mathsf{X},\mathsf{Y},\mathsf{U},\mathsf{V}$. We now propose that a D4-brane wrapping the toric surface $S_{\eta}$ associated with the phase boundary $\eta$ is determined by the phase boundary matrix as follows.


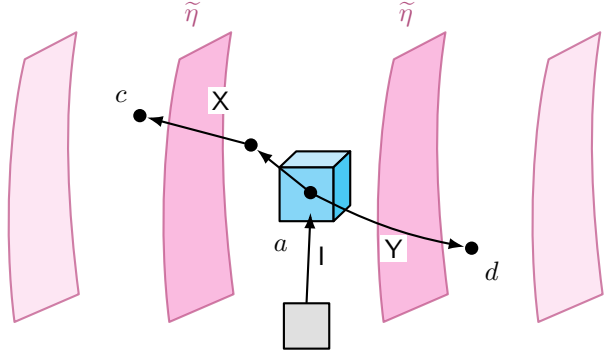
\begin{figure}[t]
\centering
\begin{tikzpicture}[
    scale=1.05,
    line cap=round,
    line join=round,
    chiral/.style={
        black,
        thick,
        -{Latex[length=2.1mm,width=1.5mm]}
    },
    gaugenode/.style={
        circle,
        fill=black,
        inner sep=0pt,
        minimum size=5pt
    },
    flavornode/.style={
        rectangle,
        draw=black,
        thick,
        fill=black!12,
        minimum width=6mm,
        minimum height=6mm,
        inner sep=0pt
    }
]


\foreach \sx/\op in {
    -3.30/0.10,
    -1.35/0.27,
     1.35/0.27,
     3.30/0.10
}{
    \begin{scope}[shift={(\sx,0)}]

    \fill[
        magenta,
        opacity=\op
    ]
        (-0.38,-1.55)
        .. controls (-0.52,-0.45) and (-0.45,0.95) ..
        (-0.25,1.76)
        --
        (0.40,2.10)
        .. controls (0.22,1.02) and (0.30,-0.20) ..
        (0.43,-1.20)
        --
        cycle;

    \draw[
        magenta!75!black,
        thick,
        opacity=0.72
    ]
        (-0.38,-1.55)
        .. controls (-0.52,-0.45) and (-0.45,0.95) ..
        (-0.25,1.76)
        --
        (0.40,2.10)
        .. controls (0.22,1.02) and (0.30,-0.20) ..
        (0.43,-1.20)
        --
        cycle;

    \end{scope}
}

\node[
    text=magenta!70!black,
    fill=white,
    inner sep=1pt
]
    at (-1.45,2.33)
    {$\widetilde{\eta}$};

\node[
    text=magenta!70!black,
    fill=white,
    inner sep=1pt
]
    at (1.25,2.33)
    {$\widetilde{\eta}$};


\filldraw[
    fill=cyan!42,
    draw=black,
    thick
]
    (-0.34,-0.28)
    rectangle
    (0.34,0.40);

\filldraw[
    fill=cyan!22,
    draw=black,
    thick
]
    (-0.34,0.40)
    --
    (-0.13,0.60)
    --
    (0.55,0.60)
    --
    (0.34,0.40)
    --
    cycle;

\filldraw[
    fill=cyan!58,
    draw=black,
    thick
]
    (0.34,-0.28)
    --
    (0.55,-0.08)
    --
    (0.55,0.60)
    --
    (0.34,0.40)
    --
    cycle;

\coordinate (a) at (0.05,0.08);

\node[
    below left=5pt
]
    at (0,-0.28)
    {$a$};


\node[flavornode] (f) at (0,-1.60) {$ $};

\draw[chiral]
    (f.north)
    --
    ($(a)+(0,-0.25)$)
    node[
        pos=0.52,
        right=2pt,
        fill=white,
        inner sep=1pt
    ]
    {$\mathsf I$};


\node[gaugenode] (xmid) at (-0.70,0.68) {};

\node[gaugenode] (c) at (-2.10,1.05) {};

\draw[chiral]
    (a)
    --
    (xmid);

\draw[chiral]
    (xmid)
    --
    (c);

\node[
    above=4pt,
    fill=white,
    inner sep=1.5pt
]
    at (-1.08,0.93)
    {$\mathsf X$};

\node[
    above left=2pt
]
    at (c)
    {$c$};


\node[gaugenode] (d) at (2.08,-0.62) {};

\draw[chiral]
    (a)
    to[bend right=8]
    (d);

\node[gaugenode] at (a){};
\node[
    below=4pt,
    fill=white,
    inner sep=1.5pt
]
    at (1.10,-0.32)
    {$\mathsf Y$};

\node[
    below right=2pt
]
    at (d)
    {$d$};

\end{tikzpicture}

\caption{
A schematic illustration of the D4-brane framing in the universal
cover. The magenta surfaces represent parallel lifts of the phase
boundary. The framing node $*$ is attached by $\mathsf I$ to the brick
$a$, from which the chiral paths $\mathsf X$ and $\mathsf Y$
originate. The field $\mathsf{X}$ is an example of a composite chiral path, while $\mathsf{Y}$ is a single chiral field.
}
\label{fig:D4-phase-boundary-crossing}
\end{figure}

\begin{proposal}[D4-brane framing from a phase boundary]
\label{prop:D4-phase-boundary-framing}
Let $\eta$ be a phase boundary of the brane brick model and $H$ be the corresponding phase boundary matrix.

\begin{enumerate}

\item
Mark all chiral fields intersected by $\eta$, namely
\bea
\Phi
\qquad\text{such that}\qquad
H_{\Phi,\eta}\neq 0.
\label{eq:D4-marked-chirals}
\eea
Every occurrence of these chiral fields in the $J$- and
$E$-polynomials is also marked.

\item
Consider a bulk Fermi multiplet $\Lambda$. It contributes an elementary
D4-brane framing if exactly one of its two polynomials
$J_{\Lambda}$ and $E_{\Lambda}$ contains marked chiral fields. We
denote the two polynomials by
\bea
Q_{\Lambda}^{\mathrm{mark}},
\qquad
Q_{\Lambda}^{\mathrm{unmark}},
\eea
where $Q_{\Lambda}^{\mathrm{mark}}$ contains the marked fields and
$Q_{\Lambda}^{\mathrm{unmark}}$ does not.

\item
Let the unmarked polynomial define a chiral path
\bea
Q_{\Lambda}^{\mathrm{unmark}}
:
b\longrightarrow a.
\eea
We introduce a framing node $*$ and two framing chiral multiplets
\bea
\mathsf I:*\longrightarrow a,
\qquad
\mathsf J:b\longrightarrow *.
\label{eq:D4-framing-chirals}
\eea
The unmarked polynomial is then deformed by
\bea
Q_{\Lambda}^{\mathrm{unmark}}
\longmapsto
Q_{\Lambda}^{\mathrm{unmark}}
+
\mathsf I\mathsf J.
\label{eq:D4-unmark-deformation}
\eea
Thus, the deformation is made in the $J$-term or the $E$-term
according to whether
$Q_{\Lambda}^{\mathrm{unmark}}=J_{\Lambda}$ or
$Q_{\Lambda}^{\mathrm{unmark}}=E_{\Lambda}$.

\item

Consider the family of parallel lifts of $\eta$ in the universal
cover of the periodic quiver, and choose the region between two
adjacent lifts that contains the framed brick $a$. The two plaquette monomials contained in $Q_{\Lambda}^{\mathrm{mark}}$ are cut when their initial chiral paths first leave this region. The corresponding
shortest paths are the composite fields $\mathsf{X},\mathsf{Y}$ discussed in \eqref{eq:D4-general-framing} (see Figure~\ref{fig:D4-phase-boundary-crossing} for an illustration).

The resulting open chiral paths are
completed by introducing two framing Fermi multiplets. The orientations of the framing Fermi multiplets and the assignment of
their $J$- and $E$-terms are fixed by closing the two cut plaquettes. In particular, an $E$-term has the same orientation as the corresponding Fermi multiplet, whereas a $J$-term has the opposite orientation.



\item
The relative signs of the new $J$- and $E$-terms are determined by the
supersymmetry condition \eqref{eq:JE-traceless}. If the cut plaquettes cannot be completed consistently, the chosen bulk
Fermi multiplet does not define an elementary D4-brane framing.

\end{enumerate}

\end{proposal}

A given phase boundary may admit more than one bulk Fermi multiplet satisfying the above conditions. Each such choice gives an elementary framing associated with the same toric surface. In the examples of $\mathcal C\times\mathbb C$ considered below, all possible choices can be read off directly from the explicit $J$- and $E$-terms. For later use, let us introduce some notations. For a phase boundary $\eta_{A}$, we denote the set of possible D4-framing nodes as $\mathcal{B}_{\eta_{A}}=\mathcal{B}_{A}$. For $\mathcal C\times\mathbb C$, the possible choices are conveniently labeled by the gauge nodes to which the framing node is attached:
\bea\label{eq:D4-admissibleflavorbranes}
&\mathcal B_{\eta_{01}}=\{2\},
\qquad
\mathcal B_{\eta_{02}}=\{1\},
\qquad
\mathcal B_{\eta_{03}}=\{2\},
\qquad
\mathcal B_{\eta_{04}}=\{1\},\\
&\mathcal B_{\eta_A}=\{1,2\},
\qquad
A\in\four^{\vee}.
\eea

Proposal~\ref{prop:D4-phase-boundary-framing} gives a combinatorial prescription for the
D4-brane framing in terms of the phase-boundary data. We do not,
however, attempt to derive it directly from the T-dual brane
configuration. In particular, the precise localization of the
T-dual image of the wrapped D4-brane, as well as the microscopic
open-string origin of the framing multiplets and their $J$- and
$E$-term interactions, will be left for future work.

Moreover, at present, it is not clear whether the above prescription extends to an arbitrary toric Calabi--Yau fourfold. For $\mathcal C\times\mathbb C$, however, it works consistently for all the phase boundaries considered below and gives the expected D4-brane framed quivers.

\paragraph{$\mathcal{N}=4$ phase boundaries}We have two inequivalent type of phase boundaries for the $\mathcal{C}\times\mathbb{C}$. We first start on the phase boundaries $\eta_{A}$ for $A\in\four$. It is enough to consider the phase boundary $\eta_{02}$, because others can be obtained by permuting the colors and the fields of the theory (see Appendix~\ref{app:JEterms-phaseboundary}).

Recall that the phase boundary matrix is
\bea
\adjustbox{valign=c}{
\tdplotsetmaincoords{60}{110}
	\begin{tikzpicture}[scale=1.5,tdplot_main_coords]
	\draw[thick,dashed,-Triangle] (0,0,0) -- node[left,pos=1] {$1$} (1.9,0,0);
	\draw[thick,dashed,-Triangle] (0,0,0) -- node[right,pos=1] {$2$} (0,1.6,0);
	\draw[thick,dashed,-Triangle] (0,0,0) -- node[above,pos=1] {$3$} (0,0,1.5);
    \node at (1,-0.1,1) {\textcolor{cyan}{$\eta_{02}$}};
	\node[draw=black,line width=1pt,circle,fill=black,minimum width=0.2cm,inner sep=1pt] (O) at (0,0,0) {};
	\node[draw=black,line width=1pt,circle,fill=black,minimum width=0.2cm,inner sep=1pt] (X1) at (1,0,0) {};
	\node[draw=black,line width=1pt,circle,fill=black,minimum width=0.2cm,inner sep=1pt] (Z) at (0,0,1) {};
	\node[draw=black,line width=1pt,circle,fill=black,minimum width=0.2cm,inner sep=1pt] (Y1) at (0,1,0) {};
	\node[draw=black,line width=1pt,circle,fill=black,minimum width=0.2cm,inner sep=1pt] (XY) at (1,1,0) {};
	\draw[line width=1pt] (O)--(X1)--(XY)--(Y1)--(O);
    	
	\draw[line width=1pt] (Z)--(O);
	\draw[line width=1pt,cyan] (Z)--(X1);
	\draw[line width=1pt] (Z)--(Y1);
	\draw[line width=1pt] (Z)--(XY);
	\end{tikzpicture}}\qquad H=\left(
\begin{array}{c|cccc|c>{\columncolor{cyan!30}}ccc}
     & \eta_{12} & \eta_{23} & \eta_{34} & \eta_{41}
     & \eta_{01} & \eta_{02} & \eta_{03} & \eta_{04} \\ \hline
   \mathsf{A}_1
     & 0&0&-1&1 & 0&0&0&-1 \\
   \mathsf{A}_2
     & -1&1&0&0 & 0&-1&0&0 \\
   \mathsf{B}_1
     & 1&0&0&-1 & -1&0&0&0 \\
   \mathsf{B}_2
     & 0&-1&1&0 & 0&0&-1&0 \\
   \mathsf{C}_1
     & 0&0&0&0 & 1&1&1&1 \\
   \mathsf{C}_2
     & 0&0&0&0 & 1&1&1&1 \\ \hline
   \Lambda^{1}_{1\rightarrow 2}
     & 0&0&-1&1 & 1&1&1&0 \\
   \Lambda^{2}_{1\rightarrow 2}
     & -1&1&0&0 & 1&0&1&1 \\
   \Lambda^{1}_{2\rightarrow 1}
     & 1&0&0&-1 & 0&1&1&1 \\
   \Lambda^{2}_{2\rightarrow 1}
     & 0&-1&1&0 & 1&1&0&1
\end{array}
\right)
\eea
and the chiral fields with non-zero elements are $\{\mathsf{A}_{2},\mathsf{C}_{1},\mathsf{C}_{2}\}$. The Fermi field whose $J,E$-terms contain these chiral fields exactly once is $\Lambda^{2}_{1\rightarrow 2}$, where for this case it is unique. In particular, the $J$-term is unmarked, giving the deformation
\bea
\begin{array}{c|c c}
  \textcolor{cyan}{\eta_{02}}   & \text{$J$-term} &  \text{$E$-term}\\ \hline
 \Lambda^{2}_{1\rightarrow 2} &  \mathsf{B}_{2}\mathsf{A}_{1}\mathsf{B}_{1}-\mathsf{B}_{1}\mathsf{A}_{1}\mathsf{B}_{2}+\textcolor{cyan}{\mathsf{I}\mathsf{J}}\quad  &  \mathsf{C}_{2}\mathsf{A}_{2}-\mathsf{A}_{2}\mathsf{C}_{1}
 \end{array}
\eea
We thus have $\mathsf{I}:\ast\rightarrow 1$, $\mathsf{J}:2\rightarrow \ast$. This is the reason why $\mathcal{B}_{02}=\{1\}$ in \eqref{eq:D4-admissibleflavorbranes}.

One can see that the two minimal oriented chiral paths that cross the universal covering $\widetilde{\eta}_{02}$ are $\mathsf{C}_{1}$ and $\mathsf{A}_{2}$. The local quiver in the periodic quiver and the resulting quiver can then be written as
\bea
\adjustbox{valign=c}{
\begin{tikzpicture}[
    scale=0.8,
    line cap=round,
    line join=round,
    chiral/.style={
        black,
        thick,
        postaction={decorate},
        decoration={
            markings,
            mark=at position 0.52 with {\arrow{Triangle}}
        }
    },
    fermi/.style={
        red!100!black,
        thick,
        postaction={decorate},
        decoration={
            markings,
            mark=at position 0.52 with {\arrow{Triangle}}
        }
    },
    qnode/.style={
        circle,
        draw=black,
        thick,
        fill=yellow!85!orange,
        minimum size=0.72cm,
        inner sep=0pt
    },
      qnode1/.style={
        circle,
        draw=black,
        thick,
        fill=blue,
        minimum size=0.72cm,
        inner sep=0pt
    },
      qnode2/.style={
        circle,
        draw=black,
        thick,
        fill=red,
        minimum size=0.72cm,
        inner sep=0pt
    },
    box/.style={
        rectangle,
        draw=black,
        thick,
        fill=black!10!white,
        minimum width=0.65cm,
        minimum height=0.65cm,
        inner sep=0pt
    }
]

\node[box] (f) at (0,3.2) {$ $};

\node[qnode1] (a) at (-3.0,0) {$ $};
\node[qnode2] (b) at ( 3.0,0) {$ $};

\node[qnode1] (c) at (-0.8, 1.05) {$ $};
\node[qnode2] (d) at ( 0.8,-1.05) {$ $};

\draw[chiral]
    ($(f.south west)+(0.02,-0.02)$)
  --
    ($(a.north)+(0,0.02)$)
    node[
        pos=0.48,
        above left=1pt,
        fill=white,
        inner sep=1.5pt
    ]
    {$\mathsf I$};

\draw[chiral]
    ($(b.north)+(0,0.02)$)
    --
    ($(f.south east)+(-0.08,-0.02)$)
    node[
        pos=0.50,
        above right=1pt,
        fill=white,
        inner sep=0.5pt
    ]
    {$\mathsf J$};

\draw[fermi]
    ($(f.south)+(-0.12,0)$)
    to[bend right=9]
    ($(c.north)+(0,0.02)$)
    node[
        pos=0.53,
        left=3pt,
        fill=white,
        inner sep=0.5pt,
        text=red!100!black
    ]
    {$ $};

\draw[chiral]
    (a)--(c)
    node[
        pos=0.46,
        above left=1pt,
        fill=white,
        inner sep=0.5pt
    ]
    {$\mathsf C_{1}$};

\draw[chiral]
    (c)--(b)
    node[
        pos=0.52,
        above right=1pt,
        fill=white,
        inner sep=0.5pt
    ]
    {$\mathsf A_{2}$};

\draw[chiral]
    (a)--(d)
    node[
        pos=0.47,
        below left=1pt,
        fill=white,
        inner sep=0.5pt
    ]
    {$\mathsf A_{2}$};

\draw[chiral]
    (d)--(b)
    node[
        pos=0.52,
        below right=1pt,
        fill=white,
        inner sep=0.5pt
    ]
    {$\mathsf C_{2}$};

\draw[fermi]
    ($(a.east)+(0.02,0)$)
    -- ($(b.west)+(-0.02,0)$)
    node[
        pos=0.50,
        below=3pt,
        fill=white,
        inner sep=1.5pt,
        text=red!100!black
    ]
    {$ $};

\draw[
    white,
    line width=4.5pt
]
    ($(d.north)+(0.05,0)$)
    to[bend right=15]
    ($(f.south)+(0.15,0)$);

\draw[fermi]
    ($(d.north)+(0.05,0)$)
    to[bend right=15]
    ($(f.south)+(0.15,0)$)
    node[
        pos=0.9,
        right=3pt,
        fill=white,
        inner sep=0.1pt,
        text=red!100!black
    ]
    {$ $};

\node at (0,-0.2) {\textcolor{red}{$\Lambda_{1\rightarrow 2}$}};

\node at (0.9,1.7) {\textcolor{red}{$\Lambda_{2}$}};

\node at (-0.3,1.7) {\textcolor{red}{$\Lambda_{1}$}};

\node[box]  at (0,3.2) {$ $};
\node[qnode2]  at ( 0.8,-1.05) {$ $};

\end{tikzpicture}}\qquad \qquad \adjustbox{valign=c}{
\begin{tikzpicture}[decoration={markings,mark=at position \arrowHeadPosition with {\arrow{latex}}}]
 \tikzset{
        box/.style={draw, minimum width=0.6cm, minimum height=0.6cm, text centered,thick},
        ->-/.style={decoration={
        markings,mark=at position #1 with {\arrow[scale=1.5]{>}}},postaction={decorate},line width=0.5mm},
        -<-/.style={decoration={
        markings,
        mark=at position #1 with {\arrow[scale=1.5]{<}}},postaction={decorate},line width=0.5mm}    
    }
\begin{scope}[xshift=4cm]

\draw[postaction={decorate},thick]
  (-1.1,2)--(-1.1,0);
\draw[postaction={decorate},red,thick]
  (-0.9,2)--(-0.9,0);
  \node[left] at (-1.1,1){$\mathsf{I}$};
  \node[right] at (-0.9,1){$\textcolor{red}{\Lambda_{1}}$};
\draw[ postaction={decorate},black,thick](1,0) to[bend right=30] (-1,2);
\draw[ postaction={decorate},red,thick](1.2,0) to[bend right=30] (-1,2.2);
\node[right] at (0.7,1.3){$\mathsf{J},\textcolor{red}{\Lambda_{2}}$};
\node[ fill=black!20!white,box] at (-1,2){};
    \draw[postaction={decorate}, black,thick,scale=1.3] (0.65,0) arc(0:-180:0.65 and 0.1) ;
    \draw[postaction={decorate}, black,thick,scale=1.3] (0.75,0) arc(0:-180:0.75 and 0.2) ;
    \draw[postaction={decorate}, black,thick,scale=1.3] (-0.65,0) arc(180:0:0.65 and 0.1) ;
    \draw[postaction={decorate}, black,thick,scale=1.3] (-0.75,0) arc(180:0:0.75 and 0.2) ;
    \draw[postaction={decorate}, black,thick,scale=1.3] (-0.8,0) arc(360:0:0.4 and 0.3) ;
    \draw[postaction={decorate}, black,thick,scale=1.3] (0.8,0) arc(-180:180:0.4 and 0.3) ;
    
    \draw[red,thick,postaction={decorate},scale=1.3] (0.65,0) arc(0:-180:0.65 and 0.3) ;
    \draw[red,thick,postaction={decorate},scale=1.3] (0.75,0) arc(0:-180:0.75 and 0.4) ;
    \draw[red,thick,postaction={decorate},scale=1.3] (-0.65,0) arc(180:0:0.65 and 0.3) ;
    \draw[red,thick,postaction={decorate},scale=1.3] (-0.75,0) arc(180:0:0.75 and 0.4) ;
    \draw[fill=blue!100!white,thick](-0.95,0) circle(0.3cm);
    \draw[fill=red!100!white,thick](0.9,0) circle(0.3cm);

\end{scope}
\end{tikzpicture}}
\eea
with the $J,E$-terms 
\bea\label{eq:D4-eta02-JE}
\begin{array}{c|c c}
  \textcolor{cyan}{\eta_{02}}   & \text{$J$-term} &  \text{$E$-term}\\ \hline
\Lambda_{ 1} \quad    & \textcolor{cyan}{\mathsf{J}\mathsf{A}_2} \quad  & \textcolor{cyan}{\mathsf{C}_1\mathsf{I}}\\ 
\Lambda_{ 2}\quad  & \textcolor{cyan}{\mathsf{A}_{2}\mathsf{I}} \quad  & \textcolor{cyan}{-\mathsf{J}\mathsf{C}_2}\\
 \Lambda^{2}_{1\rightarrow 2} &  \mathsf{B}_{2}\mathsf{A}_{1}\mathsf{B}_{1}-\mathsf{B}_{1}\mathsf{A}_{1}\mathsf{B}_{2}+\textcolor{cyan}{\mathsf{I}\mathsf{J}}\quad  &  \mathsf{C}_{2}\mathsf{A}_{2}-\mathsf{A}_{2}\mathsf{C}_{1}
 \end{array}
\eea
where the light blue terms are the new terms coming from the flavor D4-brane added to the $J,E$-terms of \eqref{eq:conifold-JEterms}.


The D4--D2--D0 systems associated with the four phase boundaries
$\eta_A$ $(A\in\four)$ admit an enhanced $\mathcal N=4$
description. Although the framed quivers are naturally formulated in
$\mathcal N=2$ language, the two framing chiral multiplets and the two
framing Fermi multiplets can be paired into two $\mathcal N=4$ chiral
multiplets. Correspondingly, their $J$- and $E$-terms can be organized
in terms of a single $\mathcal N=4$ superpotential.

Let us explain this structure for the phase boundary $\eta_{02}$. The
$\mathcal N=2$ multiplets $(\mathsf I,\Lambda_1)$ combine into an
$\mathcal N=4$ chiral multiplet, which we continue to denote by
$\mathsf I$, while $(\mathsf J,\Lambda_2)$ combine into another
$\mathcal N=4$ chiral multiplet denoted by $\mathsf J$. In this
notation, the framing contribution to the $\mathcal N=4$
superpotential is
\bea
\delta\mathsf W
=
\Tr\left(
\mathsf J\mathsf A_2\mathsf I
\right).
\label{eq:D4-eta02-N4-potential}
\eea
The $F$-term equations following from this superpotential reproduce
the $J$-term relations in \eqref{eq:D4-eta02-JE}, while the associated
$E$-terms arise from the decomposition of the $\mathcal N=4$
multiplets into $\mathcal N=2$ multiplets. Thus, the complete
$J$- and $E$-data in \eqref{eq:D4-eta02-JE} provide the
$\mathcal N=2$ presentation of the same four-supercharge system.

This enhancement also has a direct geometric interpretation. Under
dimensional reduction of $\mathcal C\times\mathbb C$ along the extra
$\mathbb C$ direction, the phase boundary $\eta_{02}$ selects the
perfect matching of the conifold brane tiling containing
$\mathsf A_2$. At the same time, the toric surface wrapped by the
D4-brane is identified with the corresponding non-compact toric
divisor of the conifold. The dimensionally reduced system is therefore
the conventional D4--D2--D0 system on a toric Calabi--Yau threefold.

In particular, the deformation
\eqref{eq:D4-eta02-N4-potential} is precisely the framed
superpotential appearing in the construction of D4--D2--D0 systems on
toric divisors developed in \cite{Nishinaka:2013mba}; see also
\cite{Nishinaka:2011is,Nishinaka:2011sv,Nishinaka:2013pua}. In this
identification, $\mathsf A_2$ is the chiral field selected by the
perfect matching associated with the divisor wrapped by the D4-brane.
A brief review of this construction, together with its application to
the conifold, is given in
Appendix~\ref{app:branetiling-conifold}.

We emphasize that the above $\mathcal N=4$ reorganization is a special
property of the four phase boundaries $\eta_A$ considered here. For a
generic phase boundary of a brane brick model, the framing chiral and
Fermi multiplets need not admit such a four-supercharge completion.
The corresponding framed theory must then be described directly in
terms of its $\mathcal N=2$ $J$- and $E$-data.




\paragraph{$\mathcal{N}=2$ phase boundaries} Let us next consider the phase boundaries $\eta_{A}$ for $A\in\four^{\vee}$. It is enough to consider the phase boundary $\eta_{34}$:
\bea
\adjustbox{valign=c}{
\tdplotsetmaincoords{60}{110}
	\begin{tikzpicture}[scale=1.5,tdplot_main_coords]
	\draw[thick,dashed,-Triangle] (0,0,0) -- node[left,pos=1] {$1$} (1.9,0,0);
	\draw[thick,dashed,-Triangle] (0,0,0) -- node[right,pos=1] {$2$} (0,1.6,0);
	\draw[thick,dashed,-Triangle] (0,0,0) -- node[above,pos=1] {$3$} (0,0,1.5);
	\node[draw=black,line width=1pt,circle,fill=black,minimum width=0.2cm,inner sep=1pt] (O) at (0,0,0) {};
	\node[draw=black,line width=1pt,circle,fill=black,minimum width=0.2cm,inner sep=1pt] (X1) at (1,0,0) {};
	\node[draw=black,line width=1pt,circle,fill=black,minimum width=0.2cm,inner sep=1pt] (Z) at (0,0,1) {};
	\node[draw=black,line width=1pt,circle,fill=black,minimum width=0.2cm,inner sep=1pt] (Y1) at (0,1,0) {};
	\node[draw=black,line width=1pt,circle,fill=black,minimum width=0.2cm,inner sep=1pt] (XY) at (1,1,0) {};
	\draw[line width=1pt] (O)--(X1)--(XY);
    \draw[line width=1pt] (Y1)--(O);
    	
	\draw[line width=1pt] (Z)--(O);
	\draw[line width=1pt,black] (Z)--(X1);
    	\draw[line width=1pt,magenta] (XY)--(Y1);
        \node at (0.5,1.3,0) {$\textcolor{magenta}{\eta_{34}}$};
	\draw[line width=1pt] (Z)--(Y1);
	\draw[line width=1pt] (Z)--(XY);
	\end{tikzpicture}}\qquad H=\left(
\begin{array}{c|cc>{\columncolor{magenta!30}}cc|cccc}
     & \eta_{12} & \eta_{23} & \eta_{34} & \eta_{41}
     & \eta_{01} & \eta_{02} & \eta_{03} & \eta_{04} \\ \hline
   \mathsf{A}_1
     & 0&0&-1&1 & 0&0&0&-1 \\
   \mathsf{A}_2
     & -1&1&0&0 & 0&-1&0&0 \\
   \mathsf{B}_1
     & 1&0&0&-1 & -1&0&0&0 \\
   \mathsf{B}_2
     & 0&-1&1&0 & 0&0&-1&0 \\
   \mathsf{C}_1
     & 0&0&0&0 & 1&1&1&1 \\
   \mathsf{C}_2
     & 0&0&0&0 & 1&1&1&1 \\ \hline
   \Lambda^{1}_{1\rightarrow 2}
     & 0&0&-1&1 & 1&1&1&0 \\
   \Lambda^{2}_{1\rightarrow 2}
     & -1&1&0&0 & 1&0&1&1 \\
   \Lambda^{1}_{2\rightarrow 1}
     & 1&0&0&-1 & 0&1&1&1 \\
   \Lambda^{2}_{2\rightarrow 1}
     & 0&-1&1&0 & 1&1&0&1
\end{array}
\right)
\eea
The chiral fields with non-zero elements of the matrix $H$ are $\{\mathsf{A}_{1},\mathsf{B}_{2}\}$. For this case, we have two Fermi fields whose $J,E$-terms contain these chiral fields once: $\Lambda^{2}_{1\rightarrow 2},\Lambda^{1}_{2\rightarrow 1}$. For both Fermi fields, the $E$-term is unmarked and we can consider two different deformations:
\bea
\begin{array}{c|c c}
   \textcolor{magenta}{\eta_{34}}  & \text{$J$-term} &  \text{$E$-term}\\ \hline
 \Lambda^{1}_{2\rightarrow 1} & \mathsf{A}_{2}\mathsf{B}_{2}\mathsf{A}_{1}-\mathsf{A}_{1}\mathsf{B}_{2}\mathsf{A}_2 \quad  &  \mathsf{C}_{1}\mathsf{B}_{1}-\mathsf{B}_{1}\mathsf{C}_{2}+\textcolor{magenta}{\mathsf{I}\mathsf{J}}\\
  \Lambda^{2}_{1\rightarrow 2} & \mathsf{B}_{2}\mathsf{A}_{1}\mathsf{B}_{1}-\mathsf{B}_{1}\mathsf{A}_{1}\mathsf{B}_2 \quad  &  \mathsf{C}_{2}\mathsf{A}_{2}-\mathsf{A}_{2}\mathsf{C}_{1}+\textcolor{magenta}{\mathsf{I}\mathsf{J}}
 \end{array}
 \eea
The first deformation determines the arrows as $\mathsf{I}:\ast\rightarrow 1,\mathsf{J}:2\rightarrow \ast$, while the second deformation gives $\mathsf{I}:\ast\rightarrow 2,\mathsf{J}:1\rightarrow \ast$, which give the two elements in $\mathcal{B}_{34}$ in \eqref{eq:D4-admissibleflavorbranes}. For later use we denote the first (second) deformation as $\eta^{(1,2)}_{34}$, respectively. 

For simplicity, let us focus on the first deformation. The two minimal oriented paths starting from $1$ that crosses $\eta_{34}$ are $\mathsf{A}_{1}$ and $\mathsf{B}_{2}\mathsf{A}_{2}$. The local quiver in the periodic quiver and the resulting quiver are then drawn as
\bea
\adjustbox{valign=c}{
\begin{tikzpicture}[
    scale=0.8,
    line cap=round,
    line join=round,
    chiral/.style={
        black,
        thick,
        postaction={decorate},
        decoration={
            markings,
            mark=at position 0.52 with {\arrow{Triangle}}
        }
    },
    fermi/.style={
        red!100!black,
        thick,
        postaction={decorate},
        decoration={
            markings,
            mark=at position 0.52 with {\arrow{Triangle}}
        }
    },
    nodeone/.style={
        circle,
        draw=black,
        thick,
        fill=blue,
        minimum size=0.72cm,
        inner sep=0pt
    },
    nodetwo/.style={
        circle,
        draw=black,
        thick,
        fill=red,
        minimum size=0.72cm,
        inner sep=0pt
    },
    box/.style={
        rectangle,
        draw=black,
        thick,
        fill=black!15!white,
        minimum width=0.65cm,
        minimum height=0.65cm,
        inner sep=0pt
    }
]


\node[] (f) at (0,3.35) {$ $};

%

\node[nodeone] (a) at (-3.4,0) {};
\node[nodetwo] (b) at ( 3.4,0) {};

\node[nodetwo] (xu) at (-2.10,1.25) {};
\node[nodeone] (c)  at (-0.65,1.25) {};

\node[nodetwo] (d)  at (-0.65,-1.25) {};
\node[nodeone] (vl) at ( 0.85,-1.25) {};

\node[
    below left=2pt
]
at (a.south)
{$ $};

\node[
    below right=2pt
]
at (b.south)
{$ $};


\draw[chiral]
    ($(f.south west)+(0.02,-0.02)$)
    .. controls
        (-1.25,2.85)
        and
        (-3.15,1.20)
    ..
    ($(a.north)+(0,0.02)$)
    node[
        pos=0.48,
        above left=3pt,
        fill=white,
        inner sep=0.5pt
    ]
    {$\mathsf I$};

\draw[chiral]
    ($(b.north)+(0,0.02)$)
    --
    ($(f.south east)+(-0.05,-0.02)$)
    node[
        pos=0.50,
        above right=1pt,
        fill=white,
        inner sep=0.5pt
    ]
    {$\mathsf J$};

%

\draw[chiral]
    (a)--(xu);

\draw[chiral]
    (xu)--(c)
    node[
        pos=0.50,
        above=2pt,
        fill=white,
        inner sep=0.5pt
    ]
    {$\mathsf B_{2}$};
\node at (-2.5,0.4) {$\mathsf{A}_{2}$};

\draw[chiral]
    (c)--(b);

\node[
    above=8pt,
    fill=white,
    inner sep=0.5pt
]
at ($(a)!0.55!(c)$)
{$ $};

\node[
    above right=5pt,
    fill=white,
    inner sep=0.5pt
]
at ($(c)!0.55!(b)$)
{$ $};

%

\draw[chiral]
    (a)--(d)
    node[
        pos=0.47,
        below left=1pt,
        fill=white,
        inner sep=0pt
    ]
    {$\mathsf A_{1}$};

\draw[chiral]
    (d)--(vl)
    node[
        pos=0.50,
        below=2pt,
        fill=white,
        inner sep=0.5pt
    ]
    {$\mathsf B_{2}$};

\draw[chiral]
    (vl)--(b);

\node[
    below left=5pt,
    fill=white,
    inner sep=0.5pt
]
at ($(a)!0.55!(d)$)
{$ $};

\node[
    below=8pt,
    fill=white,
    inner sep=0.5pt
]
at ($(d)!0.55!(b)$)
{$ $};


\draw[fermi]
    ($(b.west)+(-0.02,0)$)
    --
    ($(a.east)+(0.02,0)$)
    node[
        pos=0.50,
        below=3pt,
        fill=white,
        inner sep=0.5pt,
        text=red!100!black
    ]
    {$ $};

\node[text=red] at (0.6,-0.3){$\Lambda^{1}_{2\rightarrow 1}$};
\node[text=red] at (0.6,1.3){$\Lambda_{2}$};
\node[text=red] at (-0.3,1.9){$\Lambda_{1}$};

\draw[fermi]
    ($(f.south)+(-0.12,0)$)
    to[bend right=8]
    ($(c.north)+(0,0.02)$)
    node[
        pos=0.52,
        left=3pt,
        fill=white,
        inner sep=0.5pt,
        text=red!100!black
    ]
    {$ $};


\draw[
    white,
    line width=4.5pt
]
    ($(d.north)+(0.05,0)$)
    to[bend right=15]
    ($(f.south)+(0.15,0)$);


\draw[fermi]
    ($(d.north)+(0.05,0)$)
    to[bend right=15]
    ($(f.south)+(0.15,0)$)
    node[
        pos=0.55,
        right=3pt,
        fill=white,
        inner sep=0.5pt,
        text=red!100!black
    ]
    {$ $};
    \node[nodetwo] (xu) at (-2.10,1.25) { };
\node[box] at (0,3.25) {$ $};
\node[nodetwo]  at (-0.65,-1.25) {};

\node at (2.5,-1) {$\mathsf{A}_{2}$};
\node at (2.,0.7) {$\mathsf{A}_{1}$};

\end{tikzpicture}
}\qquad \adjustbox{valign=c}{
\begin{tikzpicture}[scale=1.3,decoration={markings,mark=at position \arrowHeadPosition with {\arrow{latex}}}]
 \tikzset{
        box/.style={draw, minimum width=0.6cm, minimum height=0.6cm, text centered,thick},
        ->-/.style={decoration={
        markings,mark=at position #1 with {\arrow[scale=1.5]{>}}},postaction={decorate},line width=0.5mm},
        -<-/.style={decoration={
        markings,
        mark=at position #1 with {\arrow[scale=1.5]{<}}},postaction={decorate},line width=0.5mm}    
    }
\begin{scope}[xshift=4cm]

\draw[postaction={decorate},thick]
  (-1.1,2)--(-1.1,0);
\draw[postaction={decorate},red,thick]
  (-0.9,2)--(-0.9,0);
  \node[left] at (-1.1,1){$\mathsf{I}$};
  \node[right] at (-0.9,1){\textcolor{red}{$\Lambda_{1} $}};
\draw[ postaction={decorate},black,thick](1,0) to[bend right=30] (-1,2);
\draw[ postaction={decorate},red,thick](1.1,0) to[bend right=30] (-0.8,2);

\node[right] at (0.7,1.3){$\mathsf{J},\textcolor{red}{\Lambda_{2}}$};
\node[ fill=black!20!white,box] at (-1,2){};
    \draw[postaction={decorate}, black,thick,scale=1.3] (0.65,0) arc(0:-180:0.65 and 0.1) ;
    \draw[postaction={decorate}, black,thick,scale=1.3] (0.75,0) arc(0:-180:0.75 and 0.2) ;
    \draw[postaction={decorate}, black,thick,scale=1.3] (-0.65,0) arc(180:0:0.65 and 0.1) ;
    \draw[postaction={decorate}, black,thick,scale=1.3] (-0.75,0) arc(180:0:0.75 and 0.2) ;
    \draw[postaction={decorate}, black,thick,scale=1.3] (-0.8,0) arc(360:0:0.4 and 0.3) ;
    \draw[postaction={decorate}, black,thick,scale=1.3] (0.8,0) arc(-180:180:0.4 and 0.3) ;
    
    \draw[red,thick,postaction={decorate},scale=1.3] (0.65,0) arc(0:-180:0.65 and 0.3) ;
    \draw[red,thick,postaction={decorate},scale=1.3] (0.75,0) arc(0:-180:0.75 and 0.4) ;
    \draw[red,thick,postaction={decorate},scale=1.3] (-0.65,0) arc(180:0:0.65 and 0.3) ;
    \draw[red,thick,postaction={decorate},scale=1.3] (-0.75,0) arc(180:0:0.75 and 0.4) ;
    \draw[fill=blue!100!white,thick](-0.95,0) circle(0.3cm);
    \draw[fill=red!100!white,thick](0.9,0) circle(0.3cm);

\end{scope}
\end{tikzpicture}}
\eea
giving the following $J,E$-terms: 
\bea\label{eq:D4-eta34-1-JE}
\begin{array}{c|c c}
   \textcolor{magenta}{\eta_{34}^{(1)}}  & \text{$J$-term} &  \text{$E$-term}\\ \hline
\Lambda_{1}\quad  &  \textcolor{magenta}{\mathsf{J}\mathsf{A}_1} \quad &\textcolor{magenta}{\mathsf{B}_2\mathsf{A}_{2}\mathsf{I}}   \\
\Lambda_{ 2} \quad    & \textcolor{magenta}{\mathsf{A}_1\mathsf{I}} \quad &\textcolor{magenta}{-\mathsf{J}\mathsf{A}_2\mathsf{B}_2}   \\ 
 \Lambda^{1}_{2\rightarrow 1} & \mathsf{A}_{2}\mathsf{B}_{2}\mathsf{A}_{1}-\mathsf{A}_{1}\mathsf{B}_{2}\mathsf{A}_2 \quad  &  \mathsf{C}_{1}\mathsf{B}_{1}-\mathsf{B}_{1}\mathsf{C}_{2}+\textcolor{magenta}{\mathsf{I}\mathsf{J}}
 \end{array}
\eea

Similarly, for the second deformation, the potential terms are determined as
\bea\label{eq:D4-eta34-2-JE}
\begin{array}{c|c c}
   \textcolor{magenta}{\eta_{34}^{(2)}}  & \text{$J$-term} &  \text{$E$-term}\\ \hline

\Lambda_{1}\quad  &  \textcolor{magenta}{-\mathsf{J}\mathsf{B}_2} \quad &\textcolor{magenta}{\mathsf{A}_1\mathsf{B}_{1}\mathsf{I}}   \\
\Lambda_{ 2} \quad    & \textcolor{magenta}{\mathsf{B}_2\mathsf{I}} \quad &\textcolor{magenta}{+\mathsf{J}\mathsf{B}_1\mathsf{A}_1}   \\ 
 \Lambda^{2}_{1\rightarrow 2} & \mathsf{B}_{2}\mathsf{A}_{1}\mathsf{B}_{1}-\mathsf{B}_{1}\mathsf{A}_{1}\mathsf{B}_2 \quad  &  \textcolor{magenta}{\mathsf{I}\mathsf{J}}
 \end{array}
\eea

For the other phase boundaries $\eta_{A}\,(A\in \four^{\vee})$, we also have two possible deformations and the matter fields, potential terms are summarized in Appendix~\ref{app:JEterms-phaseboundary}.

\subsection{The \texorpdfstring{$\mathcal N=4$}{N=4} D4--D2--D0 system: the phase boundary \texorpdfstring{$\eta_{02}$}{eta02}}\label{sec:D4-phase-02}

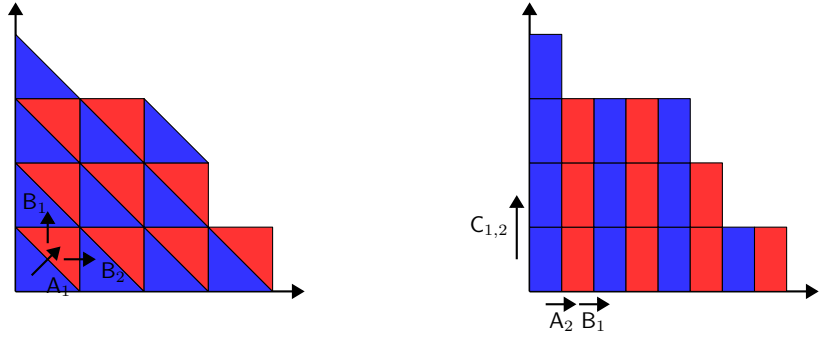
\begin{figure}[t]
\centering
\begin{tikzpicture}[
  scale=0.85,
  transform shape,
  line join=round
]

\begin{scope}

  \draw[-Triangle,thick] (0,0)--(4.5,0);
  \draw[-Triangle,thick] (0,0)--(0,4.5);

  \foreach \x in {0,1,2,3}{
    \filldraw[
      fill=blue!80!white,
      draw=black
    ]
    (\x,0)--(\x,1)--({\x+1},0)--cycle;

    \filldraw[
      fill=red!80!white,
      draw=black
    ]
    (\x,1)--({\x+1},1)--({\x+1},0)--cycle;
  }

  \foreach \x in {0,1,2}{
    \filldraw[
      fill=blue!80!white,
      draw=black
    ]
    (\x,1)--(\x,2)--({\x+1},1)--cycle;

    \filldraw[
      fill=red!80!white,
      draw=black
    ]
    (\x,2)--({\x+1},2)--({\x+1},1)--cycle;
  }

  \foreach \x in {0,1,2}{
    \filldraw[
      fill=blue!80!white,
      draw=black
    ]
    (\x,2)--(\x,3)--({\x+1},2)--cycle;
  }

  \foreach \x in {0,1}{
    \filldraw[
      fill=red!80!white,
      draw=black
    ]
    (\x,3)--({\x+1},3)--({\x+1},2)--cycle;
  }

  \filldraw[
    fill=blue!80!white,
    draw=black
  ]
  (0,3)--(0,4)--(1,3)--cycle;

  \draw[-Triangle,thick]
    (0.25,0.25)--(0.70,0.70);

  \draw[-Triangle,thick]
    (0.75,0.50)--(1.25,0.50);

  \draw[-Triangle,thick]
    (0.50,0.75)--(0.50,1.25);

  \node[below right] at (0.35,0.35)
    {$\mathsf A_{1}$};

  \node[right] at (1.20,0.30)
    {$\mathsf B_{2}$};

  \node[above] at (0.30,1.10)
    {$\mathsf B_{1}$};

\end{scope}

\begin{scope}[xshift=8cm]

  \draw[-Triangle,thick] (0,0)--(4.5,0);
  \draw[-Triangle,thick] (0,0)--(0,4.5);

  \foreach \x in {0,1,2,3}{
    \filldraw[
      fill=blue!80!white,
      draw=black
    ]
    (\x,0)--(\x,1)--({\x+0.5},1)--
    ({\x+0.5},0)--cycle;

    \filldraw[
      fill=red!80!white,
      draw=black
    ]
    ({\x+0.5},0)--({\x+0.5},1)--
    ({\x+1},1)--({\x+1},0)--cycle;
  }

  \foreach \x in {0,1,2}{
    \filldraw[
      fill=blue!80!white,
      draw=black
    ]
    (\x,1)--(\x,2)--({\x+0.5},2)--
    ({\x+0.5},1)--cycle;

    \filldraw[
      fill=red!80!white,
      draw=black
    ]
    ({\x+0.5},1)--({\x+0.5},2)--
    ({\x+1},2)--({\x+1},1)--cycle;
  }

  \foreach \x in {0,1,2}{
    \filldraw[
      fill=blue!80!white,
      draw=black
    ]
    (\x,2)--(\x,3)--({\x+0.5},3)--
    ({\x+0.5},2)--cycle;
  }

  \foreach \x in {0,1}{
    \filldraw[
      fill=red!80!white,
      draw=black
    ]
    ({\x+0.5},2)--({\x+0.5},3)--
    ({\x+1},3)--({\x+1},2)--cycle;
  }

  \filldraw[
    fill=blue!80!white,
    draw=black
  ]
  (0,3)--(0,4)--(0.5,4)--(0.5,3)--cycle;

  \draw[-Triangle,thick]
    (0.25,-0.20)--(0.73,-0.20);

  \draw[-Triangle,thick]
    (0.77,-0.20)--(1.25,-0.20);

  \draw[-Triangle,thick]
    (-0.20,0.50)--(-0.20,1.50);

  \node[below] at (0.50,-0.20)
    {$\mathsf A_{2}$};

  \node[below] at (1.00,-0.20)
    {$\mathsf B_{1}$};

  \node[left] at (-0.20,1.00)
    {$\mathsf C_{1,2}$};

\end{scope}

\end{tikzpicture}
\caption{2d BPS crystal of D4-brane wrapping phase boundary $\eta_{02}$ (left) and $\eta_{34}$ (right). Note that the fields $(\mathsf{A}_{1},\mathsf{B}_{1},\mathsf{B}_{2})$ and $(\mathsf{A}_{2},\mathsf{B}_{1},\mathsf{C}_{1},\mathsf{C}_{2})$ are associated with the equivariant parameters $(p_{1},p_{3},p_{4})$ and $(p_{2},p_{3},p_{5},p_{5})$, respectively. The coordinates of the triangles or half-boxes are determined using these equivariant parameters.}
\label{fig:D4-2dBPScrystal}
\end{figure}
In this section, we will focus on the properties of the D4--D2--D0 partition functions associated with the phase boundary $\eta_{02}$. In terms of $\mathcal{N}=2$ multiplets, the flavor charges are determined from the \eqref{eq:D4-eta02-JE} as
\bea
\mathsf{I}\rightarrow \mathsf{I},\quad \mathsf{J}\rightarrow p_{1}p_{3}p_{4}\mathsf{J},\quad \Lambda_{1}\rightarrow p_{5}\Lambda_{1},\quad \Lambda_{2}\rightarrow p_{1}p_{3}p_{4}p_{5}\Lambda_{2}.
\eea
The framing node contribution coming from the D4-brane wrapping the phase boundary $\eta_{02}$ is then
\bea
\mathcal{Z}^{\D4\tbar\D0}_{\eta_{02}}(\fra,\phi)&=\prod_{I=1}^{k_{1}}\frac{\sh(\phi_{I}^{(1)}-\fra-\eps_{5})}{\sh(\phi_{I}^{(1)}-\fra)}\prod_{J=1}^{k_{2}}\frac{\sh(\fra-\phi_{J}^{(2)}+\eps_{2})}{\sh(\fra-\phi_{J}^{(2)}+\eps_{2}+\eps_{5})}.
\eea

\begin{definition}\label{def:D4_eta02}
    The BPS partition function of the system with a D4-brane wrapping the phase boundary $\eta_{02}$ is defined as
 \bea
    \mathcal{Z}^{\D4\tbar\D0}(\eta_{02};\mathfrak{q}_{1},\mathfrak{q}_{2};p_{1,2,3,4,5})=\sum_{k_{1}=0}^{\infty}\sum_{k_{2}=0}^{\infty}\mathfrak{q}_{2}^{k_{2}}\mathfrak{q}_{1}^{k_{1}}\mathcal{Z}^{\D4\tbar\D0}_{k_{1},k_{2}}(\eta_{02})
    \eea
    where
    \bea
    \mathcal{Z}^{\D4\tbar\D0}_{k_{1},k_{2}}(\eta_{02})&=\frac{1}{k_{2}!k_{1}!}\oint_{\text{JK}}\prod_{a=1}^{2}\prod_{I=1}^{k_{a}}\frac{d\phi_{I}^{(a)}}{2\pi i } \mathcal{Z}_{\eta_{02}}^{\D4\tbar\D0}(\fra,\phi)\mathcal{Z}^{\D0\tbar\D0}(\phi).
    \eea
   
\end{definition}

Focusing on the cyclic chamber $\eta=(1,1,\ldots,1)\in\mathbb{R}^{k_{1}+k_{2}}$, the JK-residue prescription picks the same poles coming from \eqref{eq:D8-JK-poles}. Note that the anti-fundamental chiral does not contribute to the pole selection due to the JK-residue prescription. The different part compared to the magnificent four and tetrahedron instanton system is that we have two numerators
\bea
\sh(\phi_{I}^{(1)}-\fra-\eps_{5}),\quad \sh(\fra-\phi_{J}^{(2)}+\eps_{2}).
\eea
The first numerator also appeared from the D6-brane wrapping the toric divisor $\mathfrak{m}_{0}$, while the second numerator appeared from the D6 wrapping the toric divisor $\mathfrak{m}_{1}$. Imposing both numerators further restricts the growth of the crystal and actually gives a 2d BPS crystal.

\begin{theorem}
    The non-vanishing JK poles of Def.~\ref{def:D4_eta02} are in one-to-one correspondence with super partitions (see \eqref{eq:superpartition-def} and \eqref{eq:superpartition}). Let $\Lambda\in\mathsf{SP}$ be a super partition and $(\lambda^{(1)},\lambda^{(2)})$, the pairs of partitions. Then, up to permutation, the corresponding pole configuration is
    \bea
 \left\{\phi_I^{(1)}\right\}_{I=1}^{k_1}
 &=
 \left\{
 \fra
 +(\eps_1+\eps_4)(i-1)
 +(\eps_1+\eps_3)(j-1)
 \mathrel{}\middle|\mathrel{}
 (i,j,k)\in\lambda^{(1)}
 \right\},
 \\
 \left\{\phi_I^{(2)}\right\}_{I=1}^{k_2}
 &=
 \left\{
 \fra+\eps_1
 +(\eps_1+\eps_4)(i-1)
 +(\eps_1+\eps_3)(j-1)
 \mathrel{}\middle|\mathrel{}
 (i,j,k)\in\lambda^{(2)}
 \right\},
\eea
where
\bea
 k_1=|\Lambda|_1=|\lambda^{(1)}|,
 \qquad
 k_2=|\Lambda|_2=|\lambda^{(2)}|.
\eea
\end{theorem}
The proof is straightforward so we will not discuss it. Up to level two, we have
\bea
 \mathcal{Z}^{\D4\tbar\D0}(\eta_{02};\mathfrak{q}_{1},\mathfrak{q}_{2};p_{1,2,3,4,5})&=1\\
 &+\fq_1 \qquad \qquad \adjustbox{valign=c}{\begin{tikzpicture}[scale=0.3]
     \begin{scope}

  \draw[-Triangle,thick] (0,0)--(4.5,0);
  \draw[-Triangle,thick] (0,0)--(0,4.5);

    \filldraw[
      fill=blue!80!white,
      draw=black
    ]
    (0,0)--(0,1)--({0+1},0)--cycle;
\end{scope}
 \end{tikzpicture}}\\
 &+\fq_{1}\fq_{2}\frac{p_1 \left(p_2 p_3-1\right) \left(p_2 p_4-1\right)}{p_2 \left(p_1 p_3-1\right)
   \left(p_1 p_4-1\right)} \qquad \qquad \adjustbox{valign=c}{\begin{tikzpicture}[scale=0.3]
     \begin{scope}

  \draw[-Triangle,thick] (0,0)--(4.5,0);
  \draw[-Triangle,thick] (0,0)--(0,4.5);

    \filldraw[
      fill=blue!80!white,
      draw=black
    ]
    (0,0)--(0,1)--({0+1},0)--cycle;

    \filldraw[
      fill=red!80!white,
      draw=black
    ]
    (0,1)--({0+1},1)--({0+1},0)--cycle;
\end{scope}
 \end{tikzpicture}}\\
   &+\fq_{1}^{2}\fq_{2}\left(-\frac{\sqrt{p_1} \left(p_1-p_2\right) \sqrt{p_3} \left(p_1 p_2 p_3^2-1\right) \sqrt{p_4}
   \left(p_2 p_4-1\right)}{p_2^{3/2} \left(p_1 p_3-1\right) \left(p_3-p_4\right)
   \left(p_1^2 p_3 p_4-1\right)}\right.  \quad \adjustbox{valign=c}{\begin{tikzpicture}[scale=0.3]
     \begin{scope}

  \draw[-Triangle,thick] (0,0)--(4.5,0);
  \draw[-Triangle,thick] (0,0)--(0,4.5);

 \filldraw[
      fill=blue!80!white,
      draw=black
    ]
    (0,0)--(0,1)--({0+1},0)--cycle;
     \filldraw[
      fill=blue!80!white,
      draw=black
    ]
    (0,1)--(0,2)--({0+1},1)--cycle;
    
    \filldraw[
      fill=red!80!white,
      draw=black
    ]
    (0,1)--({0+1},1)--({0+1},0)--cycle;
    
\end{scope}
 \end{tikzpicture}}   \\
   &+ \left. \frac{\sqrt{p_1} \left(p_1-p_2\right) \sqrt{p_3} \left(p_2 p_3-1\right) \sqrt{p_4}
   \left(p_1 p_2 p_4^2-1\right)}{p_2^{3/2} \left(p_3-p_4\right) \left(p_1 p_4-1\right)
   \left(p_1^2 p_3 p_4-1\right)}\right)\quad \adjustbox{valign=c}{\begin{tikzpicture}[scale=0.3]
     \begin{scope}

  \draw[-Triangle,thick] (0,0)--(4.5,0);
  \draw[-Triangle,thick] (0,0)--(0,4.5);

 \filldraw[
      fill=blue!80!white,
      draw=black
    ]
    (0,0)--(0,1)--({0+1},0)--cycle;
     \filldraw[
      fill=blue!80!white,
      draw=black
    ]
    (1,0)--(1,1)--({1+1},0)--cycle;

    \filldraw[
      fill=red!80!white,
      draw=black
    ]
    (0,1)--({0+1},1)--({0+1},0)--cycle;
\end{scope}
 \end{tikzpicture}}\\
   &+\cdots 
\eea
In the unrefined limit $p_{5}\rightarrow1$, all of the factors trivialize and gives the generating function of the super partition
\bea
 \mathcal{Z}^{\D4\tbar\D0}(\eta_{02};\mathfrak{q}_{1},\mathfrak{q}_{2};p_{1,2,3,4,5})&\xrightarrow{p_{5}\rightarrow 1}1+\fq_1+\fq_1\fq_2+2\fq_1^2\fq_2+\cdots .
\eea

\subsection{The \texorpdfstring{$\mathcal N=2$}{N=2} D4--D2--D0 system: the phase boundary \texorpdfstring{$\eta_{34}$}{eta34}}\label{sec:D4-phase-34}
Let us move on to the phase boundary $\eta_{34}$. For this case, we have two possible configurations depending to which gauge node the framing node is connected.

For $\eta_{34}^{(1)}$, the flavor charges are determined from \eqref{eq:D4-eta34-1-JE} as
\bea
\mathsf{I}\rightarrow \mathsf{I},\quad \mathsf{J}\rightarrow p_{3}p_{5}\mathsf{J},\quad \Lambda_{2}\rightarrow p_{2}p_{3}p_{4}\Lambda_{2},\quad \Lambda_{1}\rightarrow p_{2}p_{4}\Lambda_{1}
\eea
giving the framing node contribution as
\bea
\mathcal{Z}^{\D4\tbar\D0}_{\eta_{34},1}(\fra,\phi)=\prod_{I=1}^{k_{1}}\frac{\sh(\phi_{I}^{(1)}-\fra-\eps_{2}-\eps_{4})}{\sh(\phi_{I}^{(1)}-\fra)}\prod_{J=1}^{k_{2}}\frac{\sh(\fra-\phi_{J}^{(2)}+\eps_{1})}{\sh(\fra-\phi_{J}^{(2)}+\eps_{1}+\eps_{2}+\eps_{4})}.
\eea
Similarly, for $\eta_{34}^{(2)}$, the framing node contribution is determined from \eqref{eq:D4-eta34-2-JE} as
\bea
\mathcal{Z}^{\D4\tbar\D0}_{\eta_{34},2}(\fra,\phi)=\prod_{I=1}^{k_{1}}\frac{\sh(\fra-\phi_{I}^{(1)}+\eps_{4})}{\sh(\fra-\phi_{I}^{(1)}+\eps_{1}+\eps_{3}+\eps_{4})}\prod_{J=1}^{k_{2}}\frac{\sh(\phi_J^{(2)}-\fra-\eps_{1}-\eps_{3})}{\sh(\phi^{(2)}_J-\fra)}.
\eea

In this section, we will mainly focus on $\eta_{34}^{(1)}$ and generalizations to $\eta_{34}^{(2)}$ is straightforward.

\begin{definition}\label{def:D4_eta34}
    The BPS partition function of the system with a D4-brane wrapping the phase boundary $\eta_{34}^{(1)}$ is defined as
 \bea
    \mathcal{Z}^{\D4\tbar\D0}(\eta^{(1)}_{34};\mathfrak{q}_{1},\mathfrak{q}_{2};p_{1,2,3,4,5})=\sum_{k_{1}=0}^{\infty}\sum_{k_{2}=0}^{\infty}\mathfrak{q}_{2}^{k_{2}}\mathfrak{q}_{1}^{k_{1}}\mathcal{Z}^{\D4\tbar\D0}_{k_{1},k_{2}}(\eta_{34}^{(1)})
    \eea
    where
    \bea
    \mathcal{Z}^{\D4\tbar\D0}_{k_{1},k_{2}}(\eta_{34}^{(1)})&=\frac{1}{k_{2}!k_{1}!}\oint_{\text{JK}}\prod_{a=1}^{2}\prod_{I=1}^{k_{a}}\frac{d\phi_{I}^{(a)}}{2\pi i } \mathcal{Z}_{\eta_{34},1}^{\D4\tbar\D0}(\fra,\phi)\mathcal{Z}^{\D0\tbar\D0}(\phi).
    \eea
   
\end{definition}

\begin{proposition}
    The non-vanishing JK-poles of Def.~\ref{def:D4_eta34} are in one-to-one correspondence with a horizontally $\mathbb{Z}_{2}$-colored partition. Let $\lambda$ be the partition, then up to permutation, the corresponding pole configuration is
    \bea
    \left\{\phi_I^{(1)}\right\}_{I=1}^{k_1}
 &=
 \left\{
 \fra
 +\frac{i-1}{2}(\eps_2+\eps_3)+(j-1)\eps_{5}
 \mathrel{}\middle|\mathrel{}
 (i,j)\in\lambda,\quad i\equiv 1\,\,(\operatorname{mod} 2)
 \right\},
 \\
 \left\{\phi_I^{(2)}\right\}_{I=1}^{k_2}
 &=
 \left\{
 \fra-\eps_{3}+\frac{i}{2}(\eps_{2}+\eps_{3})+(j-1)\eps_5
 \mathrel{}\middle|\mathrel{}
 (i,j)\in\lambda,\quad i\equiv 0\,\,(\operatorname{mod} 2)
 \right\},
    \eea
    where
    \bea
    k_{1}=\sum_{i\equiv 1}\lambda_{i},\quad k_{2}=\sum_{i\equiv 0} \lambda_{i}.
    \eea
\end{proposition}
For visualization, we group each pair of horizontally adjacent
blue and red boxes into a single unit cell, as illustrated in
Figure~\ref{fig:D4-2dBPScrystal}. It is understood as a Young diagram whose boxes are horizontally $\mathbb{Z}_{2}$-colored and for future use, we denote the set of such Young diagram as $\mathsf{ColYD}$. A physical reason for not depicting the blue and red boxes as individual unit boxes is that each of them corresponds to a fractional D0-brane, and only their combination constitutes a regular D0-brane.

The poles can also be rewritten in terms of the coordinates of the unit cell. Let $(I,j)\in\mathbb{Z}_{\geq 1}^{2}$ denote the coordinate of a unit cell. The blue half-box, corresponding to $i=2I-1$, then carries the equivariant weight
\bea
\fra+(I-1)(\eps_{2}+\eps_{3})+(j-1)\eps_{5},
\eea
whereas the red half-box, corresponding to $i=2I$, carries
\bea
\fra+I\eps_{2}+(I-1)\eps_{3}+(j-1)\eps_{5}.
\eea

In particular, up to level two, we have
\bea
 \mathcal{Z}^{\D4\tbar\D0}(\eta_{34}^{(1)};\mathfrak{q}_{1},\mathfrak{q}_{2};p_{1,2,3,4,5})&=1\\
 &+\frac{\left(p_2 p_4-1\right) \sqrt{p_5}}{\sqrt{p_2} \sqrt{p_4} \left(p_5-1\right)}\fq_1\qquad 
 \adjustbox{valign=c}{
 \begin{tikzpicture}[scale=0.3]
  \draw[-Triangle,thick] (0,0)--(3,0);
  \draw[-Triangle,thick] (0,0)--(0,3);

    \filldraw[
      fill=blue!80!white,
      draw=black
    ]
    (0,0)--(0,1)--({0+0.5},1)--
    ({0+0.5},0)--cycle;
\end{tikzpicture}}\\
 &+\frac{\left(p_2 p_4-1\right) \left(p_2 p_4-p_5\right) p_5}{p_2 p_4 \left(p_5-1\right){}^2
   \left(p_5+1\right)}\fq_1^{2} \qquad  \adjustbox{valign=c}{
 \begin{tikzpicture}[scale=0.3]
  \draw[-Triangle,thick] (0,0)--(3,0);
  \draw[-Triangle,thick] (0,0)--(0,3);

    \filldraw[
      fill=blue!80!white,
      draw=black
    ]
    (0,0)--(0,1)--({0+0.5},1)--
    ({0+0.5},0)--cycle;
     \filldraw[
      fill=blue!80!white,
      draw=black
    ]
    (0,1)--(0,2)--({0+0.5},2)--
    ({0+0.5},1)--cycle;
\end{tikzpicture}}\\
   &+\frac{\left(p_2^2 p_3 p_4-1\right) \left(p_2 p_4 p_5-1\right)}{p_2 \left(p_2 p_3-1\right)
   p_4 \left(p_5-1\right)}\fq_{1}\fq_2\qquad \adjustbox{valign=c}{
 \begin{tikzpicture}[scale=0.3]
  \draw[-Triangle,thick] (0,0)--(3,0);
  \draw[-Triangle,thick] (0,0)--(0,3);

    \filldraw[
      fill=blue!80!white,
      draw=black
    ]
    (0,0)--(0,1)--({0+0.5},1)--
    ({0+0.5},0)--cycle;
     \filldraw[
      fill=red!80!white,
      draw=black
    ]
    (0.5,0)--(0.5,1)--({0.5+0.5},1)--
    ({0.5+0.5},0)--cycle;
\end{tikzpicture}}\\\\
&+\fq_1^{3}\frac{\left(-p_2 p_4+1\right) \left(p_2 p_4-p_5\right) p_5^{3/2} \left(p_2
   p_4-p_5^2\right)}{p_2^{3/2} p_4^{3/2} \left(1-p_{5}\right) \left(1-p_{5}^{2}\right)
   \left(1-p_{5}^{3}\right)}\qquad \adjustbox{valign=c}{
 \begin{tikzpicture}[scale=0.3]
  \draw[-Triangle,thick] (0,0)--(4,0);
  \draw[-Triangle,thick] (0,0)--(0,4);

    \filldraw[
      fill=blue!80!white,
      draw=black
    ]
    (0,0)--(0,1)--({0+0.5},1)--
    ({0+0.5},0)--cycle;
 \filldraw[
      fill=blue!80!white,
      draw=black
    ]
    (0,1)--(0,2)--({0+0.5},2)--
    ({0+0.5},1)--cycle;
     \filldraw[
      fill=blue!80!white,
      draw=black
    ]
    (0,2)--(0,3)--({0+0.5},3)--
    ({0+0.5},2)--cycle;

\end{tikzpicture}}\\
&+\fq_1^2\fq_2\left(\frac{\left(p_2 p_4-1\right) \left(p_2^2 p_3 p_4-p_5\right) \sqrt{p_5} \left(p_2 p_4
   p_5-1\right) \left(p_2^2 p_3 p_4 p_5-1\right)}{p_2^2 p_4^2 \left(p_2 p_3-p_5\right)
   \left(p_5-1\right){}^2 \left(p_2 p_3 p_5-1\right)} \right.\qquad \adjustbox{valign=c}{
 \begin{tikzpicture}[scale=0.3]
  \draw[-Triangle,thick] (0,0)--(4,0);
  \draw[-Triangle,thick] (0,0)--(0,4);

    \filldraw[
      fill=blue!80!white,
      draw=black
    ]
    (0,0)--(0,1)--({0+0.5},1)--
    ({0+0.5},0)--cycle;
 \filldraw[
      fill=blue!80!white,
      draw=black
    ]
    (0,1)--(0,2)--({0+0.5},2)--
    ({0+0.5},1)--cycle;

    \filldraw[
      fill=red!80!white,
      draw=black
    ]
    (0.5,0)--(0.5,1)--({0.5+0.5},1)--
    ({0.5+0.5},0)--cycle;

\end{tikzpicture}}\\
&\left. \frac{\left(p_2 p_4-1\right) \left(p_2^2 p_3 p_4-1\right) \sqrt{p_5} \left(p_3-p_4
   p_5\right) \left(p_2^2 p_3 p_4 p_5-1\right)}{p_2 \left(p_2 p_3-1\right) p_4^2 \left(p_2
   p_3-p_5\right) \left(p_5-1\right) \left(p_2 p_3 p_5-1\right)} \right)\qquad \adjustbox{valign=c}{
 \begin{tikzpicture}[scale=0.3]
  \draw[-Triangle,thick] (0,0)--(4,0);
  \draw[-Triangle,thick] (0,0)--(0,4);

    \filldraw[
      fill=blue!80!white,
      draw=black
    ]
    (0,0)--(0,1)--({0+0.5},1)--
    ({0+0.5},0)--cycle;
  \filldraw[
      fill=blue!80!white,
      draw=black
    ]
    (1,0)--(1,1)--({1+0.5},1)--
    ({1+0.5},0)--cycle;

    \filldraw[
      fill=red!80!white,
      draw=black
    ]
    (0.5,0)--(0.5,1)--({0.5+0.5},1)--
    ({0.5+0.5},0)--cycle;

\end{tikzpicture}}\\
&+\cdots 
\eea
Compared to the previous phase boundary $\eta_{02}$, we do not have any unrefined limit. We also note that although the pole structure is the same with the instanton partition function on $\mathbb{C}/\mathbb{Z}_{2}\times \mathbb{C}$, where the orbifold action is $(z_{1},z_{2})\rightarrow(\omega z_{1},z_{2})$ with $\omega^{2}=1$, this partition function is different from it. We leave a detailed analysis for future work.

\subsection{General spiked instanton configurations}\label{sec:spiked-instanton}
We now combine the elementary D4-brane systems associated with the phase boundaries $\eta_{A}\,(A\in\eight)$ and consider their higher rank generalizations and general spiked instanton configurations. 

Similar to the tetrahedron instanton case, for $a\in\mathcal{B}_{A}$, we introduce a flavor space of rank $N_{\eta_{A}}^{(a)}$. The total number of D4-branes wrapping the toric surface ${S}_{\eta_{A}}$ is defined by
\bea
N_{\eta_{A}}=\sum_{a\in\mathcal{B}_{A}}N_{\eta_{A}}^{(a)}.
\eea
We denote the flavor parameters of the individual D4-branes by
\bea
\mathfrak{a}_{\eta_{A},\alpha}^{(a)},\quad a\in\mathcal{B}_{A},\quad \alpha=1,\ldots, N_{\eta_{A}}^{(a)}
\eea
and collectively the rank and flavor data by
\bea
\boldsymbol{ N_{\eta}}
&:=
\left(
N_{\eta_{A}}^{(a)}
\right)_{
\substack{
A\in\eight\\
a\in\mathcal B_A
}},
\quad
\boldsymbol{\mathfrak a_{\eta}}
:=
\left(
\mathfrak a_{\eta_{A},\alpha}^{(a)}
\right)_{
\substack{
A\in\eight ,\;
a\in\mathcal B_A\\
\alpha=1,\ldots,N_{\eta_{A}}^{(a)}
}}.
\eea
We assume that the flavor parameters are always generic in this paper.

\paragraph{Spiked instantons of $\mathcal{C}$}
We first consider a stack of D4-branes associated with the phase boundaries $\eta_{A}$ $(A\in\four)$, where only $\{N_{\eta_{A}}^{(a)}\mid A\in\four\}$ are non-zero. Since this system preserves $\mathcal{N}=4$ supersymmetry, let us use the $\mathcal{N}=4$ notation. For a D4-brane associated with $\eta_{A}$, a fundamental $\mathcal{N}=4$ chiral field $\mathsf{I}_{\eta_{A}}$ and an anti-fundamental $\mathcal{N}=4$ chiral field $\mathsf{J}_{\eta_{A}}$ are introduced to the quiver:

\bea
\begin{tikzpicture}[
    scale=1.05,
    gaugeone/.style={
        circle,
        minimum size=11mm,
        inner sep=0pt,
        draw=blue!70!black,
        very thick,
        fill=blue!25
    },
    gaugetwo/.style={
        circle,
        minimum size=11mm,
        inner sep=0pt,
        draw=red!75!black,
        very thick,
        fill=red!25
    },
    framing/.style={
        rectangle,
        rounded corners=2pt,
        minimum width=18mm,
        minimum height=9mm,
        draw=black,
        thick,
        fill=black!5!white
    },
    arrow/.style={
        -{Triangle[length=4pt,width=5pt]},
        thick,
        draw=black
    },
    doubletip/.style={
        postaction={decorate},
        decoration={
            markings,
            mark=at position 0.47 with
                {\arrow{Triangle[length=3.5pt,width=4.5pt]}},
            mark=at position 0.57 with
                {\arrow{Triangle[length=3.5pt,width=4.5pt]}}
        }
    }
]

\node[gaugeone] (K1) at (-1.7,0) {$k_{1}$};
\node[gaugetwo] (K2) at ( 1.7,0) {$k_{2}$};

\node[framing] (N01) at (-3.2, 3.0) {$N_{\eta_{01}}$};
\node[framing] (N03) at ( 3.2, 3.0) {$N_{\eta_{03}}$};
\node[framing] (N02) at (-3.2,-3.0) {$N_{\eta_{02}}$};
\node[framing] (N04) at ( 3.2,-3.0) {$N_{\eta_{04}}$};

\draw[thick,doubletip]
    ([yshift=5pt]K1.east)
    --
    ([yshift=5pt]K2.west);

\draw[thick,doubletip]
    ([yshift=-5pt]K2.west)
    --
    ([yshift=-5pt]K1.east);

\draw[arrow]
    ([xshift=4pt]N01.south)
    --
    node[pos=0.52,above] {$\mathsf I_{\eta_{01}}$}
    ([xshift=-4pt]K2.north);

\draw[arrow]
    ([xshift=-4pt]K1.north)
    --
    node[pos=0.48,left] {$\mathsf J_{\eta_{01}}$}
    ([xshift=-4pt]N01.south);

\draw[arrow]
    ([xshift=0pt]N03.south)
    --
    node[pos=0.5,below right] {$\mathsf I_{\eta_{03}}$}
    ([xshift=6pt]K2.north);

\draw[arrow]
    ([xshift=4pt]K1.north)
    --
    node[pos=0.52,above] {$\mathsf J_{\eta_{03}}$}
    ([xshift=-4pt]N03.south);

\draw[arrow]
    ([xshift=-2pt]N02.north)
    --
    node[pos=0.47,above left] {$\mathsf I_{\eta_{02}}$}
    ([xshift=-4pt]K1.south);

\draw[arrow]
    ([xshift=-4pt]K2.south)
    --
    node[pos=0.52,below right] {$\mathsf J_{\eta_{02}}$}
    ([xshift=3pt]N02.north);

\draw[arrow]
    ([xshift=-4pt]N04.north)
    --
    node[pos=0.52,below] {$\mathsf I_{\eta_{04}}$}
    ([xshift=4pt]K1.south);

\draw[arrow]
    ([xshift=4pt]K2.south)
    --
    node[pos=0.47,right] {$\mathsf J_{\eta_{04}}$}
    ([xshift=4pt]N04.north);

\end{tikzpicture}
\eea
where the superpotential is deformed by
\bea
\delta \mathsf{W}=\Tr\,\mathsf{J}_{\eta_{01}}\mathsf{B}_{1}\mathsf{I}_{\eta_{01}}+\Tr\,\mathsf{J}_{\eta_{02}}\mathsf{A}_{2}\mathsf{I}_{\eta_{02}}+\Tr\,\mathsf{J}_{\eta_{03}}\mathsf{B}_{2}\mathsf{I}_{\eta_{03}}+\Tr\,\mathsf{J}_{\eta_{04}}\mathsf{A}_{1}\mathsf{I}_{\eta_{04}}.
\eea

In terms of $\mathcal{N}=2$, the $J,E$-terms are obtained by introducing the $J,E$-terms for the corresponding Fermi multiplets of $\mathsf{I}_{\eta_{A}},\mathsf{J}_{\eta_{A}}$ as \eqref{eq:D4-eta02-JE} (see \eqref{eq:D4-eta01-JE}, \eqref{eq:D4-eta02-JE}, \eqref{eq:D4-eta03-JE}, \eqref{eq:D4-eta04-JE} for the full potential terms). The deformation of the superpotential deforms the $J$-terms of the Fermi multiplets $\Lambda^{1,2}_{1\rightarrow 2},\Lambda^{1,2}_{2\rightarrow 1}$ as
\bea\label{eq:4susy-spikedinst-JE}
\begin{tabular}{c|cc}
 &\text{$J$-term}&  \text{$E$-term}\\ \hline
$\Lambda^{1}_{1\rightarrow 2}$ & $\mathsf{B}_{1}\mathsf{A}_{2}\mathsf{B}_{2}-\mathsf{B}_{2}\mathsf{A}_{2}\mathsf{B}_{1}+\textcolor{cyan}{\mathsf{I}_{\eta_{04}}\mathsf{J}_{\eta_{04}}}$  &  $\mathsf{C}_{2}{\mathsf{A}_{1}}-{\mathsf{A}_{1}}\mathsf{C}_{1}$\\
$\Lambda^{2}_{1\rightarrow 2}$ & $\mathsf{B}_{2}{\mathsf{A}_{1}}\mathsf{B}_{1}-\mathsf{B}_{1}{\mathsf{A}_{1}}\mathsf{B}_{2}+\textcolor{cyan}{\mathsf{I}_{\eta_{02}}\mathsf{J}_{\eta_{02}}}$  & $\mathsf{C}_{2}\mathsf{A}_{2}-\mathsf{A}_{2}\mathsf{C}_{1}$ \\
$\Lambda^{1}_{2\rightarrow 1}$ & $\mathsf{A}_{2}\mathsf{B}_{2}{\mathsf{A}_{1}}-{\mathsf{A}_{1}}\mathsf{B}_{2}\mathsf{A}_{2}+\textcolor{cyan}{\mathsf{I}_{\eta_{01}}\mathsf{J}_{\eta_{01}}}$& $\mathsf{C}_{1}\mathsf{B}_{1}-\mathsf{B}_{1}\mathsf{C}_{2}$\\
$\Lambda^{2}_{2\rightarrow 1}$ & ${\mathsf{A}_{1}}\mathsf{B}_{1}\mathsf{A}_{2}-\mathsf{A}_{2}\mathsf{B}_{1}{\mathsf{A}_{1}}+\textcolor{cyan}{\mathsf{I}_{\eta_{03}}\mathsf{J}_{\eta_{03}}}$& $\mathsf{C}_{1}\mathsf{B}_{2}-\mathsf{B}_{2}\mathsf{C}_{2}$
\end{tabular}
\eea
Note again that the condition \eqref{eq:JE-traceless} is automatically satisfied.

The $\mathcal{N}=4$ subsector above is closely related to the
perverse-coherent-extension construction of Butson and Rap\v{c}ák
\cite{Butson:2023eid}, which provides an algebro-geometric realization
of spiked instantons on divisors in toric Calabi--Yau threefolds.
The general system below additionally involves genuinely
$\mathcal{N}=2$ framing data.

\paragraph{Spiked instantons of $\mathcal{C}\times \mathbb{C}$}
Let us now consider the most generic system with stack of D4-branes associated with all of the phase boundaries. As mentioned in section~\ref{sec:D4-phaseboundaries}, for the phase boundaries $\eta_A$ with $A\in\four^\vee$, there are two
possible admissible gauge nodes. Depending on this choice, a D4-brane
introduces a pair of $\mathcal N=2$ chiral multiplets
$\mathsf I_{\eta_A}^{(1,2)},\mathsf J_{\eta_A}^{(1,2)}$ and two framing
Fermi multiplets
$\Lambda_{\eta_A,1}^{(1,2)},\Lambda_{\eta_A,2}^{(1,2)}$. For the phase boundaries $\eta_{A}\,(A\in\four)$, the $\mathcal{N}=4$ chirals $\mathsf{I}_{\eta_{A}},\mathsf{J}_{\eta_{A}}$ decompose into a pair of $\mathcal{N}=2$ chirals $\mathsf{I}_{\eta_{A}},\mathsf{J}_{\eta_{A}}$ and $\mathcal{N}=2$ Fermi fields $\Lambda_{\eta_{A},1},\Lambda_{\eta_{A},2}$, where we abused the notation and identified the $\mathcal{N}=2,4$ chiral fields. The quiver diagram then takes the form as
\bea
\begin{tikzpicture}[
    scale=0.88,
    gaugeone/.style={
        circle,
        minimum size=11mm,
        inner sep=0pt,
        draw=blue!100!black,
        very thick,
        fill=blue!35
    },
    gaugetwo/.style={
        circle,
        minimum size=11mm,
        inner sep=0pt,
        draw=red!100!black,
        very thick,
        fill=red!35
    },
    framing/.style={
        rectangle,
        rounded corners=2pt,
        minimum width=14mm,
        minimum height=9mm,
        inner sep=2pt,
        draw=black,
        thick,
        fill=black!5!white,
        font=\small
    },
    chiral/.style={
    thick,
    draw=black,
    preaction={
        draw=white,
        line width=2pt
    },
    postaction={decorate},
    decoration={
        markings,
        mark=at position 0.58 with
        {
            \arrow{Triangle[length=3.8pt,width=4.8pt]}
        }
    }
},
fermi/.style={
    thick,
    draw=red!100!black,
    preaction={
        draw=white,
        line width=2pt
    },
    postaction={decorate},
    decoration={
        markings,
        mark=at position 0.70 with
        {
            \arrow{Triangle[length=3.8pt,width=4.8pt]}
        }
    }
},
    bulkchiral/.style={
        thick,
        draw=black,
        postaction={decorate},
        decoration={
            markings,
            mark=at position 0.47 with
                {\arrow{Triangle[length=3.5pt,width=4.5pt]}},
            mark=at position 0.57 with
                {\arrow{Triangle[length=3.5pt,width=4.5pt]}}
        }
    },
    bulkfermi/.style={
        thick,
        draw=red!100!black,
        postaction={decorate},
        decoration={
            markings,
            mark=at position 0.47 with
                {\arrow{Triangle[length=3.5pt,width=4.5pt]}},
            mark=at position 0.57 with
                {\arrow{Triangle[length=3.5pt,width=4.5pt]}}
        }
    }
]


\node[gaugeone] (K1) at (-1.4,0) {$k_1$};
\node[gaugetwo] (K2) at ( 1.4,0) {$k_2$};



\node[framing] (N121) at (-6.30,4.4)
    {$N_{\eta_{12}}^{(1)}$};
\node[framing] (N231) at (-4.50,4.4)
    {$N_{\eta_{23}}^{(1)}$};
\node[framing] (N341) at (-2.70,4.4)
    {$N_{\eta_{34}}^{(1)}$};
\node[framing] (N411) at (-0.90,4.4)
    {$N_{\eta_{41}}^{(1)}$};

\node[framing] (N122) at ( 0.90,4.4)
    {$N_{\eta_{12}}^{(2)}$};
\node[framing] (N232) at ( 2.70,4.4)
    {$N_{\eta_{23}}^{(2)}$};
\node[framing] (N342) at ( 4.50,4.4)
    {$N_{\eta_{34}}^{(2)}$};
\node[framing] (N412) at ( 6.30,4.4)
    {$N_{\eta_{41}}^{(2)}$};


\node[framing] (N01) at (-4.5,-3.7) {$N_{\eta_{01}}$};
\node[framing] (N03) at (-1.5,-3.7) {$N_{\eta_{03}}$};
\node[framing] (N02) at ( 1.5,-3.7) {$N_{\eta_{02}}$};
\node[framing] (N04) at ( 4.5,-3.7) {$N_{\eta_{04}}$};


\draw[bulkchiral]
    ([yshift=8pt]K1.east)
    --
    ([yshift=8pt]K2.west);

\draw[bulkchiral]
    ([yshift=-8pt]K2.west)
    --
    ([yshift=-8pt]K1.east);

\draw[bulkfermi]
    ([yshift=3pt]K1.east)
    --
    ([yshift=3pt]K2.west);

\draw[bulkfermi]
    ([yshift=-3pt]K2.west)
    --
    ([yshift=-3pt]K1.east);

\draw[chiral]
    (K1.145)
    to[out=150,in=210,looseness=5]
    node[left=2pt] {$\mathsf C_1$}
    (K1.215);

\draw[chiral]
    (K2.35)
    to[out=30,in=-30,looseness=5]
    node[right=2pt] {$\mathsf C_2$}
    (K2.325);


\def\UpperOne#1{%
    \draw[chiral]
        ([xshift=-7pt]#1.south)
        --
        ([xshift=-7pt]K1.north);
    \draw[fermi]
        ([xshift=-2pt]#1.south)
        --
        ([xshift=-2pt]K1.north);
    \draw[chiral]
        ([xshift=2pt]K2.north)
        --
        ([xshift=2pt]#1.south);
    \draw[fermi]
        ([xshift=7pt]K2.north)
        --
        ([xshift=7pt]#1.south);
}

\def\UpperTwo#1{%
    \draw[chiral]
        ([xshift=-7pt]K1.north)
        --
        ([xshift=-7pt]#1.south);
    \draw[fermi]
        ([xshift=-2pt]K1.north)
        --
        ([xshift=-2pt]#1.south);
    \draw[chiral]
        ([xshift=2pt]#1.south)
        --
        ([xshift=2pt]K2.north);
    \draw[fermi]
        ([xshift=7pt]#1.south)
        --
        ([xshift=7pt]K2.north);
}

\foreach \n in {N121,N231,N341,N411}{
    \UpperOne{\n}
}

\foreach \n in {N122,N232,N342,N412}{
    \UpperTwo{\n}
}


\def\LowerOne#1{%
    \draw[chiral]
        ([xshift=-7pt]#1.north)
        --
        ([xshift=-7pt]K1.south);
    \draw[fermi]
        ([xshift=-2pt]#1.north)
        --
        ([xshift=-2pt]K1.south);
    \draw[chiral]
        ([xshift=2pt]K2.south)
        --
        ([xshift=2pt]#1.north);
    \draw[fermi]
        ([xshift=7pt]K2.south)
        --
        ([xshift=7pt]#1.north);
}

\def\LowerTwo#1{%
    \draw[chiral]
        ([xshift=-7pt]K1.south)
        --
        ([xshift=-7pt]#1.north);
    \draw[fermi]
        ([xshift=-2pt]K1.south)
        --
        ([xshift=-2pt]#1.north);
    \draw[chiral]
        ([xshift=2pt]#1.north)
        --
        ([xshift=2pt]K2.south);
    \draw[fermi]
        ([xshift=7pt]#1.north)
        --
        ([xshift=7pt]K2.south);
}

\LowerTwo{N01}
\LowerTwo{N03}

\LowerOne{N02}
\LowerOne{N04}

\end{tikzpicture}
\eea
The orientations of the arrows can be determined from the previous sections by permuting the gauge nodes and the bifundamental fields in a symmetric way, but we omit the discussion. 

Let us then discuss the potential terms. The additional $J,E$-terms of the Fermi fields are simply the ones in \eqref{eq:D4-eta34-1-JE} after permuting the gauge nodes and the bifundamental fields. See Appendix~\ref{app:JEterms-phaseboundary} for the full potential terms. The extra chiral fields deform the potential terms of the $J,E$-terms of the Fermi fields $\Lambda^{1,2}_{1\rightarrow 2},\Lambda^{1,2}_{2\rightarrow 1}$ as
\bea
\begin{tabular}{c|cc}
 &\text{$J$-term}&  \text{$E$-term}\\ \hline
$\Lambda^{1}_{1\rightarrow 2}$ & $\mathsf{B}_{1}\mathsf{A}_{2}\mathsf{B}_{2}-\mathsf{B}_{2}\mathsf{A}_{2}\mathsf{B}_{1}+\textcolor{cyan}{\mathsf{I}_{\eta_{04}}\mathsf{J}_{\eta_{04}}}$  &  $\mathsf{C}_{2}{\mathsf{A}_{1}}-{\mathsf{A}_{1}}\mathsf{C}_{1}+\textcolor{magenta}{\mathsf{I}^{(2)}_{\eta_{12}}\mathsf{J}^{(2)}_{\eta_{12}}}+\textcolor{magenta}{\mathsf{I}^{(2)}_{\eta_{23}}\mathsf{J}^{(2)}_{\eta_{23}}}$\\
$\Lambda^{2}_{1\rightarrow 2}$ & $\mathsf{B}_{2}{\mathsf{A}_{1}}\mathsf{B}_{1}-\mathsf{B}_{1}{\mathsf{A}_{1}}\mathsf{B}_{2}+\textcolor{cyan}{\mathsf{I}_{\eta_{02}}\mathsf{J}_{\eta_{02}}}$  & $\mathsf{C}_{2}\mathsf{A}_{2}-\mathsf{A}_{2}\mathsf{C}_{1}+\textcolor{magenta}{\mathsf{I}^{(2)}_{\eta_{34}}\mathsf{J}^{(2)}_{\eta_{34}}}+\textcolor{magenta}{\mathsf{I}^{(2)}_{\eta_{41}}\mathsf{J}^{(2)}_{\eta_{41}}}$ \\
$\Lambda^{1}_{2\rightarrow 1}$ & $\mathsf{A}_{2}\mathsf{B}_{2}{\mathsf{A}_{1}}-{\mathsf{A}_{1}}\mathsf{B}_{2}\mathsf{A}_{2}+\textcolor{cyan}{\mathsf{I}_{\eta_{01}}\mathsf{J}_{\eta_{01}}} $& $\mathsf{C}_{1}\mathsf{B}_{1}-\mathsf{B}_{1}\mathsf{C}_{2} + \textcolor{magenta}{\mathsf{I}^{(1)}_{\eta_{34}}\mathsf{J}^{(1)}_{\eta_{34}}}+\textcolor{magenta}{\mathsf{I}^{(1)}_{\eta_{23}}\mathsf{J}^{(1)}_{\eta_{23}}}$\\
$\Lambda^{2}_{2\rightarrow 1}$ & ${\mathsf{A}_{1}}\mathsf{B}_{1}\mathsf{A}_{2}-\mathsf{A}_{2}\mathsf{B}_{1}{\mathsf{A}_{1}}+\textcolor{cyan}{\mathsf{I}_{\eta_{03}}\mathsf{J}_{\eta_{03}}}$& $\mathsf{C}_{1}\mathsf{B}_{2}-\mathsf{B}_{2}\mathsf{C}_{2}+\textcolor{magenta}{\mathsf{I}^{(1)}_{\eta_{12}}\mathsf{J}^{(1)}_{\eta_{12}}}+\textcolor{magenta}{\mathsf{I}^{(1)}_{\eta_{41}}\mathsf{J}^{(1)}_{\eta_{41}}}$
\end{tabular}
\eea

One will see that these potential terms do not obey the condition \eqref{eq:JE-traceless}. Computing the $\Tr(J\cdot E)$, the black--black terms vanish by themselves, while the black--\textcolor{magenta}{magenta} or the \textcolor{cyan}{cyan}--black terms vanish after including the $J,E$-terms of the additional Fermi fields. However, the \textcolor{cyan}{cyan}--\textcolor{magenta}{magenta} terms do not cancel out and one needs to include additional Fermi fields to compensate such contributions. From the viewpoint of the D4-branes, there are D4-branes that share point-like intersection inside the $\mathcal{C}\times \mathbb{C}$ and the additional Fermi fields come from there. This is similar to the situation in the spiked instantons of $\mathbb{C}^{4}$ \cite{Nekrasov:2016gud}.

Let us focus on the $J,E$-terms of $\Lambda^{2}_{1\rightarrow 2}$ to make it concrete (see Appendix~\ref{app:JEterms-phaseboundary} for the other cases). Generalizations to the other cases are straight forward. The additional terms of $\Tr(J\cdot E)$ that do not disappear are
\bea
\Tr\, \textcolor{cyan}{\mathsf{I}_{\eta_{02}}\mathsf{J}_{\eta_{02}}}\left(\textcolor{magenta}{\mathsf{I}^{(2)}_{\eta_{34}}\mathsf{J}^{(2)}_{\eta_{34}}}+\textcolor{magenta}{\mathsf{I}^{(2)}_{\eta_{41}}\mathsf{J}^{(2)}_{\eta_{41}}}\right).
\eea
Focusing on the first term, we can introduce an extra Fermi field with the following $J,E$-terms:
\bea
\adjustbox{valign=c}{
\tdplotsetmaincoords{60}{110}
	\begin{tikzpicture}[scale=1.2,tdplot_main_coords]
	\draw[thick,dashed,-Triangle] (0,0,0) -- node[left,pos=1] {$1$} (1.9,0,0);
	\draw[thick,dashed,-Triangle] (0,0,0) -- node[right,pos=1] {$2$} (0,1.6,0);
	\draw[thick,dashed,-Triangle] (0,0,0) -- node[above,pos=1] {$3$} (0,0,1.5);
    \node at (1,-0.1,1) {\textcolor{cyan}{$\eta_{02}$}};
	\node[draw=black,line width=1pt,circle,fill=black,minimum width=0.2cm,inner sep=1pt] (O) at (0,0,0) {};
	\node[draw=black,line width=1pt,circle,fill=black,minimum width=0.2cm,inner sep=1pt] (X1) at (1,0,0) {};
	\node[draw=black,line width=1pt,circle,fill=black,minimum width=0.2cm,inner sep=1pt] (Z) at (0,0,1) {};
	\node[draw=black,line width=1pt,circle,fill=black,minimum width=0.2cm,inner sep=1pt] (Y1) at (0,1,0) {};
	\node[draw=black,line width=1pt,circle,fill=black,minimum width=0.2cm,inner sep=1pt] (XY) at (1,1,0) {};
	\draw[line width=1pt] (O)--(X1)--(XY)--(Y1)--(O);
    	
	\draw[line width=1pt] (Z)--(O);
	\draw[line width=1pt,cyan] (Z)--(X1);
	\draw[line width=1pt] (Z)--(Y1);
	\draw[line width=1pt] (Z)--(XY);
    \draw[line width=1pt,magenta] (XY)--(Y1);
        \node at (0.5,1.3,0) {$\textcolor{magenta}{\eta_{34}}$};
	\end{tikzpicture}} \adjustbox{valign=c}{
\begin{tikzpicture}[
    scale=0.6,
    transform shape,
    gaugeone/.style={
        circle,
        minimum size=9mm,
        inner sep=0pt,
        draw=blue!0!black,
         thick,
        fill=blue!100
    },
    gaugetwo/.style={
        circle,
        minimum size=9mm,
        inner sep=0pt,
        draw=red!0!black,
         thick,
        fill=red!100
    },
    framing/.style={
        rectangle,
        rounded corners=2pt,
        minimum width=15mm,
        minimum height=8mm,
        draw=black,
        thick,
        fill=black!10!white,
        font=\small
    },
    chiral/.style={
        thick,
        draw=black,
        preaction={draw=white,line width=2pt},
        postaction={decorate},
        decoration={
            markings,
            mark=at position 0.58 with
            {\arrow{Triangle[length=3.5pt,width=4.5pt]}}
        }
    },
    fermi/.style={
        thick,
        draw=red!100!black,
        preaction={draw=white,line width=2pt},
        postaction={decorate},
        decoration={
            markings,
            mark=at position 0.70 with
            {\arrow{Triangle[length=3.5pt,width=4.5pt]}}
        }
    },
    mixedfermi/.style={
        thick,
        dashed,
        draw=red!100!black,
        postaction={decorate},
        decoration={
            markings,
            mark=at position 0.58 with
            {\arrow{Triangle[length=3.5pt,width=4.5pt]}}
        }
    },
    bulkchiral/.style={
        thick,
        draw=black,
        postaction={decorate},
        decoration={
            markings,
            mark=at position 0.45 with
            {\arrow{Triangle[length=3.2pt,width=4pt]}},
            mark=at position 0.57 with
            {\arrow{Triangle[length=3.2pt,width=4pt]}}
        }
    },
    bulkfermi/.style={
        thick,
        draw=red!100!black,
        postaction={decorate},
        decoration={
            markings,
            mark=at position 0.45 with
            {\arrow{Triangle[length=3.2pt,width=4pt]}},
            mark=at position 0.57 with
            {\arrow{Triangle[length=3.2pt,width=4pt]}}
        }
    },
    fieldlabel/.style={
        fill=white,
        inner sep=1pt,
        font=\small
    }
]

\node[gaugeone] (K1) at (-1.35,0) {} ;
\node[gaugetwo] (K2) at ( 1.35,0){} ;

\node[framing] (N34) at (1.35,2.45)
    {$N_{\eta_{34}}^{(2)}$};

\node[framing] (N02) at (-0.35,-2.45)
    {$N_{\eta_{02}}$};


\draw[bulkchiral]
    ([yshift=7pt]K1.east)
    --
    ([yshift=7pt]K2.west);

\draw[bulkchiral]
    ([yshift=-7pt]K2.west)
    --
    ([yshift=-7pt]K1.east);

\draw[bulkfermi]
    ([yshift=2pt]K1.east)
    --
    ([yshift=2pt]K2.west);

\draw[bulkfermi]
    ([yshift=-2pt]K2.west)
    --
    ([yshift=-2pt]K1.east);

\draw[chiral]
    (K1.150)
    to[out=155,in=205,looseness=8]
    node[fieldlabel,left=3pt] {$\mathsf C_1$}
    (K1.210);

\draw[chiral]
    (K2.30)
    to[out=25,in=-25,looseness=8]
    node[fieldlabel,right=3pt] {$\mathsf C_2$}
    (K2.330);


\draw[chiral]
    ([xshift=-3pt]N34.south)
    --
    node[fieldlabel,pos=0.40,left]
    {$\mathsf I_{34}^{(2)}$}
    ([xshift=-3pt]K2.north);

\draw[fermi]
    ([xshift=4pt]N34.south)
    --
    ([xshift=4pt]K2.north);

\draw[chiral]
    ([xshift=-4pt]K1.north)
    --
    node[fieldlabel,pos=0.45,above left]
    {$\mathsf J_{34}^{(2)}$}
    ([xshift=-4pt]N34.south);

\draw[fermi]
    ([xshift=3pt]K1.north)
    --
    ([xshift=3pt]N34.south);


\draw[chiral]
    ([xshift=-4pt]N02.north)
    --
    node[fieldlabel,pos=0.43,left]
    {$\mathsf I_{02}$}
    ([xshift=-4pt]K1.south);

\draw[fermi]
    ([xshift=3pt]N02.north)
    --
    ([xshift=3pt]K1.south);

\draw[chiral]
    ([xshift=3pt]K2.south)
    --
    node[fieldlabel,pos=0.43,right]
    {$\mathsf J_{02}$}
    ([xshift=3pt]N02.north);

\draw[fermi]
    ([xshift=-4pt]K2.south)
    --
    ([xshift=-4pt]N02.north);


\draw[mixedfermi]
    (N02.east)
    to[out=0,in=0,looseness=1.55]
    node[fieldlabel,pos=0.52,right=3pt]
    {$\Lambda_{\eta_{02}-\eta_{34}}$}
    (N34.east);

\end{tikzpicture}}\quad  \begin{array}{c|cc}
 \textcolor{cyan}{\eta_{02}}-\textcolor{magenta}{\eta_{34}^{(2)}} & \text{$J$-term} \qquad  & \text{$E$-term}\\\hline
\Lambda_{\eta_{02}-\eta_{34}^{(2)}}        & -\textcolor{cyan}{\mathsf{J}_{\eta_{02}}}\textcolor{magenta}{\mathsf{I}^{(2)}_{\eta_{34}}} & \textcolor{magenta}{\mathsf{J}^{(2)}_{\eta_{34}}}\textcolor{cyan}{\mathsf{I}_{\eta_{02}}}
\end{array}
    \eea
 where we illustrated the extra Fermi field in terms of dashed red lines. The choice which is $J$ or $E$ is not essential and is just a matter of choice. An observation is that the phase boundaries $\eta_{02},\eta_{34}$ do not share any intersection in the toric diagram. This also reflects the fact that the D4-branes corresponding to such phase boundaries share a point-like intersection in $\mathcal{C}\times \mathbb{C}$.

\paragraph{Relation to \cite{Nekrasov:2020qcq}}
In Appendices~B.3 and B.4 of \cite{Nekrasov:2020qcq}, Nekrasov
proposed a quiver description of instantons on $\mathcal{C}\times \mathbb{C}$. The construction is closely related to the
D4-brane systems discussed above. The single-framing system of
Appendix~B.3 agrees with one of our elementary phase-boundary
framings, whereas the crossed-instanton extension of Appendix~B.4
involves different holomorphic relations.

The dictionary between the fields in \cite{Nekrasov:2020qcq} and our
notation is
\bea
\label{eq:Nekrasov-parameter-correspondence}
(b_1,b_2,b_3,b_4,b',b'')
=
(\mathsf A_1,\mathsf A_2,\mathsf B_1,\mathsf B_2,
 \mathsf C_1,\mathsf C_2).
\eea
To compare the two descriptions, recall that the $J$- and $E$-terms
associated with a Fermi multiplet $\Lambda$ can be combined into the
Hermitian equation
\bea
s_{\Lambda}
=
J_{\Lambda}+E_{\Lambda}^{\dagger}.
\eea
The identity
\bea
\sum_{\Lambda}\|s_{\Lambda}\|^{2}
=
\sum_{\Lambda}
\left(
\|J_{\Lambda}\|^{2}+\|E_{\Lambda}\|^{2}
\right)
+
2\Re\sum_{\Lambda}\Tr(J_{\Lambda}E_{\Lambda}),
\eea
together with the supersymmetry condition
\eqref{eq:JE-traceless}, implies
\bea
s_{\Lambda}=0\ \text{for all }\Lambda
\qquad\Longleftrightarrow\qquad
J_{\Lambda}=E_{\Lambda}=0\ \text{for all }\Lambda.
\label{eq:Hermitian-holomorphic-equivalence}
\eea
This allows us to compare the Hermitian equations of
\cite{Nekrasov:2020qcq} directly with our holomorphic $J$- and
$E$-terms.

Appendix~B.3 of \cite{Nekrasov:2020qcq} introduces the framing maps
\bea
i:\ast\longrightarrow 2,
\qquad
j:1\longrightarrow\ast,
\eea
for the ADHM description of the $\mathcal{N}=2^{\ast}$ theory on the blowup
$\widehat{\mathbb{C}}^{2}$. Under the identification
\bea
(i,j)
=
(\mathsf I_{\eta_{03}},\mathsf J_{\eta_{03}}),
\eea
the Hermitian equations in Eq.~(290) of
\cite{Nekrasov:2020qcq} coincide with the equations
$s_{\Lambda}=0$ obtained from
\eqref{eq:4susy-spikedinst-JE} and
\eqref{eq:D4-eta03-JE}, after setting
$\mathsf I_{\eta_A}=\mathsf J_{\eta_A}=0$ for $A\neq03$.
Equivalently, the holomorphic equations in Eq.~(294) of
\cite{Nekrasov:2020qcq} reproduce precisely the corresponding
$J$- and $E$-term relations. Thus, the single-framing system of
Appendix~B.3 is identified with the D4-brane sector associated with
$\eta_{03}$.

Appendix~B.4 of \cite{Nekrasov:2020qcq} adds a second framing space
$\ast'$ and the maps
\bea
\widetilde{i}:\ast'\longrightarrow1,
\qquad
\widetilde{j}:2\longrightarrow\ast'.
\eea
With the Fermi-orientation conventions used in this paper, the
additional terms can be written as
\bea
\renewcommand{\arraystretch}{1.1}
\begin{array}{c|cc}
&
\text{$J$-term}
&
\text{$E$-term}
\\ \hline
\widetilde{\Lambda}_{1}
&
-\widetilde{j}\mathsf A_{1}
&
\mathsf B_{1}\mathsf A_{2}\widetilde{i}
\\
\widetilde{\Lambda}_{2}
&
\mathsf B_{1}\mathsf A_{1}\widetilde{i}
&
\widetilde{j}\mathsf A_{2}
\\
\Lambda^{2}_{2\rightarrow1}
&
\cdots
&
\cdots+\widetilde{i}\widetilde{j}
\end{array}.
\label{eq:Nekrasov-crossed-JE}
\eea
Overall signs in the first two rows depend on the choice of Fermi
representatives and can be changed by Fermi-field redefinitions.

The framing maps have the same endpoints as those of the elementary
framing associated with $\eta_{41}^{(1)}$, but the holomorphic
relations do not coincide. In the cyclic chamber,
$\widetilde{j}=0$, the additional relations following from
\eqref{eq:Nekrasov-crossed-JE} include
\bea
\mathsf B_{1}\mathsf A_{2}\widetilde{i}=0,
\qquad
\mathsf B_{1}\mathsf A_{1}\widetilde{i}=0.
\eea
By contrast, the phase-boundary framing considered above gives
\bea
\mathsf B_{1}\mathsf A_{2}\widetilde{i}=0,
\qquad
\mathsf A_{1}\widetilde{i}=0.
\eea
Therefore, the two constructions share the same underlying framing
arrows but define different superpotential relations and, in
general, different moduli spaces. The crossed-instanton system of
Appendix~B.4 of \cite{Nekrasov:2020qcq} should consequently not be
identified directly with the elementary phase-boundary framing
associated with $\eta_{41}^{(1)}$. Understanding whether it can be
realized by a more general or composite phase-boundary construction
is left for future work.

\paragraph{Spiked instanton partition functions}
The flavor charges of the additional multiplets are inherited from
the corresponding elementary D4-brane systems discussed in the
previous subsections. We denote by
\bea
\mathcal{Z}_{\eta_A,a}^{\mathrm{D4-D0}}
 \bigl(\mathfrak a_{\eta_A,\alpha}^{(a)},\phi\bigr)
\eea
the rank-one contribution of a D4-brane associated with the phase
boundary $\eta_A$, whose framing node is attached to the gauge node
$a\in\mathcal B_A$. The contributions for the remaining phase
boundaries are obtained from those of $\eta_{02}$ and $\eta_{34}$ by
the permutations of the gauge nodes, bifundamental fields and
equivariant parameters described above.

The total D4--D0 framing contribution is therefore
\bea
\mathcal{Z}_{\boldsymbol N_{\eta}}^{\mathrm{D4-D0}}
 \bigl(\boldsymbol{\mathfrak a}_{\eta},\phi\bigr)
&=
\prod_{A\in\boldsymbol{8}}
\prod_{a\in\mathcal B_A}
\prod_{\alpha=1}^{N_{\eta_A}^{(a)}}
\mathcal{Z}_{\eta_A,a}^{\mathrm{D4-D0}}
 \bigl(\mathfrak a_{\eta_A,\alpha}^{(a)},\phi\bigr).
\label{eq:spiked-framing-factor}
\eea
For $A\in\boldsymbol{4}$, the factor on the right-hand side is the
$\mathcal N=2$ decomposition of the corresponding $\mathcal N=4$
framing contribution. For $A\in\boldsymbol{4}^{\vee}$, it is the
genuine $\mathcal N=2$ framing contribution. Thus, the same expression
applies to both types of phase boundaries.

As mentioned above, in the general spiked instanton system, additional Fermi multiplets between flavor nodes may be required. Let $\mathscr F_{\mathrm{int}}$ denote the set of such Fermi
multiplets. For $\Lambda\in\mathscr F_{\mathrm{int}}$, we write
$s(\Lambda)=(\eta_A,a)$ and $t(\Lambda)=(\eta_B,b)$ for its source
and target flavor nodes, respectively, and denote its equivariant
weight by $\epsilon_{\Lambda}$. Its one-loop contribution is
independent of the gauge variables and gives
\bea
\mathcal{Z}_{\boldsymbol N_{\eta}}^{\mathrm{D4-D4}}
 \bigl(\boldsymbol{\mathfrak a}_{\eta}\bigr)
&=
\prod_{\Lambda\in\mathscr F_{\mathrm{int}}}
\prod_{\alpha=1}^{N_{s(\Lambda)}}
\prod_{\beta=1}^{N_{t(\Lambda)}}
\sh\left(
 \mathfrak a_{t(\Lambda),\beta}
-\mathfrak a_{s(\Lambda),\alpha}
-\epsilon_{\Lambda}
\right).
\label{eq:spiked-D4-D4-factor}
\eea
Here the notation for the flavor parameters includes the admissible
gauge-node labels. For example, the Fermi multiplet connecting
$N_{\eta_{02}}$ and $N_{\eta_{34}}^{(2)}$ contributes a product over
$\alpha=1,\ldots,N_{\eta_{02}}$ and
$\beta=1,\ldots,N_{\eta_{34}}^{(2)}$. The orientation of the Fermi
arrow determines the order of the two flavor parameters in
\eqref{eq:spiked-D4-D4-factor}. Reversing its orientation gives an overall factor sign factor which is non-essential in instanton computation. In particular, the contribution for the extra Fermi field coming from the phase boundaries $\eta_{02}$ and $\eta_{34}^{(2)}$ is
\bea
\sh(\fra_{34}^{(2)}-\fra_{12}-\eps_{2}-\eps_{5}).
\eea

Combining the two contributions, we define the complete framing
factor by
\bea
\mathcal{Z}_{\boldsymbol N_{\eta}}^{\mathrm{D4-D4-D0}}
 \bigl(\boldsymbol{\mathfrak a}_{\eta},\phi\bigr)
&=
\mathcal{Z}_{\boldsymbol N_{\eta}}^{\mathrm{D4-D4}}
 \bigl(\boldsymbol{\mathfrak a}_{\eta}\bigr)
\mathcal{Z}_{\boldsymbol N_{\eta}}^{\mathrm{D4-D0}}
 \bigl(\boldsymbol{\mathfrak a}_{\eta},\phi\bigr).
\label{eq:spiked-total-framing-factor}
\eea
When no additional flavor--flavor Fermi multiplet is required, we set
$\mathcal{Z}_{\boldsymbol N_{\eta}}^{\mathrm{D4-D4}}=1$. In particular, this is
the case for the purely $\mathcal N=4$ spiked instanton system
constructed only from the phase boundaries in
$\four$.

A remark is that from the viewpoint of the fractional D0-branes, the
D4--D4 contributions depend only on the flavor parameters and not on
the gauge variables $\phi$. They can therefore be interpreted as
perturbative prefactors and do not affect the JK-pole classification.
They are, however, important when one studies non-perturbative
Schwinger--Dyson equations and the associated $qq$-characters. Such
relations combine partition functions evaluated at shifted flavor
parameters, and the D4--D4 factors participate in the cancellation of
apparent poles in the corresponding spectral parameters. We leave a detailed analysis of such properties for future work.

\begin{definition}
The spiked instanton partition function on
$\mathcal C\times\mathbb C$ is defined by
\bea
\mathcal{Z}_{\boldsymbol N_{\eta}}^{\mathrm{sp}}
 \bigl(
 \boldsymbol{\mathfrak a}_{\eta};
 \fq_1,\fq_2;
 p_{1,2,3,4,5}
 \bigr)
&=
\sum_{k_1,k_2=0}^{\infty}
\fq_1^{k_1}\fq_2^{k_2}
\mathcal{Z}_{k_1,k_2;\boldsymbol N_{\eta}}^{\mathrm{sp}}
 \bigl(\boldsymbol{\mathfrak a}_{\eta}\bigr),
\\
\mathcal{Z}_{k_1,k_2;\mathbf N_{\eta}}^{\mathrm{sp}}
 \bigl(\boldsymbol{\mathfrak a}_{\eta}\bigr)
&=
\frac{1}{k_1!k_2!}
\oint_{\mathrm{JK}}
\prod_{a=1}^{2}
\prod_{I=1}^{k_a}
\frac{d\phi_I^{(a)}}{2\pi\mathrm i}\,
\mathcal{Z}^{\mathrm{D0-D0}}(\phi)\,
\mathcal{Z}_{\mathbf N_{\eta}}^{\mathrm{D4-D4-D0}}
 \bigl(\boldsymbol{\mathfrak a}_{\eta},\phi\bigr).
\label{eq:spiked-partition-function}
\eea
\end{definition}

This definition reduces to Def.~\ref{def:D4_eta02} or
Def.~\ref{def:D4_eta34} when only one of the corresponding rank-one
framing sectors is non-zero. It also includes the higher-rank
generalizations and configurations containing D4-branes associated
with different phase boundaries.

For generic flavor parameters, the flavor--flavor Fermi determinants
in \eqref{eq:spiked-D4-D4-factor} are non-zero and independent of the
gauge variables. They therefore modify the fixed-point weight but do
not introduce additional JK hyperplanes. In the cyclic chamber, the
poles are generated from the fundamental framing chirals
$\mathsf I_{\eta_A}^{(a)}$. Accordingly, the fixed-point data are
labeled by a collection
\bea
\boldsymbol{{C}}_{\eta}
&=
\left(
 {C}_{\eta_A,\alpha}^{(a)}
\right)_{
 A\in\boldsymbol{8},\,
 a\in\mathcal B_A,\,
 \alpha=1,\ldots,N_{\eta_A}^{(a)}
},
\quad 
{C}_{\eta_A,\alpha}^{(a)}
\in
\operatorname{Cryst}_{\eta_A}^{(a)}.
\label{eq:spiked-fixed-points}
\eea
Here $\operatorname{Cryst}_{\eta_A}^{(a)}$ denotes the set of
elementary two-dimensional crystal configurations generated from a
framing pole associated with $\eta_A$ and attached to the gauge node
$a$. For the phase boundaries in $\boldsymbol{4}$, these
configurations are super partitions, up to permutations of the colors
and equivariant parameters:
\bea
\operatorname{Cryst}_{\eta_A}^{(a)}
&\simeq
\mathsf{SP}_{\eta_A},
\qquad
A\in\boldsymbol{4}.
\eea
For the phase boundaries in $\boldsymbol{4}^{\vee}$, they are
$\mathbb Z_2$-colored Young diagrams: 
\bea
\operatorname{Cryst}_{\eta_A}^{(a)}
&\simeq
\mathsf{ColYD}_{\eta_A}^{(a)},
\qquad
A\in\boldsymbol{4}^{\vee}.
\eea
The two possible values of $a$ correspond to the two possible colors
of the initial box. Equivalently, they are related by exchanging the
two gauge nodes together with the corresponding colors and
equivariant parameters.

Thus, for generic flavor parameters, one elementary two-dimensional
crystal is assigned to each individual D4-brane. Nevertheless, the
contribution of a general fixed point does not factorize into a
product of independent rank-one contributions. All elementary
crystals interact through the common fractional D0--D0 sector, while
the point-like intersections of the D4-branes are encoded by the
additional D4--D4 Fermi determinants. Both effects are automatically
incorporated in the JK residue
\eqref{eq:spiked-partition-function}.

 \section{Summary and discussion}\label{sec:summary}
In this paper, we constructed the principal gauge-origami systems on
$\mathcal{C}\times\mathbb{C}$ from its brane brick model. Our main
observation is that brick matchings determine elementary D6-brane
framings, while phase boundaries determine elementary D4-brane
framings. Together with the D8--$\overline{\D8}$ system, these constructions
realize the magnificent-four, tetrahedron-instanton, and
spiked-instanton sectors within a common framed quiver quantum
mechanics.

The corresponding partition functions were defined by
Jeffrey--Kirwan residues. Their fixed points are described by solid
pyramid partitions for the magnificent four, pyramid partitions and
super plane partitions for the elementary tetrahedron instantons, and
super partitions and $\mathbb{Z}_{2}$-colored Young diagrams for the
elementary spiked instantons. General configurations are obtained by
combining these elementary crystals through the common fractional
D0--D0 sector. For intersecting D4-branes, additional Fermi
multiplets between flavor nodes are required by the supersymmetry
condition. They modify the fixed-point weights without changing the
generic JK-pole classification.

Several questions remain open. First, it is important to determine
whether the framing prescriptions proposed here extend to a general
toric Calabi--Yau fourfold. More complicated brick matchings and phase
boundaries may admit several inequivalent framing nodes, bulk Fermi
multiplets, or plaquette completions. An intrinsic characterization of
the admissible elementary framings is therefore still needed. Such a
criterion would provide the foundation for a systematic construction
of toric gauge origami beyond
$\mathcal{C}\times\mathbb{C}$. We report this in a future work \cite{Kimura-Noshita-toric}.

Second, the framing prescriptions should be derived directly from the
T-dual brane configuration. We gave a heuristic interpretation of the
D6-brane framing through tachyon condensation of a D8--$\overline{\D8}$ pair,
whose T-dual image localizes on a chiral face selected by a brick
matching. For D4-branes, the precise localization of the T-dual brane
and the microscopic origin of the framing chiral and Fermi multiplets
remain unclear. Understanding this geometry should also clarify the
role of the chosen lift of a phase boundary and the physical meaning
of the admissible framing nodes.

It would also be interesting to study wall crossing in the framed quiver quantum mechanics. Throughout this paper, we have worked in the cyclic stability chamber. Changing the FI parameters may lead to different stability conditions and crystal descriptions, in parallel with wall-crossing phenomena for Donaldson--Thomas invariants and framed quivers \cite{Kontsevich:2008fj,Joyce:2008pc,NagaoNakajima:2011,Nagao:2009rq}, as well as their Calabi--Yau fourfold counterparts \cite{Cao:2020huo,Bojko:2021cih,Bojko:2025wall}. Related chamber dependence has also appeared in recent JK-residue formulations of the DT/PT correspondence and instanton counting on the blow-up \cite{Kimura:2025lfo,Kimura:2025lig,Filoche:2026xix}.

For $\mathcal{N}=4$ quiver quantum mechanics, changes of stability chamber are closely intertwined with quiver mutations and cluster transformations \cite{Kontsevich:2008fj,Hori:2014tda,KimLeeYi:2015mutation}. In the $\mathcal{N}=2$ systems considered here, dimensional reductions of two-dimensional $(0,2)$ triality and its realization as a local transformation of brane brick models suggest that an analogous structure may exist \cite{Gadde:2013lxa,Franco:2016nwv}. It would be interesting to determine whether such transformations organize the chambers, partition functions, and BPS crystals of gauge-origami systems on $\mathcal{C}\times\mathbb{C}$.

Another aspect is the quantum-algebraic interpretation of the partition functions in terms of $qq$-characters \cite{Nekrasov:2015wsu,Nekrasov:2016qym,Nekrasov:2016ydq,Nekrasov:2017rqy,Nekrasov:2017gzb,Nekrasov:2020qcq,Bourgine:2016vsq,Bourgine:2017jsi,Kim:2016qqs} and
quiver $W$-algebras \cite{Kimura:2015rgi,Kimura:2016dys,Kimura:2017hez}, extending the gauge-origami formalism developed
for $\mathbb{C}^{4}$
\cite{Kimura:2023bxy,Kimura:2024xpr,Kimura:2024osv}. In this
description, the colored crystals found in this paper are expected to
organize the monomials of the corresponding $qq$-characters, while
the additional D4--D4 Fermi factors should play a role in non-perturbative Schwinger--Dyson equations. These algebraic structures also play an important role in quantum integrable systems, where gauge-theory partition functions and $qq$-characters provide constructions of quantum Hamiltonians, Baxter equations, and $R$-matrices \cite{Nekrasov:2009rc,Jeong:2018qpc,FrenkelKoroteevSageZeitlin:2024qopers,KoroteevZeitlin:2023toroidal,JeongLee:2025qopers}. Establishing this $qq$-character formalism, together with a comparison to K-theoretic
Donaldson--Thomas theory, would connect brane brick models,
enumerative geometry, and the BPS/CFT correspondence on general toric
Calabi--Yau fourfolds. We discuss these aspects in \cite{Kimura-Noshita1}.

A complementary perspective is provided by quiver Yangians, which
generalize the affine Yangian of $\mathfrak{gl}_{1}$ and act on BPS
crystals through operators adding and removing atoms. They encode the
BPS algebras of D-branes on toric Calabi--Yau threefolds
\cite{Li:2020rij,Galakhov:2020vyb}. Different framings and stability
chambers are organized by representations of shifted quiver Yangians
\cite{Galakhov:2021xum,Noshita:2021dgj}, while their K-theoretic
lifts are described by toroidal quiver BPS algebras
\cite{Galakhov:2021vbo,Noshita:2021ldl}.

Recent work has initiated an extension of this crystal-algebra
correspondence to two-supercharge quiver theories, including those
associated with toric Calabi--Yau fourfolds
\cite{Bao:2025hfu,Bao:2025dqs}. By analogy with quiver Yangians and
quiver quantum toroidal algebras, different choices of framing are
expected to define different modules of a common algebra. The
elementary D6- and D4-brane framings constructed in this paper provide
concrete examples of this principle: brick matchings and phase
boundaries select different crystals and should determine the
corresponding highest-weight or shift data. It would be interesting
to construct these modules explicitly, to understand how general
multi-brane configurations are assembled from them, and to develop
their shifted and K-theoretic, or toroidal, generalizations. Understanding the coproduct structure is also a perspective that deserves to be studied. This
would provide a different algebraic organization of the various
gauge-origami sectors on $\mathcal{C}\times\mathbb{C}$ and clarify
their relation to the $qq$-character and quiver $W$-algebra
description discussed above.

\acknowledgments
The authors thank Martijn Kool and Jiakang Bao for useful discussions. The work of TK was supported by EIPHI Graduate School (No.~ANR-17-EURE-0002) and Bourgogne-Franche-Comté region. The work of GN is supported by National Natural Science Foundation of China under Grant No. 12422503.

\appendix
\section{Brane tiling of the conifold}\label{app:branetiling-conifold}
In this section, we summarize the brane tiling of the conifold. Brane tilings, also known as dimer models, provide a bipartite-graph
description of the four-dimensional $\mathcal N=1$ quiver gauge
theories associated with D3-branes probing toric Calabi--Yau
threefold singularities
\cite{Hanany:2005ve,Franco:2005rj,Franco:2005sm}.
The faces, edges, and black and white nodes of the tiling encode,
respectively, the gauge groups, chiral multiplets, and superpotential
monomials. See
\cite{Kennaway:2007tq,Yamazaki:2008bt}
for comprehensive reviews.

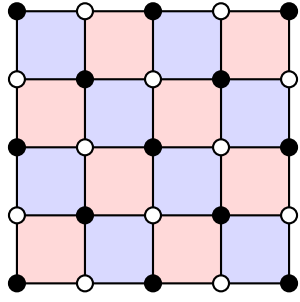
\begin{figure}[t]
\centering
\begin{tikzpicture}[
    x=0.9cm,
    y=0.9cm,
    bipartite/.style={
        circle,
        draw=black,
        thick,
        minimum size=6pt,
        inner sep=0pt
    }
]

\foreach \x in {0,...,3}{
    \foreach \y in {0,...,3}{
        \pgfmathtruncatemacro{\parity}{mod(\x+\y,2)}

        \ifnum\parity=1
            \fill[blue!15]
            (\x,\y) rectangle ({\x+1},{\y+1});
        \else
            \fill[red!15]
            (\x,\y) rectangle ({\x+1},{\y+1});
        \fi
    }
}

\foreach \x in {0,...,4}{
    \draw[black,thick]
    (\x,0)--(\x,4);
}

\foreach \y in {0,...,4}{
    \draw[black,thick]
    (0,\y)--(4,\y);
}

\foreach \x in {0,...,4}{
    \foreach \y in {0,...,4}{
        \pgfmathtruncatemacro{\parity}{mod(\x+\y,2)}

        \ifnum\parity=0
            \node[bipartite,fill=black] at (\x,\y) {};
        \else
            \node[bipartite,fill=white] at (\x,\y) {};
        \fi
    }
}

\end{tikzpicture}
\caption{The conifold brane tiling.}
\label{fig:conifold-brane-tiling}
\end{figure}

Instead of reviewing the derivation of the inverse algorithm of the brane tiling, we simply understand the brane tiling of the conifold as a projection of the third direction of the brane brick model:
\bea
\adjustbox{valign=c}{\includegraphics[width=3cm]{Conifold_brick1_orientation.pdf}}\qquad \qquad \longrightarrow \qquad  \qquad \adjustbox{valign=c}{\begin{tikzpicture}[scale=1.5]
\node[
  draw=black,
  circle,
  fill=white,
  minimum width=0.3cm,
  thick,
  inner sep=1pt
] (O) at (0,0){};

\node[
  draw=black,
  circle,
  fill=black,
  thick,
  minimum width=0.3cm,
  inner sep=1pt
] (A) at (1,0){};

\node[
  draw=black,
  circle,
  fill=black,
  thick,
  minimum width=0.3cm,
  inner sep=1pt
] (C) at (0,1){};

\node[
  draw=black,
  circle,
  fill=white,
  thick,
  minimum width=0.3cm,
  inner sep=1pt
] (E) at (1,1){};

\draw[line width=1pt]
  (O)--(A)--(E)--(C)--(O);
\draw[-Triangle,thick] (0.7,0.5)--(1.5,0.5);
\draw[-Triangle,thick] (0.3,0.5)--(-0.5,0.5);
\draw[-Triangle,thick] (0.5,1.5)--(0.5,0.7);
\draw[-Triangle,thick] (0.5,-0.5)--(0.5,0.3);
\node[right] at (1.5,0.5){$\mathsf{A}_{1}$};
\node[left] at (-0.5,0.5){$\mathsf{A}_{2}$};
\node[above] at (0.5,1.5){$\mathsf{B}_{2}$};
\node[below] at (0.5,-0.5){$\mathsf{B}_{1}$};
\end{tikzpicture}}\\
\adjustbox{valign=c}{\includegraphics[width=3cm]{Conifold_brick2_orientation.pdf}}\qquad \qquad \longrightarrow \qquad  \qquad \adjustbox{valign=c}{\begin{tikzpicture}[scale=1.5]
\node[
  draw=black,
  circle,
  fill=black,
  minimum width=0.3cm,
  thick,
  inner sep=1pt
] (O) at (0,0){};

\node[
  draw=black,
  circle,
  fill=white,
  thick,
  minimum width=0.3cm,
  inner sep=1pt
] (A) at (1,0){};

\node[
  draw=black,
  circle,
  fill=white,
  thick,
  minimum width=0.3cm,
  inner sep=1pt
] (C) at (0,1){};

\node[
  draw=black,
  circle,
  fill=black,
  thick,
  minimum width=0.3cm,
  inner sep=1pt
] (E) at (1,1){};

\draw[line width=1pt]
  (O)--(A)--(E)--(C)--(O);
\draw[-Triangle,thick] (1.5,0.5)--(0.7,0.5);
\draw[-Triangle,thick] (-0.5,0.5)--(0.3,0.5);
\draw[-Triangle,thick] (0.5,0.7)--(0.5,1.5);
\draw[-Triangle,thick] (0.5,0.3)--(0.5,-0.5);
\node[right] at (1.5,0.5){$\mathsf{A}_{2}$};
\node[left] at (-0.5,0.5){$\mathsf{A}_{1}$};
\node[above] at (0.5,1.5){$\mathsf{B}_{1}$};
\node[below] at (0.5,-0.5){$\mathsf{B}_{2}$};
\end{tikzpicture}}
\eea
which gives Figure~\ref{fig:conifold-brane-tiling}

The superpotential is
\bea
\mathsf{W}&=+\sum_{\begin{tikzpicture}
    \node[
  draw=black,
  circle,
  fill=black,
  thick,
  minimum width=0.15cm,
  inner sep=1pt
] at (0,0){};
\end{tikzpicture}} \Tr \prod_{\alpha\in \text{black node}} X_{\alpha} -\sum_{\begin{tikzpicture}
    \node[
  draw=black,
  circle,
  fill=white,
  thick,
  minimum width=0.15cm,
  inner sep=1pt
] at (0,0){};
\end{tikzpicture}} \Tr\prod_{\alpha\in \text{white node}} X_{\alpha}\\
&=\Tr\left(\mathsf{A}_{1}\mathsf{B}_{1}\mathsf{A}_{2}\mathsf{B}_{2}-\mathsf{A}_{1}\mathsf{B}_{2}\mathsf{A}_{2}\mathsf{B}_{1}\right)
\eea
where the chiral fields are written from the right to the left. Conversely, starting from the brane tiling, one perform a lifting algorithm and construct the brane brick models \cite[Sec.~7]{Franco:2015tya}.

\paragraph{Perfect matching}
A perfect matching of the brane tiling is a subset of edges such that every vertex of the brane tiling is connected to only one edge. In terms of the superpotential, it is a set of chiral fields that is contained in each monomial terms of the superpotential exactly once.

For the conifold case, we have four perfect matchings 
\bea
\mathfrak{n}_{1}=\{\mathsf{A}_1\},\quad 
\mathfrak{n}_{2}=\{\mathsf{A}_2\},\quad 
\mathfrak{n}_{3}=\{\mathsf{B}_1\},\quad 
\mathfrak{n}_{4}=\{\mathsf{B}_2\},
\eea
where each of them corresponds to the lattice points of the 2d toric diagram in \eqref{eq:toric-diagram}:
\bea
\mathfrak{n}_{i}\quad \leftrightarrow\quad \boldsymbol{n}_{i}
\eea
where $i=1,2,3,4$.

\paragraph{Zig-zag paths}
A zig-zag path is a special type of closed oriented path in a brane tiling, which forms a homology cycle on $T^{2}$. Their relation to the external legs of the toric
diagram plays a central role in the inverse construction of brane tilings \cite{Hanany:2005ss,Feng:2005gw}. Combinatorially, it follows the edges of the brane tiling, making a maximal left turn at each white node and a maximal right turn at each black node, until the path closes in $T^{2}$.

For the conifold case, we have four zig-zag paths
\bea
\begin{tikzpicture}[x=0.42cm,y=0.42cm,scale=1.2]

\DrawConifoldTiles

\draw[
    conifoldzigzag,
    orange!90!black
]
(0,0)--(1,0)--(1,1)--(2,1)
--(2,2)--(3,2)--(3,3)--(4,3);

\DrawConifoldNodes

\node[
    zigzaglabel,
    text=orange!90!black,
    below=4pt
] at (2,0)
{$\mathcal Z_1$};

\end{tikzpicture}
\hspace{0.30cm}
\begin{tikzpicture}[x=0.42cm,y=0.42cm,scale=1.2]

\DrawConifoldTiles

\draw[
    conifoldzigzag,
    purple!85!black
]
(4,4)--(3,4)--(3,3)--(2,3)
--(2,2)--(1,2)--(1,1)--(0,1);

\DrawConifoldNodes

\node[
    zigzaglabel,
    text=purple!85!black,
    below=4pt
] at (2,0)
{$\mathcal Z_2$};

\end{tikzpicture}
\hspace{0.30cm}
\begin{tikzpicture}[x=0.42cm,y=0.42cm,scale=1.2]

\DrawConifoldTiles

\draw[
    conifoldzigzag,
    green!55!black
]
(0,4)--(0,3)--(1,3)--(1,2)
--(2,2)--(2,1)--(3,1)--(3,0);

\DrawConifoldNodes

\node[
    zigzaglabel,
    text=green!55!black,
    below=4pt
] at (2,0)
{$\mathcal Z_3$};

\end{tikzpicture}
\hspace{0.30cm}
\begin{tikzpicture}[x=0.42cm,y=0.42cm,scale=1.2]

\DrawConifoldTiles

\draw[
    conifoldzigzag,
    blue!80!black
]
(4,0)--(4,1)--(3,1)--(3,2)
--(2,2)--(2,3)--(1,3)--(1,4);

\DrawConifoldNodes

\node[
    zigzaglabel,
    text=blue!80!black,
    below=4pt
] at (2,0)
{$\mathcal Z_4$};

\end{tikzpicture}
\eea
which corresponds to the dual of the edge of the toric diagram:
\bea
\adjustbox{valign=c}{\begin{tikzpicture}[scale=1.5]
\node[
  draw=black,
  circle,
  fill=black,
  minimum width=0.3cm,
  thick,
  inner sep=1pt
] (O) at (0,0){};

\node[
  draw=black,
  circle,
  fill=black,
  thick,
  minimum width=0.3cm,
  inner sep=1pt
] (A) at (1,0){};

\node[
  draw=black,
  circle,
  fill=black,
  thick,
  minimum width=0.3cm,
  inner sep=1pt
] (C) at (0,1){};

\node[
  draw=black,
  circle,
  fill=black,
  thick,
  minimum width=0.3cm,
  inner sep=1pt
] (E) at (1,1){};

\draw[line width=1pt]
  (O)--(A)--(E)--(C)--(O);
\draw[-Triangle,thick,orange] (1.1,0.5)--(2,0.5);
\draw[-Triangle,thick,purple] (-0.1,0.5)--(-1,0.5);
\draw[-Triangle,thick,blue] (0.5,1.1)--(0.5,2);
\draw[-Triangle,thick,darkgreen] (0.5,-0.1)--(0.5,-1);
\end{tikzpicture}}
\eea

\paragraph{D4-branes on toric divisor of $\mathcal{C}$ }
The quivers and superpotential of a non-compact D4-brane wrapping a corner toric divisor of a toric CY$_{3}$ $X$ were studied in \cite{Nishinaka:2013mba} (see also \cite{Nishinaka:2013pua,Nishinaka:2011sv, Nishinaka:2011is}). Note here that we are considering a toric divisor of $X$ itself but not $X\times\mathbb{C}$. We briefly summarize the construction introduced there.

Let \(X\) be a toric Calabi--Yau threefold described by a brane tiling,
and let \(Q\) and \(\mathsf{W}\) denote the associated quiver and
superpotential. A non-compact toric divisor \(D\subset X\) is associated
with an external lattice point of the toric diagram. Choosing a perfect
matching \(\mathfrak n_D\) corresponding to this lattice point specifies
the divisor wrapped by the D4-brane. Recall that the perfect matching of a corner toric divisor is unique.

Let \(z_1\) and \(z_2\) be the two zig-zag paths adjacent to the
corresponding corner of the toric diagram. Their intersection determines
a chiral field
\bea
\mathsf{X}_{\mathrm F}:a\longrightarrow b,
\eea
where the precise intersection is a part of the framing data when the
zig-zag paths intersect more than once.

The D4-brane introduces a flavor node \(*\), together with a
fundamental chiral field and an antifundamental chiral field,
\bea
\mathsf I:* \longrightarrow a,
\qquad
\mathsf J:b\longrightarrow *,
\eea
and the superpotential is deformed as
\bea
\mathsf{W}_{\mathrm{D4}}
=
\mathsf{W}+\Tr\!\left(\mathsf J \mathsf{X}_{\mathrm F}\mathsf I\right).
\label{eq:NYY-framed-superpotential}
\eea
The new $F$-term relations include
\bea
\mathsf{X}_{\mathrm F}\mathsf I=0,
\qquad
\mathsf JX_{\mathrm F}=0,
\qquad
\frac{\partial \mathsf{W}}{\partial \mathsf{X}_{\mathrm F}}
+\mathsf I\mathsf J=0.
\label{eq:NYY-framed-Fterms}
\eea
Thus, the source and target of \(X_{\mathrm F}\) determine the gauge
nodes to which \(\mathsf I\) and \(\mathsf J\) are attached.

In particular, for the conifold case, we have four perfect matchings and four possible D4-brane configurations. The possible cases are summarized in Table~\ref{tab:4susy_spiked}.

\begin{table}[ht]
\centering
\renewcommand{\arraystretch}{2}
\begin{tabular}{c|c|c|c}
\hline
Perfect matching & Point of toric diagram & Quiver & Superpotential deformation
\\
\hline


$\mathfrak n_{1}$
&
\adjustbox{valign=c}{
\begin{tikzpicture}[scale=1.5]

\node[
  draw=orange,
  circle,
  fill=orange,
  minimum width=0.2cm,
  inner sep=1pt
] (O) at (0,0){};

\node[
  draw=black,
  circle,
  fill=black,
  minimum width=0.2cm,
  inner sep=1pt
] (A) at (1,0){};

\node[
  draw=black,
  circle,
  fill=black,
  minimum width=0.2cm,
  inner sep=1pt
] (C) at (0,1){};

\node[
  draw=black,
  circle,
  fill=black,
  minimum width=0.2cm,
  inner sep=1pt
] (E) at (1,1){};

\draw[line width=1pt]
  (O)--(A)--(E)--(C)--(O);

\end{tikzpicture}
}
&
\raisebox{-0.cm}{\adjustbox{valign=c}{
\begin{tikzpicture}[
decoration={
  markings,
  mark=at position \arrowHeadPosition with {\arrow{latex}}
}
]

\tikzset{
box/.style={
  draw,
  minimum width=0.6cm,
  minimum height=0.6cm,
  text centered,
  thick
}
}

\begin{scope}

\draw[postaction={decorate},thick]
  (1,2)--(1,0);

\node[right] at (1,1){$\mathsf{I}$};

\draw[
  postaction={decorate},
  black,
  thick,
  scale=1.3
]
  (0.65,0) arc(0:-180:0.65 and 0.25);

\draw[
  postaction={decorate},
  orange,
  thick,
  scale=1.3
]
  (0.75,0) arc(0:-180:0.75 and 0.4);

\draw[
  postaction={decorate},
  black,
  thick,
  scale=1.3
]
  (-0.65,0) arc(180:0:0.65 and 0.25);

\draw[
  postaction={decorate},
  black,
  thick,
  scale=1.3
]
  (-0.75,0) arc(180:0:0.75 and 0.4);

\draw[
  postaction={decorate},
  black,
  thick
]
  (-1,0) to[bend left=30] (1,2);

\node[left] at (-0.5,1.3){$\mathsf{J}$};

\node at (0,0.8){
  $\mathsf{A}_{1},\mathsf{A}_{2}$
};

\node at (0,-0.8){
  $\textcolor{orange}{\mathsf{B}_{1}},\mathsf{B}_{2}$
};

\node[above] at (1,2.2){$ $};

\node[
  fill=black!20!white,
  box
] at (1,2){};

\node at (-1,-0.5){1};
\node at (1,-0.5){2};

\draw[
  fill=blue!100!white,
  thick
]
  (-0.95,0) circle(0.3cm);

\draw[
  fill=red!100!white,
  thick
]
  (0.9,0) circle(0.3cm);

\end{scope}
\end{tikzpicture}
}}
&
$\displaystyle
\delta\mathsf{W}
=
\Tr
\mathsf{J}\,
\textcolor{orange}{\mathsf{B}_{1}}
\mathsf{I}
$
\\[2ex]

\hline


$\mathfrak n_{2}$
&
\adjustbox{valign=c}{
\begin{tikzpicture}[scale=1.5]

\node[
  draw=black,
  circle,
  fill=black,
  minimum width=0.2cm,
  inner sep=1pt
] (O) at (0,0){};

\node[
  draw=orange,
  circle,
  fill=orange,
  minimum width=0.2cm,
  inner sep=1pt
] (A) at (1,0){};

\node[
  draw=black,
  circle,
  fill=black,
  minimum width=0.2cm,
  inner sep=1pt
] (C) at (0,1){};

\node[
  draw=black,
  circle,
  fill=black,
  minimum width=0.2cm,
  inner sep=1pt
] (E) at (1,1){};

\draw[line width=1pt]
  (O)--(A)--(E)--(C)--(O);

\end{tikzpicture}
}
&
\adjustbox{valign=c}{
\begin{tikzpicture}[
decoration={
  markings,
  mark=at position \arrowHeadPosition with {\arrow{latex}}
}
]

\tikzset{
box/.style={
  draw,
  minimum width=0.6cm,
  minimum height=0.6cm,
  text centered,
  thick
}
}

\begin{scope}

\draw[postaction={decorate},thick]
  (-1,2)--(-1,0);

\node[left] at (-1,1){$\mathsf{I}$};

\draw[
  postaction={decorate},
  black,
  thick,
  scale=1.3
]
  (0.65,0) arc(0:-180:0.65 and 0.25);

\draw[
  postaction={decorate},
  black,
  thick,
  scale=1.3
]
  (0.75,0) arc(0:-180:0.75 and 0.4);

\draw[
  postaction={decorate},
  orange,
  thick,
  scale=1.3
]
  (-0.65,0) arc(180:0:0.65 and 0.25);

\draw[
  postaction={decorate},
  black,
  thick,
  scale=1.3
]
  (-0.75,0) arc(180:0:0.75 and 0.4);

\draw[
  postaction={decorate},
  black,
  thick
]
  (1,0) to[bend right=30] (-1,2);

\node[right] at (0.5,1.3){$\mathsf{J}$};

\node at (0,0.8){
  $\mathsf{A}_{1},
  \textcolor{orange}{\mathsf{A}_{2}}$
};

\node at (0,-0.8){
  $\mathsf{B}_{1},\mathsf{B}_{2}$
};

\node[
  fill=black!20!white,
  box
] at (-1,2){};

\node at (-1,-0.5){1};
\node at (1,-0.5){2};
\node[above] at (-1,2.2){$ $};
\draw[
  fill=blue!100!white,
  thick
]
  (-0.95,0) circle(0.3cm);

\draw[
  fill=red!100!white,
  thick
]
  (0.9,0) circle(0.3cm);

\end{scope}
\end{tikzpicture}
}
&
$\displaystyle
\delta\mathsf{W}
=
\Tr 
\mathsf{J}\,
\textcolor{orange}{\mathsf{A}_{2}}
\mathsf{I}
$
\\[2ex]

\hline


$\mathfrak n_{3}$
&
\adjustbox{valign=c}{
\begin{tikzpicture}[scale=1.5]

\node[
  draw=black,
  circle,
  fill=black,
  minimum width=0.2cm,
  inner sep=1pt
] (O) at (0,0){};

\node[
  draw=black,
  circle,
  fill=black,
  minimum width=0.2cm,
  inner sep=1pt
] (A) at (1,0){};

\node[
  draw=black,
  circle,
  fill=black,
  minimum width=0.2cm,
  inner sep=1pt
] (C) at (0,1){};

\node[
  draw=orange,
  circle,
  fill=orange,
  minimum width=0.2cm,
  inner sep=1pt
] (E) at (1,1){};

\draw[line width=1pt]
  (O)--(A)--(E)--(C)--(O);

\end{tikzpicture}
}
&
\adjustbox{valign=c}{
\begin{tikzpicture}[
decoration={
  markings,
  mark=at position \arrowHeadPosition with {\arrow{latex}}
}
]

\tikzset{
box/.style={
  draw,
  minimum width=0.6cm,
  minimum height=0.6cm,
  text centered,
  thick
}
}

\begin{scope}

\draw[postaction={decorate},thick]
  (1,2)--(1,0);

\node[right] at (1,1){$\mathsf{I}$};

\draw[
  postaction={decorate},
  orange,
  thick,
  scale=1.3
]
  (0.65,0) arc(0:-180:0.65 and 0.25);

\draw[
  postaction={decorate},
  black,
  thick,
  scale=1.3
]
  (0.75,0) arc(0:-180:0.75 and 0.4);

\draw[
  postaction={decorate},
  black,
  thick,
  scale=1.3
]
  (-0.65,0) arc(180:0:0.65 and 0.25);

\draw[
  postaction={decorate},
  black,
  thick,
  scale=1.3
]
  (-0.75,0) arc(180:0:0.75 and 0.4);

\draw[
  postaction={decorate},
  black,
  thick
]
  (-1,0) to[bend left=30] (1,2);

\node[left] at (-0.5,1.3){$\mathsf{J}$};

\node at (0,0.8){
  $\mathsf{A}_{1},\mathsf{A}_{2}$
};

\node at (0,-0.8){
  $\mathsf{B}_{1},
  \textcolor{orange}{\mathsf{B}_{2}}$
};

\node[
  fill=black!20!white,
  box
] at (1,2){};

\node at (-1,-0.5){1};
\node at (1,-0.5){2};

\draw[
  fill=blue!100!white,
  thick
]
  (-0.95,0) circle(0.3cm);

\draw[
  fill=red!100!white,
  thick
]
  (0.9,0) circle(0.3cm);

\node[above] at (1,2.2){$ $};
\end{scope}
\end{tikzpicture}
}
&
$\displaystyle
\delta\mathsf{W}
=
\Tr 
\mathsf{J}\,
\textcolor{orange}{\mathsf{B}_{2}}
\mathsf{I}
$
\\[2ex]

\hline


$\mathfrak n_{4}$
&
\adjustbox{valign=c}{
\begin{tikzpicture}[scale=1.5]

\node[
  draw=black,
  circle,
  fill=black,
  minimum width=0.2cm,
  inner sep=1pt
] (O) at (0,0){};

\node[
  draw=black,
  circle,
  fill=black,
  minimum width=0.2cm,
  inner sep=1pt
] (A) at (1,0){};

\node[
  draw=orange,
  circle,
  fill=orange,
  minimum width=0.2cm,
  inner sep=1pt
] (C) at (0,1){};

\node[
  draw=black,
  circle,
  fill=black,
  minimum width=0.2cm,
  inner sep=1pt
] (E) at (1,1){};

\draw[line width=1pt]
  (O)--(A)--(E)--(C)--(O);

\end{tikzpicture}
}
&
\adjustbox{valign=c}{
\begin{tikzpicture}[
decoration={
  markings,
  mark=at position \arrowHeadPosition with {\arrow{latex}}
}
]

\tikzset{
box/.style={
  draw,
  minimum width=0.6cm,
  minimum height=0.6cm,
  text centered,
  thick
}
}

\begin{scope}

\draw[postaction={decorate},thick]
  (-1,2)--(-1,0);

\node[left] at (-1,1){$\mathsf{I}$};

\draw[
  postaction={decorate},
  black,
  thick,
  scale=1.3
]
  (0.65,0) arc(0:-180:0.65 and 0.25);

\draw[
  postaction={decorate},
  black,
  thick,
  scale=1.3
]
  (0.75,0) arc(0:-180:0.75 and 0.4);

\draw[
  postaction={decorate},
  black,
  thick,
  scale=1.3
]
  (-0.65,0) arc(180:0:0.65 and 0.25);

\draw[
  postaction={decorate},
  orange,
  thick,
  scale=1.3
]
  (-0.75,0) arc(180:0:0.75 and 0.4);

\draw[
  postaction={decorate},
  black,
  thick
]
  (1,0) to[bend right=30] (-1,2);

\node[right] at (0.5,1.3){$\mathsf{J}$};

\node at (0,0.8){
  $\textcolor{orange}{\mathsf{A}_{1}},
  \mathsf{A}_{2}$
};

\node at (0,-0.8){
  $\mathsf{B}_{1},\mathsf{B}_{2}$
};

\node[
  fill=black!20!white,
  box
] at (-1,2){};

\node at (-1,-0.5){1};
\node at (1,-0.5){2};

\draw[
  fill=blue!100!white,
  thick
]
  (-0.95,0) circle(0.3cm);

\draw[
  fill=red!100!white,
  thick
]
  (0.9,0) circle(0.3cm);

\node[above] at (-1,2.2){$ $};
\end{scope}
\end{tikzpicture}
}
&
$\displaystyle
\delta\mathsf{W}
=
\Tr
\mathsf{J}\,
\textcolor{orange}{\mathsf{A}_{1}}
\mathsf{I}
$
\\[2ex]

\hline

\end{tabular}

\caption{
Quivers and superpotential deformation for D4-branes wrapping toric divisors of $\mathcal{C}$.
}
\label{tab:4susy_spiked}

\end{table}

\paragraph{The two-dimensional crystal.}
In the cyclic stability chamber, the representation is generated by the
framing vector through oriented paths starting at the flavor node.
The $F$-term relations imply that the chiral fields belonging to the
perfect matching \(\mathfrak n_D\) vanish at the torus-fixed points.
Consequently, the D4-brane moduli space can be regarded as a subspace of
the parent D6--D2--D0 moduli space obtained by imposing
\bea
\mathsf{X}_{\mathrm{F}}=0,
\qquad
\mathsf{X}_{\mathrm{F}}\in\mathfrak n_D.
\label{eq:NYY-perfect-matching-zero}
\eea

Combinatorially, the surviving paths form a two-dimensional crystal.
Its boundary is determined by the two zig-zag paths \(z_1\) and \(z_2\),
and the torus-fixed points are in one-to-one correspondence with finite
molten configurations of this crystal. Equivalently, the D4-brane
crystal is a two-dimensional slope face of the three-dimensional crystal
of the parent D6--D2--D0 system.

For example, the conifold construction gives the super partitions in \eqref{eq:superpartition}, while more general toric singularities lead to other two-dimensional
crystals, such as oblique or orbifold partitions.

\paragraph{Relation to the phase-boundary prescription.}
The construction above provides the three-dimensional precursor of the
D4-brane prescription used in the main text. The relevant correspondence
may be summarized schematically as
\bea
\begin{array}{c|c}
\text{Toric CY}_{3}
&
\text{Toric CY}_{4}
\\ \hline
\text{external lattice point}
&
\text{extremal toric edge}
\\
\text{toric divisor }D
&
\text{toric surface }S_{\eta}
\\
\text{perfect matching }\mathfrak m_D
&
\text{phase boundary }\eta
\\
\text{adjacent zig-zag paths}
&
\text{a lift of }\eta\text{ in the universal cover}
\\
\text{distinguished chiral field }\mathsf{X}_{\mathrm F}
&
\text{admissible bulk Fermi multiplet}
\\
\mathsf{W}\longmapsto \mathsf{W}+\mathsf J\mathsf{X}_{\mathrm F}\mathsf I
&
J_\Lambda\text{ or }E_\Lambda
 \longmapsto J_\Lambda+\mathsf I\mathsf J
 \text{ or }E_\Lambda+\mathsf I\mathsf J
\\
\text{two-dimensional slope face}
&
\text{effective crystal bounded by }\eta
\end{array}
\label{eq:NYY-phase-boundary-dictionary}
\eea

In the brane-tiling construction, the two zig-zag paths adjacent to the
chosen corner select the chiral field \(\mathsf{X}_{\mathrm F}\) appearing in
\eqref{eq:NYY-framed-superpotential}. In a brane brick model, the
analogous information is distributed among the \(J\)- and \(E\)-terms
of the Fermi multiplets. We therefore use the phase-boundary matrix to
mark the chiral fields intersected by a chosen phase boundary. A bulk
Fermi multiplet for which only one of its \(J\)- and \(E\)-polynomials
is marked then plays the role analogous to \(\mathsf{X}_{\mathrm F}\).

The unmarked polynomial is deformed by the mesonic path
\(\mathsf I\mathsf J\), while the marked polynomial determines the
minimal chiral paths crossing the chosen lift of the phase boundary.
Completing the corresponding plaquettes requires two additional framing
Fermi multiplets. For a generic two-supercharge phase boundary, this plaquette
completion is a genuinely Calabi--Yau fourfold feature. For the four phase boundaries admitting an $\mathcal N=4$ completion, however, the framing Fermi multiplets recombine with the framing chirals into $\mathcal N=4$ multiplets.


\section{Superpotential terms of phase boundaries}\label{app:JEterms-phaseboundary}
In this section, we summarize the notation of the $J,E$-terms of all of the phase boundaries of $\mathcal{C}\times\mathbb{C}$. 
\paragraph{$\mathcal{N}=4$ phase boundaries}For the $\mathcal{N}=4$ phase boundaries $\eta_{A}$ ($A\in\four$), one can either use Proposal~\ref{prop:D4-phase-boundary-framing} or decompose the superpotential in Table~\ref{tab:4susy_spiked} to obtain the $J,E$-terms. They are summarized as
\begin{align}
\begin{array}{c|c c}
  \textcolor{cyan}{\eta_{01}}   & \text{$J$-term} &  \text{$E$-term}\\ \hline
\Lambda_{ \mathsf{I}} \quad    & \textcolor{cyan}{\mathsf{J}\mathsf{B}_1} \quad  & \textcolor{cyan}{\mathsf{C}_2\mathsf{I}}\\ 
\Lambda_{ \mathsf{J}}\quad  & \textcolor{cyan}{\mathsf{B}_{1}\mathsf{I}} \quad  & \textcolor{cyan}{-\mathsf{J}\mathsf{C}_1}\\
 \Lambda^{1}_{2\rightarrow 1} &  \cdots +\textcolor{cyan}{\mathsf{I}\mathsf{J}}\quad  &  \mathsf{C}_{1}\mathsf{B}_{2}-\mathsf{B}_{2}\mathsf{C}_{2}
 \end{array}\label{eq:D4-eta01-JE}\\
 \begin{array}{c|c c}
  \textcolor{cyan}{\eta_{02}}   & \text{$J$-term} &  \text{$E$-term}\\ \hline
\Lambda_{ 1} \quad    & \textcolor{cyan}{\mathsf{J}\mathsf{A}_2} \quad  & \textcolor{cyan}{\mathsf{C}_1\mathsf{I}}\\ 
\Lambda_{ 2}\quad  & \textcolor{cyan}{\mathsf{A}_{2}\mathsf{I}} \quad  & \textcolor{cyan}{-\mathsf{J}\mathsf{C}_2}\\
 \Lambda^{2}_{1\rightarrow 2} &  \cdots+\textcolor{cyan}{\mathsf{I}\mathsf{J}}\quad  &  \mathsf{C}_{2}\mathsf{A}_{2}-\mathsf{A}_{2}\mathsf{C}_{1}
 \end{array}\\
 \begin{array}{c|c c}
  \textcolor{cyan}{\eta_{03}}   & \text{$J$-term} &  \text{$E$-term}\\ \hline
\Lambda_{ \mathsf{I}} \quad    & \textcolor{cyan}{\mathsf{J}\mathsf{B}_2} \quad  & \textcolor{cyan}{\mathsf{C}_2\mathsf{I}}\\ 
\Lambda_{ \mathsf{J}}\quad  & \textcolor{cyan}{\mathsf{B}_{2}\mathsf{I}} \quad  & \textcolor{cyan}{-\mathsf{J}\mathsf{C}_1}\\
 \Lambda^{1}_{2\rightarrow 1} & \cdots + \textcolor{cyan}{\mathsf{I}\mathsf{J}}\quad  &  \mathsf{C}_{1}\mathsf{B}_{1}-\mathsf{B}_{1}\mathsf{C}_{2}
 \end{array}\label{eq:D4-eta03-JE}\\
 \begin{array}{c|c c}
  \textcolor{cyan}{\eta_{04}}   & \text{$J$-term} &  \text{$E$-term}\\ \hline
\Lambda_{ \mathsf{I}} \quad    & \textcolor{cyan}{\mathsf{J}\mathsf{A}_1} \quad  & \textcolor{cyan}{\mathsf{C}_1\mathsf{I}}\\ 
\Lambda_{ \mathsf{J}}\quad  & \textcolor{cyan}{\mathsf{A}_{1}\mathsf{I}} \quad  & \textcolor{cyan}{-\mathsf{J}\mathsf{C}_2}\\
 \Lambda^{2}_{1\rightarrow 2} &  \cdots + \textcolor{cyan}{\mathsf{I}\mathsf{J}}\quad  &  \mathsf{C}_{2}\mathsf{A}_{1}-\mathsf{A}_{1}\mathsf{C}_{1}
 \end{array}\label{eq:D4-eta04-JE}
\end{align}
where $\cdots$ mean the original potential terms of the unframed quiver.

\paragraph{$\mathcal{N}=2 $ phase boundaries}For the $\mathcal{N}=2$ phase boundaries $\eta_{A}\,(A\in\four^{\vee})$, the $J,E$-terms are determined as follows. For this case, we have two cases for each phase boundaries and we denote them as $\eta_{A}^{(1,2)}$.
 
\begin{align}\label{eq:D4-eta12-JE}
\begin{array}{c|c c}
   \textcolor{magenta}{\eta_{12}^{(1)}}  & \text{$J$-term} &  \text{$E$-term}\\ \hline

\Lambda_{1}\quad  &  \textcolor{magenta}{\mathsf{J}\mathsf{A}_2} \quad &\textcolor{magenta}{\mathsf{B}_1\mathsf{A}_{1}\mathsf{I}}   \\
\Lambda_{ 2} \quad    & \textcolor{magenta}{\mathsf{A}_2\mathsf{I}} \quad &\textcolor{magenta}{-\mathsf{J}\mathsf{A}_1\mathsf{B}_1}   \\ 
 \Lambda^{2}_{2\rightarrow 1} & \mathsf{A}_{1}\mathsf{B}_{1}\mathsf{A}_{2}-\mathsf{A}_{2}\mathsf{B}_{1}\mathsf{A}_1 \quad  &\cdots +  \textcolor{magenta}{\mathsf{I}\mathsf{J}}
 \end{array}\qquad \begin{array}{c|c c}
   \textcolor{magenta}{\eta_{12}^{(2)}}  & \text{$J$-term} &  \text{$E$-term}\\ \hline

\Lambda_{1}\quad  &  \textcolor{magenta}{-\mathsf{J}\mathsf{B}_1} \quad &\textcolor{magenta}{\mathsf{A}_2\mathsf{B}_{2}\mathsf{I}}   \\
\Lambda_{ 2} \quad    & \textcolor{magenta}{\mathsf{B}_1\mathsf{I}} \quad &\textcolor{magenta}{+\mathsf{J}\mathsf{B}_2\mathsf{A}_2}   \\ 
 \Lambda^{1}_{1\rightarrow 2} & \mathsf{B}_{1}\mathsf{A}_{2}\mathsf{B}_{2}-\mathsf{B}_{2}\mathsf{A}_{2}\mathsf{B}_1 \quad  &  \cdots +\textcolor{magenta}{\mathsf{I}\mathsf{J}}
 \end{array}\\
 \begin{array}{c|c c}
   \textcolor{magenta}{\eta_{23}^{(1)}}  & \text{$J$-term} &  \text{$E$-term}\\ \hline

\Lambda_{1}\quad  &  \textcolor{magenta}{-\mathsf{J}\mathsf{A}_2} \quad &\textcolor{magenta}{\mathsf{B}_2\mathsf{A}_{1}\mathsf{I}}   \\
\Lambda_{ 2} \quad    & \textcolor{magenta}{\mathsf{A}_2\mathsf{I}} \quad &\textcolor{magenta}{\mathsf{J}\mathsf{A}_1\mathsf{B}_2}   \\ 
 \Lambda^{1}_{2\rightarrow 1} & \mathsf{A}_{2}\mathsf{B}_{2}\mathsf{A}_{1}-\mathsf{A}_{1}\mathsf{B}_{2}\mathsf{A}_2 \quad  &  \cdots +\textcolor{magenta}{\mathsf{I}\mathsf{J}}
 \end{array}\qquad \begin{array}{c|c c}
   \textcolor{magenta}{\eta_{23}^{(2)}}  & \text{$J$-term} &  \text{$E$-term}\\ \hline
   \Lambda_{1}\quad  &  \textcolor{magenta}{\mathsf{J}\mathsf{B}_2} \quad &\textcolor{magenta}{\mathsf{A}_2\mathsf{B}_{1}\mathsf{I}}   \\
\Lambda_{ 2} \quad    & \textcolor{magenta}{\mathsf{B}_2\mathsf{I}} \quad &\textcolor{magenta}{-\mathsf{J}\mathsf{B}_1\mathsf{A}_2}   \\ 
 \Lambda^{1}_{1\rightarrow 2} & \mathsf{B}_{1}\mathsf{A}_{2}\mathsf{B}_{2}-\mathsf{B}_{2}\mathsf{A}_{2}\mathsf{B}_1 \quad  &  \cdots +\textcolor{magenta}{\mathsf{I}\mathsf{J}}
 \end{array}\label{eq:D4-eta23-JE}\\
\begin{array}{c|c c}
   \textcolor{magenta}{\eta_{34}^{(1)}}  & \text{$J$-term} &  \text{$E$-term}\\ \hline
\Lambda_{1}\quad  &  \textcolor{magenta}{\mathsf{J}\mathsf{A}_1} \quad &\textcolor{magenta}{\mathsf{B}_2\mathsf{A}_{2}\mathsf{I}}   \\
\Lambda_{ 2} \quad    & \textcolor{magenta}{\mathsf{A}_1\mathsf{I}} \quad &\textcolor{magenta}{-\mathsf{J}\mathsf{A}_2\mathsf{B}_2}   \\ 
 \Lambda^{1}_{2\rightarrow 1} & \mathsf{A}_{2}\mathsf{B}_{2}\mathsf{A}_{1}-\mathsf{A}_{1}\mathsf{B}_{2}\mathsf{A}_2 \quad  &  \cdots+\textcolor{magenta}{\mathsf{I}\mathsf{J}}
 \end{array}\qquad \begin{array}{c|c c}
   \textcolor{magenta}{\eta_{34}^{(2)}}  & \text{$J$-term} &  \text{$E$-term}\\ \hline

\Lambda_{1}\quad  &  \textcolor{magenta}{-\mathsf{J}\mathsf{B}_2} \quad &\textcolor{magenta}{\mathsf{A}_1\mathsf{B}_{1}\mathsf{I}}   \\
\Lambda_{ 2} \quad    & \textcolor{magenta}{\mathsf{B}_2\mathsf{I}} \quad &\textcolor{magenta}{+\mathsf{J}\mathsf{B}_1\mathsf{A}_1}   \\ 
 \Lambda^{2}_{1\rightarrow 2} & \mathsf{B}_{2}\mathsf{A}_{1}\mathsf{B}_{1}-\mathsf{B}_{1}\mathsf{A}_{1}\mathsf{B}_2 \quad  &  \cdots +\textcolor{magenta}{\mathsf{I}\mathsf{J}}
 \end{array}\\
 \begin{array}{c|c c}
   \textcolor{magenta}{\eta_{41}^{(1)}}  & \text{$J$-term} &  \text{$E$-term}\\ \hline

\Lambda_{1}\quad  &  \textcolor{magenta}{-\mathsf{J}\mathsf{A}_1} \quad &\textcolor{magenta}{\mathsf{B}_1\mathsf{A}_{2}\mathsf{I}}   \\
\Lambda_{ 2} \quad    & \textcolor{magenta}{\mathsf{A}_1\mathsf{I}} \quad &\textcolor{magenta}{\mathsf{J}\mathsf{A}_2\mathsf{B}_1}   \\ 
 \Lambda^{2}_{2\rightarrow 1} & \mathsf{A}_{1}\mathsf{B}_{1}\mathsf{A}_{2}-\mathsf{A}_{2}\mathsf{B}_{1}\mathsf{A}_1 \quad  &  \cdots +\textcolor{magenta}{\mathsf{I}\mathsf{J}}
 \end{array}\qquad \begin{array}{c|c c}
   \textcolor{magenta}{\eta_{41}^{(2)}}  & \text{$J$-term} &  \text{$E$-term}\\ \hline

\Lambda_{1}\quad  &  \textcolor{magenta}{\mathsf{J}\mathsf{B}_1} \quad &\textcolor{magenta}{\mathsf{A}_1\mathsf{B}_{2}\mathsf{I}}   \\
\Lambda_{ 2} \quad    & \textcolor{magenta}{\mathsf{B}_1\mathsf{I}} \quad &\textcolor{magenta}{-\mathsf{J}\mathsf{B}_2\mathsf{A}_1}   \\ 
 \Lambda^{2}_{1\rightarrow 2} & \mathsf{B}_{2}\mathsf{A}_{1}\mathsf{B}_{1}-\mathsf{B}_{1}\mathsf{A}_{1}\mathsf{B}_2 \quad  &  \cdots +\textcolor{magenta}{\mathsf{I}\mathsf{J}}
 \end{array}\label{eq:D4-eta41-JE}
\end{align}

\paragraph{Spiked instanton of $\mathcal{C} \times\mathbb{C}$}
The potential terms deformed by the framing multiplets are
\bea
\begin{tabular}{c|cc}
 &\text{$J$-term}&  \text{$E$-term}\\ \hline
$\Lambda^{1}_{1\rightarrow 2}$ & $\mathsf{B}_{1}\mathsf{A}_{2}\mathsf{B}_{2}-\mathsf{B}_{2}\mathsf{A}_{2}\mathsf{B}_{1}+\textcolor{cyan}{\mathsf{I}_{\eta_{04}}\mathsf{J}_{\eta_{04}}}$  &  $\mathsf{C}_{2}{\mathsf{A}_{1}}-{\mathsf{A}_{1}}\mathsf{C}_{1}+\textcolor{magenta}{\mathsf{I}^{(2)}_{\eta_{12}}\mathsf{J}^{(2)}_{\eta_{12}}}+\textcolor{magenta}{\mathsf{I}^{(2)}_{\eta_{23}}\mathsf{J}^{(2)}_{\eta_{23}}}$\\
$\Lambda^{2}_{1\rightarrow 2}$ & $\mathsf{B}_{2}{\mathsf{A}_{1}}\mathsf{B}_{1}-\mathsf{B}_{1}{\mathsf{A}_{1}}\mathsf{B}_{2}+\textcolor{cyan}{\mathsf{I}_{\eta_{02}}\mathsf{J}_{\eta_{02}}}$  & $\mathsf{C}_{2}\mathsf{A}_{2}-\mathsf{A}_{2}\mathsf{C}_{1}+\textcolor{magenta}{\mathsf{I}^{(2)}_{\eta_{34}}\mathsf{J}^{(2)}_{\eta_{34}}}+\textcolor{magenta}{\mathsf{I}^{(2)}_{\eta_{41}}\mathsf{J}^{(2)}_{\eta_{41}}}$ \\
$\Lambda^{1}_{2\rightarrow 1}$ & $\mathsf{A}_{2}\mathsf{B}_{2}{\mathsf{A}_{1}}-{\mathsf{A}_{1}}\mathsf{B}_{2}\mathsf{A}_{2}+\textcolor{cyan}{\mathsf{I}_{\eta_{01}}\mathsf{J}_{\eta_{01}}} $& $\mathsf{C}_{1}\mathsf{B}_{1}-\mathsf{B}_{1}\mathsf{C}_{2} + \textcolor{magenta}{\mathsf{I}^{(1)}_{\eta_{34}}\mathsf{J}^{(1)}_{\eta_{34}}}+\textcolor{magenta}{\mathsf{I}^{(1)}_{\eta_{23}}\mathsf{J}^{(1)}_{\eta_{23}}}$\\
$\Lambda^{2}_{2\rightarrow 1}$ & ${\mathsf{A}_{1}}\mathsf{B}_{1}\mathsf{A}_{2}-\mathsf{A}_{2}\mathsf{B}_{1}{\mathsf{A}_{1}}+\textcolor{cyan}{\mathsf{I}_{\eta_{03}}\mathsf{J}_{\eta_{03}}}$& $\mathsf{C}_{1}\mathsf{B}_{2}-\mathsf{B}_{2}\mathsf{C}_{2}+\textcolor{magenta}{\mathsf{I}^{(1)}_{\eta_{12}}\mathsf{J}^{(1)}_{\eta_{12}}}+\textcolor{magenta}{\mathsf{I}^{(1)}_{\eta_{41}}\mathsf{J}^{(1)}_{\eta_{41}}}$
\end{tabular}
\eea
The additional Fermi fields are
\bea
 \begin{array}{c|cc}
 \textcolor{cyan}{\eta_{04}}-\textcolor{magenta}{\eta_{12}^{(2)}} & \text{$J$-term} \qquad  & \text{$E$-term}\\\hline
\Lambda_{\eta_{04}-\eta_{12}^{(2)}}        & -\textcolor{cyan}{\mathsf{J}_{\eta_{04}}}\textcolor{magenta}{\mathsf{I}^{(2)}_{\eta_{12}}} & \textcolor{magenta}{\mathsf{J}^{(2)}_{\eta_{12}}}\textcolor{cyan}{\mathsf{I}_{\eta_{04}}}
\end{array}\qquad  \begin{array}{c|cc}
 \textcolor{cyan}{\eta_{04}}-\textcolor{magenta}{\eta_{23}^{(2)}} & \text{$J$-term} \qquad  & \text{$E$-term}\\\hline
\Lambda_{\eta_{04}-\eta_{23}^{(2)}}        & -\textcolor{cyan}{\mathsf{J}_{\eta_{04}}}\textcolor{magenta}{\mathsf{I}^{(2)}_{\eta_{23}}} & \textcolor{magenta}{\mathsf{J}^{(2)}_{\eta_{23}}}\textcolor{cyan}{\mathsf{I}_{\eta_{04}}}
\end{array}
 \\
 \begin{array}{c|cc}
 \textcolor{cyan}{\eta_{02}}-\textcolor{magenta}{\eta_{34}^{(2)}} & \text{$J$-term} \qquad  & \text{$E$-term}\\\hline
\Lambda_{\eta_{02}-\eta_{34}^{(2)}}        & -\textcolor{cyan}{\mathsf{J}_{\eta_{02}}}\textcolor{magenta}{\mathsf{I}^{(2)}_{\eta_{34}}} & \textcolor{magenta}{\mathsf{J}^{(2)}_{\eta_{34}}}\textcolor{cyan}{\mathsf{I}_{\eta_{02}}}
\end{array}\qquad  \begin{array}{c|cc}
 \textcolor{cyan}{\eta_{02}}-\textcolor{magenta}{\eta_{41}^{(2)}} & \text{$J$-term} \qquad  & \text{$E$-term}\\\hline
\Lambda_{\eta_{02}-\eta_{41}^{(2)}}        & -\textcolor{cyan}{\mathsf{J}_{\eta_{02}}}\textcolor{magenta}{\mathsf{I}^{(2)}_{\eta_{41}}} & \textcolor{magenta}{\mathsf{J}^{(2)}_{\eta_{41}}}\textcolor{cyan}{\mathsf{I}_{\eta_{02}}}
\end{array}\\
 \begin{array}{c|cc}
 \textcolor{cyan}{\eta_{01}}-\textcolor{magenta}{\eta_{34}^{(1)}} & \text{$J$-term} \qquad  & \text{$E$-term}\\\hline
\Lambda_{\eta_{01}-\eta_{34}^{(1)}}        & -\textcolor{cyan}{\mathsf{J}_{\eta_{01}}}\textcolor{magenta}{\mathsf{I}^{(1)}_{\eta_{34}}} & \textcolor{magenta}{\mathsf{J}^{(1)}_{\eta_{34}}}\textcolor{cyan}{\mathsf{I}_{\eta_{01}}}
\end{array}\qquad  \begin{array}{c|cc}
 \textcolor{cyan}{\eta_{01}}-\textcolor{magenta}{\eta_{23}^{(1)}} & \text{$J$-term} \qquad  & \text{$E$-term}\\\hline
\Lambda_{\eta_{01}-\eta_{23}^{(1)}}        & -\textcolor{cyan}{\mathsf{J}_{\eta_{01}}}\textcolor{magenta}{\mathsf{I}^{(1)}_{\eta_{23}}} & \textcolor{magenta}{\mathsf{J}^{(1)}_{\eta_{23}}}\textcolor{cyan}{\mathsf{I}_{\eta_{01}}}
\end{array}\\
 \begin{array}{c|cc}
 \textcolor{cyan}{\eta_{01}}-\textcolor{magenta}{\eta_{34}^{(1)}} & \text{$J$-term} \qquad  & \text{$E$-term}\\\hline
\Lambda_{\eta_{03}-\eta_{12}^{(1)}}        & -\textcolor{cyan}{\mathsf{J}_{\eta_{03}}}\textcolor{magenta}{\mathsf{I}^{(1)}_{\eta_{12}}} & \textcolor{magenta}{\mathsf{J}^{(1)}_{\eta_{12}}}\textcolor{cyan}{\mathsf{I}_{\eta_{03}}}
\end{array}\qquad  \begin{array}{c|cc}
 \textcolor{cyan}{\eta_{03}}-\textcolor{magenta}{\eta_{41}^{(1)}} & \text{$J$-term} \qquad  & \text{$E$-term}\\\hline
\Lambda_{\eta_{03}-\eta_{41}^{(1)}}        & -\textcolor{cyan}{\mathsf{J}_{\eta_{03}}}\textcolor{magenta}{\mathsf{I}^{(1)}_{\eta_{41}}} & \textcolor{magenta}{\mathsf{J}^{(1)}_{\eta_{41}}}\textcolor{cyan}{\mathsf{I}_{\eta_{03}}}
\end{array}
\eea

\bibliographystyle{amsalpha_mod}
\bibliography{Worigami}
\end{document}